\documentclass[onecolumn,numbers,sort&compress]{els-mrw}    % For numbered references Style 
\usepackage{hyperref}
\usepackage{mathrsfs}
\usepackage{amsmath,amssymb,amsfonts,amsthm,makeidx,graphicx}
\usepackage{slashed}
\usepackage{txfonts}
\usepackage{helvet}
\usepackage{mathrsfs}
\usepackage[normalem]{ulem}

\usepackage{xcolor,wasysym,tikzsymbols}
\newcommand{\uvec}[1]{{\vec #1}}
\newcommand{\po}{$\phantom{.0}$}
\newcommand{\be}{\begin{equation}}
\newcommand{\ee}{\end{equation}}
\newcommand{\ba}{\begin{eqnarray}}
\newcommand{\ea}{\end{eqnarray}}
\newcommand{\bi}{\begin{itemize}}
\newcommand{\ei}{\end{itemize}}
\newcommand{\la}{\langle}
\newcommand{\ra}{\rangle}

\newcommand{\SFF}[1]{F_{#1}^{{}_{\rm SF}}}
\newcommand{\SFG}[1]{G_{#1}^{{}_{\rm SF}}}

\usepackage{comment}
\usepackage{diagbox}
\begin{document}

%%%%%%%%%%%%%%%%%%%%%%%%%%%%%%%%%%%%%%%%%%%%%%%%%%%%%%%%%%%%%%%%
%% the following items are mandatory: 
%% - title
%% - author names
%% - affiliation details
%% - abstract
%% - keywords

%% Precise, concise, and informative description of the focus of this work. Avoid abbreviations and formulae in the title
\chapter{Parton Distribution Functions and their Generalizations}
\label{chap1}

%% All author names and affiliations, and email address for corresponding author
\author[1]{Cédric Lorcé}%
\author[2]{Andreas Metz}%
\author[3,4]{Barbara Pasquini}
\author[5]{Peter Schweitzer}

\address[1]{\orgname{CPHT, CNRS, \'{E}cole polytechnique}, 
    \orgdiv{Institut Polytechnique de Paris}, \orgaddress{91120 Palaiseau, France}}
\address[2]{\orgname{Temple University}, 
    \orgdiv{Department of Physics, SERC}, \orgaddress{Philadelphia, PA 19122, U.S.A.}}
\address[3]{\orgname{Università degli Studi di Pavia}, 
    \orgdiv{Dipartimento di Fisica "A. Volta"}, \orgaddress{27100 Pavia, Italy}}
\address[4]{\orgname{Istituto Nazionale di Fisica Nucleare}, 
    \orgdiv{Sezione di Pavia}, \orgaddress{27100 Pavia, Italy}}
\address[5]{\orgname{University of Connecticut}, 
    \orgdiv{Department of Physics}, \orgaddress{Storrs, CT 06269, U.S.A.}}

\articletag{Chapter Article tagline: update of previous edition, reprint.}

\maketitle

%%%%%%%%%%%%%%%%%%%%%%%%%%%%%%%%%%%%%%%%%%%%%%%%%%%%%%%%%%%%%%%%
%% the following item is mandatory: 
%% 100-150 word summary of the chapter
\begin{abstract}[Abstract]
        This article is an introduction to parton distribution functions and their generalizations which describe the quark and gluon structure of hadrons, and can be measured in various high-energy scattering processes. 
        We provide the theoretical background, highlight both  historical and recent developments, explain the connections between the different functions, and expose in which processes these functions can be accessed and what we can learn from them about hadron structure.
\end{abstract}

%% 5-10 words that embody the key topics in the chapter. What terms would someone put into a search engine if they were looking for a chapter like this?
\begin{keywords}
%% 	    please enter 5 keywords as follows:
%% 	    keyword1\sep keyword2\sep keyword3 \sep keyword4 \sep keyword5
        parton distribution functions and generalizations\sep
        high-energy deep-inelastic reactions\sep
        hard exclusive processes \sep
        quark-gluon dynamics inside hadrons
        \sep form factors of the energy-momentum tensor
\end{keywords}

%%%%%%%%%%%%%%%%%%%%%%%%%%%%%%%%%%%%%%%%%%%%%%%%%%%%%%%%%%%%%%%%
%% the following item is optonal: 
%% - Single figure visually illustrating the key topic/method/outcome described in the chapter
\begin{figure}[h!]
    \centering
    \includegraphics[height=9.1cm]{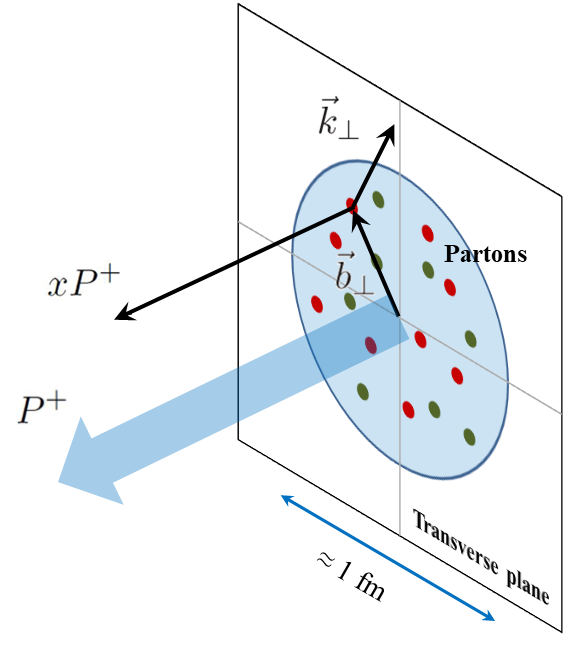}
    \caption{\footnotesize \label{fig:titlepage} A nucleon moving with the 
    momentum $P_z \to \infty$ along the $z$-axis has a large light-front 
    plus momentum $P^+=\frac{1}{\sqrt{2}}(P^0+P_z) \to \infty$. 
    In this so-called infinite-momentum frame, the nucleon looks 
    like a two-dimensional pancake due to extreme Lorentz contraction. 
    Each of its constituents, called partons, carries a certain fraction $x$
    of the nucleon's momentum, has some transverse momentum $\vec{k}_\perp$, 
    and is located at a transverse position $\vec{b}_\perp$ relative to the 
    center of the nucleon. The distributions of the partons in these variables, 
    which further depend on the nucleon and parton polarizations, are described 
    by parton distribution functions and their generalizations.}
    \label{f:pancake}
\end{figure}
 
\

%%%%%%%%%%%%%%%%%%%%%%%%%%%%%%%%%%%%%%%%%%%%%%%%%%%%%%%%%%%%%%%%
%% the following item is optional: 
%% - System of abbreviations/terms/symbols used in the specific field of study/community. List and define
\begin{comment}
\begin{glossary}[Nomenclature]
	\begin{tabular}{@{}lp{34pc}@{}}
            QCD     & quantum chromodynamics \\
            QED     & quantum electrodynamics \\
            FF      & form factor\\
            PDF     & parton distribution function \\
            GPD     & generalized parton distribution function\\
            TMD     & transverse momentum dependent parton distribution function\\
            GTMD    & generalized transverse momentum dependent parton distribution function\\
            DIS     & deep-inelastic lepton-nucleon scattering\\
            LO & leading order\\
            NLO & next-to-leading order\\
            NNLO &next-to-next-to-leading order
  %          EIC     & Electron-Ion Collider \\
		% LHC   & Large Hadron Collider (?) \\
	\end{tabular}
\end{glossary}
\end{comment}

%%%%%%%%%%%%%%%%%%%%%%%%%%%%%%%%%%%%%%%%%%%%%%%%%%%%%%%%%%%%%%%%
%% the following item is mandatory: 
%% List of the key points and topics a reader can expect to learn from this chapter 
\section*{Objectives}
\begin{itemize}
	\item Section 1: 
                A brief historical overview of the nucleon structure research field, the emergence of different types of parton distribution functions,  quantum chromodynamics, and major experimental milestones.
	\item Section 2: 
                Deep-inelastic scattering in the Bjorken limit, the parton model, the definition and interpretation of parton distribution functions (PDFs), global fit analysis, and the behavior of PDFs at small and large $x$.
        \item Section 3: 
                Introduction of transverse parton momenta and transverse momentum-dependent distributions (TMDs), including theoretical aspects such as naive time-reversal symmetry and universality properties, process dependence, and phenomenological applications.
	\item Section 4: 
                Generalized parton distributions (GPDs), their definition and properties, polynomiality, interpretation in impact-parameter space, form factors of the energy-momentum tensor, and experimental observables sensitive to GPDs in nucleons and other hadrons.
	\item Section 5: 
                Generalized transverse momentum-dependent distributions (GTMDs), their connection to Wigner distributions and orbital angular momentum, and discussion of processes that could probe them.
	\item Section 6: 
                Theoretical approaches to studying hadronic structure, including insights from the large-$N_c$ limit, phenomenological hadron models, and results from lattice QCD.
\end{itemize}
%\clearpage

%%%%%%%%%%%%%%%%%%%%%%%%%%%%%%%%%%%%%%%%%%%%%%%%%%%%%%%%%%%%%%%%
%% the following items are mandatory: 
%% - Section: Introduction 
%% - further sections
%% - Section: Conclusion

\section{Introduction}\label{Sec-1:intro}

Hadrons are described in terms of quantum chromodynamics (QCD) and its
fundamental colored quark, antiquark and gluon (parton) degrees of freedom.
The nucleon (proton and neutron), the by far most studied hadron,
is not a simple color-neutral bound state of three quarks but a complex,
extended many-body system with an intricate internal structure
which emerges from the parton interactions in~QCD.
Due to color confinement, free partons have never been experimentally
observed, and yet their existence is well established. This is partly based
on studies of high-energy reactions with hadrons, where we can ``observe'' 
partons and study the structure of hadrons. The key focus of this article is 
the description of hadron structure in terms of parton distributions. 
In Sec.~\ref{Sec-1.1-pre-parton} we briefly review how hadron structure research began.
In Sec.~\ref{sect.1.2} we give a qualitative overview of the different types of parton distributions, some of whose variables are shown in Fig.~\ref{f:pancake}.  
This is followed by a first account of their definition in QCD in Sec.~\ref{sect.1.3}.

\subsection{Magnetic moments and electromagnetic form factors of the nucleon}
\label{Sec-1.1-pre-parton}

If protons and neutrons were pointlike spin-$\frac12$ fermions 
like electrons, their magnetic moments would be predicted by the
Dirac equation as $\mu_p=\mu_N$ and $\mu_n=0$, where $\mu_N = e\hbar/(2M)$ is the nuclear magneton with $M$ the 
nucleon mass. 
Starting in the 1930s, it was discovered that $\mu_p$ and $\mu_n$ exhibit significant deviations from these values~\cite{Frisch-Stern,Alvarez:1940zz}, known as the anomalous magnetic moments and defined as $\mu_p = (1+\kappa_p)\,\mu_N$ and $\mu_n=\kappa_n\, \mu_N$, with the modern values $\kappa_p = 1.79$ and $\kappa_n = -1.91$~\cite{ParticleDataGroup:2022pth}.  
These were the first indications that the nucleon is not pointlike. 

\begin{figure}[b!] 
    \begin{center}
    \includegraphics[height=0.22\textwidth]{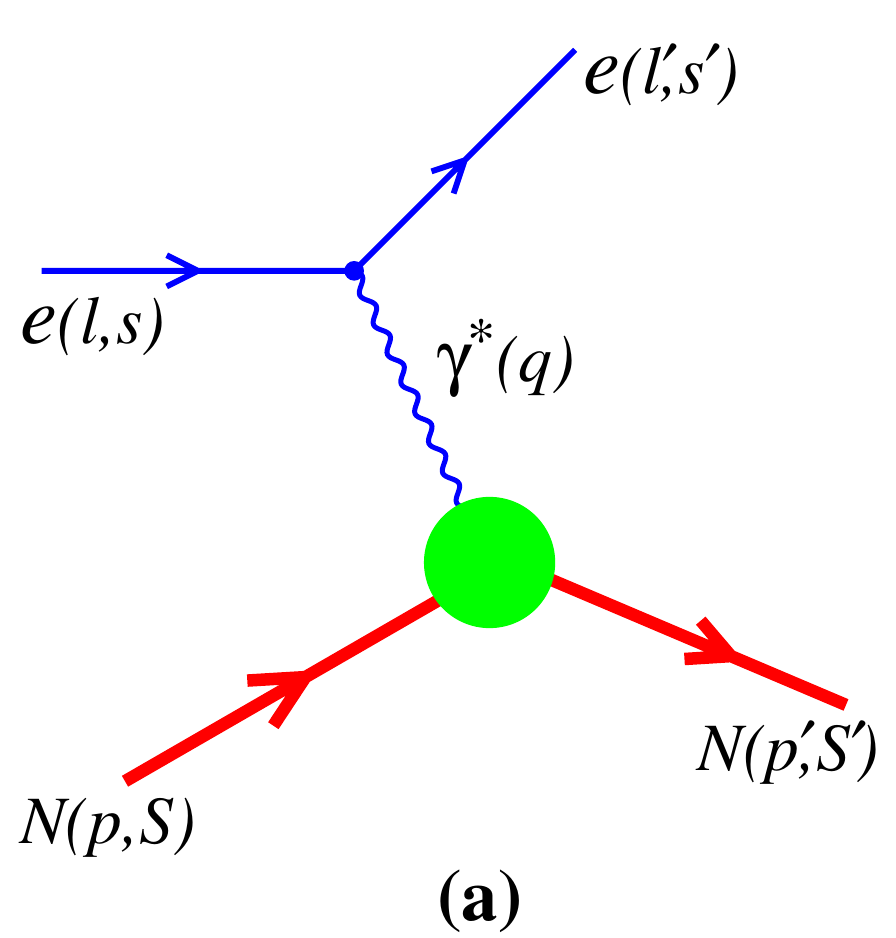}     \hspace{5mm}
    \includegraphics[height=0.22\textwidth]{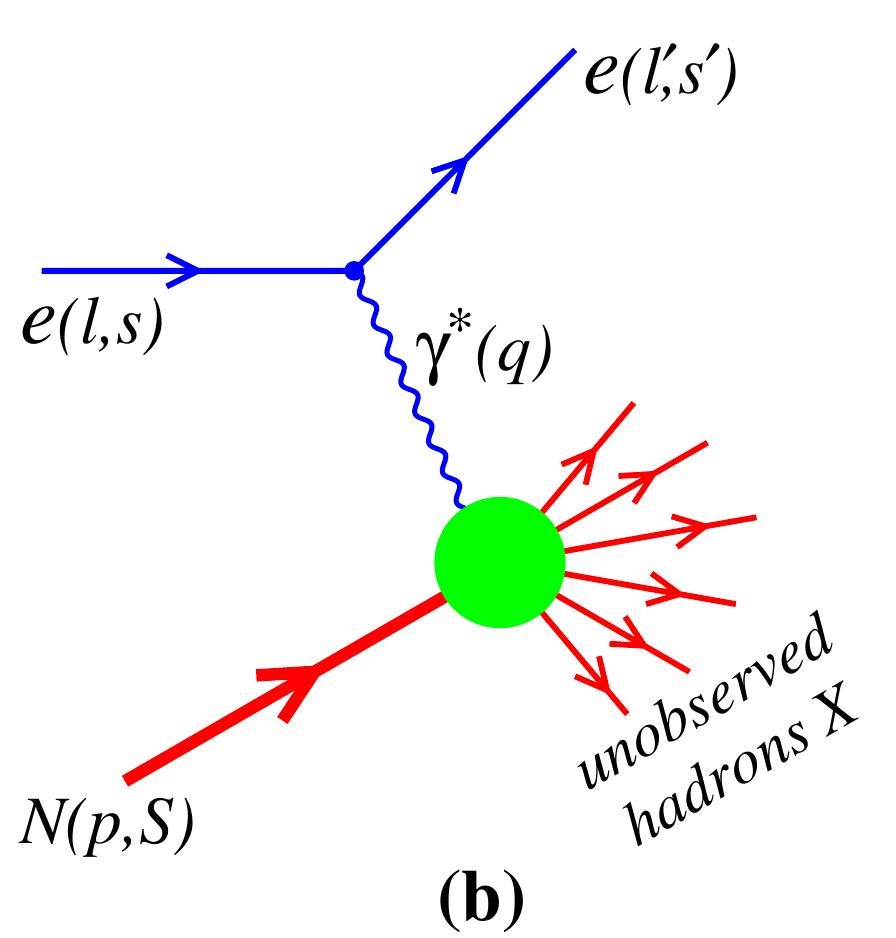}    \hspace{5mm}
    \includegraphics[height=0.22\textwidth]{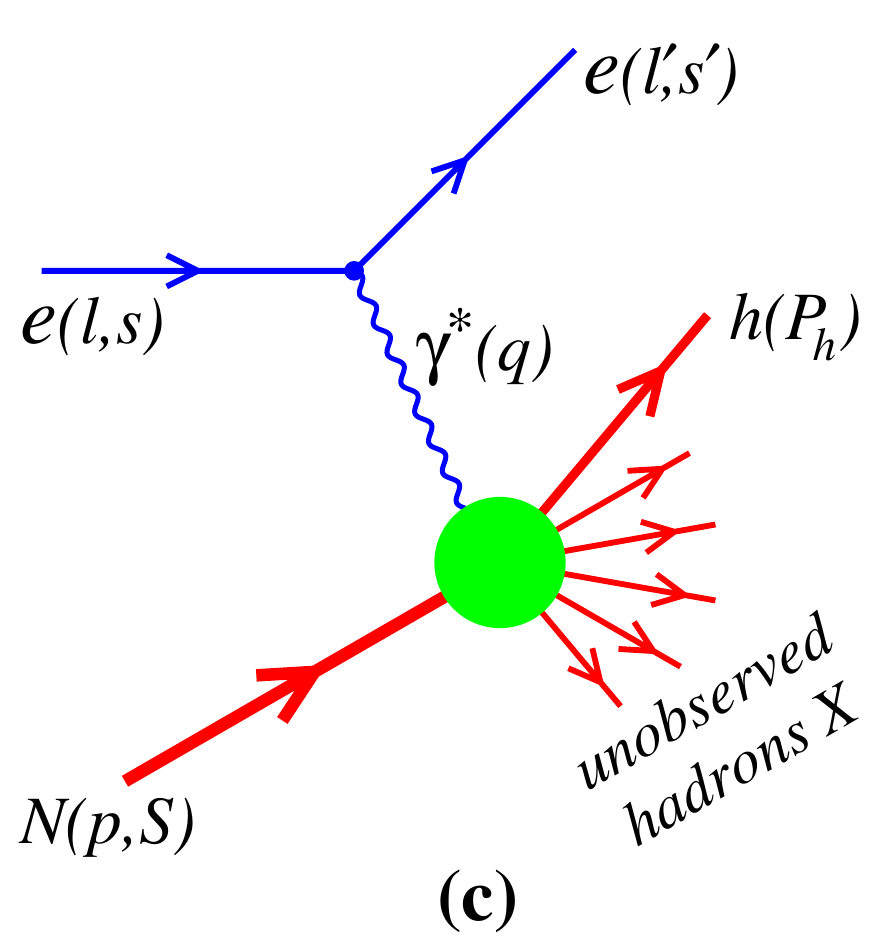}  \hspace{5mm}
    \includegraphics[height=0.22\textwidth]{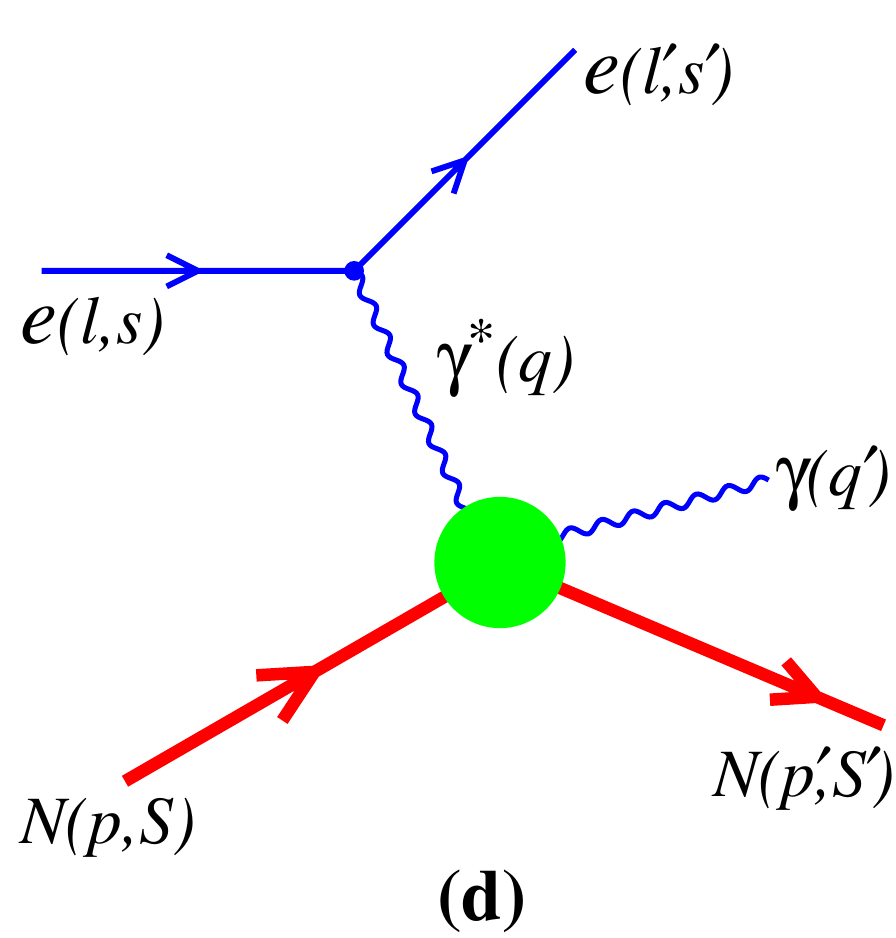}   \hspace{5mm}
    \end{center}
    \caption{\footnotesize\label{Fig-2:FF-DIS-DVCS}
    Amplitudes of 
    (a) elastic electron-nucleon scattering, $eN\to e'N'$, where electromagnetic form factors $F_i(t)$ are measured;
    (b) inclusive deep-inelastic scattering (DIS), $eN\to e'X$, from which PDFs are extracted; 
    (c) semi-inclusive DIS, $eN\to e' h X$, where a specific hadron $h$ is detected and TMDs can be studied; 
    (d) deeply virtual Compton scattering, $eN\to e'N'\gamma$, where GPDs are accessed.}
\end{figure}

In the 1950s elastic electron-proton scattering experiments provided the first determination of the proton charge radius~\cite{Mcallister:1956ng,Hofstadter:1956qs}.
In the one-photon exchange approximation, the amplitude of the elastic electron-proton scattering is given by
\begin{equation}\label{elastic_amplitude}
    {\cal M}   =    \frac{e^2}{q^2}\,
    \langle l',s'|J_{\mu\, {\rm em}}(0)|l,s\rangle\,
    \langle p',S'|J_{\rm em}^{\mu}(0)|p,S\rangle \,,
\end{equation}
with the definitions of momenta and polarizations shown in Fig.~\ref{Fig-2:FF-DIS-DVCS}(a). 
For convenience, the space-time position of the electromagnetic current operators can be chosen at the origin. 
Evaluating the matrix element $\langle l',s'|J_{\rm em}^{\mu}(0)|l,s\rangle$ of the electron in lowest-order perturbative quantum electrodynamics (QED) typically provides sufficient accuracy.
For the nucleon matrix element $\langle p',S'|J_{\rm em}^{\mu}(0)|p,S\rangle$ the situation is different. 
The composite nature of hadrons manifests itself in a nontrivial coupling to the photon which, for a spin-$J$ hadron, is described by $2J+1$ electromagnetic form factors (FFs). 
For the nucleon, this is expressed in terms of the Dirac FF $F_1(t)$ and Pauli FF $F_2(t)$ as
\begin{equation}\label{Eq:def-em-FF}
    \langle p',S'|J_{\rm em}^{\mu}(0)|p,S\rangle
    =e\,\bar u(p',S')\left[\gamma^\mu\, F_1(t)
    +\frac{i\sigma^{\mu\nu}\Delta_\nu}{2M}\,F_2(t)\right]u(p,S) \,,
\end{equation}
where $\Delta =p^\prime-p$ is the four-momentum transfer to the nucleon, $t = \Delta^2<0$, $\gamma^\mu$ and $\sigma^{\mu\nu}=\frac{i}2[\gamma^\mu,\gamma^\nu]$ are Dirac matrices,
and the spinor normalization is $\bar u(p,S')u(p,S) = 2M\delta_{SS'}$.
Notice that in this elastic process $q=l-l^\prime=\Delta$.
The form of Eq.~(\ref{Eq:def-em-FF}) follows from Poincar\'e 
symmetry and electromagnetic current conservation. 
The FFs are dimensionless Lorentz-invariant functions normalized such that $F_1(0)$ yields the electric charge in units of the elementary charge $e > 0$, i.e.,~$F_1(0)$ is 1 
for proton and 0 for neutron, while $F_2(0)$ is the anomalous magnetic moment $\kappa$ discussed above. 
The nucleon FFs exhibit nontrivial $t$-dependences indicating that they are extended particles.

A three-dimensional interpretation is obtained based on the Sachs FFs  $G_E(t)=F_1(t)+\frac{t}{4M^2}\,F_2(t)$ and $G_M(t)=F_1(t)+F_2(t)$~\cite{Sachs:1962zzc}. In the Breit frame defined by $\vec{p}^{\,\prime}=-\uvec p$ and under the assumption $|t|\ll M^2$, one has $G_E(t) = \int d^3r\,\rho_{e}(r)\,e^{i\uvec\Delta\cdot\uvec r}$ where $\rho_{e}(r)$ is the electric charge distribution in the nucleon. The mean square charge radius is defined as $\langle r^2_{\rm el} \ra =\int d^3r\,\rho_{\rm el}(r)\,r^2= 6\left[dG_E/dt\right]_{t=0}$ and, in the proton case, is commonly used as a proxy for the ``proton radius'', leading to the value 
 $\sqrt{\langle r^2_{\rm el} \ra_p^{ }} = 0.84\,\rm fm$~\cite{ParticleDataGroup:2022pth}. 
%$\langle r^2_{\rm el} \ra_p^{1/2} = 0.84\,\rm fm$~\cite{ParticleDataGroup:2022pth}. 
In the neutron case, the mean square charge radius is negative, $\langle r^2_{\rm el}\ra_n^{ }=-0.11\,\rm fm^2$~\cite{ParticleDataGroup:2022pth}, and does not characterize the size of the neutron but reflects the fact that its charge distribution is positive in the center and negative outside. The FF $G_M(t)$ has a similar interpretation in terms of magnetization density inside the nucleon. The interpretations in terms of 3D spatial distributions suffer from relativistic recoil corrections. Exact interpretations in terms of probabilistic 2D spatial densities can, however, be formulated in terms of light-front densities in the impact-parameter space $\vec{b}_\perp$~\cite{Burkardt:2000za,Miller:2010nz}. We will come back to this topic in more detail in Sec.~\ref{Sec-4:GPD}. For comprehensive review articles on electromagnetic nucleon FFs, we refer to~\cite{Punjabi:2015bba,Gao:2021sml}.

\subsection{Insights into the partonic structure}
\label{sect.1.2}
\label{Sec:1.2-partonic-functions-overview}

The late 1960s brought the first insights into nucleon structure through inclusive electron-nucleon deep-inelastic scattering (DIS) experiments, characterized by a so-called hard scale given by the virtuality of the exchanged photon as $Q^2 = - q^2 \gg M_{\rm had}^2$, where $M_{\rm had}$ is a typical hadronic scale set, e.g., by the nucleon mass. With increasing $Q^2$ the elastic electron-nucleon scattering in Fig.~\ref{Fig-2:FF-DIS-DVCS}(a) becomes less and less likely, while  it becomes more and more likely that the nucleon breaks up producing a large number of hadrons; see Fig.~\ref{Fig-2:FF-DIS-DVCS}(b). To interpret such high-energy experiments, it is convenient to think of the proton as moving with a large momentum, e.g., in the $\gamma^\ast$-proton center-of-mass frame. The DIS cross section can be described by assuming that the electron scatters (not off the entire, extended nucleon, but) off an electrically charged, pointlike, spin-$\frac12$ constituent dubbed parton (nowadays identified as quark or antiquark). The probabilities to encounter a parton carrying some fraction $x\in[0,1]$ of the nucleon's momentum are described by parton distribution functions (PDFs) \cite{Feynman:1972original}.  
Another important reaction was the Drell-Yan process where partons and antipartons (quarks and antiquarks) annihilate to form virtual photons that eventually produce $e^+e^-$ or $\mu^+\mu^-$ pairs. (At larger energies, also annihilation into the electroweak boson $Z^0$ becomes important, or a quark and an antiquark of different flavors may annihilate to form a $W^\pm$ boson. These processes are also considered Drell-Yan in a broader sense.)
The quantitative understanding of DIS, Drell-Yan and similar processes paved the path to QCD; see Sec.~\ref{Sec-2:PDF}.

In the late 1970s and early 1980s \cite{Ralston:1979ys,Collins:1981uw}, the concept of transverse momentum dependent PDFs, or TMDs for short, was established in QCD because partons carry not only longitudinal momenta (described by the fraction $x$) but also momenta $\uvec{k}_\perp$ perpendicular to the nucleon's momentum. TMDs are therefore functions of $x$ and $k_\perp=|\uvec{k}_\perp|$. The $k_\perp$-dependence of TMDs provides new insights into the hadron structure, beyond what can be learned from the $x$-dependence of PDFs. TMDs enter the description of, for instance, semi-inclusive DIS (SIDIS) where, in addition to the electron, one of the hadrons produced in a DIS event is detected; see Fig.~\ref{Fig-2:FF-DIS-DVCS}(c). We will discuss TMDs in Sec.~\ref{Sec-3:TMD}.

A different extension of the concept of PDFs emerged in the 1990s, known as generalized PDFs, or GPDs for short \cite{Muller:1994ses,Ji:1996ek,Ji:1996nm,Radyushkin:1996nd,Radyushkin:1996ru,Collins:1996fb}. These functions enter the description of certain hard exclusive reactions, such as deeply virtual Compton scattering (DVCS) shown in Fig.~\ref{Fig-2:FF-DIS-DVCS}(d). GPDs depend on $x$, the longitudinal momentum transfer to the nucleon $\xi$, and the four-momentum transfer $t = \Delta^2$. Among other things, GPDs allow one to access the angular momentum of partons as well as other so-called mechanical properties of hadrons, which are encoded in the FFs of the energy-momentum tensor. Moreover, their dependence on the transverse momentum transfer $\Delta_\perp=|\uvec \Delta_\perp|$ provides information about the spatial distributions of the partons inside the hadron. We will discuss GPDs in Sec.~\ref{Sec-4:GPD}.

A further generalization introduced in the new millennium~\cite{Ji:2003ak,Belitsky:2003nz,Meissner:2008ay,Meissner:2009ww} bears the name of generalized TMDs, or GTMDs for short. They are basically the overarching functions that contain information on both GPDs and TMDs, as well as unique new information. GTMDs depend on five variables, namely $x$, $k_\perp$, $\xi$, $\Delta_\perp$, and $\uvec k_\perp\cdot\uvec\Delta_\perp$. These functions are still the subject of intensive theoretical research and the first processes through which they may be measured have been proposed only recently. GTMDs will be discussed in Sec.~\ref{Sec-5:GTMDs}.

\subsection{Overview of QCD and description of hadron structure}
\label{sect.1.3}
The properties of hadrons are described by QCD discovered about 50 years ago following the pioneering DIS measurements; see~\cite{Gross:2022hyw} for a review of the historical developments. 
The QCD Lagrangian is given by
\be\label{Eq:Lagrangian}
      {\cal L} = \sum_q \bar{\psi}_q(i\slashed{D}-m_q)\psi_q
      - \tfrac{1}{4}\,F^a_{\mu\nu}F^{a\,\mu\nu} \,,
\ee 
where $\psi_q$ are quark fields of flavor $q$, $m_q$ are the current quark masses, and the summation is over the $N_f$ 
quark flavors $q\in\{u,d,s,c,b,t\}$. 
The covariant derivative is $D_\mu = \partial_\mu - i g A_\mu^aT^a$ and the field-strength tensor is $F^a_{\mu\nu}=\partial_\mu A^a_\nu-\partial_\nu A^a_\mu+g f^{abc}A^b_\mu A^c_\nu$, where $A_\mu^a$ are the gauge (gluon) fields and $T^a$ are the generators in the fundamental representation of the group SU$(N_c)$ with $a\in\{1,\cdots,N_c^2-1\}$, $f^{abc}$ are the structure constants of SU$(N_c)$, and $N_c=3$ is the number of colors. 
Non-Abelian gauge theories like QCD are renormalizable \cite{tHooft:1972tcz}.
In field theory, the coupling constants depend in general on the renormalization scale $\mu$. 
Assuming the QCD coupling constant $\alpha_{\rm s}(\mu_0)=g^2(\mu_0)/(4\pi)$ to be small at an initial scale $\mu_0$, its dependence on scales $\mu>\mu_0$ can be determined perturbatively and, to lowest order, can be expressed as
\begin{equation}
    \alpha_{\rm s}(\mu) 
    = \frac{\alpha_{\rm s}(\mu_0)}{1 + \frac{\beta_0}{4\pi} \alpha_{\rm s}(\mu_0) \ln \frac{\mu^2}{\mu_0^2}} \,,
\label{e:as_running}
\end{equation}
where $\beta_0 = \frac{11}{3}N_c - \frac23 N_f$. 
Increasing $\mu$ for fixed $\mu_0$ in Eq.~\eqref{e:as_running}  
shows that $\alpha_{\rm s}(\mu)$ gets smaller, reaching the value 
$\alpha_{\rm s}(91\rm{~GeV})\approx 0.12$ at the scale of the $Z$-boson mass. 
Note that $\alpha_{\rm s}(\mu) \to 0$ for $\mu \to \infty$, a property 
known as asymptotic freedom, one of the key features of QCD.
In principle, one can also use Eq.~(\ref{e:as_running}) to go to 
renormalization scales $\mu<\mu_0$, but in this case $\alpha_{\rm s}(\mu)$ 
increases implying, sooner or later, a breakdown of 
perturbative QCD (pQCD) from which Eq.~(\ref{e:as_running}) was derived.
For a review on $\alpha_{\rm s}(\mu)$ see \cite{Deur:2025rjo}.
When addressing the question of how quarks and gluons form 
the nucleon, the scale is $\mu \sim M_{\rm had}\sim 1\,\rm GeV$ and 
$\alpha_{\rm s}(\mu)$ is of the order of unity. The interaction is 
then strong and non-perturbative techniques must be used to solve QCD. 
The PDFs and their generalizations are non-perturbative objects, which 
means they cannot be computed in pQCD. 
The second key feature of QCD is the confinement hypothesis describing 
the empirical fact that free color charges have never been observed in nature. 
The theoretical explanation of confinement remains an outstanding open question.
The third key feature of QCD is spontaneous breaking of chiral symmetry
which is present in the Lagrangian (\ref{Eq:Lagrangian}) to a good approximation
(due to $m_q\ll M_{\rm had}$ for the light flavors ${q=u,\,d,\,s}$), but is not realized 
in the hadron spectrum.

\begin{figure}
\centering
\includegraphics[width=0.3\textwidth]{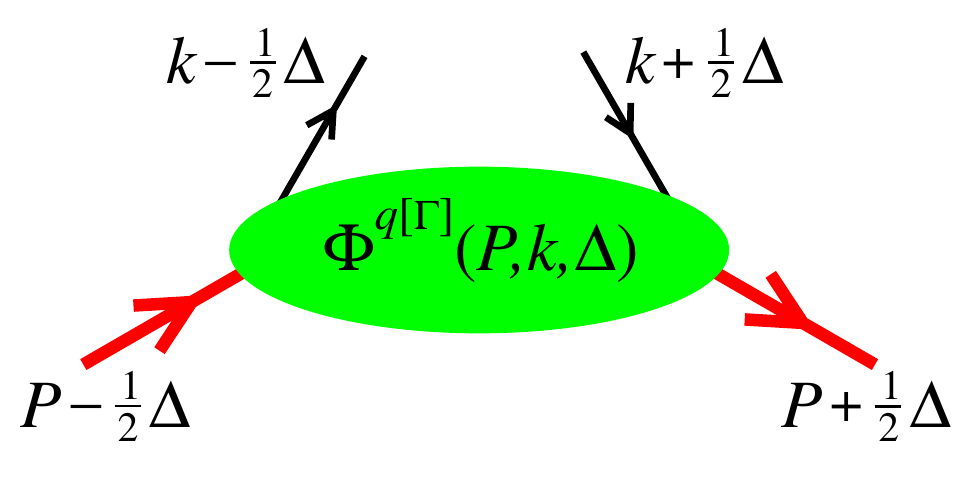}
\caption{\footnotesize Graphical representation of the quark correlation function 
    $\Phi^{q[\Gamma]}(P,k,\Delta)$ in Eq.~(\ref{GPCF}) with the definitions of nucleon and quark momenta. 
    For brevity we have omitted polarization labels for the initial-state and final-state nucleons.
    \label{f:correlator}}
\end{figure}

After this brief overview of QCD and before sketching the definitions of FFs, PDFs, TMDs, GPDs  and GTMDs, we introduce light-front coordinates, which are best suited to describe high-energy processes characterized, generically, by a ``hard momentum flow'' along a certain spatial direction. 
Without loss of generality, we can choose the latter as the 
3-direction (i.e.,\ $z$-axis), and describe four-vectors 
$a^\mu = (a^0, \, a^1, \, a^2, a^3)$  in terms of the light-front components $a^\mu = (a^+, \, a^-, \, \vec{a}_\perp)$ with
\be\label{Eq:LF-coordinates}
    a^\pm = \frac1{\sqrt{2}}(a^0\pm a^3) \,,
    \quad \uvec{a}_\perp = (a^1,\,a^2)\,.
\ee
Now we can also state more precisely how the variables $x$ and $\xi$ mentioned in Sec.~\ref{Sec:1.2-partonic-functions-overview} are defined --- namely, in the notation of Fig.~\ref{f:correlator}, 
\be\label{Eq:x-xi-in-LF}
      x = \frac{k^+}{P^+}\, , \quad
    {\xi=\frac{p^+-p^{\prime +}}{p^++p^{\prime +}}=-\frac{\Delta^+}{2P^+}}
   \,,
 \ee
where $P = \frac{1}{2}(p + p')$ is the average four-momentum of the hadron. 
In the following, we will, for convenience, restrict ourselves to frames with vanishing average transverse momentum, i.e.,
\begin{equation}
    P^\mu=(P^+,P^-,\uvec 0_\perp)\quad \text{with} \quad P^-=\frac{P^2}{2P^+}=\frac{M^2+t/4}{2P^+} \,.
\end{equation}

In order to state the formal QCD definitions of the non-perturbative functions describing the quark structure of hadrons, we introduce the following correlator, which is diagrammatically represented in Fig.~\ref{f:correlator} along with the definitions of the momenta,
\be\label{GPCF}
    \Phi^{q[\Gamma]}(P,k,\Delta)= \frac{1}{2}
    \int \frac{d^4z}{(2\pi)^4}\;\mathrm{e}^{i k\cdot z}\,
    \langle p',S'|\,\bar{\psi}_q(-\tfrac{z}{2})\;\Gamma\,{\cal W}[- \tfrac{z}{2},\tfrac{z}{2}]\;\psi_q(\tfrac{z}{2})\,
    |p,S\rangle\,.
\ee
Here the Wilson line ${\cal W}[- \tfrac{z}{2},\tfrac{z}{2}]$ 
connects the positions of the quark fields along a certain path, and $\Gamma$ is some Dirac matrix.
Generally, we specify a Wilson line connecting two points $a$ and $b$ with a straight line by the exponential of the line integral of the gluon field according to
\begin{align}
    {\cal W}[a,b] = [a^+, a^-, \uvec{a}_\perp; b^+, b^-, \uvec{b}_\perp]
= {\cal P} \, {\rm exp} \, \Bigg(  i g \int_b^a dy^\mu A_\mu(y) \Bigg) \,,
\label{e:Wline_def}
\end{align}
where $A_\mu = A_\mu^a T^a$, and ${\cal P}$ indicates path-ordering.
For the bilocal quark operator in Eq.~\eqref{GPCF} a summation over color is implicit.
The FFs, PDFs, TMDs, GPDs, and GTMDs of quarks are defined through certain 
limits or projections of the correlator in Eq.~\eqref{GPCF}, i.e., 
\begin{subequations}\label{Eq:definitions-overview}
\begin{align}
    \Phi_{\rm FF}^{q[\Gamma]}(P,\Delta)             
    & =  \int d^4 k \;\Phi^{q[\Gamma]}(P,k,\Delta) \,, & \label{FF-correlator}\\
    \Phi_{\rm PDF}^{q[\Gamma]}(P,x)                
    & =  \int d^4 k \;\Phi^{q[\Gamma]}(P,k,0)\;\delta(k^+-xP^+) \,, \label{PDF-correlator}\\
    \Phi_{\rm TMD}^{q[\Gamma]}(P,x,\uvec k_\perp)    
    & =  \int d k^+d k^-\; \Phi^{q[\Gamma]}(P,k,0)\;\delta(k^+-xP^+) \,, \label{TMD-correlator}\\
    \Phi_{\rm GPD}^{q[\Gamma]}(P,x,\xi,\uvec\Delta_\perp)          
    & =  \int d^4 k \;\Phi^{q[\Gamma]}(P,k,\Delta)\;\delta(k^+-xP^+) \,,\label{GPD-correlator}\\
    \Phi_{\rm GTMD}^{q[\Gamma]}(P,x,\uvec k_\perp,\xi,\uvec\Delta_\perp) 
    & =  \int d k^+d k^-\; \Phi^{q[\Gamma]}(P,k,\Delta)\;\delta(k^+-xP^+) \,. \label{GTMD-correlator}
\end{align} \end{subequations}
More precisely, e.g.,  $2\sum_q e_q\Phi_{\rm FF}^{q[\gamma^\mu]}(P,\Delta)$ corresponds to the matrix elements of the electromagnetic current $J^\mu_{\rm em}(0)=\sum_qe_q\bar{\psi}_q(0)\gamma^\mu\psi_q(0)$ and defines the electromagnetic FFs as shown in Eq.~(\ref{Eq:def-em-FF}). 
Similarly, all the above correlators are parameterized in terms of appropriate non-perturbative functions. 
Gluon distributions are defined by 
similar correlators, which, however, feature gluon field-strength tensors instead of quark fields.

In the field-theoretical treatment, the operators defining the correlators 
in~\eqref{FF-correlator}--\eqref{GTMD-correlator} may exhibit various types of 
divergences requiring regularization and renormalization. 
This introduces dependence on one or more (renormalization) scales which we do 
not indicate for brevity. We will comment on these aspects in the following sections. 
Only for certain FFs there is no scale dependence, namely when one deals with an external 
(from the point of view of QCD) conserved current, as is the case, e.g., for the electromagnetic 
FFs in Eq.~(\ref{Eq:def-em-FF}).

\begin{figure}
\begin{center}
\includegraphics[width=0.9\textwidth]{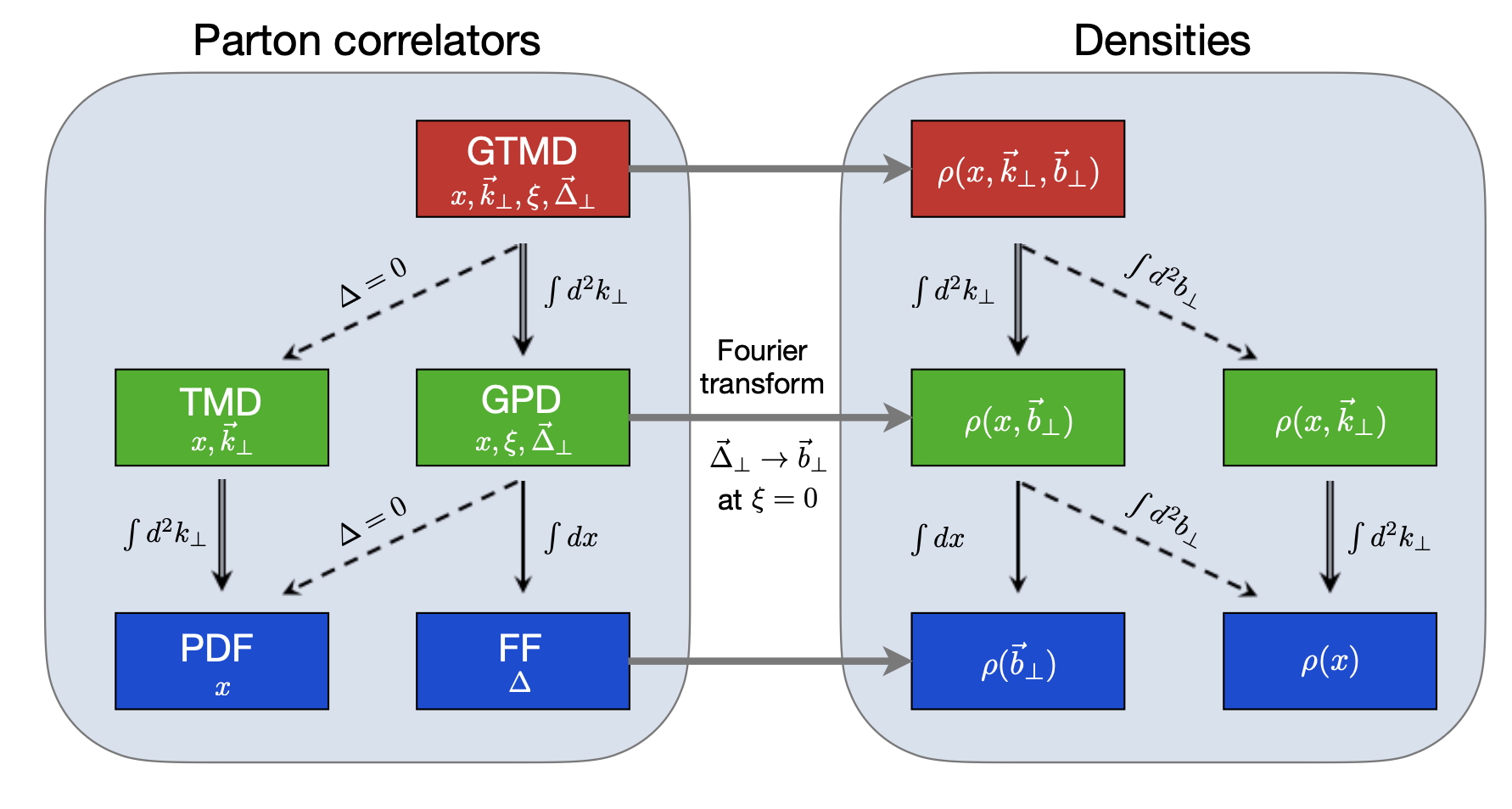}
\end{center}
\caption{\footnotesize Experimentally observable functions (left panel), 
            their interpretations (right panel) and inter-relations 
            (indicated by the arrows).  Note that the PDFs and the TMDs are densities in momentum space, indicated in the right panel as   $\rho(x)$ and $\rho(x,\vec k_\perp)$, respectively. \label{Fig:functions+interpretation}}
\end{figure}

The definitions in~\eqref{FF-correlator}--\eqref{GTMD-correlator} make the inter-relations between the different non-perturbative functions apparent, as illustrated in the left panel of Fig.~\ref{Fig:functions+interpretation}.
For instance, in the forward limit $\Delta^\mu\to 0$ the correlator $\Phi_{\rm GTMD}^{q[\Gamma]}$ reduces to $\Phi_{\rm TMD}^{q[\Gamma]}$, implying that all TMDs arise from the forward limit of some of the GTMDs.
Furthermore, integrating $\Phi_{\rm GTMD}^{q[\Gamma]}$ over $\uvec k_\perp$ yields $\Phi_{\rm GPD}^{q[\Gamma]}$, implying relations between GPDs and some of the GTMDs, and so on. 
We point out though that in QCD the relations in Fig.~\ref{Fig:functions+interpretation} arising from the integral over $\uvec k_\perp$ have to be considered with care as we will discuss in more detail in Sec.~\ref{Sec-3:TMD}.

Interpretations of the non-perturbative functions are of importance for our understanding of the partonic structure of hadrons. 
PDFs and TMDs were introduced as partonic probability densities. 
We also already saw the interpretations of FFs in terms of spatial distributions. 
For GPDs, we set $\xi=0$ and take the Fourier transform 
$\int \frac{d^2\Delta_\perp}{(2\pi)^2}\, e^{-i\vec{\Delta}_\perp\cdot\vec{b}_\perp}\,\Phi_{\rm GPD}^{q[\Gamma]}(P,x,0,\uvec\Delta_\perp)$ that yields impact-parameter 
space distributions $\rho(x,\uvec b_\perp)$ describing the probability densities to find partons with momentum fraction $x$ at the position $\vec{b}_\perp$ in the transverse plane.
Carrying out similar steps for GTMDs yields the so-called relativistic Wigner functions $\rho(x,\uvec k_\perp,\uvec b_\perp)$, which are phase-space distributions, whereby $\vec{b}_\perp$ and $\vec{k}_\perp$ are not Fourier-conjugate variables. 
The right panel of Fig.~\ref{Fig:functions+interpretation} shows an overview of the various densities and their interpretations.

In the following sections, we will successively discuss PDFs, TMDs, GPDs, and GTMDs.
We will make their definitions more precise, give an overview of their theoretical properties, elaborate on their interpretations, discuss to some extent how we can learn about them from experiments, and briefly highlight some of the latest developments. 
This article is not intended as a review.  
In fact, many important results and aspects are not covered.
We have rather aimed at an introduction and (incomplete) overview of the vast field of parton distributions and their generalizations. In particular, we will discuss only single-parton distributions in this contribution. For an account of the more recently developing field of double-parton distributions, we refer to~\cite{Diehl:2017wew,Bartalini:2018qje}.

\subsection{Overview of selected experiments}

The theoretical understanding and phenomenological insights
presented in this article rely on decades of dedicated 
medium- and high-energy experiments carried out world wide.
We will not be able, not even close, to give full justice to
the tremendous experimental~effort. We will not even try to 
cite all data which played a role in acquiring the current 
knowledge of the parton structure of hadrons. Rather we restrict 
ourselves here to a brief overview, see Table~\ref{Table-experiments}, 
and will refer in the following to selected publications or review articles.

The pioneering DIS experiments that laid out the foundation for QCD
and led to the 1990 Nobel Prize of Friedman, Kendall, and~Taylor 
were carried out at SLAC (Stanford Linear Accelerator Center). 
The facility operated  from 1967-1999, with interruptions for upgrades,
using electron beams of initially 8 GeV, reaching up to 50 GeV in the 
1995–1999 period. The first experiment E1 dealt with elastic 
electron-proton scattering. The groundbreaking DIS experiments, conducted 
in 1968 \cite{Bloom:1969kc,Breidenbach:1969kd}, used a 20 GeV beam. 
Since 1976, beam and target polarizations became available.

In the Gargamelle experiment at CERN 
(Conseil Europ\'een pour la Recherche Nucl\'eaire), $\nu_\mu$ and $\bar{\nu}_\mu$ 
beams with energies between 1 and 11 GeV impinged on a heavy freon CF$_3$Br 
target that filled a bubble chamber and simultaneously served as detector 
material. Neutrino induced DIS events like $\nu_\mu p \to \mu^-X$ were 
observed, analogous to Fig.~\ref{Fig-2:FF-DIS-DVCS}(b), except that the 
process was mediated by the exchange of a virtual $W^-$ boson in this example. 
The highlights of this experiment include the discovery that the partons 
observed in DIS carry only half of the nucleon momentum \cite{Eichten:1973cs}.
Combining the SLAC and Gargamelle data, proved that the partons 
(participating in DIS event with electron and neutrino beams) 
carry fractional electric charges \cite{GargamelleNeutrino:1974exc}.

The Drell-Yan process was first observed in 1970 with the AGS 
(Alternating Gradient Synchrotron, built in 1960) at BNL 
(Brookhaven National Lab) by using 22 to 29.5 GeV proton beams 
on a uranium target and detecting $\mu^+\mu^-$ pairs 
\cite{Christenson:1970um}. The charm quark 
was discovered at AGS in October 1974 and confirmed by SLAC in 
November 1974. Early Drell-Yan measurements were also carried out at 
CERN  (R108, R209) exploring the ISR (Intersecting Storage Ring, 1971-1984),
the world's first collider running two counter-circulating proton 
beams each with up to 31.4$\,$GeV and $\sqrt{s}$ of $62.8\,$GeV
\cite{CERN-Columbia-Oxford-Rockefeller:1979gxw,Antreasyan:1980yb}.

At CERN, the SPS (Super Proton Synchrotron) delivered since 1976
proton beams of 400$\,$GeV which were either directly used for Drell-Yan 
experiments (NA3) or used to produce secondary $\pi^\pm$ beams with 
beam energies between 40 and 286 GeV which were then used for 
Drell-Yan and other measurements (NA10, WA11, WA39, WA70)
\cite{NA3:1983ejh,NA10:1986fgk,NA10:1987sqk,WA70:1987vvj}.
The experiments were conducted over a period of 1976-1985 using 
unpolarized proton or nuclear targets. The abbreviations 
stand for North Area (NA) and West Area (WA) experiments.

The Fermilab Drell-Yan experiments used proton beams with 
400-800$\,$GeV impinging on fixed targets leading in 1977 to 
the first observation of $\Upsilon$ and $b$-quark discovery. 
Secondary pion and antiproton beams were also used for Drell-Yan 
experiments. The E866/NuSea experiment ran with 800 GeV proton beams, 
followed by the E906/SeaQuest experiment with a 120$\,$GeV proton 
beam. The ongoing E906/SpinQuest experiment uses a polarized 
120$\,$GeV proton beam and polarized proton and deuteron targets.   

The EMC (European Muon Collaboration, NA2, NA9, NA28) 
and BCDMS (Bologna-CERN-Dubna-Munich-Saclay, NA4) experiments 
at CERN ran in parallel (1978-1985) using 100-280$\,$GeV muon 
beams impinging on proton, deuterium and other nuclear targets 
making DIS measurements up to $Q^2=260\,\rm GeV^2$ possible.
Polarized ($p$, D) targets became available at EMC in 1984. 
The 400$\,$GeV SPS proton beam was used to produce pion beams
which then decayed into muons. Longitudinal beam polarizations were 
obtained by selecting muons of different energies from pion beams of 
a fixed energy by exploiting parity violation in the weak pion decays. 
EMC was superseded by NMC (New Muon Collaboration, 1989) 
and SMC (Spin Muon Collaboration, 1991-1996) experiments.
These muon experiments discovered 
the EMC effect \cite{EuropeanMuon:1983wih} and demonstrated that a 
significant fraction of nucleon spin is not due to the spins of 
quarks and antiquarks \cite{EuropeanMuon:1989yki,SpinMuonSMC:1994met}.

\begin{table}[t!]
\begin{tabular}{llllrr}
\hline
facility - experiments  & years     &   beam    & target            & $\sqrt{s}/$GeV & $Q_{\rm max}^2/\rm GeV^2$ \\
\hline
SLAC - E1...E155        & 1967-1999 & $e^-$                         & $p$, D, $^3$He &  9.7	 & 35     \\
CERN - Gargamelle       & 1970-1976 & $\nu_\mu,\,\bar\nu_\mu$    
                                                & CF$_3$Br          & 4.4   & 3   \\
CERN - NA3, NA10, WA11, WA39, WA70
                        & 1976-1985 & $\pi^\pm$ & $p$, nuclei       & 27.4  & 72  \\
FNAL - E288...E1039     
                        & 1977-now  & $p$, $\bar p$, $\pi^\pm$     
                                                & $p$, nuclei       & 38.7  & 324     \\
CERN - BCDMS, EMC, NMC, SMC  
                        & 1978-1996 & $\mu^+$   & $p$, D, nuclei    & 19.4  & 200   \\
FNAL - E616, E701, E744, E770 
                        & 1979-1988 & $\nu_\mu$, $\bar{\nu}_\mu$
                                                & iron              & 27.4         & 100   \\ 
FNAL - TEVATRON (CDF, D$\slashed{0}$)
                        & 1983-2011 & $\bar{p}$ & $p$ (beam)        & 1,960\po     & 13,500 \\
DESY - HERA (H1, ZEUS)   & 1992-2007 & $e^\pm$   & $p$ (beam)        & 318\po       & 50,000  \\ 
DESY - HERMES 		    & 1995-2007 & $e^\pm$   & $p$, D, $^3$He    & 7.2	       & 15      \\
JLab - Hall A, B, C, D  & 1997-now  & $e^-$     & $p$, D, $^3$He     & 4.8	       & 10          \\
CERN - COMPASS          & 2002-2022  & $\mu^\pm,\,\pi^-$  
                                                & $p$, D             & 19.4	       & 70         \\
RHIC - BRAHMS, PHENIX, STAR     
                        & 2002-now  & $p$       & $p$ (beam)	    & 510\po       & 400       \\
LHC  - ATLAS, CMS, LHCb & 2010-now  & $p$       & $p$ (beam)	    & 13,000\po    & 2,250,000 \\
\end{tabular}
\caption{\footnotesize \label{Table-experiments}
Overview of selected hadron structure experiments.
Shutdown periods for upgrades are not indicated. Only the 
highest available center of mass energies $\sqrt{s}$ 
are shown. The experiments covered the kinematic  
range $Q_{\rm min}^2< Q^2 < Q_{\rm max}^2$ 
with typically $Q_{\rm min}^2=1\,{\rm GeV}^2$ in fixed-target experiments 
and often much higher in collider experiments. For all experiments, 
$Q^2_{\rm max}$ is specified in the last column.}
\end{table}

The Fermilab experiments E616 (1979) and E701 (1982)
explored neutrino beams with beam energies up to 300$\,$GeV 
produced by 400$\,$GeV proton beams from Fermilab's Main Injector.
The E744 (1985) and E770 (1987-1988) used neutrino beams up to
600$\,$GeV produced from the 800$\,$GeV proton beam at Tevatron.
The E815 NuTeV experiment (1996-1997) used 150$\,$GeV neutrino 
beams produced by the Main Injector proton beam. These experiments 
provided important DIS data.
Tevatron at Fermilab operated from 1983 to 2011 colliding 
$p$ and $\bar{p}$ beams with $\sqrt{s}=1.96\,{\rm TeV}$
with the first detection of the top quark in 1995. The
CDF and D$\slashed{0}$ experiments provided $W^\pm$ and $Z^0$
as well as jet production data which play an important role
in fits of the unpolarized parton distributions
\cite{D0:1999jba,CDF:1999bpw,D0:2010dbl,CDF:2012brb}.

HERA (Hadron–Electron Ring Accelerator) at DESY 
(Deutsches Elektronen-Synchrotron) was the only (so far)
electron-proton collider. Operating from 1992 to 2007, HERA collided
920$\,$GeV proton beams and 27.5$\,$GeV $e^\pm$ beams at $\sqrt{s}$
of 318$\,$GeV. The two experiments H1 and ZEUS took DIS data 
(1994-2000) in the kinematic range $0.005 < x < 0.65$ and
$200\,\rm GeV^2 < Q^2 < 30,000\,GeV^2$ including structure functions
due to exchange of $W^\pm$ and $Z^0$ as well as exclusive reactions
\cite{H1:2000kis,ZEUS:2001mhd,ZEUS:2002wfj,H1:2005dtp,H1:2009pze,H1:2015ubc}.
The polarized HERA $e^\pm$ beams were used in the HERMES 
experiment on polarized proton, deuterium and $^3$He fixed-targets  
to investigate deep-inelastic and exclusive reactions
\cite{HERMES:1997hjr,HERMES:1999ryv,HERMES:2004mhh,HERMES:2006jyl,HERMES:2008abz,HERMES:2009lmz,HERMES:2012uyd,HERMES:2020ifk}.

The Jefferson Lab (JLab) experiments in Halls A, B, C started data taking in 1997
with initially $4\,\rm GeV$ polarized electron beams which were upgraded over the 
years to 12$\,$GeV exploring $p$, D, $^3$He and heavier nuclear targets. In the
Hall D experiment, which began data taking in 2017, the electron beam is used to
produce a beam of real photons with energies up to 11.4$\,$GeV. Thanks to the
high-luminosities of the CEBAF (Continuous Electron Beam Accelerator Facility) 
the JLab experiments deliver high accuracy data on deep-inelastic and exclusive 
reactions in electro-production as well as photo-production
\cite{CLAS:2001wjj,CLAS:2003qum,JeffersonLabHallA:2003joy,CLAS:2007clm,JeffersonLabHallA:2011ayy,GlueX:2023pev}. 

The COMPASS (Common Muon and Proton Apparatus for Structure and Spectroscopy) 
experiment at CERN used 160$\,$GeV $\mu^-$ or 190$\,$GeV $\pi^-$ beams produced
by the SPS. The muon beam was longitudinally polarized. The proton and deuteron 
targets in that experiment could be longitudinally or transversely polarized.
The experiment was taking data in the period 2002-2022, and is currently in the
analysis stage.
Many important results related to hadron structure were obtained in this experiment
\cite{COMPASS:2005qpp,COMPASS:2008isr,COMPASS:2010wkz,COMPASS:2010hbb,COMPASS:2013bfs,
COMPASS:2014bze,COMPASS:2017jbv,COMPASS:2017mvk,COMPASS:2023cgk}.

RHIC (Relativistic Heavy Ion Collider) is an intersecting storage ring accelerator
which can accelerate heavy ions up to 100$\,$GeV per nucleon and protons up to
500$\,$GeV. It is the only proton collider built so far with polarized proton
beams. The RHIC experiments delivered important results 
on heavy-ion collisions including the observation of quark-gluon plasma in 2010. 
The experiments BRAHMS, PHENIX, and STAR at RHIC 
delivered important data related to proton spin physics 
\cite{PHENIX:2007kqm,BRAHMS:2008doi,STAR:2012ljf,STAR:2014wox,STAR:2014afm,
STAR:2015vmv,STAR:2017wsi,PHENIX:2018dwt,STAR:2023jwh}.

At LHC (Large Hadron Collider) proton beams were collided with center-of-mass 
energies $\sqrt{s}$ of initially 7$\,$TeV and meanwhile 13$\,$TeV leading to 
the Higgs boson discovery in 2012. LHC can also collide heavy ions. The ATLAS, 
CMS, and LHCb experiments yield valuable data for proton structure studies 
\cite{LHCb:2015mad,ATLAS:2019zci,CMS:2022ubq}.

In the Belle (1999-2010, KEK) and BaBar (1999-2008, SLAC) experiments
$e^+e^-$ beams were collided at $\sqrt{s}$ of 10.58$\,$GeV, i.e.,\
the peak of the $\Upsilon(4S)$ resonance which decays into $B$-mesons.
These experiments were designed to study CP-violation and other phenomena~\cite{BaBar:2014omp}.
The data taken in these experiments allow us to study, among others,
fragmentation functions 
\cite{Belle:2005dmx,Belle:2008fdv,Belle:2011cur,BaBar:2013jdt,Belle:2017rwm}
which are indispensable for the determinations of PDFs or TMDs 
from certain processes; see Sec.~\ref{Sec-3.3:SIDIS-and-DY}.
After detector and luminosity upgrades, in 2019 the Belle II 
experiment was launched.

\section{Parton distribution functions}
\label{Sec-2:PDF}

PDFs describe the structure of hadrons in high-energy processes such as DIS, 
which was historically of paramount importance for the development of QCD, 
leading to the Nobel Prizes for the experimental work by Friedman, Kendall, 
and Taylor (1990), and theoretical work by Gross, Wilczek, and Politzer (2004); 
see the review \cite{Gross:2022hyw} and the historical account~\cite{Parisi:2025nob}.
In the following, we briefly review the DIS process before introducing the parton 
model and the QCD definition of PDFs along with their properties, interpretation, 
certain limits, as well as some results extracted from experimental data.

\subsection{Inclusive deep-inelastic electron-nucleon scattering and Bjorken scaling} 
\label{Sec-2.1:DIS+scaling}

The process $e\,N\to e^\prime X$, depicted in Fig.~\ref{Fig-2:FF-DIS-DVCS}(b) in the 
one-photon exchange approximation, can be described in terms of the kinematic variables 
\begin{equation}
        s = (P+l)^2 \,, \quad
        Q^2 = - q^2 \,, \qquad
        x_B = \frac{Q^2}{2P\cdot q} \,, \qquad
        y = \frac{P\cdot q}{P\cdot l} \,,
        \label{Eq:DIS-variables}
\end{equation}
where $s$ is the square of the center-of-mass energy, 
$x_B$ the Bjorken scaling variable satisfying $0\leq x_B\leq 1$,
while $y$ describes the energy loss of the electron in the nucleon rest frame. 
The variables in~\eqref{Eq:DIS-variables} are related by $x_B y\,(s-M^2-m_e^2) = Q^2$, 
where the electron mass $m_e$ is (here and in the following) negligible. 
Also the nucleon mass $M$ can often be neglected.
Notice that in Eq.~(\ref{Eq:DIS-variables}) we denote the four-momentum of the 
incoming nucleon by $P$ instead of $p$. ($P$ and $p$ coincide in the language of the 
correlator in Eq.~(\ref{GPCF}) in the case of PDFs where $\Delta=0$.)

DIS refers to the regime where both $2P\cdot q \gg M_{\rm had}^2$ and 
$Q^2 \gg M_{\rm had}^2$. In this kinematics, the nucleon typically 
breaks up and produces a final state $X$ containing many hadrons 
(and possibly other particles), which remains unresolved in the inclusive process. 
The DIS scattering amplitude ${\cal M}$ in the one-photon exchange approximation 
is similar to the elastic one in Eq.~\eqref{elastic_amplitude} and reads
\begin{align}
    {\cal M}   =    \frac{e^2}{q^2}\,
    \langle l',s'|J_{\mu\, {\rm em}}(0)|l,s\rangle\,
    \langle X|J_{\rm em}^{\mu}(0)|p,S\rangle \,,
\label{e:DIS_amp}
\end{align}
with the states labeled by momenta and polarization vectors of the particles as 
defined in Fig.~\ref{Fig-2:FF-DIS-DVCS}(b). The DIS cross section is proportional to
\begin{equation}
    \frac{|{\cal M}|^2}{(4\pi)^2} = 
    \frac{\alpha_{\rm em}^2}{Q^4}\,L_{\mu \nu} W^{\mu \nu} \,,
\end{equation}
with the leptonic tensor $L^{\mu\nu}$ given, to the lowest-order in QED perturbation theory, by
\begin{equation}
    L^{\mu \nu} 
    = \sum\limits_{s'}[\bar{u}(l',s') \gamma^\mu u(l,s) ]^*[\bar{u}(l',s') \gamma^\nu u(l,s) ]
    = 2 (l^\mu l'^\nu + l^\nu l'^\mu - l \cdot l' g^{\mu \nu} 
    - i\lambda_e\epsilon^{\mu\nu\alpha\beta} l_\alpha l^\prime_\beta ) \,,  \quad \mbox{with} \quad\epsilon_{0123}=+1 \,,\label{leptonictensor}
\end{equation}
where $\lambda_e = \uvec{s}\cdot\uvec{l}/(|\uvec{s}|\,|\uvec{l}|)$ denotes the helicity of 
the incoming (highly relativistic) electron. The hadronic tensor $W^{\mu\nu}$ defined as
\begin{equation}\label{Eq:hadronic-tensor}
        W^{\mu \nu}  
    =   \frac{1}{4 \pi} \sum_X \langle P,S | J_{\rm em}^{\mu \dagger}(0) | X \rangle 
        \langle X | J_{\rm em}^\nu(0) | P,S \rangle \,(2 \pi)^4 \delta^{(4)}(P_X - P - q)
    =   \frac{1}{4 \pi} \int d^4z \,e^{i q \cdot z} \,
        \langle P,S | J_{\rm em}^{\mu}(z) J_{\rm em}^{\nu}(0)  | P,S \rangle 
\end{equation}
contains the non-perturbative information about the nucleon structure which is encoded 
in terms of two spin-independent structure functions, $\SFF{1}$ and $\SFF{2}$, and two 
spin-dependent ones, $\SFG{1}$ and $\SFG{2}$, according to
\begin{align}
    W^{\mu \nu} = \left( \frac{q^\mu q^\nu}{q^2} - g^{\mu \nu} \right) \SFF{1} 
    + \frac{\hat P^\mu \hat P^\nu}{P \cdot q}  \SFF{2}
    +i\epsilon^{\mu\nu\alpha\beta}\,\frac{q_\alpha M}{P\cdot q}
    \left(S_\beta \,\SFG{1}+\left[S_\beta-\frac{(S\cdot q)}{(P\cdot  q)}P_\beta\right]\, \SFG{2}\right) \,,
    \quad \mbox{with} \quad
    \hat P^\mu = P^\mu - \frac{P\cdot q}{q^2}\,q^\mu \,.
    \label{Eq:hadr-tensor-decomposition}
\end{align}
The nucleon polarization vector satisfies $P\cdot S=0$ and $S^2=-1$.  
The Lorentz decomposition in Eq.~\eqref{Eq:hadr-tensor-decomposition} 
follows from the symmetries of the strong interaction and the conservation 
of the electric current, implying $q_\mu W^{\mu\nu}=0$ and $q_\nu W^{\mu\nu}=0$. 
Neglecting terms of ${\cal O}(M^2/Q^2)$, the unpolarized DIS cross section 
takes the form
\begin{equation}\label{DIS_cross_section}
    \frac{d^2\sigma_{\rm DIS}}{dx_B\, dQ^2}\biggl|_{\rm unp} 
    \ \simeq
    \frac{4\pi\alpha^2_\textrm{em}}{Q^4}\left[y^2\SFF{1}+\left(1-y\right)\frac{\SFF{2}}{x_B}\right] \,.
\end{equation}
The expression for the polarized DIS cross section can be found, e.g., 
in Sec.~18 of Ref.~\cite{ParticleDataGroup:2024cfk}. 

The DIS structure functions are Lorentz-invariant functions of 
$x_B$ and $Q^2$, i.e.,\ $\SFF{i}(x_B,Q^2)$ and $\SFG{i}(x_B,Q^2)$. 
Based on current algebra and dispersion relation techniques, 
Bjorken predicted that, in the limit $2P\cdot q\to\infty$ and 
$Q^2\to\infty$ with their ratio $x_B$ fixed, the structure 
functions should merely depend on $x_B$
\cite{Bjorken:1967fb,Bjorken:1968dy}, a phenomenon called Bjorken scaling. 
Soon after, experiments conducted at SLAC confirmed (within the energy range 
and error bars of that time) the Bjorken scaling of the unpolarized 
DIS structure functions \cite{Bloom:1969kc}. 
Bjorken's prediction greatly helped pave the path towards QCD; 
see \cite{Gross:2005nobel} for the historical context.

\subsection{Feynman's parton model}
\label{Sec-2.2:parton-model}

\begin{figure}[t!] 
\begin{center}
\includegraphics[height=.22\textwidth]{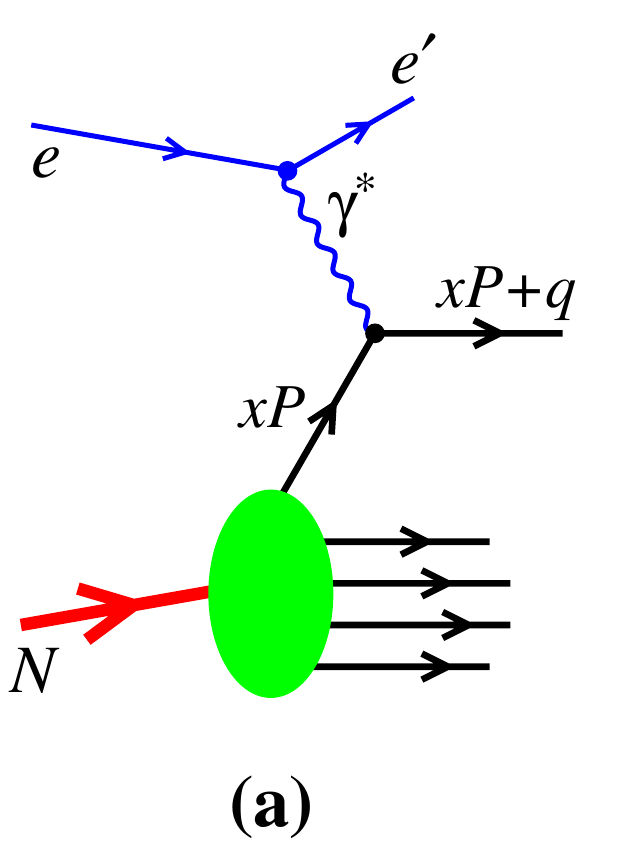}        \hspace{5mm}
\includegraphics[height=.22\textwidth]{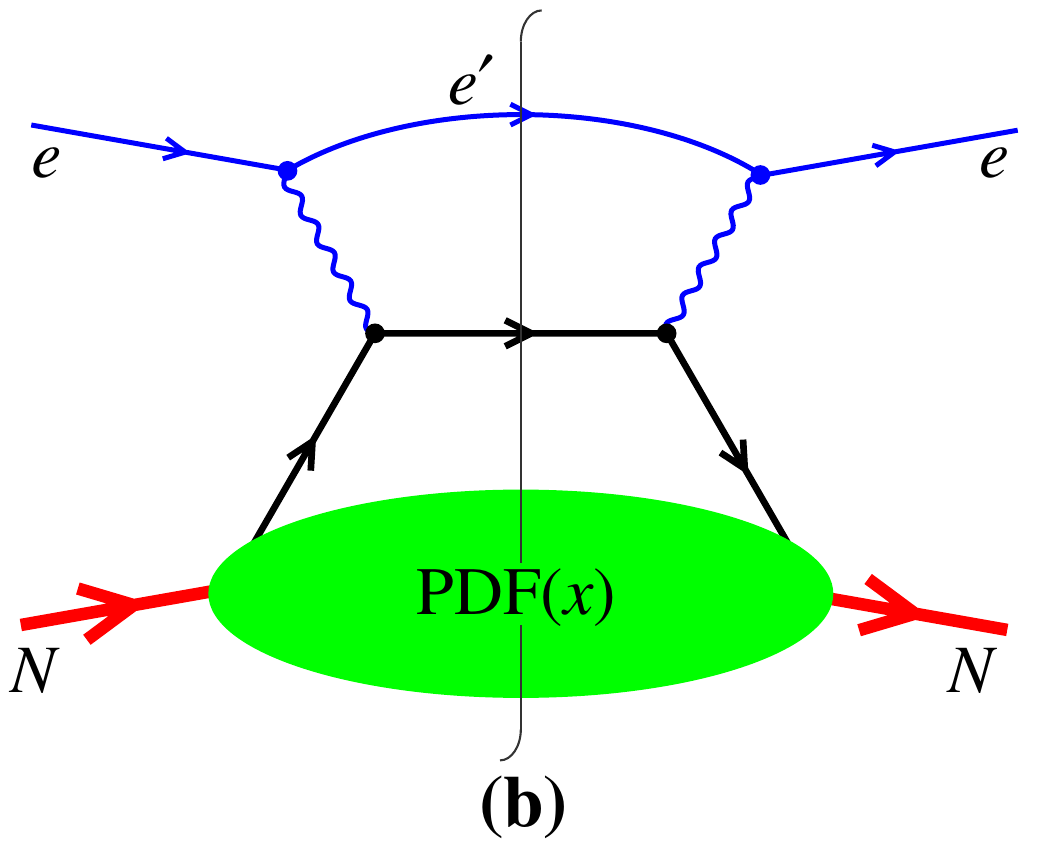} \hspace{5mm}
\includegraphics[height=.22\textwidth]{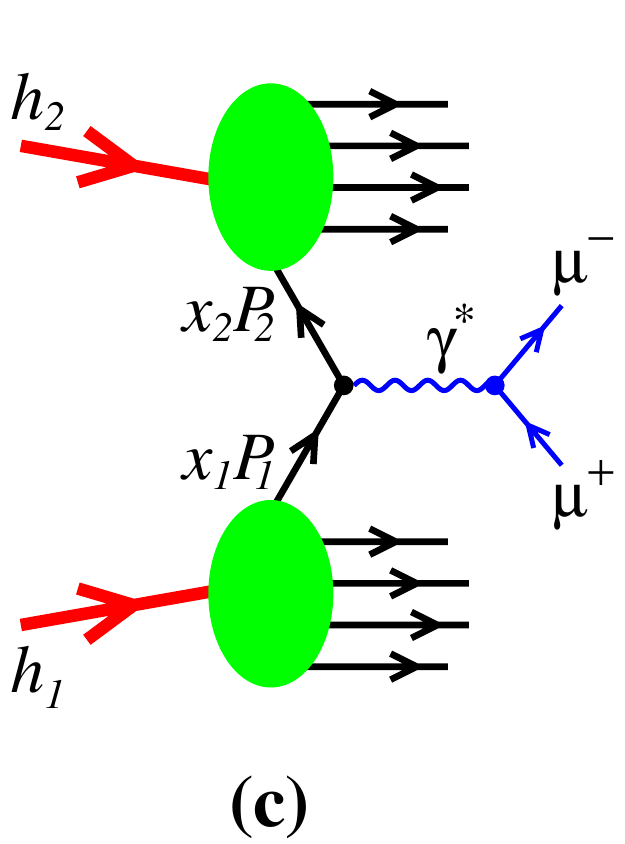}         \hspace{5mm}
\includegraphics[height=.22\textwidth]{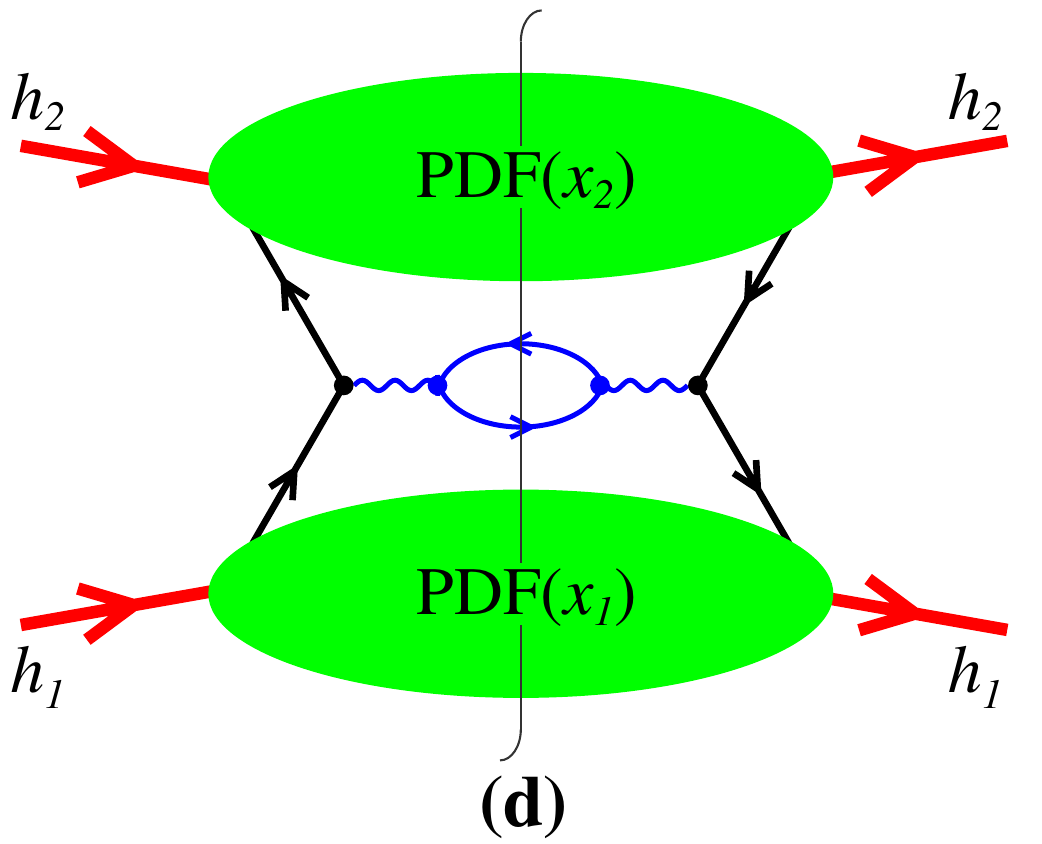}  
\end{center}
\caption{\footnotesize \label{Fig-03:DIS+DY}
(a) 
The amplitude of deep-inelastic scattering (DIS) in the Bjorken-limit,
where the electron scatters elastically off a charged parton carrying 
the momentum fraction $x \simeq x_B$ of the nucleon's momentum $P$.
(b) 
The DIS cross section is proportional to the ``handbag'' diagram representing the product of the DIS amplitude (left of vertical line) and its complex conjugate (right of vertical line).
The non-perturbative blob labeled PDF$(x)$ is the diagrammatical representation of a PDF and includes a summation over a complete set of states.
(c) 
Amplitude for the Drell-Yan process, where a parton and antiparton from two colliding hadrons annihilate to produce a virtual photon. The photon momentum $q^\mu$ is reconstructed from its decay into a $\mu^+\mu^-$ pair in the limit $s=(P_1+P_2)^2\to\infty$ and $Q^2=q^2\to\infty$, with their ratio equal 
to $x_1 x_2 = Q^2/s$ and fixed.
(d) 
The square of the Drell-Yan amplitude.}
\end{figure}

A physically appealing, heuristic explanation of the Bjorken scaling in 
DIS was given by Feynman in his parton model \cite{Feynman:1972original,Feynman:1969ej}
and further formalized in \cite{Bjorken:1969ja}.
In a frame where the proton momentum is large (e.g.\ electron-proton or $\gamma^\ast$-proton center-of-mass frame), Feynman envisaged ``the proton of momentum $P$ as being made of partons of momenta $x_i\,P$ all sharing in various proportions
the momentum of the proton'' with $0< x_i < 1$ and $\sum_i x_i =1$, whereby the partons are ``practically free''~\cite{Feynman:1972original}.
If the parton momentum is given by $k^\mu$ in the initial state and, after 
absorbing the virtual photon, by $k^\mu + q^\mu$ in the final state, then 
the onshell assumption of the initial and final partons yields
\be\label{Eq:parto-on-shell-assumption}
    0=(k + q)^2-k^2 = q^2 + 2 k \cdot q 
    \quad \quad \Rightarrow \quad \quad
    x \simeq\frac{Q^2}{2P\cdot q} = x_B \,,
    \ee
where we set $k^\mu\simeq xP^\mu$ and neglected transverse parton momenta $k_\perp$. 
Notice that in the parton model $k_\perp^2$ and off-shellness $\delta^2=k^2-m_i^2\neq0$, where $m_i$ is the parton mass, are not assumed to be exactly zero, but merely small compared to $Q^2$.
Thus, in the parton model, the Bjorken variable $x_B$ is equal to the fraction of nucleon momentum carried by the initial-state parton; see Fig.~\ref{Fig-03:DIS+DY}(a).
To describe the partons inside the proton, Feynman introduced 
the concept of parton distribution function $f^i(x)$, which represents a number density. 
Specifically, $f^i(x) dx$ is the number of partons of type $i$ 
with momentum fraction in the interval $[x, x+dx]$.
%
% BEGIN FIGURE 03: DIS WITH DETAILS AND DRELL-YAN
%

Except for being electrically charged, it was a priori not clear which 
properties (including spin) the partons should have. Depending on whether 
one assumes the partons to be elementary spin-0 or spin-$\frac12$ particles, 
the parton model makes different predictions, namely \cite{Feynman:1972original}, 
\be\label{Eq:spin0-spin12-predictions}
     \mbox{(i)  spin-0         partons:} \quad \SFF{T} \ll \SFF{L} \,, \quad \quad
     \mbox{(ii) spin-$\frac12$ partons:} \quad \SFF{T} \gg \SFF{L} \,,   
\ee
where $\SFF{T}=2x_B\SFF{1}$ and $\SFF{L}=\SFF{2}-2x_B\SFF{1}$ are the contributions 
associated with transversely and longitudinally polarized virtual photons, respectively. 
Experimentally, $\SFF{L}\ll\SFF{T}$ was found, indicating that the charged partons have 
spin $\frac12$. The predictions in Eq.~(\ref{Eq:spin0-spin12-predictions}) 
have been derived, prior to the parton model, by means of current algebra and 
dispersion relation techniques~\cite{Callan:1969uq}, and the experimental result 
$\SFF{L}\ll\SFF{T}$ is equivalent to $\SFF{2}\approx2x_B\SFF{1}$ which is known 
as the Callan-Gross relation~\cite{Callan:1969uq}. 
It is instructive to quote how the experimentally supported prediction (ii) 
in~\eqref{Eq:spin0-spin12-predictions} arises in the parton model, namely 
\cite{Feynman:1972original}
\be\label{Eq:Callan-Gross-in-parton-model}
    \frac{\SFF{L}}{\SFF{T}} = 4\,\frac{\langle k_\perp^2\rangle+m_i^2\pm\delta^2}{Q^2} \simeq 0 \,,
\ee
with average squared transverse parton momenta $\la k_\perp^2\ra$, $m_i^2$, and 
off-shellness corrections $\pm\delta^2$ assumed to be small compared to $Q^2$.

If $e_i$ denotes the electric charge of partons of type $i$ in units of the 
elementary charge, the parton model results for the unpolarized DIS structure 
functions are $\SFF{1}=\frac12\sum_i e_i^2 f^i(x_B)$ and $\SFF{2} = 2x_B\SFF{1}$, 
and the DIS cross section can be written as 
\be\label{factorized_DIS_cross_section}
    d\sigma_{\rm DIS} \, \big|_{\rm unp} 
    \;\stackrel{ {\rm parton}\atop{\rm \phantom{_1}model\phantom{_1}}}{=}\;
    \sum_i \int_0^1dx\,f^i(x)\,
    d\hat\sigma^i_{\rm elast} \,,
    \quad \quad
    \frac{d^2\hat\sigma^i_{\rm elast}}{dx_B\, dQ^2}
    =e^2_i\,\frac{2\pi\alpha^2_{\rm em} }{Q^4}\left[1+(1-y)^2\right]\,\delta(x-x_B) \, ,
\ee
with $y\simeq Q^2/(x_Bs)$ and fixed large $s$.
The left equation in~\eqref{factorized_DIS_cross_section}   
can be considered differential in any set of independent DIS 
variables, such as $x_B$ and $Q^2$. 
The parton model relates the DIS cross section to the calculable 
cross section $\hat{\sigma}^i_{\rm elast}$ for elastic electron-parton 
scattering, shown in~\eqref{factorized_DIS_cross_section} to leading order (LO) in QED.
The fact that not amplitudes but cross sections for electron-parton scattering 
are added in Eq.~(\ref{factorized_DIS_cross_section}) follows from the parton 
model assumption that DIS is an incoherent process.

DIS events can also be mediated by electroweak bosons (instead of virtual photons), 
e.g.\ when neutrino beams are used (instead of electron beams). 
Then different structure functions contribute, but they are expressed in the parton 
model in terms of the same PDFs, albeit weighted by electroweak charges of the partons 
rather than electric charges.
This feature can be, and has been, used to separate 
the PDFs of different partons.

The parton model can be applied to other deeply inelastic reactions as well. 
One prominent example is the Drell-Yan process in Fig.~\ref{Fig-03:DIS+DY}(c)~\cite{Drell:1970wh}. 
This implies an important property, namely the PDFs in the parton model are universal in the sense that the same PDFs describe different high-energy processes. 
This is a key ingredient which allows one, in principle, to extract 
PDFs from experimental data for one process and make predictions for other processes.

While the parton model had a lot of success, it also gave rise to open questions. 
(i) What is the nature of the spin-$\frac12$ partons seen in DIS?
(ii) It was a natural speculation that the partons could correspond to the quarks used by Gell-Mann and Zweig in the early 1960s as building blocks in order to describe the quantum numbers of hadrons; see \cite{Entem:2025bqt} and references therein.
However, how is it then possible that quarks are so tightly bound in hadrons that they can never be observed as free particles (confinement hypothesis) and at the same time appear as nearly free in DIS?
(iii) The partons must carry 100$\%$ of the nucleon's momentum, i.e.,\ 
$\sum_i \int_0^1 dx\,x\,f^i(x) = 1$, while already early experiments indicated that the partons seen in DIS carry only about half of the proton momentum~\cite{GargamelleNeutrino:1974exc}. So, who carries the rest? 
The quest to find answers to these and other questions paved the way to the developments of QCD.

\subsection{QCD description} 
\label{Sec-2.3-QCD-description-of-DIS}

Because of the asymptotic freedom of QCD (see Sec.~\ref{sect.1.3}), for $Q \gg M_{\rm had}$ the parton model result in Eq.~\eqref{factorized_DIS_cross_section} can potentially be considered a first reasonable approximation, and corrections might be calculable in pQCD. 
However, a large scale is not sufficient since most high-energy processes, including DIS, are also sensitive to non-perturbative QCD dynamics.
The additional key element is factorization, stating that the cross section can be separated into two parts: (i) a perturbatively calculable partonic cross section and (ii) a non-perturbative (universal) function~\cite{Collins:1989gx, CTEQ:1993hwr, Collins:2011zzd}.
In the parton model expression in Eq.~\eqref{factorized_DIS_cross_section}, the perturbative part is given by the cross section for elastic electron-quark scattering calculated in lowest-order QED, while the non-perturbative part is given by the PDFs.
For the DIS cross section, this factorization still holds once radiative corrections in QCD are taken into account.

\begin{figure}[b!] 
\begin{center}
\includegraphics[height=.18\textwidth]{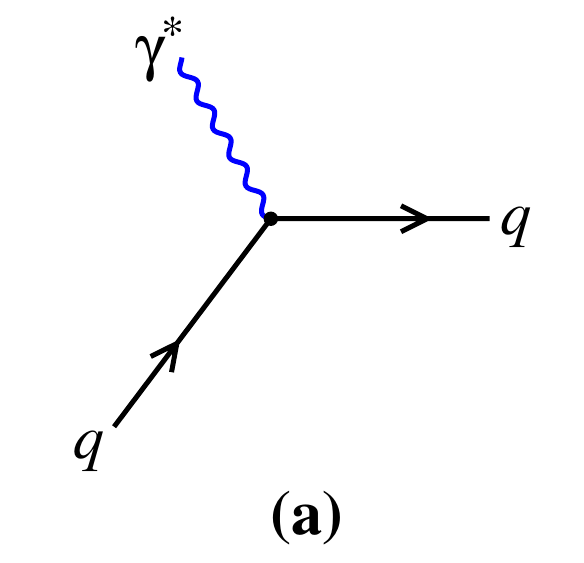}                 \hspace{5mm}
\includegraphics[height=.18\textwidth]{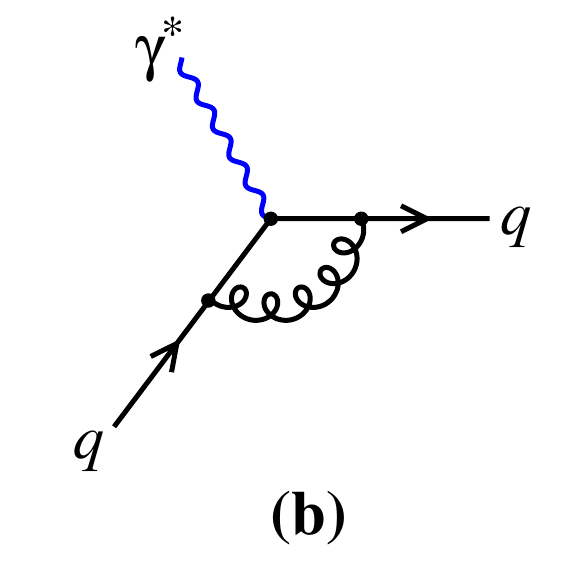}   \hspace{5mm}
\includegraphics[height=.18\textwidth]{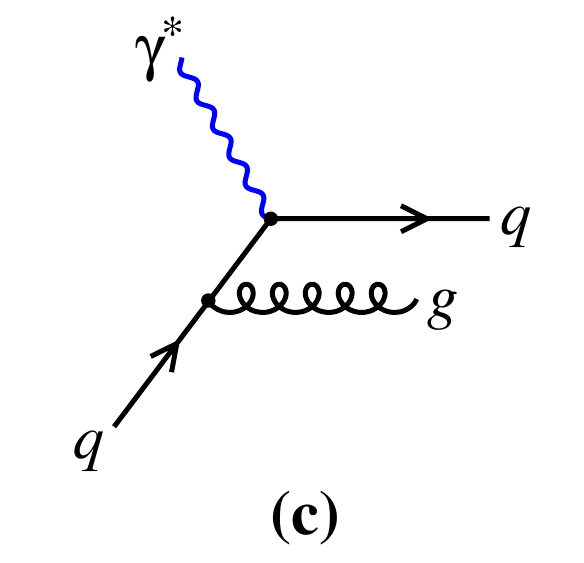} \hspace{5mm}
\includegraphics[height=.18\textwidth]{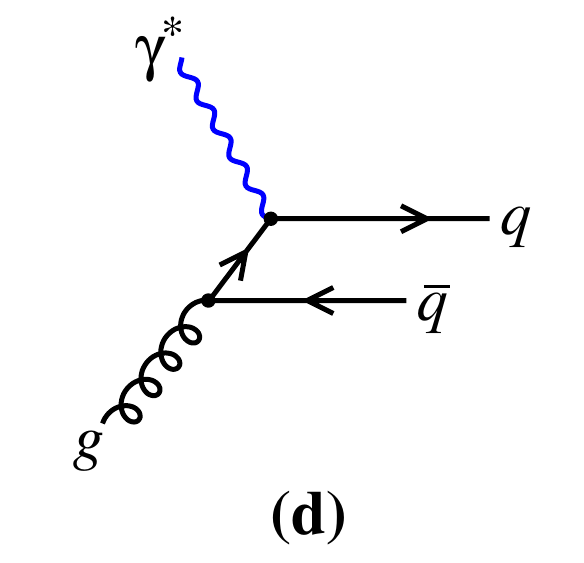}  
\end{center}
\caption{\footnotesize \label{Fig-NEW-DGLAP:DIS}
QCD diagrams relevant for DIS.
(a) The LO (parton model) diagram $\gamma^\ast q \to q$.
The NLO corrections to this process:
(b) one of the virtual one-loop corrections;
(c) one of the real radiative corrections for the QCD Compton process $\gamma^\ast q \to g q$;  
(d) one of the photon-gluon fusion diagrams.}
\end{figure}

Before writing down the QCD factorization formula for DIS, we briefly discuss the ${\cal O}(\alpha_{\rm s})$ corrections that go beyond the parton model, which are referred to as one-loop or next-to-leading order (NLO) contributions.
If one sets aside the electron, the lowest-order perturbative process (parton model approximation) 
is $\gamma^\ast q \to q$ in Fig.~\ref{Fig-NEW-DGLAP:DIS}(a) and the 
corresponding reaction $\gamma^\ast \bar{q} \to \bar{q}$ for antiquarks.
At ${\cal O}(\alpha_{\rm s})$, there exist virtual one-loop corrections 
to the process $\gamma^\ast q \to q$ in Fig.~\ref{Fig-NEW-DGLAP:DIS}(b), 
as well as real radiative corrections in the form of the so-called QCD 
Compton process, $\gamma^\ast q \to g q$ in Fig.~\ref{Fig-NEW-DGLAP:DIS}(c), 
and the photon-gluon fusion process, $\gamma^\ast g \to q \bar{q}$, 
in Fig.~\ref{Fig-NEW-DGLAP:DIS}(d).
The boson-gluon fusion process provides sensitivity to the gluon PDF in the nucleon.
In fact, DIS experiments have played a very important role in revealing the gluon structure of the nucleon.

%
% BEGIN NEW FIGURE "25" (DIS DIAGRAMS FOR DGLAP)
%

In the discussion of the parton model we used the generic expression $f^i$ for PDFs. 
In QCD we denote the unpolarized (spin-averaged) PDF by $f_1^a$ with $a=u,\bar u,d, \bar d,\cdots,g$. 
The QCD factorization formula for the unpolarized DIS cross section can then be written as 
\be
d \sigma_{\rm DIS}\bigl|_{\rm unp} \ = \sum_a \int_{x_B}^1 \frac{dx}{x}\, f_1^a(x,\mu) \, d\hat{\sigma}^a \left(\frac{x_B}{x},\frac{Q}{\mu}, \alpha_{\rm s}(\mu) \right) + {\cal O}\left(\frac{M_\text{had}}{Q}\right) \,.
\label{e:QCD_factorization}
\ee
The partonic cross section $d\hat{\sigma}^a$ in~\eqref{e:QCD_factorization} can be computed order-by-order in perturbation theory, where the zeroth order (``tree-level'') term coincides with the parton model result in Eq.~\eqref{factorized_DIS_cross_section}.
Note that the lower limit of the integral in~\eqref{e:QCD_factorization} is $x_B$, which ensures that the final state has positive energy once radiative corrections are included; see, e.g., Sec.8.2.2 in~\cite{Collins:2011zzd}.
The DIS cross section factorizes to all orders in perturbation theory (see~\cite{Collins:1989gx, CTEQ:1993hwr, Collins:2011zzd} and references therein), with explicit results fully available up to three loops~\cite{Moch:2004pa,Vogt:2004mw}, and partly worked out at four-loop accuracy; see, e.g., Ref.~\cite{Moch:2017uml}.
The PDFs entering in the factorized expression in Eq.~\eqref{e:QCD_factorization} correspond to the leading contribution in an expansion of the cross section in powers of $M_\text{had}/Q$; see Sec.~\ref{sect:ope}.
Similar expressions can be written for 
other processes, and in some cases one may even write a factorization formula for subleading contributions.  
When this occurs, the non-perturbative part typically involves multi-parton correlators. 
These objects are qualitatively different from the simpler single-parton densities on which we focus in this contribution.

Both the partonic cross section and the PDFs in Eq.~\eqref{e:QCD_factorization} depend on a scale denoted by $\mu$.
In the case of the PDFs, $\mu$ is referred to as the renormalization scale, which is needed to regularize the ultraviolet (UV) divergences that appear in QCD.
At ${\cal O}(\alpha_{\rm s})$ the most important UV divergence arises when integrating over the transverse quark momentum all the way up to infinity, because in pQCD the integrand behaves like $1/k_\perp^2$ for large $k_\perp$. 
Leaving aside problems with gauge invariance in QCD, $\mu$ can be considered as a cutoff for the transverse momentum integral.
In this case, factorization can readily be understood as the separation of the physics at low transverse momenta (smaller than $\mu$) contained in the PDF and at large transverse momenta (larger than $\mu$) contained in the partonic cross section.
This also motivates why often in a factorization formula like in Eq.~\eqref{e:QCD_factorization}, the scale of the PDF is called factorization scale.
In principle, one could distinguish the factorization scale at which 
perturbative and non-perturbative contributions are separated and the 
renormalization scale. In practice, both scales are typically set equal. 
In dimensional regularization commonly used in pQCD, the renormalization 
scale $\mu$ enters when divergences are tamed  by evaluating loop integrals 
in $4 - 2 \varepsilon$ dimensions instead of 4 dimensions~\cite{tHooft:1972tcz,Bollini:1972ui,Cicuta:1972jf}.
To keep the QCD coupling dimensionless in this step, one also makes the replacement $g \to g \mu^{\varepsilon}$, introducing the mass scale $\mu$ in dimensional regularization.
The results of such calculations are given by a series in $\varepsilon$ with the leading terms proportional to $1/\varepsilon$ (or even higher powers of $1/\varepsilon$), which need to be subtracted via a renormalization procedure.
Since this subtraction is not unique, a dependence on the renormalization scheme arises, where mostly the so-called modified minimal subtraction ($\overline{\rm MS}$) scheme~\cite{Bardeen:1978yd} is used in the community.
(In the $\overline{\rm MS}$ scheme, the singularity is removed along with certain finite terms that appear in dimensional regularization.) 
As a result, PDFs and partonic cross sections are both scale- and scheme-dependent.
However, these dependences must drop out for the full DIS cross section which is a physical observable.

Like in the case of $\alpha_{\rm s}(\mu)$, pQCD determines the scale dependence of the PDFs through the so-called DGLAP (Dokshitzer, Gribov, Lipatov, Altarelli, Parisi) evolution equations~\cite{Dokshitzer:1977sg, Gribov:1972ri, Altarelli:1977zs}. 
The structure of these equations looks like
\be
\frac{d}{d \ln \mu}f_1^a(x,\mu) = 2 \sum_b \int_x^1 \frac{dx'}{x'}\, P_{ab}\left(\frac{x}{x'}, \alpha_{\rm s}(\mu)\right) f_1^b(x',\mu) \,,
\label{e:DGLAP}
\ee
where the splitting functions  
\be
P_{ab}(x, \alpha_{\rm s}) = \frac{\alpha_{\rm s}}{2 \pi}\, P_{ab}^{(1)}(x) + \bigg( \frac{\alpha_{\rm s}}{2 \pi} \bigg)^2 P_{ab}^{(2)}(x) +\cdots
\ee
can be computed order-by-order in pQCD.
Because of the quark-gluon coupling and the gluon self-coupling, a parton can split into other partons. 
The $P_{ab}$ are a measure of the probability of finding a parton $a$ inside a parton $b$.
Note the summation over the parton type on the r.h.s.~of Eq.~\eqref{e:DGLAP}, which generally introduces mixing between different partons.
Solving the DGLAP equations then, in principle, allows one to find the PDFs at a given scale $\mu$ if they are known at another scale $\mu_0$. 
For this to work, both $\mu$ and $\mu_0$ must be large enough for pQCD to be applicable.
For brevity, throughout the document we do not display the dependence of the PDFs and their generalizations on the renormalization scale, unless it is important for the context.

We now briefly discuss the question of the appropriate numerical value for $\mu$ on the r.h.s.~of Eq.~\eqref{e:QCD_factorization}.
Although in principle $\mu$ is arbitrary, it should in practice be chosen to be similar to the hard scale $Q$ of the DIS process.
In a one-loop analysis, a term proportional to $\alpha_{\rm s}(\mu) \ln(Q/\mu)$ appears, at two loops a term proportional to $(\alpha_{\rm s}(\mu) \ln(Q/\mu))^2$, and so on.
If $\mu$ and $Q$ are very different, then $\ln(Q/\mu)$ is large so that even the product $\alpha_{\rm s}(\mu) \ln(Q/\mu)$ may not be small, leading to a poor convergence of the perturbative expansion of the factorized expression in Eq.~\eqref{e:QCD_factorization}. 
The same discussion applies to other high-energy scattering processes, that is, $\mu$ must be of the order of the hard scale of a given process to avoid large logarithms that endanger the stability of the pQCD expansion.

The pQCD analysis of DIS at ${\cal O}(\alpha_{\rm s})$ leads to two qualitative changes compared to the lowest-order (parton model) results.
First, the scale dependence of the PDFs implies in practice 
a nontrivial (logarithmic) dependence of the structure functions on $Q^2$, meaning that the Bjorken scaling is violated.
It was considered a triumph of QCD when such scaling violations were first observed in the DIS data, and found to be in quantitative agreement with the predictions of pQCD.
Second, at this order in perturbation theory the structure function $F_{L}^{\rm SF}$ becomes nonzero.
DIS data have confirmed this result as well.

Before discussing the field-theoretic definition of the unpolarized PDF, we want to 
emphasize that also the Drell-Yan process played an important role in establishing QCD.
Within the framework of the parton model, the cross section for this process was strongly 
underestimated based on information on PDFs from DIS. But the NLO pQCD corrections for the 
Drell-Yan process turned out to be sizable and, once taken into account, resolved the 
issue~\cite{Altarelli:1979ub}.
This development also provided further phenomenological evidence that PDFs are universal.
Proving all-order factorization for Drell-Yan is a challenging task which was addressed in the 1980s~\cite{Bodwin:1984hc, Collins:1985ue}.
We also note that QCD factorization involving PDFs and related non-perturbative parton correlation functions has been proven for a variety of processes in lepton-nucleon scattering, hadron-production in electron-positron annihilation, and in hadronic collisions~\cite{Collins:1989gx}. 
Factorization together with the universality of PDFs provides a powerful framework for extracting PDFs from global data analysis involving a variety of processes and for making predictions for observables. 

With the understanding that the QCD operator $\bar{\psi}_q(-\tfrac{z}{2})\;\gamma^+ \, {\cal W}[- \tfrac{z}{2},\tfrac{z}{2}]\,\psi_q(\tfrac{z}{2})$ is renormalized at a given order in $\alpha_{\rm s}$ within a certain scheme at the scale $\mu$ and the Wilson line ${\cal W}$ specified in Eq.~\eqref{e:Wline_def},
the unpolarized PDF for quarks and antiquarks is defined through~\cite{Collins:1981uw}
\begin{equation}\label{e:f1_def}
    \Phi^{q[\gamma^+]}_\text{PDF}(P,x,\mu)= \int \frac{dz^-}{4\pi} \, \mathrm{e}^{i xP^+ z^-} \,
    \langle P,S|\,\bar{\psi}_q(-\tfrac{z}{2})\;\gamma^+ \, 
    {\cal W}[- \tfrac{z}{2},\tfrac{z}{2}] \, \psi_q(\tfrac{z}{2})\,|P,S\rangle 
    \big|_{z^+ = |\vec{z}_\perp| = 0}=f_1^q(x,\mu) \,.
\end{equation}
This expression has the support $x \in [-1, 1]$ with the positive-$x$ region describing, as usual, the distribution of quarks. The meaning of $f_1^q(x,\mu)$ at negative $x$ is as follows: as a consequence of charge conjugation symmetry, the distribution of antiquarks is given by $f^{\bar q}_1(x,\mu)=-f^q_1(-x,\mu)$ (here $x$ is positive). 
It is also customary to introduce the valence quark distribution as $f_1^{q,\text{val}}(x,\mu)= f_1^q(x,\mu)-f_1^{\bar q}(x,\mu)$ with $x>0$, so as to exclude the contribution from the  $q\bar q$-pairs in the sea. 
For the definition of the unpolarized gluon PDF $f_1^g$ we refer to the literature~\cite{Collins:1981uw}.
As already mentioned above, PDFs for different partons can mix under DGLAP evolution.
Specifically, the flavor non-singlet distributions, defined as differences between PDFs of different quark flavors, $f_1^q - f_1^{q'}$, do not mix, while the flavor singlet combination $\sum_q (f_1^q + f_1^{\bar q})$ mixes with $f_1^g$.
Furthermore, in Sec.~\ref{sect-2.4} we will discuss how different gamma matrices $\Gamma$ in $\Phi^{q[\Gamma]}_\text{PDF}$ define unpolarized or polarized PDFs, or even correlation functions that cannot be interpreted as single-parton densities.

The QCD definition in Eq.~\eqref{e:f1_def} can be obtained by computing the lowest-order diagram for the DIS cross section in Fig.~\ref{Fig-03:DIS+DY}(b) for a nucleon with a large $P^+$ and making the parton-model approximations, which means neglecting the (small) minus and transverse momentum components of the initial-state quark which interacts with the virtual photon.
This approximation, in particular, provides a correlation function in which the two quark-field operators are separated along the light-front minus direction, as is the case in Eq.~\eqref{e:f1_def}.
The Wilson line in Eq.~\eqref{e:f1_def} arises when taking into account the (all-order) gluon exchange between the struck (final-state) quark and the remnants of the nucleon, and keeping the leading terms of the expansion in powers of $M_{\rm had}/Q$. 

\subsection{Operator product expansion and twist}\label{sect:ope}

Originally, DIS was not treated in the diagrammatic approach described above, 
but by using operator product expansion, a powerful method in quantum field theory \cite{Hollands:2023txn}; for applications in QCD see \cite{Collins:1984xc} and the pedagogical exposition \cite{Jaffe:1996zw}. It is instructive to sketch the main steps of this method. In the last step in Eq.~(\ref{Eq:hadronic-tensor}), the product of currents $J(z)J(0)$ can be replaced by the commutator $[J(z),J(0)]$ because the opposite ordering does not contribute to the hadronic tensor. (Lorentz indices and subscripts of the currents are not essential for the generic argument, and are therefore omitted.)  When the exponential $e^{iq\cdot z}$ is expressed in light-front coordinates, one sees that the integral over $d^4z$ in (\ref{Eq:hadronic-tensor}) is dominated by the region of $z^2\approx 0$, suggesting a ``Taylor expansion'' around $z^2=0$ of the form~\cite{Itzykson:1980rh}
\be\label{Eq:OPE}
    [J(z),J(0)]\sim\sum_n C_n(z^2)\,z_{\alpha_1}\cdots z_{\alpha_{k_n}}\hat{\cal X}_n^{\alpha_1\cdots\alpha_{k_n}}(0) \,,
\ee
with c-number (singular) Wilson coefficients $C_n(z^2)$ and local operators $\hat{\cal X}_n^{\alpha_1\cdots\alpha_{k_n}}(0)$. These operators can be organized in a basis of symmetric traceless tensors, i.e.,\ $\hat{\cal X}_n^{\alpha_1\cdots\alpha_{k_n}}g_{\alpha_i\alpha_j}=0$ for any $1\le i,j\le k_n$, with matrix elements
\begin{equation}\label{MELO}
    \la P|\hat{\cal X}_n^{\alpha_1\cdots\alpha_{k_n}}(0)|P\ra = P^{\alpha_1}\cdots P^{\alpha_{k_n}} M_{\rm had}^{d_n-k_n-2} f_n +\cdots\,.
\end{equation}
The $f_n$ are dimensionless numbers and the dots indicate Lorentz structures like ${g^{\alpha_1\alpha_2}P^{\alpha_3}\cdots P^{\alpha_{k_n}}}$ that are suppressed in the Bjorken limit. The powers of $M_{\rm had}$ are introduced for dimensional reasons, $d_n$ is the mass dimension of the operator $\hat{\cal X}_n^{\alpha_1\cdots\alpha_{k_n}}(0)$ and $k_n$ its ``spin'' (i.e.,\ number of Lorentz indices). Inserting the series expansion (\ref{Eq:OPE}) into the hadronic tensor in Eq.~(\ref{Eq:hadronic-tensor}) yields an infinite series in which each term is weighted by a prefactor $(M_{\rm had}/Q)^{d_n-k_n-2}$. In the limit $Q^2\to \infty$, the series is dominated by an infinite tower of local operators with increasing values of $k_n$ and $d_n$, such that the difference $d_n-k_n$ is the lowest. For unpolarized DIS, the infinite tower of leading operators in the quark sector is $\bar{\psi}(0)\gamma^+D^+\cdots D^+\psi(0)$. (These operators mix under renormalization with similar leading operators in the gluonic sector.) The first operator in this tower is $\bar{\psi}(0)\gamma^+\psi(0)$ with  mass dimension 3 (due to $\bar{\psi}\psi$) and spin 1 (one Lorentz index), i.e.,\ one has $d_n-k_n=2$ for $n=1$. Insertions of the $n-1$ operators $D^+$ (each with mass dimension 1 and spin 1) satisfy all $d_n-k_n=2$. One defines the twist $t_n$ of a local operator $\hat{\cal X}_n^{\alpha_1\cdots\alpha_{k_n}}(0)$ as $t_n=d_n-k_n$. This means that the lowest possible value in DIS is twist-2. Other operators may be of higher twists. For example, the operator $\bar{\psi}(0)\psi(0)$ is twist-3 (mass dimension 3 and spin 0; the operator does not contribute to DIS though). The matrix elements $f^q_n$ of the infinite tower of leading (i.e.,\ lowest) twist quark operators in DIS correspond to the Mellin moments $f^q_n=\int dx\,x^{n-1} f_1^q(x)$ of the unpolarized PDF, and similarly in the gluon sector. The Mellin transform can be uniquely inverted to obtain from the infinitely many $f^a_n$ the PDF $f_1^a(x)$. The latter is said to be twist-2, even though it is defined through a non-local operator. (Keep in mind that here twist is defined for local operators.) For a pedagogical exposition see \cite{Jaffe:1996zw}.

Twist only tells us about the leading $Q^2$-dependence associated with $\hat{\cal X}_n^{\alpha_1\cdots\alpha_{k_n}}(0)$, as trace terms represented by the dots in Eq.~\eqref{MELO} are suppressed by additional powers of $M_{\rm had}/Q$. This means that if the matrix element of a local operator is suppressed by some power $(M_{\rm had}/Q)^{t-2}$, the twist of that operator is at most $t$. Matrix elements with no suppression factor are therefore twist-2, whereas those suppressed by one power of $M_{\rm had}/Q$ generally involve both twist-2 and twist-3 operators. It is important to note
that the operator product expansion can be applied to some processes like DIS, 
but not to all processes (with Drell-Yan being one counter-example). 
In such cases, only the diagrammatic approach sketched in the previous section can be
used to separate leading from subleading contributions (and prove factorization).  
The concept of twist can be applied not only to PDFs but also to other hadronic 
properties including GPDs; see Sec.~\ref{Sec-4:GPD}.
However, in the case of TMDs and GTMDs it is in general not
possible to apply the notion of twist in the above strict sense
because such quantities are defined in terms of non-local operators
with no simple representation in terms of towers of local operators.
Nevertheless, one can still distinguish such functions according to 
the powers $(M_{\rm had}/Q)^{t-2}$ with which 
they contribute to cross sections. In this sense, one can speak of  
``leading'' or ``leading-power'' functions (for $t=2$)  and distinguish them from 
``subleading'' functions (for $t= 3$); see Secs.~\ref{Sec-3:TMD}~and~\ref{Sec-5:GTMDs}.

\subsection{Interpretation and properties of PDFs in QCD}
\label{sect-2.4}

Similar to the unpolarized PDF in Eq.~\eqref{e:f1_def}, 
there exist at leading twist two polarized quark PDFs defined as
\begin{equation}\label{Eq:def-g1-h1}
    \Phi^{q[\gamma^+\gamma_5]}_\text{PDF}(P,x)=S_L\, g_1^q(x) \,, \qquad  \Phi^{q[i\sigma^{j+}\gamma_5]}_\text{PDF}(P,x)=S^j_\perp\, h_1^q(x) \,,
\end{equation}
with the transverse index $j=1,2$ and the nucleon polarization 
characterized by 
$S^\mu=(0,\uvec S_\perp,S_L)$ in the nucleon rest frame.
As a result of charge conjugation symmetry, the corresponding antiquark distributions are given by $g^{\bar q}_1(x)=g^q_1(-x)$ and $h^{\bar q}_1(x)=-h^q_1(-x)$.
The PDFs $f^q_1(x)$ and $g^q_1(x)$ are chiral-even, in the sense 
that chirality $\psi_{R,L}=\frac{1}{2}(1\pm\gamma_5)\psi$ is 
preserved, while $h^q_1(x)$ is chiral-odd~\cite{Jaffe:1991ra}, i.e., 
\begin{equation}\label{Eq:psiR-psiL-1}
    \bar \psi\gamma^+\psi=\bar\psi_R\gamma^+\psi_R+\bar\psi_L\gamma^+\psi_L \,,\qquad \bar \psi\gamma^+\gamma_5\psi=\bar\psi_R\gamma^+\gamma_5\psi_R-\bar\psi_L\gamma^+\gamma_5\psi_L \,,\qquad \bar\psi i\sigma^{j+}\gamma_5\psi=\bar\psi_Li\sigma^{j+}\gamma_5\psi_R-\bar\psi_Ri\sigma^{j+}\gamma_5\psi_L \,.
\end{equation}
Chiral-odd PDFs are suppressed in inclusive DIS by a factor 
$m_q/Q$, but can be accessed in other high-energy processes like, 
e.g., SIDIS and Drell-Yan, where they can appear in combination 
with another chiral-odd non-perturbative partonic function.

When considered in a matrix element with no momentum transfer, these operators can be re-expressed at the bare level in terms of sums and differences of light-front quark number operators~\cite{Jaffe:1996zw}. 
For example, for the unpolarized PDF one finds
\begin{equation}\label{PDF-parton-interpretation}
f_1^q(x)=\frac{1}{\langle P,S|P,S\rangle}\int  \frac{d^2k_\perp}{2|x|(2\pi)^3}
       \sum_\lambda\begin{cases}\,\langle P,S|\,b^\dagger_{q,\lambda}(xP^+, \uvec k_\perp) b_{q,\lambda}(xP^+,\uvec k_\perp)\,|P,S\rangle&{\rm for}\quad x> 0\vspace{.2cm}\\
        \,-\langle P,S|\,d^\dagger_{q,\lambda}(-xP^+, \uvec k_\perp) d_{q,\lambda}(-xP^+,\uvec k_\perp)\,|P,S\rangle&{\rm for}\quad x< 0\end{cases}\quad,
        \end{equation}
where $\lambda=\pm$ is the quark light-front helicity, i.e., the projection of the light-front quark polarization along the $z$-direction~\cite{Soper:1972xc}. 
This shows that $f^q_1(x)$ counts the number of quarks 
(and $f^{\bar q}_1(x)$ that of antiquarks) with a given momentum fraction $x$, irrespective of their polarization.
Since the notions of chirality and light-front helicity are essentially equivalent for leading-twist operators, it follows from Eq.~\eqref{Eq:psiR-psiL-1} that the so-called helicity PDF $g^q_1(x)$ measures the longitudinal polarization of quarks with momentum fraction $x$. Similarly, since quarks polarized along the transverse direction $\uvec n_\perp=(\cos\varphi,\sin\varphi)$ are defined via the linear combination $b^\dag_{q,\uvec n_\perp}=(b^\dag_{q,+}\pm e^{i\varphi}b^\dag_{q,-})/\sqrt{2}$, it can be shown that the so-called transversity PDF $h^q_1(x)$ measures the transverse polarization of quarks with momentum fraction $x$. We stress that, in a relativistic theory, a transversely-polarized quark state is not an eigenstate of the transverse spin operator. Transverse spin is a higher-twist property associated with $\gamma^j\gamma_5$, whereas transverse polarization (or transversity) is a leading-twist property associated with $i\sigma^{j+}\gamma_5$~\cite{Jaffe:1991ra}. The corresponding quark bilinear operators therefore behave in a different way under Lorentz transformations. Based on the partonic density interpretation, leading-twist quark PDFs are expected to satisfy inequality relations~\cite{Jaffe:1991ra}, namely 
\begin{equation}\label{Eq:PDF-inequalities}
    f_1^q(x)  \ge 0 \,, \qquad
    |g^q_1(x)| \le f^q_1(x) \,, \qquad
    |h^q_1(x)| \le f^q_1(x) \,, \qquad 
    |h^q_1(x)| \le \frac12\Bigl(f^q_1(x)+g^q_1(x)\Bigr) \,,
\end{equation}
with the latter known as the Soffer bound~\cite{Soffer:1994ww}. These inequalities are not rigorously valid in QCD due to the subtraction of UV divergences \cite{Ralston:2008sm,Collins:2021vke}. However, provided that they are valid at some (high enough) renormalization scale, they are known in some schemes to be preserved under evolution to higher scales
\cite{Vogelsang:1997ak,Candido:2023ujx}. 
According to~\eqref{Eq:psiR-psiL-1}, $g_1^q$ and $h_1^q$ are given by differences of two densities and, therefore, can become negative.
Note also that a gluon helicity PDF $g_1^g$ can be defined~\cite{Collins:1981uw} as the counterpart of $g_1^q$, while there is no gluon transversity PDF for spin-$\frac{1}{2}$ hadrons because of angular momentum conservation.

Integrals of the leading-twist PDFs $f_1^q$, $g_1^q$ and $h_1^q$ give the 
number of valence quarks, axial charge and tensor charge, respectively, 
\begin{equation}
    \int_{-1}^1d x\,f^q_1(x,\mu)=N_q \,,\qquad \int_{-1}^1d x\,g^q_1(x,\mu)=g^q_A(\mu) \,,\qquad \int_{-1}^1d x\,h^q_1(x,\mu)=g^q_T(\mu) \,,
\label{e:moments}
\end{equation}
with 
$\int_{-1}^1d x\,f^q_1(x,\mu)=\int_0^1d x\,(f^q_1-f_1^{\bar q})(x,\mu)$ and likewise for 
$h_1^q$, while $\int_{-1}^1d x\,g^q_1(x,\mu)=\int_0^1d x\,(g^q_1+g_1^{\bar q})(x,\mu)$; 
see the definitions following Eqs.~\eqref{e:f1_def} and ~\eqref{Eq:def-g1-h1}.
One has $N_u=2$ and $N_d=1$ for the proton, and vice versa for the neutron.
The axial charge $g^q_A$ is also denoted as $\Delta q$ or $\Delta\Sigma_q$. 
It is interpreted as twice the total intrinsic angular momentum (or spin) carried by quarks and antiquarks. 
The tensor charge $g^q_T$ is also denoted as $\delta q$. 
Contrary to $N_q$, both the axial and tensor charges depend on the 
renormalization scheme and scale,
except  for the flavor non-singlet combination 
$\int_{-1}^1 dx\,(g_1^u-g_1^d)(x,\mu)=g_A^u(\mu)-g_A^d(\mu)=g_A$ with the 
axial coupling constant $g_A=1.275$ 
known from neutron decays \cite{ParticleDataGroup:2024cfk},
due to the conservation of the flavor non-singlet axial 
vector current in the chiral limit (or more generally in the limit of exact isospin symmetry)~\cite{Goldberger:1958vp,Adler:1965ka,Adler:1969gk}.
Of particular importance is the Bjorken sum rule for the polarized structure 
functions \cite{Bjorken:1966jh,Bjorken:1969mm}, whose expression in QCD reads 
${\int_0^1 dx\,(\SFG{1p}-\SFG{1n})(x,Q^2)=\frac16\,g_A\left(1-\frac{\alpha_{\rm s}(Q)}{\pi}+\cdots\right)}$
with higher-order corrections known up to ${\cal O}(\alpha_{\rm s}^4)$
\cite{Baikov:2010je}, making it one of the most precisely calculated 
quantities in pQCD \cite{Boer:2011fh}. This fundamental sum 
rule can be tested experimentally \cite{Altarelli:1996nm,Accardi:2012qut}
and used as an independent method to determine $\alpha_{\rm s}(\mu)$ from 
DIS data \cite{Deur:2025rjo}. 

For the unpolarized PDFs also the second moment is of distinct importance, as it defines the (scale-dependent) momentum fractions $\langle x \rangle_a (\mu)$ 
carried by the partons according to
\begin{equation}
\int_{-1}^1 d x \, x \,f^q_1(x, \mu) = \langle x \rangle_q (\mu) \, , \qquad
\int_{0}^1 d x \, x \,f^g_1(x, \mu) = \langle x \rangle_g (\mu) \,.
\label{e:momentum_fractions}
\end{equation}
Summing over all partons provides the momentum sum rule~\cite{Gross:1973zrg,Politzer:1974fr,Georgi:1974wnj,Collins:1981uw} 
\begin{equation}
\sum_a \, \langle x \rangle_a (\mu) = 1 \,.
\end{equation}
The valence and momentum sum rules are protected, respectively, by the conservation of the vector-current and the energy-momentum tensor, and therefore hold even though UV divergences spoil the density interpretation of the PDFs in QCD.

In this section, we so far have focused on the twist-2 PDFs, which for quarks are related to the operator $\bar{\psi} \Gamma \psi$, with the two quark fields separated along the light-front and $\Gamma \in \{\gamma^+, \gamma^+ \gamma_5, i \sigma^{j+} \gamma_5\}$.
For any other gamma matrix, $\bar{\psi} \Gamma \psi$ involves higher-twist PDFs.
As discussed above, twist-2 PDFs can be understood as single-parton number densities.
On the other hand, higher-twist PDFs no longer have a density interpretation.
They rather have an intimate relation with multi-parton correlations, the simplest of which are (3-parton) quark-gluon-quark correlations, symbolically denoted by $\langle  \bar{\psi} A \psi \rangle $ with $A$ representing the gluon field; see~\cite{Ellis:1982cd, Jaffe:1983hp, Balitsky:1987bk, Kanazawa:2015ajw} and references therein.
Generally, such multi-parton correlations are of high interest as they encode quantum interference effects.
They are also important for quantifying the partonic structure of hadrons in QCD.
However, because multi-parton correlations typically appear suppressed in observables, their present knowledge is poor compared to the leading-twist PDFs.
In this work, we will not discuss these objects in any detail.

\subsection{Extractions of PDFs from experimental data} 
\label{Sec-2.5-global-fits}

Modern efforts to extract PDFs leverage data from a broad range of experiments. 
However, the input data sets differ substantially among the three leading-twist PDFs  and the hadron being analyzed. 
In this section we concentrate on the unpolarized and helicity PDFs, while we discuss fits of the transversity PDFs below in Sec.~\ref{Sec-3.6-TMD-extractions}, since exploiting transverse parton momenta has played a crucial role for our present understanding of this function.
In the proton case, the PDFs $f_1^a$ and $g_1^a$ are constrained, though to varying degrees, by data obtained through DIS and SIDIS, Drell-Yan, and various proton-proton and proton-antiproton scattering processes, such as those involving weak gauge boson and quarkonium production. 
Thousands of data points are currently available to constrain the unpolarized PDFs, while the helicity PDFs are typically determined using only hundreds of data points.
A recent compilation of the kinematic coverage of the hadronic cross-section data for the processes commonly included in modern analyses can be found in Ref.~\cite{Gross:2022hyw}.
The disparity in data coverage clearly highlights the challenges of achieving comparable precision for the unpolarized and helicity PDFs.
However, the situation is expected to evolve in the coming years~\cite{Caldwell:2025fjg}, driven by new measurements from JLab~\cite{Accardi:2023chb} and eventually by future colliders such as the Large Hadron-electron Collider (LHeC) and the future circular collider (FCC)~\cite{Ahmadova:2025vzd},
and electron-ion colliders in the US (EIC)~\cite{AbdulKhalek:2021gbh} and in China (EIcC)~\cite{Anderle:2021wcy}.

The (global) data analyses rely on QCD factorization theorems such as the one for DIS in Eq.~\eqref{e:QCD_factorization}.
To describe  PDFs at an initial renormalization scale, traditional approaches employ physics-driven parameterizations for the unknown PDFs, while alternative methods adopt neural network-based inputs to reduce the dependence on specific functional forms.
The DGLAP  equations are then used to evolve these PDFs to higher  scales. 
Predictions for cross sections are compared with experimental measurements, and the best-fit PDF parameters are extracted using statistical techniques, assessing the uncertainties in the extracted PDFs, which reflect experimental errors, model assumptions, and theoretical uncertainties.
Several groups have been generating nucleon PDF sets for decades, constantly releasing updated versions. 
The LHAPDF library~\cite{Buckley:2014ana} offers access to a comprehensive collection of unpolarized PDF sets for the nucleon.

\begin{figure}[t]
\begin{center}
\begin{tabular}{cc}
    \includegraphics[width=0.46\textwidth]{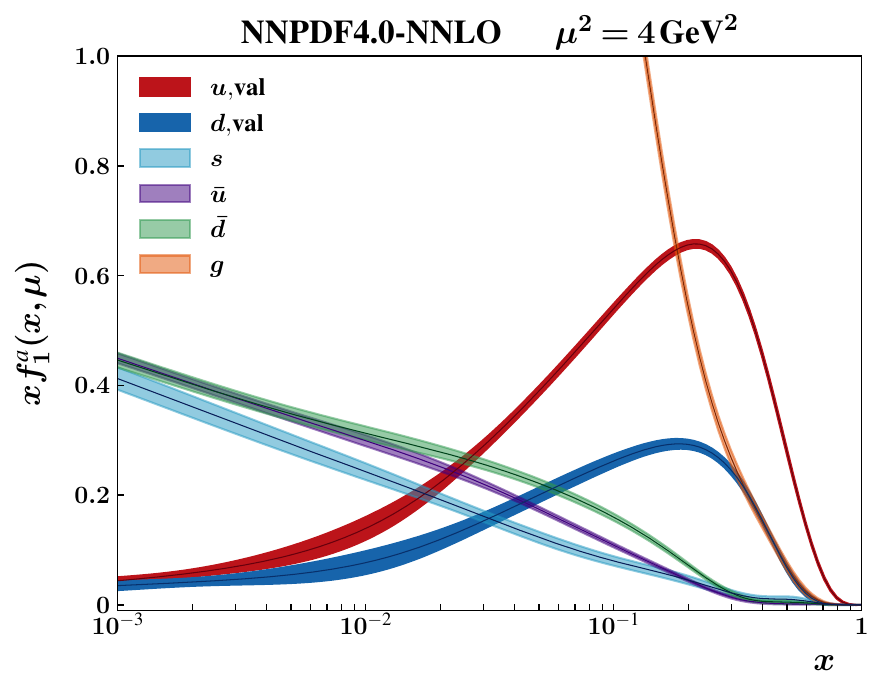}    &
    \includegraphics[width=0.46\textwidth]{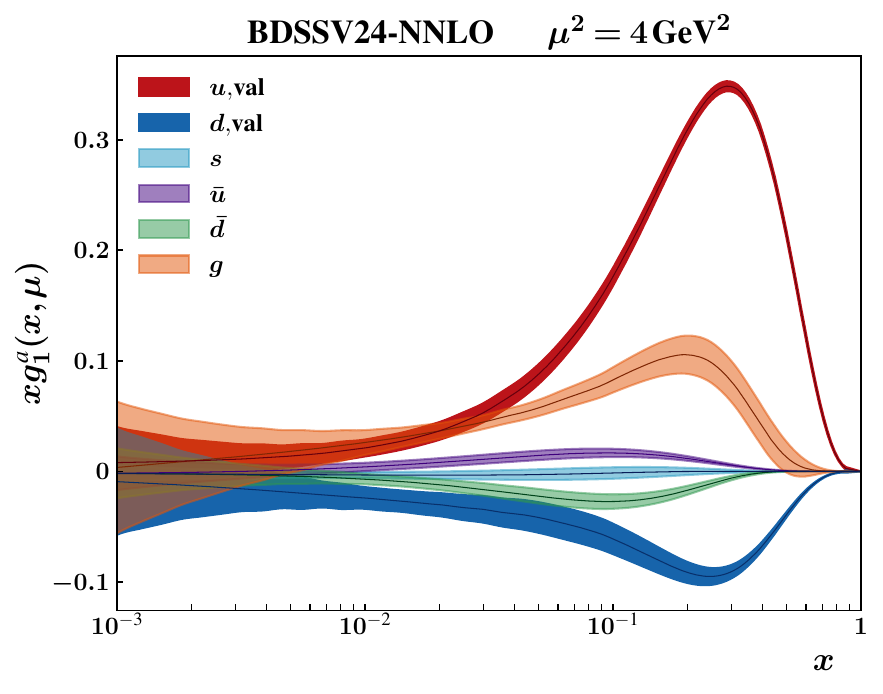}\vspace{-5mm}\\
    {\bf (a)} & \hspace{5mm} {\bf (b)} 
\end{tabular}
\end{center}
\caption{\footnotesize \label{fig3bis} 
Unpolarized (a) and helicity (b) parton distribution functions (PDFs) of the proton at the scale $\mu^2=4 \, \textrm{GeV}^2$ from, respectively, the  NNPDF4.0~\cite{NNPDF:2021njg} and BDSSV24~\cite{Borsa:2024mss} NNLO analyses. Here and in the following figures, the uncertainty bands correspond to 68\% confidence level.}
\end{figure}
The precision of unpolarized proton PDF fits has considerably advanced over time, driven by significant improvements in both experimental data and theoretical modeling; see, for example, Refs.~\cite{NNPDF:2021njg,Bailey:2020ooq,Hou:2019qau}. 
Next-to-next-to-leading order (NNLO) calculations have become the benchmark for unpolarized proton PDFs, with ongoing efforts focused on incorporating higher-order corrections to further minimize theoretical uncertainties.
In the following, we summarize the key features of the most recent global determinations of unpolarized PDFs; see also Ref.~\cite{Ethier:2020way} for more details.
Results from various unpolarized PDF sets show very good agreement for the up-valence distribution, which exhibits a relative uncertainty of only a few percent within the range covered by the experimental data. 
In contrast, the down-valence distribution shows slightly greater variation among the sets.
Unlike valence distributions, sea quark distributions are significantly suppressed at large $x$ and exhibit a steep increase at small $x$, primarily due to gluon splitting. 
These distributions exhibit more pronounced variations between different PDF sets compared to their valence counterparts.
The unpolarized gluon PDF of the proton has different shapes depending on the PDF set, especially at large $x$, where the uncertainties are significant, and remains largely unconstrained at very small $x$,
where techniques other than DGLAP may apply \cite{Ball:2017otu}; see next section.
As an example, we show in Fig.~\ref{fig3bis}(a) the unpolarized proton PDFs from the NNPDF4.0 extraction~\cite{NNPDF:2021njg}.
One remarkable observation is that $f_1^{\bar u}(x)\neq f_1^{\bar d}(x)$, contrary to what one could naively think, imagining the nucleon as a ``perturbative bound state'', i.e., a system of 3 valence quarks bound by the exchange of perturbative gluons. In such a picture, sea quarks would emerge only from gluon fluctuations into $q\bar q$ pairs. The negligible mass difference between the light $u$- and $d$-flavors would then imply, e.g., for the Gottfried ``sum rule'' $\int_0^1\frac{dx}{x}\,(\SFF{2p}-\SFF{2n})(x,Q^2) = \frac13+\frac23\int_0^1 dx \, (f_1^{\bar u}-f_1^{\bar d})(x,Q)$ the value $\frac13$. This naive expectation is not a fundamental sum rule and is clearly
violated as observed starting in the 1980s; see Ref.~\cite{Kumano:1997cy} 
for a review  and the recent study \cite{Cocuzza:2021cbi}.
This observation is naturally explained in chiral models and illustrates the
significance of chiral symmetry breaking for the understanding of nucleon
structure; see Sec.~\ref{Sec6-models}.
At a scale of $4\,\rm GeV^2$, only $f_1^u(x,\mu)$ and only for $x\gtrsim 0.2$ is larger than $f_1^g(x,\mu)$, which strongly dominates the nucleon structure for $x<0.1$; see Sec.~\ref{Sec-small-large-x}.
Contributions to the PDFs from heavy quarks, i.e., quarks other than the light quarks $q = u, \, \bar u, \, d, \, \bar d,\, s,\, \bar  s$, 
can be separated  into a perturbative and
a non-perturbative component. 
The DGLAP evolution perturbatively generates heavy-quark PDFs when the evolution scale crosses the corresponding quark mass thresholds. 
On the other hand, heavy flavors, especially charm, may also have an intrinsic component in the proton’s wave function, associated with   non-perturbative dynamics. 
The notion of intrinsic charm has been a long-standing subject of investigation~\cite{Brodsky:1980pb}, and phenomenological indications for intrinsic charm in the proton~\cite{Ball:2022qks} are still controversial~\cite{Guzzi:2022rca}. 

\begin{figure}[t]
\centering
\includegraphics[width=0.7\textwidth]{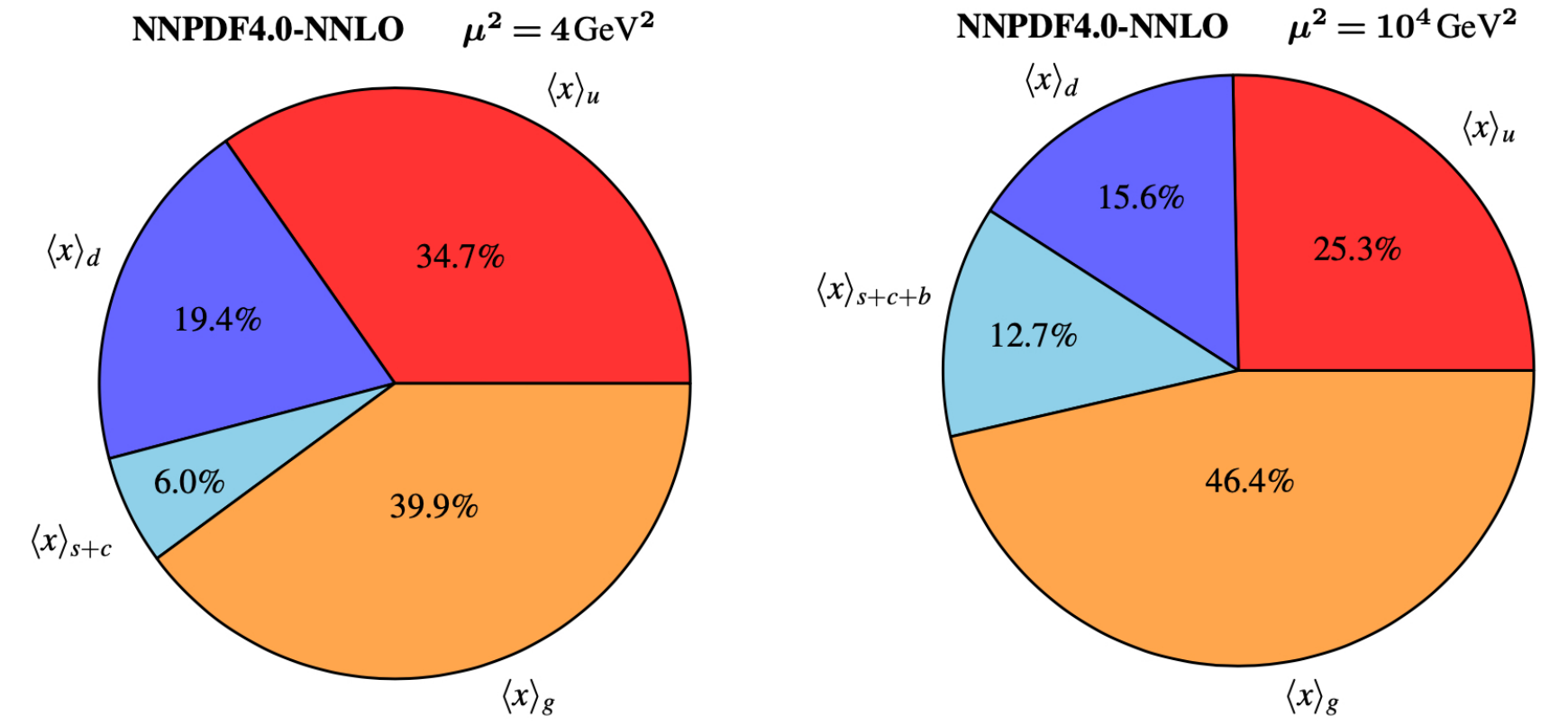} 
\caption{\footnotesize Fraction of the proton momentum carried by the partons for $\mu^2 = 4 \, \textrm{GeV}^2$ (left panel) and $\mu^2 = 10^4 \, \textrm{GeV}^2$ (right panel) for the NNPDF4.0 NNLO set of PDFs~\cite{NNPDF:2021njg}.
The uncertainties are below $1\%$ for all the numbers shown in these pie charts.
Recall that the $\langle x \rangle_q$ include the quark and antiquark contributions.
Here we also quote some values for the momentum fractions $\langle x \rangle_{\bar{q}} (\mu) = \int_0^1 dx \, x \, f_1^{\bar{q}}(x, \mu)$ carried by the light antiquarks: 
$\langle x \rangle_{\bar{u}} = 3.1\%$ ($3.6\%$) and
$\langle x \rangle_{\bar{d}}= 4.2\%$ ($4.3\%$) at $\mu^2=4 \, \textrm{GeV}^2$ ($10^4 \, \textrm{GeV}^2$).
\label{fig_momentum_fractions}}
\end{figure}

Modern extractions of the unpolarized PDFs provide rather precise values for the parton momentum fractions defined in Eq.~\eqref{e:momentum_fractions}.
In Fig.~\ref{fig_momentum_fractions}, we show results for the $\langle x \rangle_a (\mu)$ based on the NNPDF4.0 set of PDFs for two different renormalization scales.
As the scale increases, the momentum fractions for the up and down quarks decrease while those of the heavier quarks and gluons increase.
Asymptotically, the exact result 
$\lim_{\mu \to \infty} \langle x \rangle_q (\mu) / \langle x \rangle_g (\mu) = 3/16$ holds for each quark flavor~\cite{Gross:1973zrg}, 
which means that $\langle x \rangle_q (\infty) \approx 8.8\%$ and $\langle x \rangle_g (\infty) \approx 47.1\%$.
At $\mu^2 = 10^4 \, \textrm{GeV}^2$, the quark momentum fractions still differ significantly from the asymptotic value, but the one for the gluon is already quite close.
The approach to the asymptotic limit is controlled by $\ln \mu$, and therefore quite slow.

Extractions of the helicity PDFs of the proton have also reached a mature level; see, for instance, Refs.~\cite{Bertone:2024taw, Borsa:2024mss, Cruz-Martinez:2025ahf,Cocuzza:2025qvf}.
As an example, in Fig.~\ref{fig3bis}(b) we show the results from the BDSSV24 fit~\cite{Borsa:2024mss}, one of the two available NNLO analyses~\cite{Bertone:2024taw, Borsa:2024mss}.
The valence distributions are considered fairly well constrained and show good agreement across different sets.
The sea quark helicity PDFs are smaller than the valence distributions and less well constrained, with differences among the available extractions primarily arising from the specific measurements included in each PDF set~\cite{Ethier:2020way}. 
Nevertheless, one very clearly emerging feature is the flavor asymmetry 
of the light sea quark helicity distributions; see Fig.~\ref{fig3bis}(b) and the dedicated 
study \cite{Cocuzza:2022jye}.
This finding was predicted in a chiral model providing further illustration of the impact of chiral symmetry 
on the nucleon structure; see Sec.~\ref{Sec6-models}. %XXX
The extraction of the gluon helicity distribution relies on data covering a limited $x$-range, specifically $0.02\lesssim x \lesssim 0.4$, where the distribution is consistently found to be sizable in all available extractions, suggesting that approximately half of the nucleon spin comes from the gluon polarization~\cite{Borsa:2024mss,Cruz-Martinez:2025ahf,Cocuzza:2025qvf}. 
However, extrapolations beyond the currently measured $x$-range are affected by large uncertainties and impact studies indicate that this picture could be significantly refined with the extended kinematic reach achievable by the  EIC~\cite{Aschenauer:2015ata,Ball:2013tyh}.

\begin{figure}[b]
\centering
\includegraphics[width=0.55\textwidth]{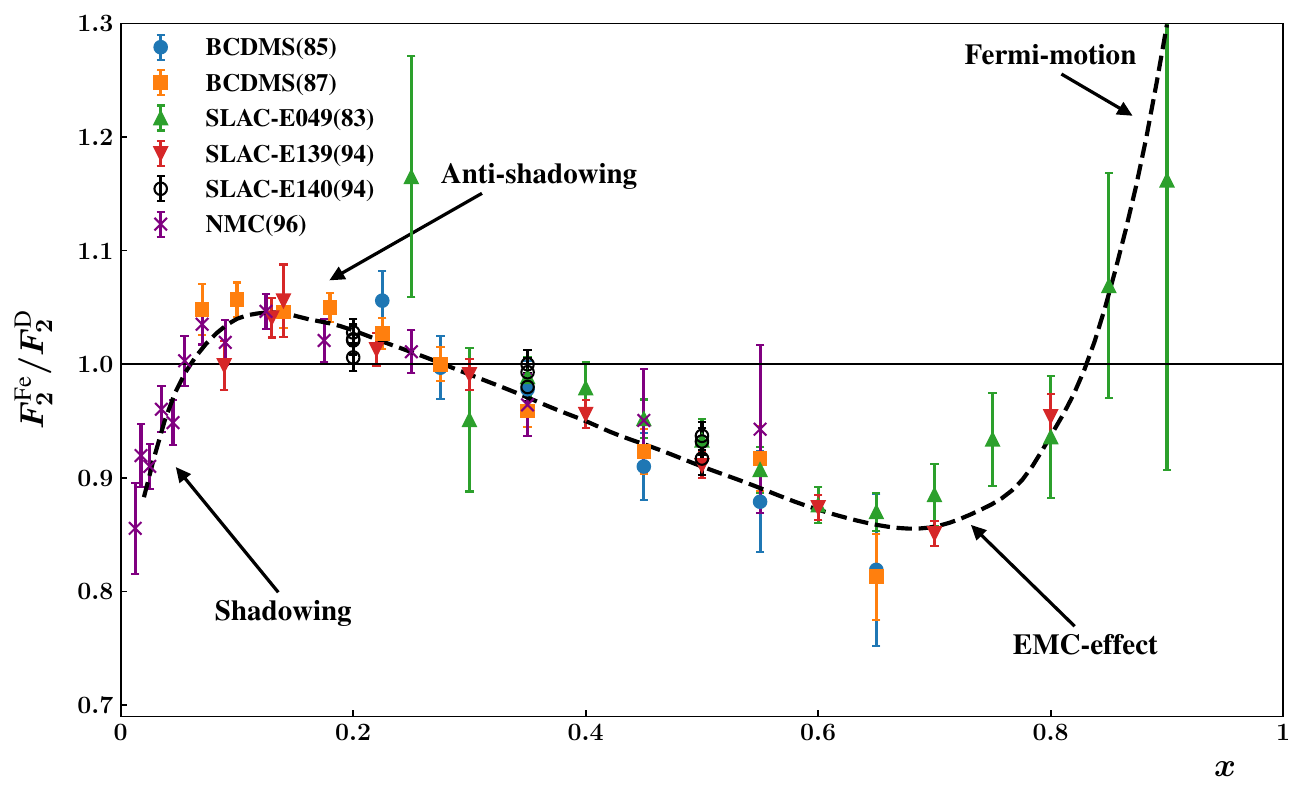}
\caption{\footnotesize \label{fig4} Parameterization for the structure function ratio $F_2^{\rm Fe}/F_2^{\rm D}$ in comparison with the experimental results from the BCDMS collaboration (BCDMS-85~\cite{BCDMS:1985dor}, BCDMS-87~\cite{BCDMS:1987upi}), from experiments at SLAC (E049~\cite{Bodek:1983qn}, E139~\cite{Gomez:1993ri}, and E140~\cite{Dasu:1993vk}), and from the NMC collaboration (NMC~\cite{NewMuon:1995cua,NewMuon:1996yuf}).}
\end{figure}

The analysis of PDFs in nucleons bound within nuclei, referred to as nuclear PDFs, has reached a level of sophistication  comparable to that of nucleon PDFs.
Historically, the majority of data on nuclear PDFs has come from DIS with charged leptons or neutrinos, and from fixed-target Drell-Yan. 
In the past decade,  collider data from the LHC and RHIC  have further advanced our understanding of unpolarized nuclear PDFs.
The extraction of nuclear PDFs is based on factorization theorems~\cite{Barshay:1975zz}, with global analyses commonly employing NLO perturbative calculations.
Although NNLO accuracy has been explored in some studies, such efforts have typically been restricted to a narrower dataset; see~\cite{Klasen:2023uqj,Ruiz:2023ozv,Ethier:2020way,Schienbein:2007fs} for reviews of the datasets and the methodologies employed by various groups in the nuclear PDF extractions. 
Nuclear modifications of PDFs, referring to the differences in parton distributions between bound and free nucleons, are classified according to the $x$-region. 
These effects are usually illustrated by the nuclear modification 
factor, defined as the ratio of per-nucleon structure functions of a nucleus with mass number $A$ and deuteron, $R= F^A_2/F^D_2$.
Figure~\ref{fig4} shows the characteristic pattern of the nuclear corrections for the iron (Fe) nucleus.
At small $x$ (i.e., $x\lesssim 0.05$), we observe a depletion in the nuclear structure function $F_2^{{\rm Fe}}(x)$,  commonly referred to as shadowing.
This is followed by a mild enhancement in the region $0.05\le x \le 0.3$, known as antishadowing. 
In the intermediate range $0.3\le x\le 0.7$, again a depletion appears, known as the EMC effect (named after its discovery by the European Muon Collaboration~\cite{EuropeanMuon:1983wih}). 
Finally, for $x\ge 0.7$, the ratio increases again due to Fermi motion, reflecting the high-momentum tail of nucleons bound within the nucleus.
Despite steady progress in global fitting efforts, the extraction of nuclear PDFs remains more complex than that of nucleon PDFs, involving more free parameters and using smaller data sets, resulting in larger uncertainties and significant differences among the available global analyses. 
This situation is expected to improve with upcoming measurements at future facilities.

The quark-gluon structure of light mesons is poorly known, mainly because pertinent experimental data are sparse.
Global analyses of pion  PDFs have relied on data for pion-induced Drell-Yan,
%~\cite{NA10:1985ibr,Conway:1989fs}, 
$J/\psi$  or direct photon production,
%~\cite{WA70:1987bai}, 
and leading-neutron electroproduction.
%~\cite{H1:2010hym,ZEUS:2002gig}.
The latest extractions are performed at NLO accuracy. Additionally,  following the pioneering analysis of Ref.~\cite{Aicher:2010cb},  threshold-resummation corrections at high $x$ have attracted significant interest, particularly for their impact on the valence-quark distributions~\cite{Barry:2021osv,Roberts:2021nhw}.
While the valence PDFs of the pion are relatively well constrained, the available dataset lacks the sensitivity needed to unambiguously determine the sea-quark and gluon distributions; see, for example, Refs.~\cite{Kotz:2023pbu,Pasquini:2023aaf} for a comparison of the results from the most recent extractions.
The situation is even more challenging for kaon PDFs, as the only available information comes from a forty-year-old Drell-Yan measurement of the $K^-/\pi^-$ structure function ratio~\cite{Saclay-CERN-CollegedeFrance-EcolePoly-Orsay:1980fhh},  consisting of just eight data points with limited statistical precision.  
Recent analyses~\cite{Bourrely:2023yzi,Chang:2024rbs} have shown the potential of kaon induced $J/\psi$-production data to better constrain the valence and gluon kaon PDFs.  
A deeper understanding of the parton structure of light mesons is anticipated with data from upcoming experiments, including COMPASS++/AMBER~\cite{Adams:2018pwt} and the SuperBigbite program at JLab~\cite{Arrington:2021alx}, as well as future measurements at the EIcC and EIC~\cite{Arrington:2021biu}.

\subsection{\boldmath Small-$x$ and large-$x$ limits} 
\label{Sec-small-large-x}

The small-$x$ and large-$x$ limits 
are of interest because the experimentally accessible $x$-range in DIS 
is limited, from below by the available accelerator energies and from 
above by the requirement to take measurements above the resonance region.

As $x$ decreases, the nucleon structure is more and more gluon dominated, 
see Sec.~\ref{Sec-2.5-global-fits}, and eventually enters a different regime. 
Recalling that $x_B\simeq Q^2/(y\,s)$ for $s\gg M_{\rm had}^2$, 
we can distinguish two relevant limits for fixed $y$:
(i) the Bjorken limit, where both $Q^2\to\infty$ and $s\to\infty$ 
 at fixed $x$, leading to DGLAP evolution;
(ii) the Regge-Gribov limit, where $Q^2$ is kept large enough for 
pQCD to apply while $s\rightarrow \infty$,  which implies 
$x\to0$. In this regime, the dominant expansion parameter in the 
evolution of PDFs becomes  $\alpha_{\rm s}\ln\frac1x$, rather than 
$\alpha_{\rm s}$.
As $x$ decreases, one reaches the point where $\alpha_{\rm s}\ln\frac1x$ is 
no longer small, making the neglect of higher-order terms  no longer justified.
 The DGLAP evolution equations (truncated at fixed order in $\alpha_{\rm s}$) break down and a resummation of the leading 
logs $\alpha_{\rm s}^n\,\ln^n\frac1x$ becomes necessary. This is 
accomplished by the Balitsky-Fadin-Kuraev-Lipatov (BFKL) evolution 
equation \cite{Kuraev:1977fs,Balitsky:1978ic} which replaces DGLAP 
at small $x$. 
The BFKL framework predicts that the gluon PDF behaves in
LO as $xf_1^g(x) \sim x^{-\lambda}$ for $x \lll 1$ 
with $\lambda = 4N_c\frac{\alpha_{\rm s} }{\pi}\ln 2$.
NLO results for the BFKL equation were first presented in 
Refs.~\cite{Fadin:1998py, Ciafaloni:1998gs}. 
The more complicated small-$x$ asymptotics of helicity and transversity PDFs were studied in \cite{Kovchegov:2015pbl,Kovchegov:2017lsr,Kovchegov:2018zeq,Cougoulic:2022gbk} and incorporated as theory constraints in recent extractions of $g_1^a(x)$ and $h_1^a(x)$ from experimental data~\cite{Adamiak:2023yhz, Adamiak:2025dpw, Cocuzza:2023oam, Cocuzza:2023vqs}.

The strong rise of $xf_1^g(x)$ at small $x$ is due to the growing phase space 
available for emitting more and more soft (i.e., low-$x$) gluons 
as $s\to\infty$ which, if it continued indefinitely, would lead to 
a violation of the Froissart unitarity bound~\cite{Froissart:1961ux}. 
A proposed solution to this problem is that the system should enter a 
new regime where the growth of the gluon density should slow down and 
eventually saturate at some scale $Q^2_s$~\cite{Gribov:1983ivg} 
due to non-perturbative finite density effects.
In this regime, the nucleon becomes a dense many-body system of gluons 
where non-linear effects, like gluon recombination, start to play a 
significant role. 
The physics of the dense gluon medium can be described by the color 
glass condensate effective theory (see~\cite{McLerran:1993ni,McLerran:1993ka}
and the reviews~\cite{Gelis:2010nm,Morreale:2021pnn}), where a separation is 
made between fast gluons with $k^+>\mu$ and slow gluons with $k^+<\mu$. 
The former are long-lived configurations which can therefore be 
treated as static color sources, hence the term ``glass'' in analogy 
to disordered systems in condensed matter physics. 
The latter are short-lived quantum fluctuations. 
The evolution of PDFs 
w.r.t.\ the arbitrary separation scale $\mu$ is governed by the JIMWLK 
(Jalilian-Marian, Iancu, McLerran, Weigert, Leonidov, Kovner) 
renormalization group equation 
\cite{Jalilian-Marian:1997qno,Jalilian-Marian:1997jhx,Weigert:2000gi,Iancu:2000hn,Iancu:2001ad}.
The Balitsky-Kovchegov equation \cite{Balitsky:1987bk,Kovchegov:1999yj} 
is a numerically more easily tractable mean-field approximation to JIMWLK 
in the limit of a large number of colors $N_c$. The saturation scale 
depends on both the value of $x$ and the target as 
$Q^2_s(x,A)\sim A^{1/3}x^{-\lambda_s}$, where $A$ denotes the mass number 
of the nucleus, with $\lambda_s\sim 0.2$--0.3 suggested by phenomenological 
studies~\cite{Kowalski:2007rw,Kowalski:2006hc,Kowalski:2003hm}.
Saturation effects are theoretically well motivated and supported by 
phenomenological models, and direct experimental evidence is still 
under active investigation. 

In the opposite limit $x\to 1$, it is intuitively clear that 
the probability for a parton to carry nearly the entire nucleon 
momentum vanishes, though the manner in which it vanishes 
depends on the parton type. This can be understood in terms of 
light-front wave functions of hadrons which describe the 
probability amplitudes for the hadron to be found in a specific 
$n$-parton Fock state configuration with free invariant mass ${\cal M}_n$ given by 
${\cal M}_n^2 = \sum_{i=1}^n(\vec{k}_{\perp i}^2+m_i^2)/x_i$
where $\sum_{i=1}^n\vec{k}_{\perp i}=\uvec 0_\perp$ and $\sum_{i=1}^nx_i=1$.
For the nucleon, the lowest values $n=3,4,5,\cdots$ correspond 
schematically to the Fock states $|qqq\ra$, $|qqqg\ra$, $|qqqq\bar{q}\ra$, etc.
When some parton $a$ carries nearly the entire nucleon momentum, 
i.e.,\ $x_a\to 1$, the momentum fractions of the other partons
($x_i$ with $i\neq a$) become small, making the free invariant mass 
${\cal M}_n$ 
large and the configuration far off-shell.
Assuming that the dominant hadronic light-front wavefunction 
corresponds to the lowest-mass (minimal ${\cal M}_n$) Fock state, 
the far off-shellness of the configuration can be obtained only 
through the exchange of hard gluons.
Under these assumptions, one may use pQCD 
techniques to derive quark counting rules which 
predict $f_1^a(x) \propto (1-x)^{2n_a-1}$, where $n_a$ is 
the minimal number of spectators, i.e., those partons that 
are not involved in the hard scattering subprocess of
parton $a$.
For the proton $n_u=n_d=2$, $n_g=3$, $n_a=4$ for 
$a=\bar{u}, \, \bar{d}, \,s,\,\bar{s}$ \cite{Matveev:1973ra,Brodsky:1973kr},  
see \cite{Brodsky:1994kg} for a pedagogical exposition.
These predictions are compatible with modern PDF parameterizations
within the uncertainties of global fits \cite{Ball:2016spl}.

Studies of PDFs in the large-$x$ region require to remain in the DIS regime. 
However, when approaching the limit $x_B\to1$ in experiments, one inevitably
leaves the DIS continuum, enters the resonance region, and ultimately approaches
the elastic limit. Thus, the large-$x_B$ limit naturally connects to the physics of quark-hadron duality, where one observes that structure functions measured in the resonance region, when averaged appropriately over 
resonances, correspond to those measured in the DIS region~\cite{Bloom:1970xb}. 
Quark–hadron duality is a rich topic in its own right;
see, for example, Refs.~\cite{Accardi:2009br, Accardi:2016qay} for further discussions.
At large $x_B$, an additional complication arises due to the need of resumming large threshold logarithms in order to have a robust framework for extracting PDFs from Drell-Yan~\cite{Sterman:1986aj,Catani:1989ne,Aicher:2011ai} and DIS~\cite{Corcella:2005us}.

\section{Transverse momentum dependent parton distributions}
\label{Sec-3:TMD}

The notion of transverse parton momenta $k_\perp=|\uvec k_\perp|$ was present already in the 
earliest parton model applications to DIS \cite{Feynman:1972original}, as
we saw in Eq.~(\ref{Eq:Callan-Gross-in-parton-model}). After early parton 
model investigations about $k_\perp$ effects in SIDIS
and Drell-Yan~\cite{Ravndal:1973kt,Cahn:1978se, Ralston:1979ys},
a systematic description of TMDs in QCD began with the works \cite{Collins:1981uw,Collins:1984kg} and was explored to predict
and describe new effects  \cite{Sivers:1989cc, Collins:1992kk, Anselmino:1994tv, Tangerman:1994eh,
Kotzinian:1994dv, Mulders:1995dh, Boer:1997nt}. 
In this section, we explain how $k_\perp$ effects 
offer new insights into nucleon structure. We introduce TMDs, their
interpretation and properties in QCD. We also review how TMDs can be 
accessed experimentally and what we know about them phenomenologically. 
A comprehensive review of TMD physics can be found in \cite{Boussarie:2023izj}.

\subsection{Transverse parton momenta and new effects}
\label{Sec-3.1:TMD-introductory-part}

Transverse parton momenta introduce new effects such as 
azimuthal and single-spin asymmetries (SSAs); see, for instance, 
Ref.~\cite{DAlesio:2007bjf} for a review. One of the earliest 
examples, known as the Cahn effect~\cite{Cahn:1978se}, 
is an azimuthal $\cos \phi_h$ (see Fig.~\ref{Fig:SIDIS-kin-and-more}(a)) 
modulation of the distribution of hadrons produced in unpolarized SIDIS 
with transverse momenta $|\vec{P}_{h\perp}|\ll Q$ that arises already in 
the parton model due to the probability (density) $f_1^a(x,k_\perp)\neq0$ 
to encounter partons with transverse momenta $k_\perp$.  
Although the Cahn effect is subleading, i.e.,~suppressed like 
$M_{\rm had}/Q$, it did play an important role in the 
development of the TMD field~\cite{Anselmino:2005nn}. 

Transverse momentum effects were also studied early in the Drell-Yan 
process~\cite{Ralston:1979ys,Collins:1984kg}. If the annihilating 
quark and antiquark were collinear, then the momentum $\vec{q}$ of 
the virtual photon (reconstructed from the observed lepton pair) 
would point, in LO, along the collision axis. 
However, non-zero $q_\perp=|\uvec q_\perp|$ are clearly measured, 
and the cross section behaves like 
$d\sigma_{\rm DY}/dq_\perp^2 \propto \exp(-q_\perp^2/\la q_\perp^2\ra)$
up to $q_\perp^2 \lesssim 3$--$4\,\rm GeV^2$ with 
$\la q_\perp^2\ra \sim 1$--$2\,\rm GeV^2$, depending on the 
kinematics of the experiment~\cite{DAlesio:2004eso, Schweitzer:2010tt}. 
For $q_\perp \ll Q$, where $Q$ is the invariant mass of the 
lepton pair in Drell-Yan, the process is described by TMDs.

Another phenomenon that can be explained with the help of 
transverse parton momenta are transverse SSAs. Here we consider 
the generic process $A(p_A)\,N(p_N, S) \to B(p_B)\, X$ with 
unpolarized particles $A$ and $B$ and a polarized nucleon. 
Since parity is conserved in strong interactions, the only
allowed correlation involving $S$ (and the four momenta of the particles) is 
$\varepsilon_{\alpha \beta \gamma \delta} S^\alpha p_A^\beta p_N^\gamma p_B^\delta$. 
Evaluating this correlation in the center-of-mass frame with particles $A$ and $N$ 
having no transverse momenta leads to an expression proportional to
$\uvec{S}_\perp \cdot( \uvec{p}_N \times \uvec{p}_{B\perp})$. 
First, we observe that for this process single-spin effects are always
associated with transverse polarization, unless contributions from the weak
interaction are taken into account.
Second, the vectors $\uvec{S}_\perp$, $\uvec{p}_N$, and $\uvec{p}_{B\perp}$ 
flip sign under time-reversal. Therefore, SSAs are often said to be
``odd under naive time-reversal transformations'' (naive \textsf{T}-odd), 
where the term ``naive'' is used to remind that a true time-reversal 
transformation would also require to interchange in- and out-states 
in the S-matrix, which is not done here. Third, transversely polarized
states along, say, the $y$-axis can be expressed as linear combinations
of helicity states $|N^\pm\rangle$ according to
$|N^{\uparrow,\downarrow}\rangle = \frac{1}{\sqrt{2}}(|N^+\rangle \pm i|N^-\rangle)$. 
If ${\cal M}^i$ denote the amplitudes for scattering off protons in transverse polarization 
($i=\{\uparrow,\downarrow\}$) or helicity (or $i=\{+,-\}$) states, then
\be\label{Eq:transverse-SSA-and-phases}
    {\rm SSA} = 
       \frac{d\sigma^\uparrow-d\sigma^\downarrow}
            {d\sigma^\uparrow+d\sigma^\downarrow}
    =  \frac{|{\cal M}^{\uparrow}|^2-|{\cal M}^{\downarrow}|^2}
            {|{\cal M}^{\uparrow}|^2+|{\cal M}^{\downarrow}|^2}
    =  \frac{2\,{\rm Im}[{\cal M}^+({\cal M}^-)^\ast]}
            {|{\cal M}^+|^2+|{\cal M}^-|^2} \,,
\ee
which reveals that transverse SSAs require non-trivial phases.
Early on, it was found that, in collinear factorization using twist-2 PDFs only, 
such effects are very small because the imaginary part requires 
a loop correction, and the result is proportional to the (current) quark 
mass~\cite{Kane:1978nd}. This was in stark contrast to the large experimental 
results for transverse target SSAs first observed in the 1970s in 
$pp^\uparrow\to\pi X$ \cite{Dick:1975ty}. 
A promising solution to this puzzle later emerged 
within the TMD approach \cite{Sivers:1989cc}.

Our last example concerns the Lam-Tung relation~\cite{Lam:1980uc}. 
The angular distribution of lepton pairs in unpolarized Drell-Yan is given by
\be\label{Eq:unpol-DY-and-Lam-Tung}
    \frac{1}{\sigma}\,\frac{d\sigma}{d\Omega} = \frac{3}{4\pi}\,
    \frac1{\lambda+3}\left(1
    +\lambda\,\cos^2\theta
    +\mu\,\sin^2\theta\,\cos\phi
    +\nu\,\frac12\,\sin^2\theta\,\cos2\phi\right)
\ee
in the Collins-Soper frame, a particular dilepton rest frame 
\cite{Collins:1977iv}.
The coefficients $\lambda$, $\mu$, $\nu$ depend on the kinematical 
variables of the process and, in collinear factorization, satisfy 
the Lam-Tung relation ${\lambda+2\nu = 1}$, which is exact through 
${\cal O}(\alpha_{\rm s})$. As first observed in pion-nucleus Drell-Yan 
experiments~\cite{NA10:1986fgk}, this relation is strongly violated 
due to a sizeable coefficient $\nu$.
It was then shown that the TMD factorization approach can serve as 
a very promising explanation for this observation~\cite{Boer:1999mm}, 
although using collinear factorization to ${\cal O}(\alpha_{\rm s}^2)$ 
also provides a good description of the data~\cite{Peng:2015spa,Lambertsen:2016wgj}.

\begin{figure}
\centering
\includegraphics[width=0.6\textwidth]{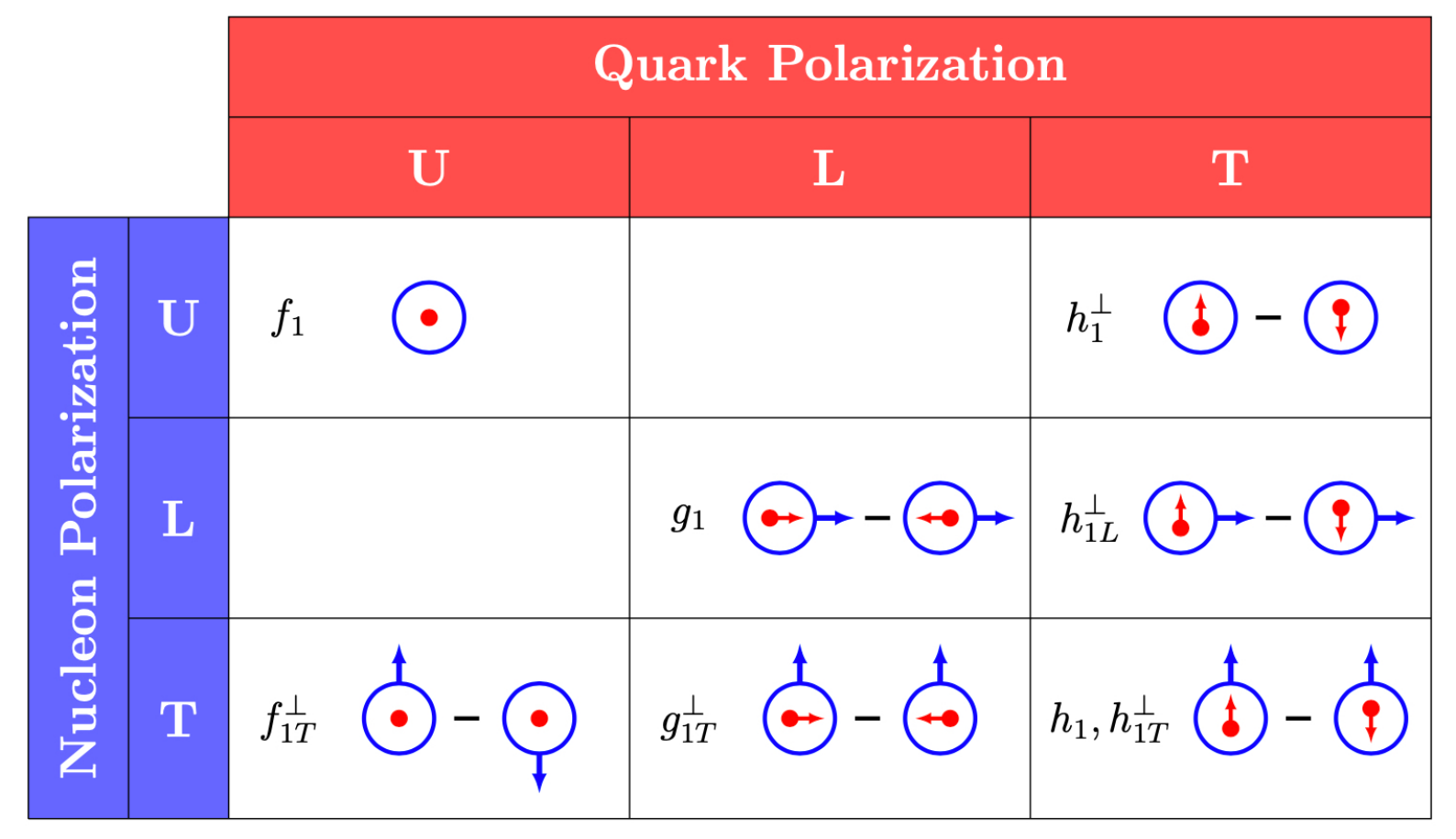}
\caption{\footnotesize Leading-power quark TMDs for a spin-$\frac{1}{2}$ hadron. 
The blue circle and red dot represent the hadron and quark, respectively, while 
the blue and red arrows represent the directions of their polarizations.
\label{Fig-09-TMD-TABLE}}
\end{figure}

\subsection{Definition, properties, and partonic interpretation}
\label{Sec-3.2:TMD-interpretation-etc}

TMDs offer more insights into the partonic picture and a richer 
phenomenology than PDFs, because transverse parton momenta lead to
several additional correlations. As in the PDF case, the gamma matrices
$\gamma^+$, $\gamma^+\gamma_5$, $i\sigma^{j+}\gamma_5$ are associated with unpolarized,
longitudinally and transversely polarized quarks and antiquarks, respectively. 
However, due to the presence of $\vec{k}_\perp$, eight leading-power TMDs exist,
compared to the three leading-twist PDFs.
(Leading-power TMDs appear in observables at leading order in $M_{\rm had}/Q$.  
Although they are sometimes referred to as twist-2, in the case of TMDs there is no one-to-one correspondence between the order at which a
particular TMD enters observables and the notion of twist; see Sec.~\ref{sect:ope}.
More details on this point can also be found, for example, in Ref.~\cite{Boussarie:2023izj}.)
More precisely, the TMD correlator in Eq.~\eqref{TMD-correlator}, for a specific choice of the gauge link path (discussed further in Sec.~\ref{Sec-3.4-TMD-unversality}), is parameterized in terms of leading quark TMDs as
\begin{subequations}
\label{e:TMD_example}
\begin{align}
    \Phi_{\rm TMD}^{q[\gamma^+]}(P,x,\uvec k_\perp)    
    & = f_1^q(x,k_\perp) 
    - \frac{\varepsilon_\perp^{jl}k_\perp^{\,j} S_\perp^l}{M}\,f_{1T}^{\perp q}(x,k_\perp) \,, 
    \label{e:TMD_example-f}\\
    \Phi_{\rm TMD}^{q[\gamma^+\gamma_5]}(P,x,\uvec k_\perp)    
    & = S_L\, g_1^q(x,k_\perp) 
    + \frac{\vec{k}_\perp\cdot\vec{S}_\perp}{M}\,g_{1T}^{\perp q}(x,k_\perp) \,,
    \label{e:TMD_example-g}\\
    \Phi_{\rm TMD}^{q[i\sigma^{j+}\gamma_5]}(P,x,\uvec k_\perp)    
    & = S_\perp^j\, h_1^q(x,k_\perp) 
    + \frac{S_L k_\perp^j}{M}\,h_{1L}^{\perp q}(x,k_\perp)
    + \frac{(k_\perp^j k_\perp^l-\frac12\delta_\perp^{jl}\vec{k}_\perp^{\,2})\,S_\perp^l}{M^2}\,h_{1T}^{\perp q}(x,k_\perp)
    + \frac{\varepsilon_\perp^{jl}k_\perp^l}{M}\,h_1^{\perp q}(x,k_\perp) \,,
    \label{e:TMD_example-h}
\end{align}
\end{subequations}
where $\delta_\perp^{jl}$ and $\varepsilon_\perp^{jl}$ denote the Kronecker 
symbol and the totally antisymmetric tensor in the transverse plane with 
$\varepsilon_\perp^{12}=+1$, respectively; see~\cite{Bacchetta:2006tn} 
and references therein for more details. The expressions in 
Eqs.~(\ref{e:TMD_example-f})--(\ref{e:TMD_example-h}) define 
TMDs of antiquarks according to 
\be
    {\rm TMD}^{\bar q}(x,k_\perp) = \pm {\rm TMD}^q(-x,k_\perp)
    \qquad \mbox{with} \qquad
    \begin{cases}   
    \, + & \mbox{for\quad $g_1^{\bar q}$, $f_{1T}^{\perp\bar q}$, $h_{1L}^{\perp\bar q}$, $h_1^{\perp\bar q}$} \\ 
    \, - & \mbox{for\quad $f_1^{\bar q}$, $\,h_1^{\bar q}$, $\,g_{1T}^{\perp\bar q}$, $\,h_{1T}^{\perp\bar q}$.}
    \end{cases}
\ee
We also note that a total of eight leading-power 
gluon TMDs exist~\cite{Mulders:2000sh, Meissner:2007rx}.

The partonic interpretation of the leading quark TMDs is as follows.
The distribution of unpolarized quarks is described by $f_1^q$ if the 
nucleon is unpolarized, and by $f_{1T}^{\perp q}$ if it is transversely 
polarized (Fig.~\ref{Fig-09-TMD-TABLE}, first column).
Similarly, longitudinally polarized quarks are described by $g_1^q$ if
the nucleon is longitudinally polarized, and by $g_{1T}^{\perp q}$ if it 
is transversely polarized (Fig.~\ref{Fig-09-TMD-TABLE}, second column).
The distributions of transversely polarized partons are described by 
$h_1^{\perp q}$ if the nucleon is unpolarized, by $h_{1L}^{\perp q}$ if 
it is longitudinally polarized, and by both $h_1^q$ and $h_{1T}^{\perp q}$ 
if it is transversely polarized (Fig.~\ref{Fig-09-TMD-TABLE}, third column).
The latter case is described by two TMDs because one can have a monopole 
(proportional to $\delta_\perp^{jl}$ and associated with $h_1^q$) and quadrupole 
(proportional to $k_\perp^j k_\perp^l-\frac12\delta_\perp^{jl}\vec{k}_\perp^{\,2}$ 
and associated with $h_{1T}^{\perp q}$) structure in the transverse plane.
The matrix elements on the l.h.s.~of Eqs.~(\ref{e:TMD_example-f})--(\ref{e:TMD_example-h})
actually represent the distributions of partons inside the nucleon with a fixed polarization.
Therefore, like in the case of PDFs, also TMDs correspond to averages or differences of distributions.
The empty fields in Fig.~\ref{Fig-09-TMD-TABLE} would correspond to
distributions of unpolarized (longitudinally polarized) quarks 
inside a longitudinally polarized (unpolarized) nucleon 
which are forbidden due to parity. 

Furthermore, the Sivers function $f_{1T}^{\perp q}$~\cite{Sivers:1989cc} and 
Boer-Mulders function $h_1^{\perp q}$~\cite{Boer:1997nt} are naive \textsf{T}-odd.
The structure $\varepsilon_\perp^{jl}k_\perp^{\,j} S_\perp^l$ in front of
$f_{1T}^{\perp q}$ in Eq.~\eqref{e:TMD_example-f} is directly proportional to 
the correlation $\uvec{S}_\perp \cdot( \uvec{p}_N \times \uvec{k}_{\perp})$. 
This means that the Sivers function generates a transverse SSA at the level
of the parton correlator $\Phi_{\rm TMD}^{q[\gamma^+]}$. 
Keeping transverse parton momenta in the description of the process 
$A \,N(S_\perp) \to B \, X$ (see Sec.~\ref{Sec-3.1:TMD-introductory-part}) 
can lead to a transverse SSA given by the correlation 
$\uvec{S}_\perp \cdot( \uvec{p}_N \times \uvec{p}_{B\perp})$~\cite{Sivers:1989cc}.
We point out that the TMD approach for such observables must be considered a model, 
while collinear factorization involving twist-3 quark-gluon-quark correlations 
provides a rigorous framework in QCD; 
see~\cite{Efremov:1984ip, Qiu:1991pp, Kouvaris:2006zy, Koike:2009ge, Kang:2010zzb, Metz:2012ct} 
and references therein.
Nevertheless, successful phenomenology using the TMD approach exists as 
reported, for example, in Refs.~\cite{Anselmino:1994tv, Anselmino:2005sh}.
Even more importantly, the proposal of Ref.~\cite{Sivers:1989cc} 
to address the physics of such SSAs by means of transverse parton 
momenta had a tremendous impact on the development of the TMD field.

The inequalities for PDFs in Eq.~\eqref{Eq:PDF-inequalities} hold
with the obvious generalization for the corresponding TMDs. 
The TMDs satisfy furthermore the inequalities~\cite{Bacchetta:1999kz}
\ba\label{Eq:TMD-inequalities}
    \bigl|h_{1T}^{\perp(1)a}\bigr| \le \frac12\biggl(f_1^a-g_1^a\biggr) \,, \hspace{4mm}
    \bigl[f_{1T}^{\perp(1)a}\bigr]^2+\bigl[g_{1T}^{\perp(1)a}\bigr]^2 \le 
    \frac{\vec{k}^{\,2}_\perp}{4M^2}\biggl(\bigl[f_1^a\bigr]^2-\bigl[g_1^a\bigr]^2\biggr) \,, \hspace{4mm}
    \bigl[h_{1L}^{\perp(1)a}\bigr]^2+\bigl[h_{1 }^{\perp(1)a}\bigr]^2 \le 
    \frac{\vec{k}^{\,2}_\perp}{4M^2}\biggl(\bigl[f_1^a\bigr]^2-\bigl[g_1^a\bigr]^2\biggr) \,,  \hspace{4mm}
    a = q, \bar{q}
\ea
with transverse TMD moments defined as
$h_{1T}^{\perp(1)q}(x,k_\perp) = \frac{\uvec k_\perp^2}{2M^2}\,h_{1T}^{\perp q}(x,k_\perp)$ 
and arguments $x,\,k_\perp$ omitted for brevity. As in the PDF case, the  inequalities in 
Eq.~(\ref{Eq:TMD-inequalities}) cannot be rigorously proven for renormalized TMDs in QCD.

\subsection{Observables for TMDs}
\label{Sec-3.3:SIDIS-and-DY}

\begin{figure}[t!]
\centering
\begin{tabular}{cc}
\includegraphics[height=4cm]{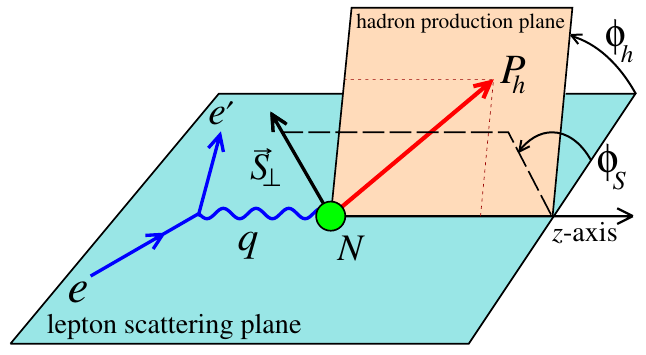}  &  
\hspace{12mm} \includegraphics[height=4cm]{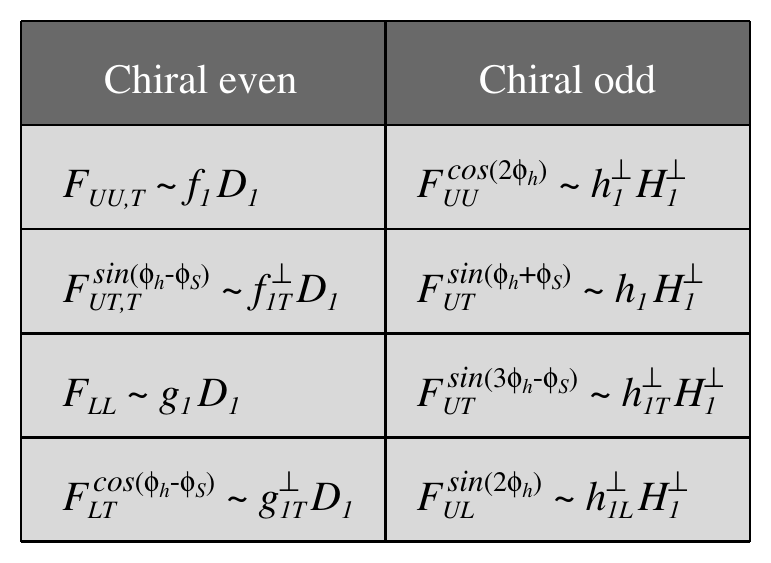} \\
{\bf (a)} \hspace{12mm} & \hspace{12mm} {\bf (b)}
\end{tabular}
\caption{\label{Fig:SIDIS-kin-and-more} \footnotesize 
(a) Definitions of azimuthal angles in semi-inclusive deep-inelastic 
electron-nucleon scattering (SIDIS), $eN\to e^\prime hX$. 
The transverse nucleon polarization component (shown) is defined with respect to the momentum of
$\gamma^\ast$. The longitudinal nucleon polarization component (not shown) 
is by convention along the $z$-axis. 
(b) Schematic expressions of the leading SIDIS structure functions.}
\end{figure}

Sensitivity to TMDs requires the measurements of some transverse 
momenta in the final state, e.g.\ $\vec{P}_{h\perp}$ of a hadron $h$ 
produced in SIDIS; see Fig.~\ref{Fig:SIDIS-kin-and-more}(a). The SIDIS 
cross section is, in addition to DIS variables, also differential in 
$z_h=(P_h\cdot q)/(P\cdot q)$ and $d^2P_{h\perp}=P_{h\perp}d\phi_h dP_{h\perp}$. 
Focusing on the production of spin-zero (or unpolarized) hadrons, 
the SIDIS cross section is described in terms of 18 structure functions 
denoted by $F_{XY,Z}^{w(\phi)}(x_B,z_h, P_{h\perp},Q^2)$, where $X=U,\,L$
denotes the electron polarization and $Y=U,\,L,\,T$ denotes the nucleon
polarization, while $Z=T,\,L$ (when needed) denotes the
polarization of the virtual photon, analogously to the DIS
structure functions in Eq.~(\ref{Eq:spin0-spin12-predictions}).
The superscripts $w(\phi_h)$ indicate the type of azimuthal distribution 
and may, in the case of transverse nucleon polarization, depend on the 
azimuthal angle $\phi_S$ of the nucleon polarization vector; see 
Fig.~\ref{Fig:SIDIS-kin-and-more}(a). The absence of a 
$w(\phi_h)$-superscript signals an isotropic distribution 
of produced hadrons. 
Out of the 18 SIDIS structure functions, 8 are leading and
another 8 are subleading, while 2 are subsubleading in an 
expansion in $M_{\rm had}/Q$ \cite{Bacchetta:2006tn}.

\begin{figure}
\centering
\includegraphics[height=0.25\textwidth]{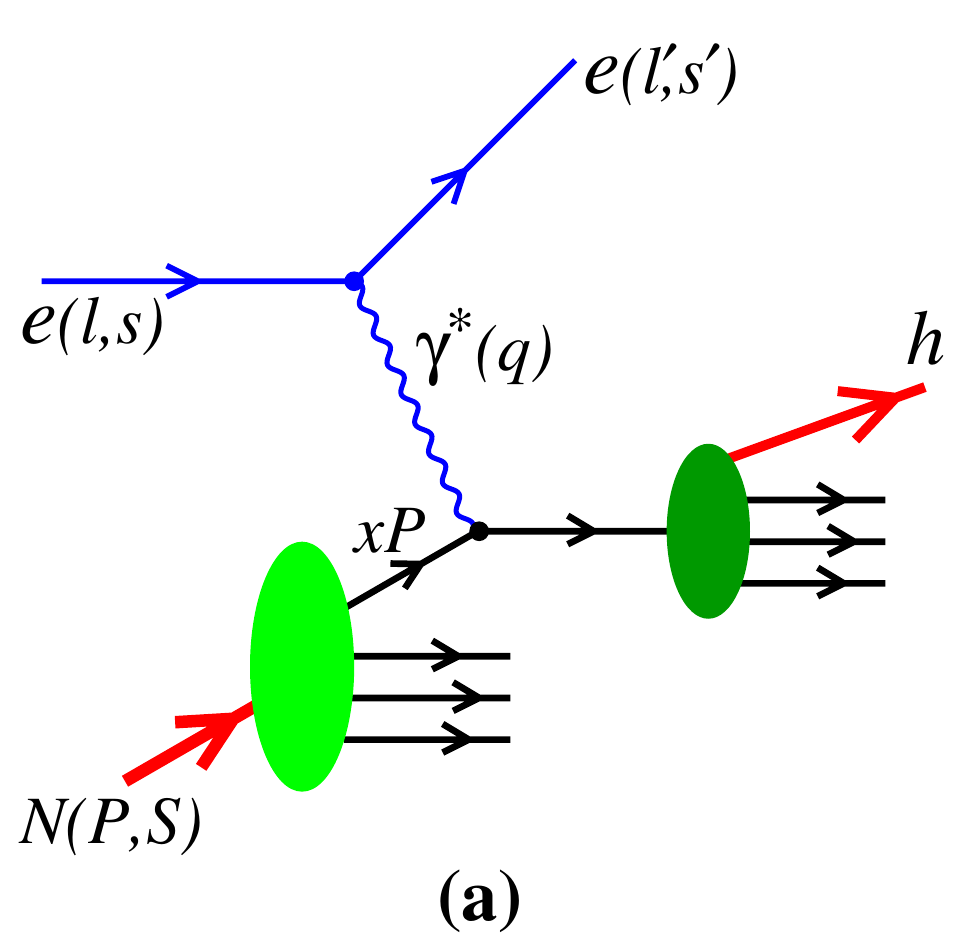} \hspace{5mm}
\includegraphics[height=0.25\textwidth]{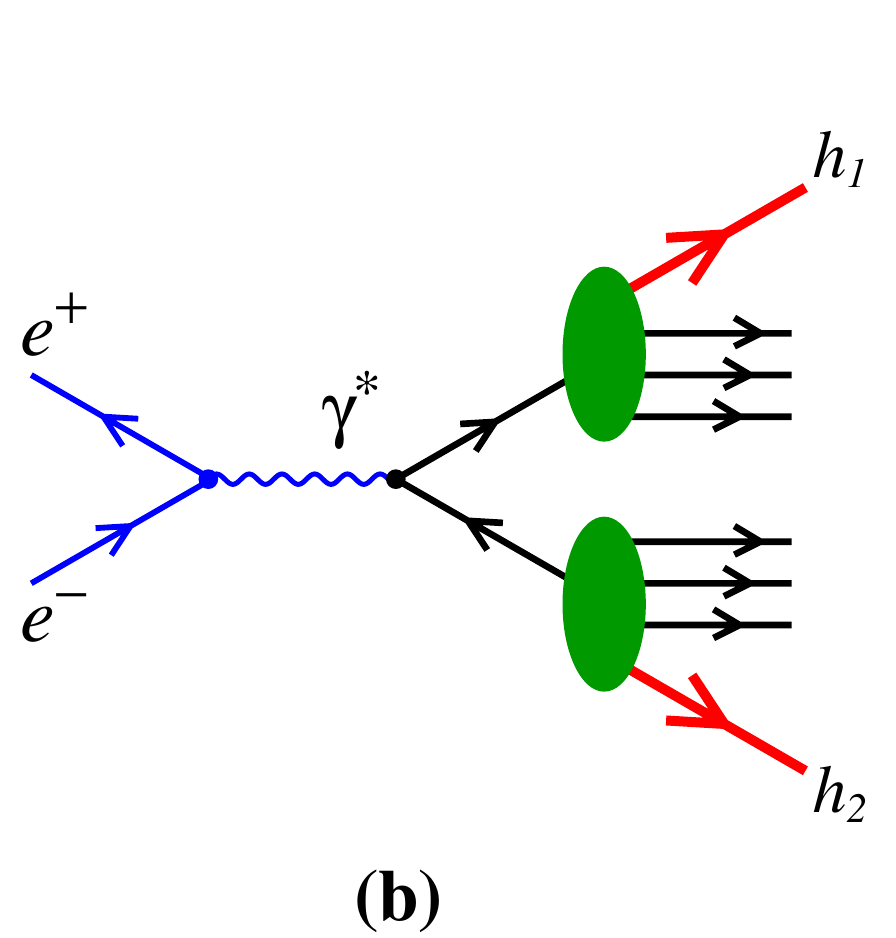}
\caption{\footnotesize \label{Fig-11:TMD-processes}
Lowest-order diagrams for the amplitudes of 
     (a) semi-inclusive deep inelastic production 
     of hadrons in electron-nucleon scattering, and
     (b) production of hadrons in electron-positron 
     annihilation. The dark shaded non-perturbative 
     blobs represent the fragmentation process.}
\end{figure}

In the Bjorken limit (which in SIDIS includes $P\cdot P_h\to\infty$ 
with $z_h$ fixed), each  leading structure function is described in 
terms of a different TMD combined with either the unpolarized 
fragmentation function $D_1^q(z,K_\perp)$ or the Collins function 
$H_1^{\perp q}(z,K_\perp)$~\cite{Collins:1992kk},
as sketched in Fig.~\ref{Fig:SIDIS-kin-and-more}(b).
The LO SIDIS amplitude is shown in Fig.~\ref{Fig-11:TMD-processes}(a).
$D_1^q(z,K_\perp)$ is chiral-even and describes the probability density 
that a quark with momentum $k^\mu$ undergoes fragmentation $q\to h\,X$ 
into a hadron with momentum $P^\mu_h = z k^\mu + K_\perp^\mu$, up to 
terms negligible in the Bjorken limit. $H_1^{\perp q}(z,K_\perp)$ is 
a chiral-odd and naive \textsf{T}-odd fragmentation function describing 
an asymmetry in the fragmentation into unpolarized hadrons sensitive to 
the transverse polarization of partons. In this sense, the Collins effect 
provides an analyzer for transverse parton polarization \cite{Collins:1992kk} 
that is related to the TMDs $h_1^q$, $h_{1L}^{\perp q}$, $h_{1T}^{\perp q}$, 
$h_1^{\perp q}$, as featured in Fig.~\ref{Fig-09-TMD-TABLE}.
The fragmentation functions can be studied independently of TMDs in 
hadron production from $e^+e^-$ annihilation. 
At large center-of-mass energies and away from resonances, the LO
annihilation process is $e^+e^-\to\gamma^\ast\to \bar{q}q$, where the quark 
and antiquark subsequently hadronize into jets of hadrons;
see Fig.~\ref{Fig-11:TMD-processes}(b).
By measuring the cross sections for the production of specific hadrons 
(in one jet, or two or more hadrons from different jets) one obtains information 
about different types of fragmentation functions. For a comprehensive review 
of fragmentation functions see~\cite{Metz:2016swz}.

The transverse momenta $\uvec{k}_\perp$ and $\uvec K_\perp$ appearing in 
the TMDs and fragmentation functions in SIDIS cannot be measured directly, 
but instead appear in convolution integrals. For example, 
the LO (parton-model) expression for the $F_{UU,T}$ structure function reads
\be\label{Eq:conv-integral-SIDIS-FUU}
    F_{UU,T}(x_B,z_h,P_{h\perp},Q^2) \stackrel{\rm LO}{=} x_B \int d^2k_\perp \int d^2K_\perp \,
    \delta^{(2)}(z_h\vec{k}_\perp+\vec{K}_\perp-\vec{P}_{h\perp})\,
    \sum_{a=q,\bar{q}} e_a^2 \, f_1^a(x_B,k_\perp)\,D_1^a(z_h,K_\perp)\,.
\ee
Note that, as for the DIS structure functions, 
the LO result for $F_{UU,T}$ does not explicitly depend on $Q^2$.
Apart from the delta function which ensures the conservation of transverse momentum, the hard electron-(anti)quark scattering only provides the factor $e_a^2$.
The $F_{LL}$ structure function is given by the same expression as 
in Eq.~\eqref{Eq:conv-integral-SIDIS-FUU} with $f_1^a$ replaced by 
$g_1^a$, while the convolution integrals in the other cases in
Fig.~\ref{Fig:SIDIS-kin-and-more}(b) include specific weighting 
functions that project out the azimuthal correlation of interest; 
see~\cite{Mulders:1995dh, Bacchetta:2006tn} for full details. 
We also point out that there exist some connections between SIDIS and DIS structure functions~\cite{Bacchetta:2006tn}, such as $\sum_h\int dz_h \int d^2P_{h\perp} F_{UU,T}(x_B,z_h,P_{h\perp},Q^2)=\SFF{T}(x_B,Q^2)$.  

In the TMD case, the Drell-Yan amplitude in the parton model looks exactly as in
Fig.~\ref{Fig-03:DIS+DY}(c), except that the transverse momenta $\vec{q}_\perp$ 
of the produced lepton pairs with respect to the collision axis of the two 
hadrons, $h_1$ and $h_2$, are measured. 
For the scattering of polarized nucleons off spin-0 or unpolarized hadrons, 
there are 12 structure functions, 6 of which are at leading order in 
$M_{\rm had}/Q$, and many more in the case of Drell-Yan with two 
polarized nucleons; see, for example, Refs.~\cite{Tangerman:1994eh, Arnold:2008kf}. 
The Drell-Yan structure functions correspond to convolutions analogous to 
Eq.~(\ref{Eq:conv-integral-SIDIS-FUU}), where this time the TMD of a quark 
is convoluted with the TMD of an antiquark; see Fig.~\ref{Fig-03:DIS+DY}(d). 
For example, the 6 leading structure functions in nucleon-pion Drell-Yan 
are given by the convolutions 
$f^q_1 f_1^{\bar q}$, $f^{\perp q}_{1T} f_1^{\bar q}$,
$h^q_1 h_1^{\perp \bar q}$, $h^{\perp q}_{1L} h_1^{\perp \bar q}$,
$h^{\perp q}_{1T} h_1^{\perp \bar q}$, $h^{\perp q}_1 h_1^{\perp\bar q}$ 
(plus the ones with $q$ and $\bar q$ interchanged),
where the first TMD refers to the nucleon and the second to the pion. 
These structure functions are uniquely distinguished by their angular
dependences. Noteworthy, the Drell-Yan structure function 
$F_{UU}^{\cos(2\phi)}\sim h^{\perp q}_1 h_1^{\perp\bar q}$
(plus the one with $q$ and $\bar q$ interchanged)
gives rise to the coefficient $\nu$ in Eq.~\eqref{Eq:unpol-DY-and-Lam-Tung}; 
see the discussion in Sec.~\ref{Sec-3.1:TMD-introductory-part}.

At hadron colliders, TMDs can also be studied in $W^\pm$ and $Z^0$ production.
Further processes include lepton-jet correlations in DIS~\cite{Liu:2018trl}.
Observables have also been proposed, e.g., 
in Higgs production or quarkonium production in proton-proton 
collisions~\cite{Boer:2011kf,Echevarria:2015uaa,Echevarria:2019ynx},
which can give access to TMDs describing unpolarized or linearly 
polarized gluons, with the latter playing a particularly important 
role at small $x$~\cite{Metz:2011wb, Dominguez:2011br}.
For a comprehensive overview about further  
TMD processes we refer to~\cite{Boussarie:2023izj}.

\subsection{Universality of TMDs}
\label{Sec-3.4-TMD-unversality}

\begin{figure}[b]
\centering
\includegraphics[height=3.4cm]{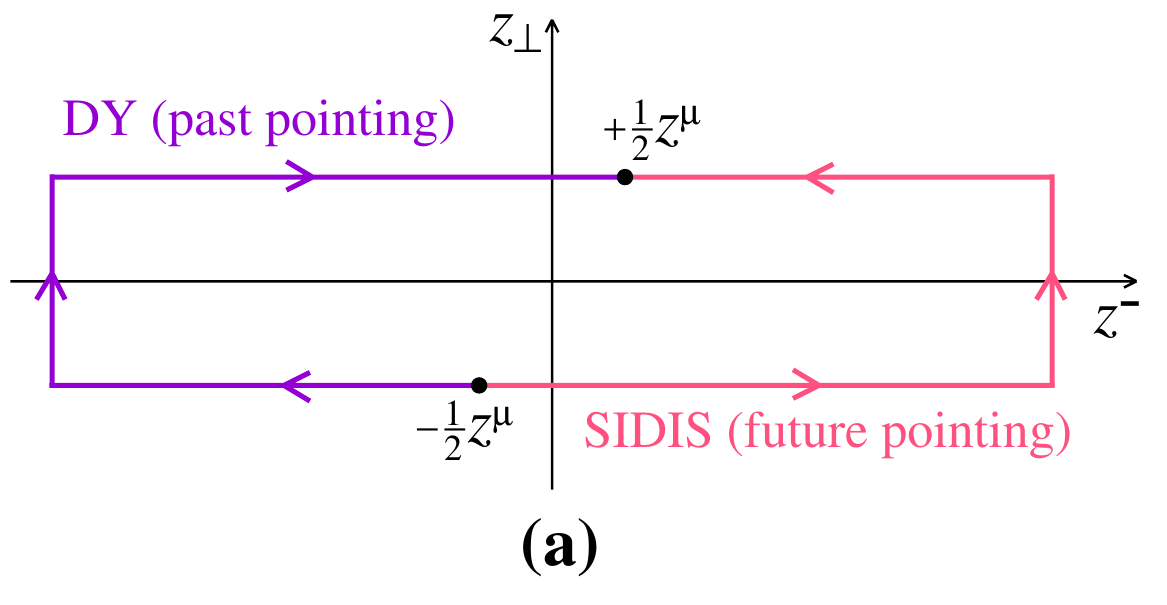} \hspace{5mm}
\includegraphics[height=3.4cm]{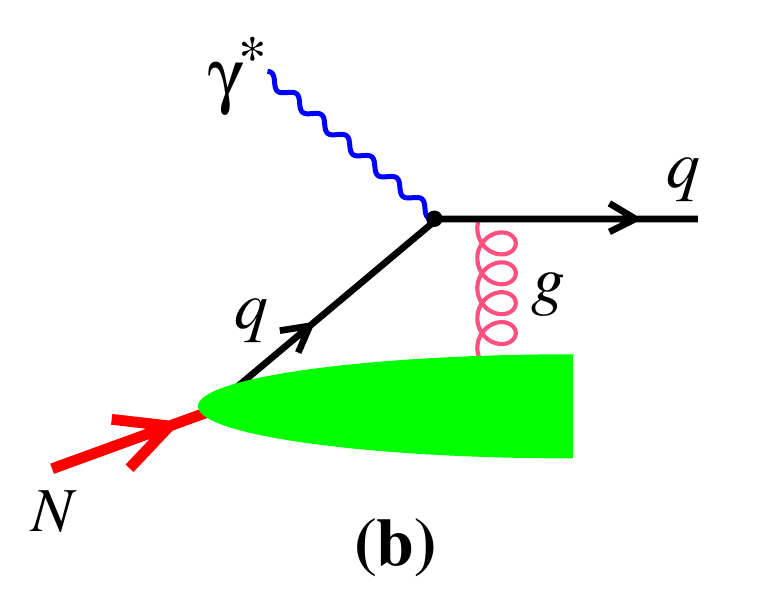}           \hspace{5mm}
\includegraphics[height=3.4cm]{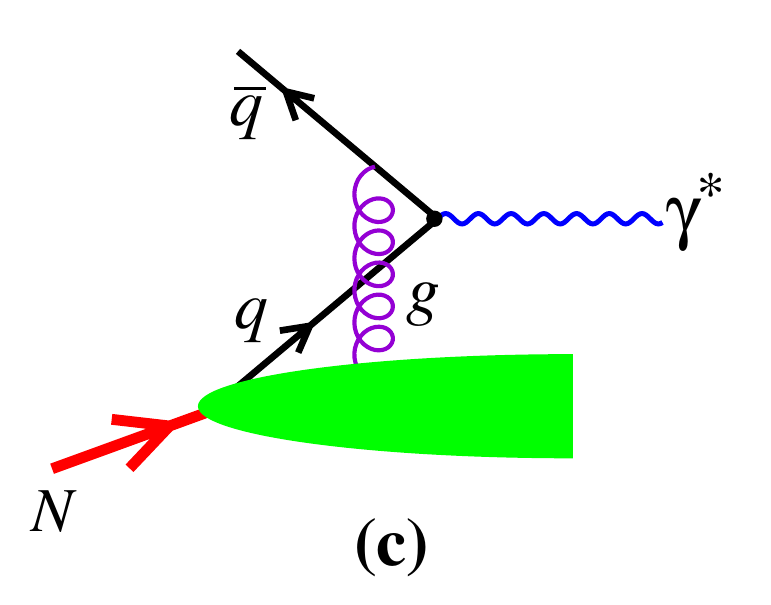}
\caption{\footnotesize
(a) Paths of Wilson lines in the non-local operators $\bar{\psi}_q(-\frac{z}2)\Gamma{\cal W}\psi_q(+\frac{z}2)|_{z^+=0}$ in semi-inclusive deep-inelastic scattering (SIDIS) and Drell-Yan (DY).  
The future-pointing Wilson line in SIDIS goes to the positive light-front infinity, $+\infty^-$, while the past-pointing Wilson line for Drell-Yan goes to the negative light-front infinity, $-\infty^-$. 
(b) One-gluon exchange between the active parton and target remnant in SIDIS
(final-state interactions).
(c) One-gluon exchange between the active parton of one hadron (not indicated)
and the spectator system of another hadron in Drell-Yan (initial-state interactions).
The gluon lines in panels (b) and (c) are colored to indicate their correspondence with 
the paths in panel (a).
\label{Fig:sign_change}}
\end{figure}

Defining TMDs in QCD requires extra care.
Some of the TMD-specific features in that regard will be discussed in the following section, while here we just concentrate on the gauge link ${\cal W}_{\rm TMD}$.
Like for the PDFs discussed above, this gauge link is generated through the exchange of gluons between the active partons in a process and the remnants of the target.
However, in contrast to the PDF case, ${\cal W}_{\rm TMD}$ does not connect the two quark fields in Eq.~\eqref{TMD-correlator} via a straight line, but is given by a structure shown in Fig.~\ref{Fig:sign_change}(a) that follows from the QCD factorization of the SIDIS and Drell-Yan processes~\cite{Collins:2002kn}.
Since the two quark fields in the TMD correlator have a nonzero transverse separation, the final-state interaction of the active quark in SIDIS, Fig.~\ref{Fig:sign_change}(b), and the initial-state interaction in Drell-Yan, Fig.~\ref{Fig:sign_change}(c), give rise to different gauge links ${\cal W}_{\rm TMD}^{\rm SIDIS}$ and ${\cal W}_{\rm TMD}^{\rm DY}$, respectively.
(The one-gluon exchange between the active quark and the target remnants is shown in Figs.~\ref{Fig:sign_change}(b) and \ref{Fig:sign_change}(c).)
To be specific, the Wilson line for the SIDIS process is given by
\begin{align}
{\cal W}_{\rm TMD}^{\rm SIDIS} = 
\Big[ 0^+, -\tfrac{1}{2} z^-, -\tfrac{1}{2}\uvec{z}_\perp; 0^+, \infty^-, -\tfrac{1}{2}\uvec{z}_\perp \Big] \times
\Big[ 0^+, \infty^-, -\tfrac{1}{2}\uvec{z}_\perp; 0^+, \infty^-, \tfrac{1}{2}\uvec{z}_\perp \Big] \times
\Big[ 0^+, \infty^-, \tfrac{1}{2}\uvec{z}_\perp; 0^+, \tfrac{1}{2} z^-, \tfrac{1}{2}\uvec{z}_\perp \Big] \,.
\label{e:Wline_SIDIS_def}
\end{align}
This is called a future-pointing (staple-like) gauge link, while for Drell-Yan a past-pointing gauge link occurs.  
Integrating the TMD correlator in Eq.~\eqref{TMD-correlator} upon $\uvec{k}_\perp$ implies $\uvec{z}_\perp = \uvec 0_\perp$, that is, the two quark fields of the bilocal operator are separated along the light cone only, as is the case for PDFs. 
One readily sees from Fig.~\ref{Fig:sign_change}(a) that the Wilson lines for SIDIS and Drell-Yan become identical in this case.

Because the Wilson lines are different, the TMDs in SIDIS and Drell-Yan, a priori, appear to be different as well.
However, by applying the time-reversal and parity transformations, they can be related to each other~\cite{Collins:2002kn}.
It turns out that the 6 naive \textsf{T}-even quark TMDs are the same for the two processes, while the naive \textsf{T}-odd Sivers and Boer-Mulders functions exhibit a sign change, i.e., 
\begin{align}
f_{1T}^{\perp q} \big|_{\rm DY} = - f_{1T}^{\perp q} \big|_{\rm SIDIS} \,,
\qquad
h_1^{\perp q} \big|_{\rm DY} = - h_1^{\perp q} \big|_{\rm SIDIS} \,.
\label{e:sign_change}
\end{align}
This remarkable breakdown of universality, put forward in 2002~\cite{Collins:2002kn}, is a prediction based on TMD factorization.
It also reflects the maximal impact of the final-state/initial-state interactions of the active quark.
For more than a decade, there was no direct experimental check of the sign change in Eq.~\eqref{e:sign_change}. However, there is now strong evidence supporting it; see Sec.~\ref{Sec-3.6-TMD-extractions}.

In the case of hadronic collisions, the gauge links % of TMDs
can be even more complicated than the staple-shaped Wilson lines discussed above~\cite{Bomhof:2006dp}.
An example is dihadron production with the two hadrons being nearly back-to-back, where both initial-state and final-state interactions are present. 
The resulting ``non-standard'' Wilson lines prevent one from relating TMDs in different processes~\cite{Collins:2007nk}.
Interestingly, it has been shown that for such a process TMD factorization actually fails due to color entanglement caused by overlaps of initial-state and final-state interactions~\cite{Rogers:2010dm}.

It is also important to explore whether transverse momentum dependent fragmentation functions are universal.
Factorization leads, a priori, to past-pointing (future-pointing) Wilson lines for fragmentation functions in SIDIS (electron-positron annihilation).
Unlike for TMDs, fragmentation functions defined with past-pointing and future-pointing Wilson lines cannot be related by time-reversal and parity transformations.
However, the specific kinematics for parton fragmentation ensures the universality of fragmentation functions, as was found in a model calculation~\cite{Metz:2002iz} and shown in a more general analysis~\cite{Collins:2004nx}.
Further work in this area, part of which considers transverse-momentum moments of fragmentation functions, has also been put forward~\cite{Yuan:2008yv, Gamberg:2008yt, Meissner:2008yf}.

\subsection{TMDs in QCD}
\label{Sec-3.5-TMD-QCD}

As discussed in Sec.~\ref{Sec-3.3:SIDIS-and-DY}, SIDIS at low transverse momentum of the produced hadron, $P_{h \perp} \ll Q$, and Drell-Yan for low transverse momentum of the dilepton, $q_T \ll Q$, are key observables for studying TMDs.
While the parton model analysis of leading-power effects in these processes is relatively simple, all-order QCD factorization is more involved \cite{Collins:1981uk, Collins:1984kg, Ji:2004wu, Collins:2011zzd, Boussarie:2023izj}.
Here we largely follow the formulation presented in Ref.~\cite{Collins:2011zzd}.
We only sketch key results and refer to the literature for more details; see, in particular, Refs.~\cite{Collins:2011zzd, Boussarie:2023izj} and references therein.

As an example, we consider the leading-power SIDIS structure function $F_{UU,T}$ for which the all-order factorization formula can be written as
\begin{align}
F_{UU,T}(x_B, z_h, P_{h \perp}, Q^2) =
x_B \sum_{a,b} \, {\cal H}_{ab} \left( \frac{Q}{\mu}, \alpha_{\rm s}(\mu) \right) \, \int d^2 z_\perp \, e^{-i \uvec{q}_\perp \cdot \uvec{z}_\perp} \,
f_1^a (x_B, z_\perp, \mu, \zeta) \,
D_1^b (z_h, z_\perp, \mu, \tilde{\zeta}) + Y \,,
\label{e:TMD_factorization}
\end{align}
where $\uvec{q}_\perp = - \uvec{P}_{h\perp} / z_h$ is the transverse momentum of the virtual photon in a frame in which both % the target particle and the 
the incoming and detected hadrons have no transverse momentum. 
The longitudinal momentum fractions in the TMDs and fragmentation functions in Eq.~\eqref{e:TMD_factorization} are given by the external variables $x_B$ and $z_h$, respectively, since in the region $P_{h\perp} \ll Q$ radiation of unobserved particles from hard sub-processes is suppressed.
In Eq.~\eqref{e:TMD_factorization} the position-space representation is used for the TMDs, 
\begin{align}
f_1^a (x, z_\perp, \mu, \zeta) = \int d^2 k_\perp \, e^{i \uvec{k}_\perp \cdot \uvec{z}_\perp} \, f_1^a (x, k_\perp, \mu, \zeta) \,,
\end{align}
and similarly for the fragmentation functions, where $\uvec{z}_\perp$ is the transverse separation between the two quark fields in Eq.~\eqref{TMD-correlator}.
The reason for switching to position space is that the convolution in transverse momenta, which we already have seen in the parton model expression~\eqref{Eq:conv-integral-SIDIS-FUU}, simplifies into a product of functions.
The Wilson lines of the TMDs and fragmentation functions in Eq.~\eqref{e:TMD_factorization} follow the discussion in Sec.~\ref{Sec-3.4-TMD-unversality}.
The TMD factorization formula for the other leading SIDIS structure functions looks similar to Eq.~\eqref{e:TMD_factorization}, while additional complexities arise at subleading power~\cite{Ebert:2021jhy, Gamberg:2022lju, Rodini:2023plb}.

The factorization formula in~\eqref{e:TMD_factorization} shows three differences compared to the parton-model result: First, including higher-order terms introduces non-trivial hard factors ${\cal H}_{ab}$, as in inclusive DIS.
Second, the TMDs depend on two auxiliary scales, $\mu$ and $\zeta$, and likewise for the fragmentation functions.
(Note that $\zeta$ has the dimension mass$^2$.)
Third, the $Y$-term describes the structure function at large $P_{h \perp} \gg M_{\rm had}$.
The $Y$-term is computed in fixed-order pQCD using PDFs and (collinear) fragmentation functions, with a subtraction to remove the large-$P_{h\perp}$ asymptotic contribution  already captured by the first term in Eq.~\eqref{e:TMD_factorization}. 
This subtraction is needed to avoid double-counting.
Note also that resummation techniques involving collinear non-perturbative functions, first applied to the Drell-Yan process~\cite{Dokshitzer:1978yd, Parisi:1979se}, can be used to describe the SIDIS cross section in the region $M_{\rm had} \ll P_{h_\perp} \ll Q$.
More details on the description of SIDIS from low to high transverse momenta can be found elsewhere~\cite{Ji:2006br, Bacchetta:2008xw, Boglione:2014oea, Collins:2016hqq, Gonzalez-Hernandez:2023iso}.

We now briefly comment on the additional scale $\zeta$ of TMDs, with analogous comments applying to $\tilde{\zeta}$ in fragmentation functions.
(Note that in the formulation of Ref.~\cite{Collins:2011zzd}, $\zeta \tilde{\zeta} = Q^4$ is used, leading to hard factors that do not depend explicitly on these scales.)
In Sec.~\ref{Sec-2.3-QCD-description-of-DIS} we have seen that PDFs depend on a renormalization scale introduced to regularize UV divergences.
TMDs show in addition so-called rapidity divergences.
These divergences become explicit in low-order calculations of TMDs in pQCD~\cite{Collins:2003fm}, where they arise when the rapidity of certain partons goes to infinity.
(The rapidity of a particle with four-momentum $p$ is defined as $\frac{1}{2} \ln \frac{p^+}{p^-}$.)
In the case of PDFs, the rapidity divergences cancel between real and virtual diagrams, leading to a significant simplification.
The parameter $\zeta$ in the TMDs appears as a regulator of the rapidity divergences.
The $\zeta$-dependence of the TMDs is governed by the Collins-Soper evolution equation~\cite{Collins:1981uk, Collins:2011zzd, Aybat:2011zv},
\begin{align}
\frac{d}{d \ln \! \sqrt{\zeta}} \, f_1^a (x, z_\perp, \mu, \zeta)
= K(z_\perp, \mu) \, f_1^a (x, z_\perp, \mu, \zeta)  
\qquad \textrm{with} \qquad 
\frac{d K(z_\perp, \mu)}{d \ln \mu} = - \gamma_{K} \big( \alpha_{\rm s}(\mu) \big) \,.
\label{e:CS_evolution}
\end{align}
The r.h.s.~of the evolution equation for $f_1^a$ in Eq.~\eqref{e:CS_evolution} is a simple product since we work with functions in position rather than momentum space.
An important feature of the Collins-Soper evolution equation is that the kernel $K(z_\perp, \mu)$ receives non-perturbative contributions, thus complicating the phenomenology of TMDs.
The evolution of the TMDs in $\mu$ is given by 
\begin{align}
\frac{d}{d \ln \mu} \, f_1^a (x, z_\perp, \mu, \zeta) = \left[ \gamma_a \big( \alpha_{\rm s}(\mu) \big) - \frac{1}{2} \gamma_K \big( \alpha_{\rm s}(\mu) \big) \ln \frac{\zeta}{\mu^2} \right] \,  f_1^a (x, z_\perp, \mu, \zeta) \,.
\label{e:TMD_evol_mu}
\end{align}
For explicit expressions of the anomalous dimensions $\gamma_a$ and $\gamma_K$, we refer to the literature~\cite{Collins:2011zzd, Boussarie:2023izj}. 
We emphasize that, in contrast to the DGLAP equation~\eqref{e:DGLAP}, the evolution of TMDs in both $\zeta$ and $\mu$ are for fixed $x$.
Like for factorization formulas involving PDFs, the renormalization scale $\mu$ of the TMDs and fragmentation functions in Eq.~\eqref{e:TMD_factorization} are to be taken similar to the hard scale $Q$ to avoid large logarithms.

At small $z_\perp$, or equivalently at large $k_\perp$, TMDs can be computed in pQCD~\cite{Collins:1984kg, Collins:2011zzd, Boussarie:2023izj},
\begin{align}
f_1^a (x, z_\perp, \mu, \zeta) = \sum_b \int_x^1 \frac{d x'}{x'} \,
C_{ab} \left( \frac{x}{x'}, z_\perp, \mu, \zeta, \alpha_{\rm s}(\mu) \right) \, f_1^b(x', \mu) \,,
\label{e:TMD_small_zperp}
\end{align}
with the perturbative coefficients $C_{ab}$ and PDFs appearing on the r.h.s~of that equation.
Results for the $C_{ab}$ at ${\cal O}(\alpha_{\rm s}^2)$ have been presented in Ref.~\cite{Gehrmann:2014yya}. 
We point out that in Eq.~\eqref{e:TMD_small_zperp} only the leading term of an expansion in powers of $M_{\rm had} \, z_\perp$ is shown.
The small-$z_\perp$ expansion, which exists for all quark and gluon TMDs, is nowadays routinely incorporated in analysis frameworks for extracting TMDs from experimental data. 
Not all leading TMDs can be related to twist-2 PDFs like, e.g., the  Sivers function $f_{1T}^{\perp q}$ whose small-$z_\perp$ expansion contains the ${\mbox{twist-3}}$ Qiu-Sterman function~\cite{Kang:2011mr, Aybat:2011ge}, 
which parameterizes a collinear twist-3 quark-gluon-quark correlator; see also the last paragraph in Sec.~\ref{sect-2.4}. 
We refrain from discussing the details of a complete field-theoretic definition of TMDs in QCD that is free of divergences but rather refer to the literature~\cite{Collins:2011zzd, Boussarie:2023izj}.
However, we emphasize that a properly defined TMD does not integrate to the corresponding PDF,
\begin{align}
\int d^2 k_\perp \, f_1^a (x, k_\perp, \mu, \zeta) \neq f_1^a(x, \mu) \,,
\label{e:TMD_PDF_rel}
\end{align}
and similarly for the other TMDs.
In model calculations where the TMDs do not exhibit divergences,~\eqref{e:TMD_PDF_rel} holds with the equal sign.
But in QCD the l.h.s.~and r.h.s.~of that equation differ by terms of ${\cal O}(\alpha_{\rm s})$ and power corrections; see~\cite{Ebert:2022cku, delRio:2024vvq} and references therein for detailed studies of this point.

\subsection{Extractions of TMDs from experimental data}
\label{Sec-3.6-TMD-extractions}

\begin{figure}[b!]
\centering
\includegraphics[height=5cm]{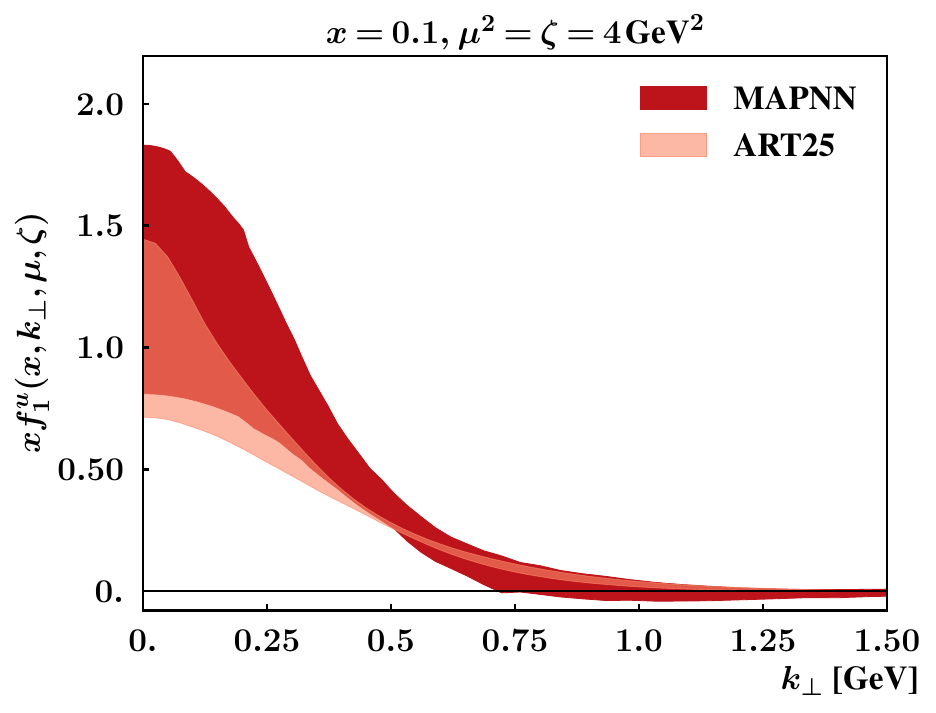} \hspace{1cm}
\includegraphics[height=5cm]{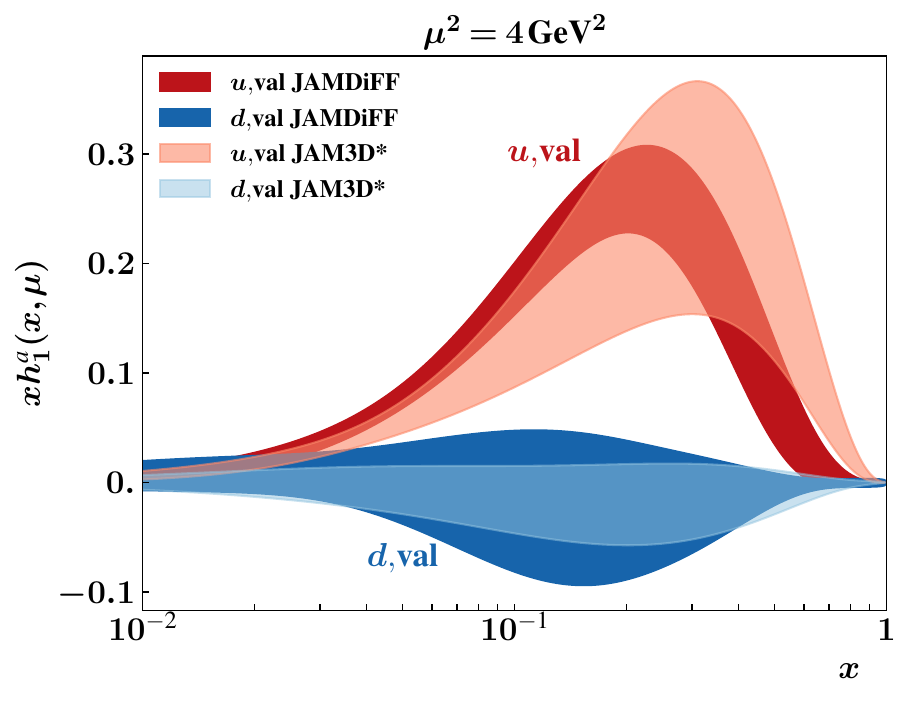}
\caption{\label{Fig-13:TMD-extractions-f1-h1}\footnotesize
(a) Results for $f_1^u(x,k_\perp,\mu,\zeta)$ as a function of $k_\perp$ at $x=0.1$ and $\mu^2=\zeta = 4 \, \textrm{GeV}^2$  from the neural network framework (MAPNN)~\cite{Bacchetta:2025ara} 
and ART25~\cite{Moos:2025sal}.
(b) Results for valence transversity distributions $xh_1^{u,\text{val}}(x, \mu)$ (red bands) 
and $xh_1^{d,\text{val}}(x, \mu)$ (blue bands) as a function of $x$ at $\mu^2=4 \, \textrm{GeV}^2$. 
The results correspond to the extraction from dihadron production (JAMDiFF, darker bands) ~\cite{Cocuzza:2023oam, Cocuzza:2023vqs} and single-hadron production (JAM3D$^*$, lighter bands)~\cite{Cammarota:2020qcw, Gamberg:2022kdb, Cocuzza:2023oam}.}
\end{figure}

Dedicated efforts to determine $f_1^q(x,k_\perp)$ began in the 1980s, 
initially based on Drell-Yan data from unpolarized proton-proton or 
proton-nucleus collisions measured in fixed-target experiments 
at CERN and Fermilab with $\sqrt{s}$ up to $60\,\rm GeV$. 
In the 1990s, data on $W^\pm$ and $Z^0$ production from 
collider experiments at Tevatron with $\sqrt{s}=1.8\,\rm TeV$ 
became available \cite{D0:1999jba,CDF:1999bpw,D0:2010dbl,CDF:2012brb}. 
More recently Drell-Yan and weak-boson data became available 
from BNL at $\sqrt{s}=200\,\rm GeV$ and $510 \, \textrm{GeV}$
\cite{PHENIX:2018dwt,STAR:2023jwh}, as well as the LHC with 
$\sqrt{s}$ up to $13\,\rm TeV$ 
\cite{LHCb:2015mad,ATLAS:2019zci,CMS:2022ubq}.
This amounts to a volume of ${\cal O}(500-600)$ data points as used 
in the analyses of the Ref.~\cite{Bacchetta:2025ara,Moos:2025sal} 
that we shall discuss as examples for recent extractions below.
These experiments include data for $Q^2$ from $16\,{\rm GeV}^2$
in fixed-target Drell-Yan experiments all the way up to 
$10^6\,\rm GeV^2$ in collider experiments at the LHC and 
$0 < q_\perp \le 40\,{\rm GeV}$. 
A comparable data volume with ${\cal O}(600)$ data points
was used in the analysis of~\cite{Moos:2025sal} 
from the SIDIS 
experiments at HERMES and COMPASS for $\pi^\pm$ and $K^\pm$ production 
from proton and deuteron targets covering overall a range of $Q^2$ from 
$1\,\rm GeV^2$ to $81\,\rm GeV^2$ and $0 < P_{h\perp} < 0.68\,{\rm GeV}$
\cite{HERMES:2012uyd,COMPASS:2017mvk}.
In Fig.~\ref{Fig-13:TMD-extractions-f1-h1}(a) we show as 
an example the extractions of $f_1^u(x,k_\perp)$ as a function 
of $k_\perp$ for $x=0.1$ at ${\mu^2 = \zeta =  4\,{\rm GeV}^2}$
from \cite{Bacchetta:2025ara,Moos:2025sal}.
The analysis of \cite{Bacchetta:2025ara} used neural-network 
techniques and was carried out at next-to-next-to-next-to-leading 
logarithmic (N$^3$LL) accuracy fitting Drell-Yan data only.
The analysis of \cite{Moos:2025sal} was carried out at N$^4$LL 
accuracy using a flexible fit Ansatz  
(for the TMDs Fourier transformed to $z_\perp$-space) 
and fitting in addition to Drell-Yan data also the SIDIS data. 
The neural network fit of Ref.~\cite{Bacchetta:2025ara} 
(labeled MAPNN in Fig.~\ref{Fig-13:TMD-extractions-f1-h1}(a))
has larger and more conservative uncertainty bands. In contrast,
the fit of \cite{Moos:2025sal} has narrower uncertainty bands, 
in part due to fitting a larger data set and in part due to
working with a fixed fit Ansatz. Despite differences in 
technical details, the extractions of $f_1^u(x,k_\perp)$ 
from \cite{Bacchetta:2025ara,Moos:2025sal} agree well.

We now proceed to the chiral-odd transversity PDFs.
Chiral-odd functions cannot be measured as a single function due to the helicity conservation of interactions between fermions and gauge bosons in the Standard Model. 
Rather, they enter observables always in combination with another chiral-odd function. It is therefore impossible to measure their absolute signs, although their relative signs can be determined. 
The sign of $h_1^u(x)$ is informed by theoretical predictions from models and lattice QCD.
Transversity can be extracted as TMD in TMD-factorization or as PDF in collinear factorization. 
Both approaches are sometimes combined. 
Data from Drell-Yan with both protons transversely polarized, the first process proposed to measure $h_1^a(x)$ \cite{Ralston:1979ys, Jaffe:1991ra},
is not available yet. 
Phenomenological analyses exist for both single-hadron and dihadron production.
Single-hadron processes include, in TMD factorization, an azimuthal transverse SSA in SIDIS~\cite{Collins:1992kk}; see Fig.~\ref{Fig:SIDIS-kin-and-more}(b).
The needed information on $H_1^{\perp q}(z,K_\perp)$ is obtained from $e^+e^-$ data for a specific angular modulation in the production of back-to-back hadrons~\cite{Boer:1997mf}. 
Another single-hadron observable is the aforementioned transverse SSA for $p^\uparrow p\to \pi X$ described in collinear twist-3 factorization, where $h_1^q(x)$ couples to (collinear) twist-3 fragmentation functions~\cite{Metz:2012ct, Kang:2010zzb}.
For dihadron production, the transversity can couple to a naive T-odd (collinear) dihadron fragmentation function in both lepton-nucleon~\cite{Collins:1993kq, Bianconi:1999cd} and proton-proton scattering~\cite{Collins:1993kq, Bacchetta:2004it}.
This dihadron fragmentation function is constrained by measuring pertinent correlations of dihadron pairs in opposite jets in $e^+e^-$ annihilation \cite{Artru:1995zu, Boer:2003ya, Matevosyan:2018icf}.
In dihadron production, the relative momentum of the two measured hadrons plays the role of an analyzer for the transverse quark spin encoded in $h_1^q(x)$, making this method amenable to collinear factorization.
On the other hand, the pertinent dihadron fragmentation functions also depend on the invariant mass of the hadrons~\cite{Collins:1993kq, Bianconi:1999cd, Pitonyak:2023gjx, Rogers:2010dm}, leading to extra challenges for their phenomenology.
In Fig.~\ref{Fig-13:TMD-extractions-f1-h1}(b) we compare two recent extractions of $h_1^{q,\text{val}}(x)$ from single-hadron 
production~\cite{Cammarota:2020qcw, Gamberg:2022kdb, Cocuzza:2023oam} 
based on the data from Refs.~\cite{COMPASS:2008isr,COMPASS:2014bze,HERMES:2020ifk,Belle:2008fdv} 
and dihadron production~\cite{Cocuzza:2023oam, Cocuzza:2023vqs} 
based on the data from Refs.~\cite{HERMES:2008mcr,STAR:2017wsi,Belle:2017rwm,COMPASS:2023cgk}.
The uncertainties are sizable because much less data is available than, e.g., for unpolarized TMDs. 
The agreement of the transversity valence distribution from the single-hadron and dihadron methods in 
Fig.~\ref{Fig-13:TMD-extractions-f1-h1}(b) is good.
In fact, it was shown that presently all experimental information on transversity, theory constraints on the transversity and lattice-QCD results for the tensor charges are compatible with each other~\cite{Cocuzza:2023oam}.

\begin{figure}
\centering
    \begin{tabular}{ccc}
            \begin{minipage}[b]{0.305\textwidth} 
            \vspace{6.5mm}
      \includegraphics[width=1.03\textwidth]{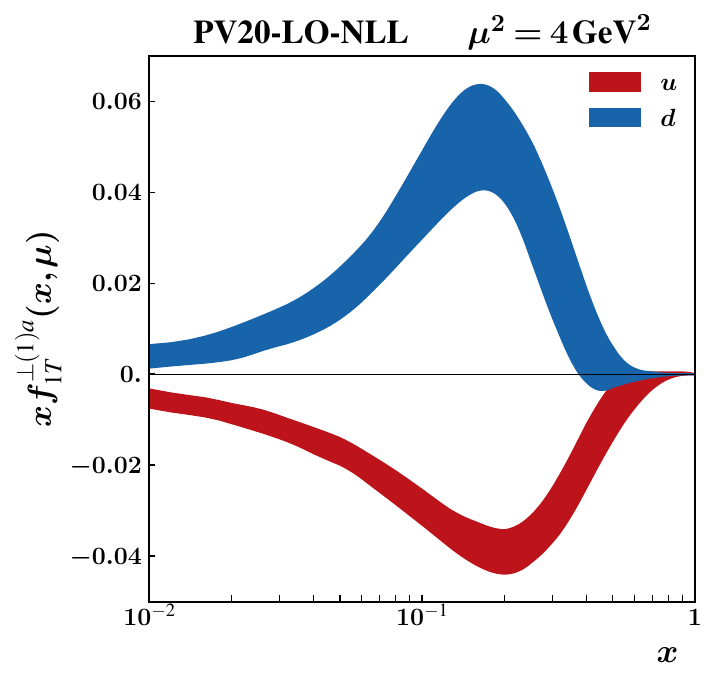}
            \vspace{-5.mm}
            \end{minipage}
    & \includegraphics[width=0.345\textwidth]{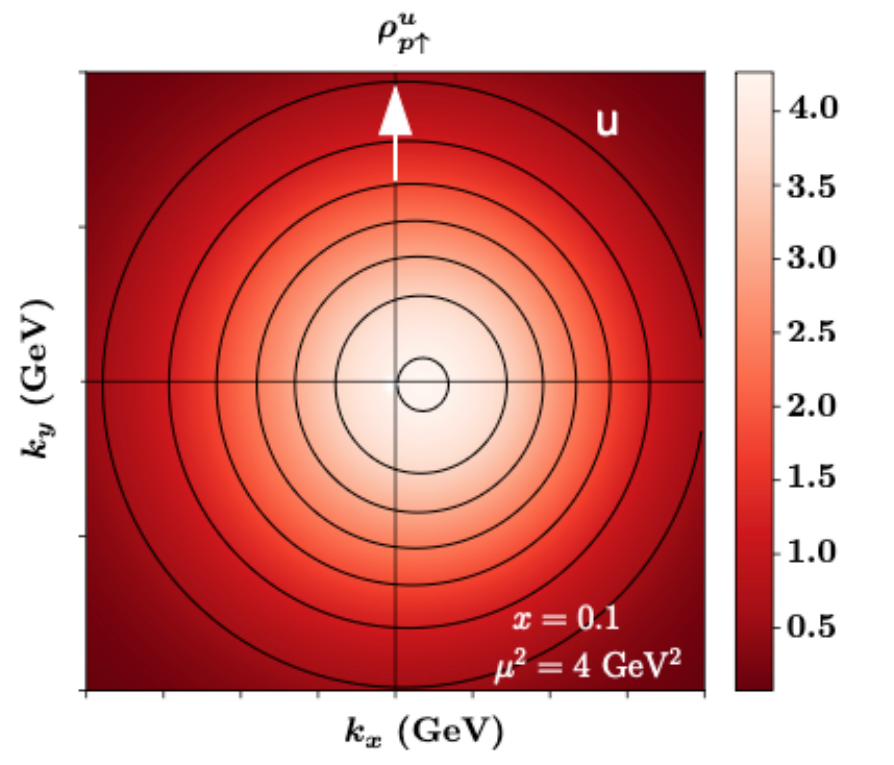}
    &  \includegraphics[width=0.33\textwidth]{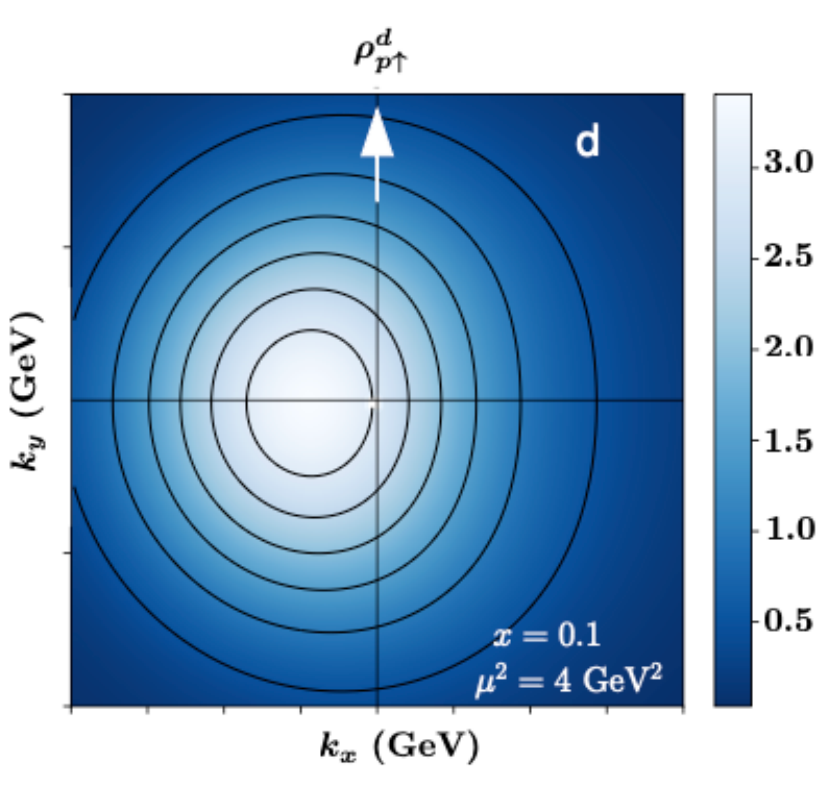} \cr
    \ \hspace{5mm}  \textbf{ (a)} & 
    \ \hspace{-5mm} \textbf{(b)} & 
    \ \hspace{-5mm} \textbf{(c)}
    \end{tabular}
\caption{\label{Fig:14-or-15-Sivers-extraction}\footnotesize 
(a) The transverse moment $xf_{1T}^{\perp(1)q}(x,\mu)$ 
of the Sivers function for $u$- and $d$-quarks as functions of $x$ 
at $\mu^2=4$ GeV$^2$. 
(b,~c) The density distributions $\rho_{p\uparrow}$ of unpolarized up and down quarks in a proton polarized along $+y$ direction and moving towards the reader, as functions of $(k_x, k_y)$ at $x=0.1$ 
and $\mu^2 = 4$ GeV$^2$. Adapted from Ref.~\cite{Bacchetta:2020gko}.}
\end{figure}

The last example we will cover is that of the quark Sivers function, a TMD with no collinear twist-2 counterpart in contrast to $f_1^q(x,k_\perp)$ 
and $h_1^q(x,k_\perp)$.
This naive \textsf{T}-odd TMD generates transverse SSAs in SIDIS in combination with the fragmentation function $D_1^q(z,K_\perp)$; see Fig.~\ref{Fig:SIDIS-kin-and-more}(b).
It also generates a transverse SSA in Drell-Yan from $\pi p^\uparrow$ in combination with the unpolarized TMDs of pions, and in $W^\pm$ and $Z^0$ production from $p^\uparrow p$  collisions. 
Based on SIDIS data from HERMES \cite{HERMES:2009lmz}
and
COMPASS \cite{COMPASS:2008isr}, the prediction of the 
sign reversal of the Sivers function in Eq.~(\ref{e:sign_change}) 
received first support from RHIC data on $W^{\pm}$ and $Z^0$ production
in $p^\uparrow p$ collisions \cite{STAR:2015vmv} and then stronger
support through Drell-Yan measurements from $\pi p^\uparrow$ collisions
at COMPASS~\cite{COMPASS:2017jbv}. Closely related earlier studies
can be found elsewhere~\cite{Metz:2012ui, Gamberg:2013kla}. 
The prediction of the $k_\perp$ factorization of the sign reversal of the
Sivers function in SIDIS and Drell-Yan \cite{Collins:2002kn} is nowadays 
considered as confirmed, and Eq.~(\ref{e:sign_change}) is routinely
used in phenomenological analyses.
In Fig.~\ref{Fig:14-or-15-Sivers-extraction}(a) we show the results 
for $f_{1T}^{\perp(1)q}(x)$ with the sign corresponding to SIDIS from
Ref.~\cite{Bacchetta:2020gko}; transverse moments of TMDs are defined 
below Eq.~(\ref{Eq:TMD-inequalities}). 
The extraction is based on SIDIS and collider data 
\cite{HERMES:2009lmz,COMPASS:2008isr,JeffersonLabHallA:2011ayy,STAR:2015vmv}.
The interesting observation that the up and down quark Sivers functions have nearly the same magnitude and opposite signs was predicted theoretically; 
see Sec.~\ref{Sec-6.2-large-Nc}. The Sivers antiquark distributions are not
constrained by available data.
Based on information on the unpolarized and Sivers TMDs, 
one can visualize the distribution of unpolarized quarks 
in momentum space for a transversely polarized proton. 
The associated parton density,   
see Fig.~\ref{Fig:functions+interpretation}, 
for a proton polarized along the $y$-axis corresponds to 
$\rho_{p\uparrow}^q(x,\vec{k}_\perp) = \Phi_{\rm TMD}^{q[\gamma^+]}(P,x,\uvec k_\perp)    
 = f_1^q(x,k_\perp)-\frac{k_x}{M}\,f_{1T}^{\perp q}(x,k_\perp)$
and is shown in Figs.~\ref{Fig:14-or-15-Sivers-extraction}(b) 
and \ref{Fig:14-or-15-Sivers-extraction}(c), where $x=0.1$ is 
fixed and the proton moves towards the reader. We see that 
the proton polarization in conjunction with the final-state
interaction causes a left-right asymmetry of the quarks in 
the $(k_x,k_y)$-plane, with the contours indicating 
lines of equal probability density. Overall, the unpolarized 
$u$ quarks appear shifted to the right, and $d$ quarks to the 
left. The shifts are opposite and stronger for $d$ quarks than for $u$ quarks because, very roughly at this $x$, we have $f_1^u(x,k_\perp)\approx 2f_1^d(x,k_\perp)$ while
$f_{1T}^{\perp u}(x,k_\perp)\approx -f_{1T}^{\perp d}(x,k_\perp)$.

Compared to the PDF $f_1^a(x)$, much less data is available for $f_1^a(x,k_\perp)$ and still less than that for transversity and the Sivers function. 
These are the three presently most studied quark TMDs. 
Much less is known about the other quark TMDs, mainly due to even fewer available data~\cite{Barone:2009hw,Lefky:2014eia,Bhattacharya:2021twu,Horstmann:2022xkk,Yang:2024bfz,Bacchetta:2024yzl}.

\section{Generalized parton distributions}  \label{Sec-4:GPD}

GPDs have attracted growing interest over more than three decades as a powerful tool for obtaining unique insights into the origin of certain fundamental properties of the nucleon in terms of its constituents, which are not readily accessible through other observables.
GPDs were initially recognized as playing a role in the description of high-energy exclusive processes, as noted in Ref.~\cite{Muller:1994ses}, and their importance in describing the internal structure of hadrons was highlighted in Ref.~\cite{Ji:1996ek}. Additionally, they were identified as key quantities in the description of DVCS~\cite{Muller:1994ses,Ji:1996ek,Ji:1996nm,Radyushkin:1996nd}, hard diffractive electroproduction of vector mesons~\cite{Radyushkin:1996ru}, and hard exclusive electroproduction of mesons \cite{Collins:1996fb}. 
A pivotal development in understanding GPDs and their role in unraveling the quark and gluon structure of the nucleon has been their interpretation in impact-parameter space~\cite{Soper:1976jc,Burkardt:2000za,Burkardt:2002hr,Ralston:2001xs,Diehl:2002he}.
This framework allows for the mapping of the nucleon constituents in both longitudinal momentum and transverse coordinate space, providing a unified description that bridges the properties of FFs and PDFs. Furthermore, the link between GPDs and the energy-momentum tensor (EMT) FFs~\cite{Ji:1996ek,Polyakov:2002yz} has emerged as a key motivation for studying GPDs and has largely driven experimental efforts to extract them.

In the following, we begin by introducing  the definition of GPDs and describe their partonic interpretation and key properties, including their relationship to collinear PDFs and FFs. We then discuss the physical significance of GPDs in impact-parameter space and explore their connection to the EMT FFs, emphasizing  the role of the latter in unraveling the quark and gluon structure of the nucleon. Finally, we summarize some of the major experimental initiatives and phenomenological analysis aimed at extracting these quantities from observables. 
More technical details on the theoretical aspects of GPDs can be found in review papers~\cite{Diehl:2003ny,Ji:1998pc,Goeke:2001tz,Belitsky:2005qn,Boffi:2007yc} and lecture notes~\cite{Mezrag:2022pqk}, while a comprehensive account of the phenomenological and experimental developments is provided in Refs.~\cite{Guidal:2013rya,Favart:2015umi,Kumericki:2016ehc,Diehl:2023nmm}.

\subsection{Definition and properties}
\label{Sec-GPD-definition-and-all-that}

GPDs parameterize the parton correlator in~\eqref{GPD-correlator} corresponding to the off-forward  matrix elements of the same bilocal quark (or gluon) operator that defines PDFs.
For a hadron of spin $J$, the number of independent leading-twist GPDs is $2(2J + 1)^2$~\cite{Diehl:2003ny}. 
Focusing on the proton in the quark sector, the leading-twist GPDs describe the correlator in Eq.~\eqref{GPD-correlator} with $\Gamma=\{\gamma^+,\gamma^+\gamma^5, i\sigma^{j+}\gamma_5\}$. For example, the twist-2 GPD correlator for unpolarized quarks, corresponding to $\Gamma=\gamma^+$, is parameterized as follows: 
\begin{equation}
    \Phi_{\rm GPD}^{q[\gamma^+]}(P,x,\xi,\uvec\Delta_\perp)=\frac{1}{2P^+}\,\bar u(p',S')\left[\gamma^+\, H^q(x,\xi,t)
    +\frac{i\sigma^{+\mu}\Delta_\mu}{2M}\,E^q(x,\xi,t)\right]u(p,S) \,.
\end{equation}
As for the matrix elements of the electromagnetic current in Eq.~\eqref{Eq:def-em-FF}, we need two GPDs, $H^q$ and $E^q$. A similar parameterization for $\Gamma=\gamma^+\gamma^5$ requires two helicity GPDs, $\tilde{H}^q$ and $\tilde{E}^q$. $E^q$ and $\tilde E^q$ are associated with nucleon helicity-flip amplitudes, whereas helicity-conserving amplitudes involve a $\xi$-dependent combination of $H^q$ and $E^q$, or of $\tilde H^q$ and $\tilde E^q$. For $\Gamma=i\sigma^{j+}\gamma_5$ with $j=1,2$, we need four transversity GPDs, $H^q_T$, $E^q_T$, $\tilde H^q_T$, and $\tilde E^q_T$. Combining hermiticity and time-reversal symmetries, GPDs are defined as real-valued functions that are even or odd in $\xi$.

As sketched in Fig.~\ref{Fig:functions+interpretation} , 
GPDs integrate information from both FFs and PDFs.
PDFs are obtained by taking the forward limit $\Delta\to 0$, which implies $\xi,t\to0$, and leads to
\begin{align}
H^q(x,0,0)=f_1^q(x) \,, \qquad \tilde H^q(x,0,0)=g_1^q(x) \,, \qquad H_T^q(x,0,0)=h_1^q(x) \,,
%, \, \qquad{\rm for}\, x>0\\
%H^q(x,0,0)&=-f_1^{\bar q}(x), & \tilde H^q(x,0,0)&=g_1^{\bar q}(x), & H_T^q(x,0,0)&=-h_1^{\bar q}(x), \quad{\rm for}\, x<0,
\end{align}
with analogous relations for gluons. Note that $E^q,\tilde E^q, \tilde H^q_T$, and $E^q_T$ do not vanish in this limit but instead disappear from the correlator, as they are multiplied by kinematic factors proportional to $\Delta$. On the other hand, $\tilde E^q_T(x,0,0)=0$ since it is $\xi$-odd, though one may have $\tilde E^q_T(x,\xi,0)/\xi\neq0$ for $\xi\to0$.
FFs are obtained through integration over $x$. For example, we have
\begin{align}\label{gpd-ff-relation}
    F_1^q(t)=\int_{-1}^1dx \, H^q(x,\xi,t) \,,\qquad
     F_2^q(t)=\int_{-1}^1dx\,  E^q(x,\xi,t) \,,\qquad
     G_A^q(t)=\int_{-1}^1dx\, \tilde{H}^q(x,\xi,t) \,,\qquad
       G_P^q(t)=\int_{-1}^1dx\,\tilde{E}^q(x,\xi,t) \,,
\end{align}
where $F_1^q$ and $F_2^q$ are the usual electromagnetic FFs associated with the quark flavor $q$, $G^q_A$ is the axial-vector FF (with $G_A^q(0)=g_A^q$ introduced in Sec.~\ref{sect-2.4}) and $G^q_P$ is the induced pseudoscalar FF. Another fundamental relation, known as Ji's spin sum rule~\cite{Ji:1996nm}, provides a means to assess the parton contributions to the longitudinal total angular momentum of the nucleon $\frac{1}{2}=\sum_q J^q+J^g$ with
\begin{align}\label{ji-sumrule}
   J^{q}=\frac{1}{2}\int_{-1}^1dx\,x\,[H^q(x,0,0)+E^q(x,0,0)] \quad  {\rm and} \quad
J^{g}=\frac{1}{2}\int_{0}^1dx\,x\,[H^g(x,0,0)+E^g(x,0,0)]\,. 
\end{align}
The quark contribution in Eq.~\eqref{ji-sumrule} can be further split into spin and orbital angular momentum components as $J^q=\frac{1}{2} \Delta q + L^q$, where $\Delta q$ is the axial charge defined in Eq.~\eqref{e:moments} and $L^q$ can be determined by a specific twist-3 GPD~\cite{Penttinen:2000dg,Kiptily:2002nx,Hatta:2012cs}.
However, a similar decomposition does not readily apply to the gluon contribution.
It is worth noting that there exist alternative ways to decompose the total angular momentum into quark and gluon contributions, differing by terms involving the gluon field. For a more detailed discussion on this topic, we refer to the review in Ref.~\cite{Leader:2013jra} and the contribution~\cite{Ji-encyclopedia} in this volume. The relations in Eqs.~\eqref{gpd-ff-relation} and \eqref{ji-sumrule} are special cases of the so-called polynomiality property, which asserts that the $n$-Mellin moments $\int dx\, x^{n-1}$ of GPDs are even 
polynomials in $\xi$ (except for $\tilde E_T^q$, where the polynomials are odd) of order $n$ or less, whose coefficients correspond to the FFs of local currents \cite{Ji:1998pc,Hagler:2004yt}.
The second (i.e., $n=2$) Mellin moments are of particular interest, since they can be related to the EMT FFs; see Sec.~\ref{Sec-EMT-FFs}.

\begin{figure}[t]
\begin{center}
\includegraphics[width=0.3\textwidth]{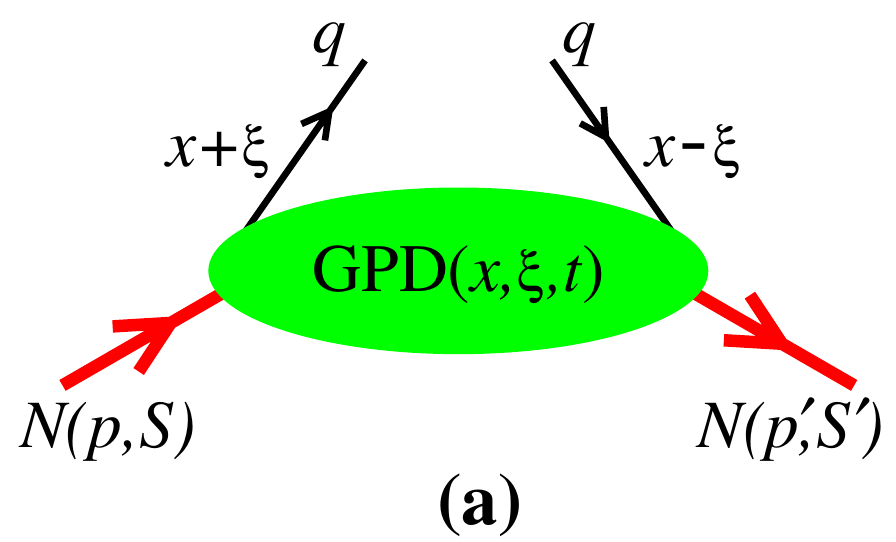}    
\hspace{0.04\textwidth}
\includegraphics[width=0.3\textwidth]{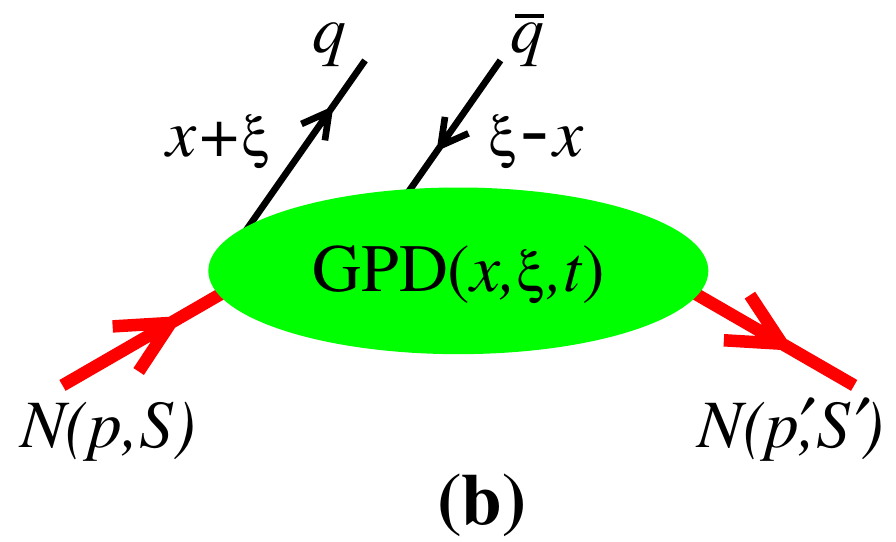}  \hspace{0.04\textwidth}
\includegraphics[width=0.3\textwidth]{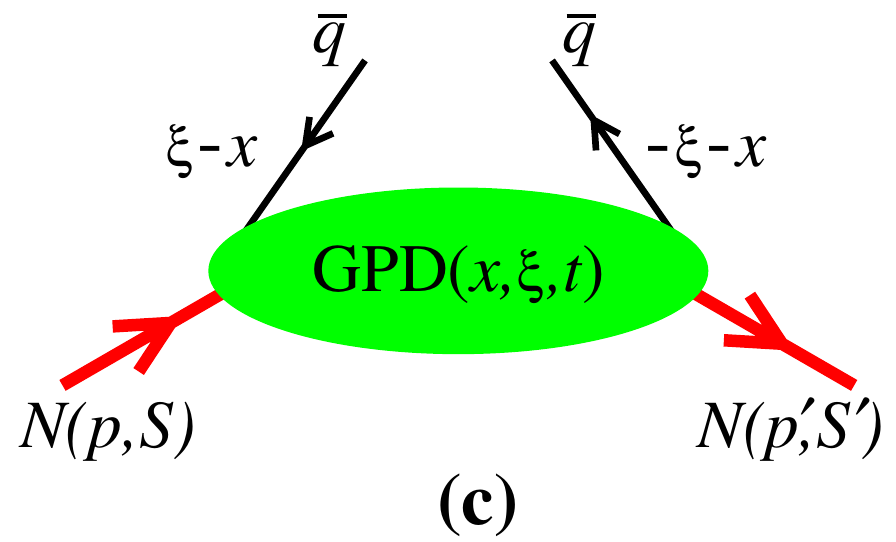}   
\caption{\label{Fig-GPD-DGLAP-ERBL} \footnotesize
Partonic representation of generalized parton 
distributions (GPDs) in the three intervals 
(a) $x\in[\xi,1]$,
(b) $x\in[-\xi,\xi]$, and
(c) $x\in[-1,-\xi]$ for 
$\xi > 0$.
}
\end{center}
\end{figure}

GPDs are not parton densities.
Their partonic representation depends on the 
values of $x$ and $\xi$, as illustrated in Fig.~\ref{Fig-GPD-DGLAP-ERBL} 
for $\xi>0$ (the same description applies to the experimentally 
inaccessible region $\xi<0$ by replacing $\xi\rightarrow -\xi$).
In the region $\xi < x < 1$ (see Fig.~\ref{Fig-GPD-DGLAP-ERBL}(a)),
GPDs describe the emission from the initial hadron
of a quark carrying light-front plus momentum 
$(x+\xi)P^+$, followed by the absorption of a quark with 
plus momentum $(x-\xi)P^+$ that recombines with the spectator system to form the final state.
In the central region  $-\xi<x<\xi$ (see Fig.~\ref{Fig-GPD-DGLAP-ERBL}(b)), 
GPDs describe the emission a quark-antiquark pair with total plus momentum 
$2\xi P^+$. The $-1<x<-\xi$ region (see Fig.~\ref{Fig-GPD-DGLAP-ERBL}(c))
corresponds to the emission and absorption of an antiquark.
The distributions in the outer regions $\xi<|x|<1$ generalize the situation illustrated for the ordinary PDFs in sect.~\ref{sect-2.4}
and evolve according to modified DGLAP equations. 
In contrast, the central region has no analogue in ordinary PDFs. GPDs here resemble meson distribution amplitudes and, accordingly, evolve under the modified Efremov-Radyushkin-Brodsky-Lepage (ERBL) equations~\cite{Efremov:1979qk,Lepage:1980fj}.
This ERBL region provides genuinely new information on hadron structure, as it is absent in DIS (which corresponds to $\xi= 0$).

\subsection{Impact-parameter distributions} 
\label{sec:IPD}
As mentioned in Sec.~\ref{Sec-1.1-pre-parton}, a 3D
density interpretation in position space is hampered by relativistic recoil corrections. In this discussion, we will disregard the hadron spin because it only makes the situation more involved.
If we consider, e.g., the charge density operator $j^0(\uvec x)$, the Fourier transform of its expectation value in a normalized state $|\Psi\rangle$ is given by
\begin{equation}
    \tilde\rho_\Psi(\uvec\Delta)=\int d^3x\,e^{i\uvec\Delta\cdot\uvec x}\,\langle\Psi|j^0(\vec x)|\Psi\rangle=\int\frac{d^3p}{(2\pi)^3}\,\tilde\Psi^*(\uvec p+\uvec\Delta)\tilde\Psi(\uvec p)\,\frac{\langle p+\Delta|j^0(\uvec 0)|p\rangle}{2\sqrt{E_{p+\Delta}E_{p}}},
\end{equation}
where $\tilde\Psi(\uvec p)=\langle p|\Psi\rangle/\sqrt{2E_p}$ is the momentum-space wave packet with 
the energy $E_p=\sqrt{\uvec p^2+M^2}$, and the momentum eigenstates normalized as $\langle p'|p\rangle=(2\pi)^3\,2E_p\,\delta^{(3)}(\uvec p'-\uvec p)$. For a spin-$0$ state,
Lorentz covariance implies that $\langle p+\Delta|j^0(\uvec 0)|p\rangle=(E_{p+\Delta}+E_p)\,F(-\uvec\Delta^2)$ with $F(-\uvec\Delta^2)$ the electric FF. 
In the non-relativistic limit, we have $E_{p+\Delta}\approx E_p\approx M$ and so $\tilde\rho_\Psi(\uvec\Delta)\approx F(-\uvec\Delta^2)\int\frac{d^3p}{(2\pi)^3}\,\tilde\Psi^*(\uvec p+\uvec\Delta)\tilde\Psi(\uvec p)$. 
Now, if the wave packet is very flat in momentum space (i.e.,~very narrow in position space), it is justified to identify $F(-\uvec\Delta^2)$ with $\tilde\rho_\Psi(\uvec\Delta)$. 
In contrast, in the relativistic regime the kinematical factor $(E_{p+\Delta}+E_p)/(2\sqrt{E_{p+\Delta}E_p})\not\approx 1$ entangles the $p$- and $\Delta$-dependences in a way that prevents us 
from expressing the 3D Fourier transform $\int\frac{d^3\Delta}{(2\pi)^3}\,e^{-i\uvec\Delta\cdot\uvec x}\,F(-\uvec\Delta^2)$ as a genuine charge density $\langle\Psi|j^0(\vec x)|\Psi\rangle$. Note, however, that $\int\frac{d^3\Delta}{(2\pi)^3}\,e^{-i\uvec\Delta\cdot\uvec x}\,F(-\uvec\Delta^2)$ is a legitimate quasi-density from Wigner's phase-space perspective~\cite{Lorce:2020onh}. Therefore, it remains a meaningful and useful concept in the relativistic theory. More details on the phase-space approach will be presented in Sec.~\ref{sec.-Wigner_distribution}. 

If one insists on a genuine density interpretation, a key property to satisfy is that the system's inertia (usually identified with energy in relativity) does not appreciably depend on the momentum transfer. A simple solution in the relativistic theory is to work in infinite-momentum frame $p_z\to\infty$ so that $E_{p+\Delta}\approx E_p$, which naturally leads to 2D densities in position space as a result of extreme Lorentz contraction~\cite{Fleming:1974af,Burkardt:2000za}. Equivalently, we can switch to the light-front formalism, where the role of inertia is played by the momentum component $p^+=(p^0+p_z)/\sqrt{2}$. In the Drell-Yan frame, characterized by the condition $\Delta^+=0$, the 2D spatial charge distribution $\int\frac{d^2\Delta_\perp}{(2\pi)^2}\,e^{-i\uvec\Delta_\perp\cdot\uvec b_\perp}\,F(-\uvec\Delta^2_\perp)$ admits a proper density interpretation, provided that the wave packet is well-localized in both $p^+$ and transverse-position spaces~\cite{Burkardt:2002hr,Miller:2010nz}. The so-called impact-parameter variable $\uvec b_\perp$ represents the position relative to the center of light-front inertia of the system, i.e.,~from the center of $P^+$ ($p^+$ and $P^+$ are the same in the Drell-Yan frame). For recent developments regarding the notion of relativistic densities, see e.g.~\cite{Jaffe:2020ebz,Freese:2021mzg,Epelbaum:2022fjc}.

Moving on to GPDs, the fact that they are defined in terms of non-local light-front operators  singles out the 2D interpretation in impact-parameter space. 
If we consider $\Phi^{q[\Gamma]}_{\rm GPD}$ in the Drell-Yan frame, which amounts to setting $\xi=0$, and Fourier transform it to impact-parameter space $\vec{b}_\perp$, 
we obtain the so-called impact-parameter distributions~\cite{Burkardt:2000za,Burkardt:2002hr}. When the initial and final nucleon polarizations are the same, impact-parameter distributions admit a density interpretation: 
they give the average number of partons with momentum fraction $x$ at the transverse position $\uvec b_\perp$ relative to the nucleon’s center of $P^+$, as illustrated in Fig.~\ref{f:pancake}.
In that sense, they provide a tomographic picture of the nucleon internal structure in a hybrid momentum-position space. Since longitudinal momentum and transverse position coordinates are not conjugate variables, the quantum-mechanical uncertainty principle does not impose a constraint in this context. For a discussion of the meaning of the Fourier transform of GPDs in the more general $\xi\neq 0$ case, we refer to~\cite{Diehl:2002he,Miller:2019ysh}.  Similar to PDFs and TMDs, some positivity constraints are expected to impose upper bounds on GPDs in terms of PDFs~\cite{Pire:1998nw,Kirch:2005in,Pobylitsa:2002gw}. These constraints have been established in the DGLAP regions $x\in [\xi,1]$ and $x\in[-\xi,-1]$, while their extension to the ERBL region $|x|\le \xi$  is unknown. Further positivity constraints based on the impact-parameter representation have also been derived in~\cite{Diehl:2005jf}.

If we compare the parameterization of the correlator~\eqref{TMD-correlator} in momentum space, expressed in terms of TMDs, with the parameterization of the Fourier-transformed correlator~\eqref{GPD-correlator} in impact-parameter space at $\xi=0$ given in terms of GPDs, we find a corresponding structure with $\uvec k_\perp$ interchanged with $\uvec b_\perp$.
The correspondence in terms of distributions reads~\cite{Diehl:2005jf,Meissner:2007rx}
\begin{equation}
\begin{aligned}
f^q_1 &\leftrightarrow {\mathscr H}^q ,\quad & 
f_{1T}^{\perp q} &\leftrightarrow - {\mathscr E}^{q\prime} ,\quad &
g_1^q &\leftrightarrow \tilde{{\mathscr H}}^q ,\\
h_1^q&\leftrightarrow {\mathscr H}^q_T - \Delta_b \tilde{{\mathscr H}}^q_T /(4M^2), \quad &
h_1^{\perp q} &\leftrightarrow - ({\mathscr E}^{q\prime}_T + 2\tilde{{\mathscr  H}}^{q\prime}_T),\quad &
h_{1T}^{\perp q} &\leftrightarrow 2 \tilde{{\mathscr H}}^{q\prime\prime}_T ,
\end{aligned}
\end{equation}
where, for the GPDs in impact-parameter space, we used the notation $\mathscr{  F}(x,\uvec b_\perp^2)=\int\tfrac{d^2 \Delta_\perp}{(2\pi)^2}\, e^{-i\uvec \Delta_\perp\cdot\uvec b_\perp}\, F(x,0,-\uvec\Delta^2_\perp)$ and we introduced the shorthand ${\mathscr F}'=\partial {\mathscr F}/\partial b^2_\perp$ and ${\mathscr F}''=\partial^2
{\mathscr F}/(\partial b^2_\perp)^2$. Furthermore, $\Delta_b$ stands for the usual Laplace operator in $\uvec b_\perp$-space. The GPD amplitudes corresponding to $g_{1T}^{\perp q}$ and $h_{1L}^{\perp q}$ are $\xi$-odd~\cite{Lorce:2011dv} and therefore disappear at $\xi=0$.
 
As a result, the densities in momentum space $(x,\uvec k_\perp)$, described by TMDs, are mirrored by densities in the mixed space $(x,\uvec b_\perp)$, described by GPDs, where the same multipole patterns emerge, modulated by different types of distributions. However, beyond this formal connection, model-independent relations between GPDs and TMDs generally cannot be established. Indeed, $\uvec b_\perp$ is the Fourier conjugate variable to the transverse momentum transfer $\uvec\Delta_\perp$, whereas $\uvec k_\perp$is the Fourier conjugate variable to the transverse separation $\uvec z_\perp$ of the fields in Eq.~\eqref{GPCF}. Nonetheless, non-trivial relations can arise in  specific  models, such as those leading to the lensing mechanism that connect T-odd TMDs and GPDs in impact parameter space~\cite{Burkardt:2003je,Burkardt:2003uw}.
Although these relations hold only under particular model assumptions~\cite{Meissner:2007rx,Pasquini:2019evu}, they correctly predict the absolute signs of the Sivers TMDs~\cite{Burkardt:2003uw}.

\begin{figure}[ht]
\center
\includegraphics[width=0.8\textwidth]{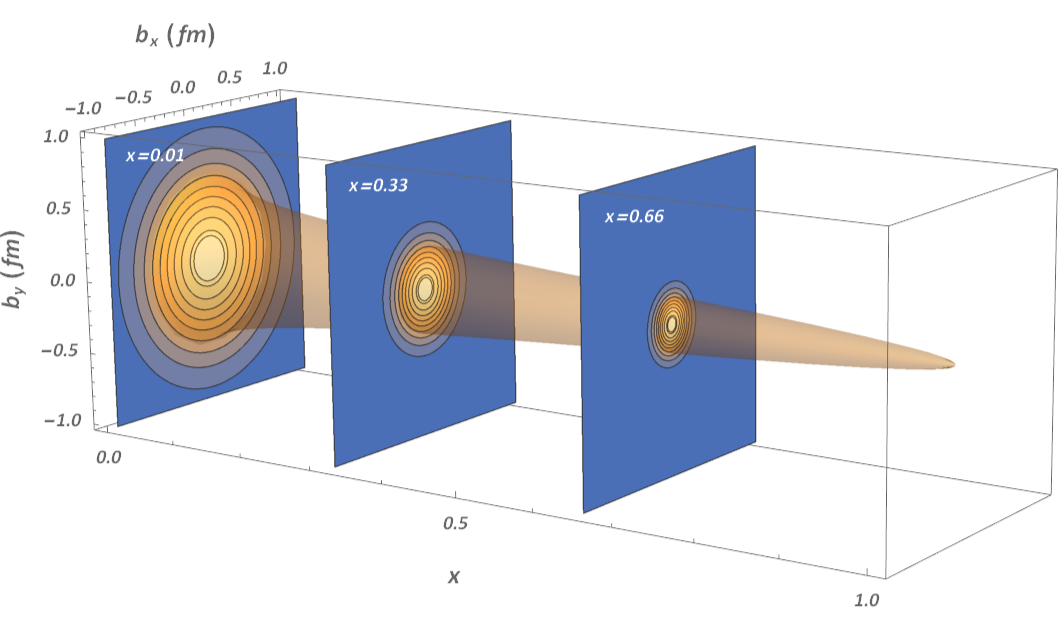}
\caption{\footnotesize Spatial extension of the unpolarized quark distribution in an unpolarized nucleon at various values of $x$, defined as ${\mathscr H}^q(x,\uvec b_\perp^2)/\int d^2b_\perp\,{\mathscr H}^q(x,\uvec b_\perp^2)$. The transparent cone represents the mean transverse radius $\sqrt{\langle b_\perp^2\rangle(x)}$. Adapted from Fig. 25 of Ref.~\cite{Dupre:2017hfs}.\label{fig-ipd}}
\end{figure}

The first attempt to reconstruct the GPDs in impact-parameter space from experimental data was performed in Ref.~\cite{Dupre:2017hfs}, with the result for the GPD ${\mathscr H}^q$ shown in Fig.~\ref{fig-ipd}.
This map corresponds to unpolarized quarks in an unpolarized proton, assuming flavor independence. 
It was constructed using model assumptions to extrapolate the GPD to $\xi=0$, a kinematic point that is not experimentally accessible, and to a region beyond the measured $(x,t)$-range. Nevertheless, the mapping of the quark in impact-parameter space exhibits a clear trend: the transverse width shrinks as $x\rightarrow 1$, in line with expectations since, in this limit, the active quark carries the entire longitudinal momentum of the proton and defines therefore the position of the center of $P^+$.

\subsection{Form factors of the energy-momentum tensor}
\label{Sec-EMT-FFs}

The notion of electromagnetic FF can be generalized to other local probes. For example, weak FFs describe the response of a hadron to a weak probe, like $W$ and $Z$ bosons, that couples also to the axial-vector current $j^\mu_5(x)=\bar\psi(x)\gamma^\mu\gamma_5\psi(x)$. Similarly, EMT FFs describe the response to a probe that couples to the EMT~\cite{Kobzarev:1962wt,Pagels:1966zza}. Since the EMT is the source of gravity according to general relativity, these FFs are also often called gravitational FFs. Although appealing, this name may be a bit misleading for three reasons. First, it may give an incorrect impression that these FFs are in practice determined via a gravitational probe. Second, the EMT in quantum field theory does not necessarily coincide with the gravitational EMT of general relativity. In particular, the former is not necessarily symmetric due to the quantum notion of intrinsic spin, unlike the latter; see e.g.~\cite{Leader:2013jra} for a detailed discussion. Last but not least, it follows from the equivalence principle that gravity does not distinguish between quarks and gluons, whereas quark and gluon contributions to the EMT make perfect sense in quantum field theory.

For a spin-$\frac{1}{2}$ hadron, one can parameterize the matrix elements of the quark and gluon contributions to the EMT as~\cite{Kobzarev:1962wt,Pagels:1966zza,Ji:1996ek,Bakker:2004ib}
\begin{equation}\label{EMTparameterization}
    \langle p',S'|T^{\mu\nu}_a(0)|p,S\rangle=\bar u(p',S')\left[P^{\{\mu} \gamma^{\nu\}}\,A_a(t)+\frac{P^{\{\mu}i\sigma^{\nu\}\lambda}\Delta_\lambda}{2M}\,B_a(t)+Mg^{\mu\nu}\,\bar C_a(t)+\frac{\Delta^\mu\Delta^\nu-g^{\mu\nu}\Delta^2}{4M}\,D_a(t)- \frac{P^{[\mu}i\sigma^{\nu]\lambda}\Delta_\lambda}{M}\,S_a(t)\right]u(p,S) \,,
\end{equation}
where $a^{\{\mu}b^{\nu\}}=\frac{1}{2}(a^\mu b^\nu+a^\nu b^\mu)$ and $a^{[\mu}b^{\nu]}=\frac{1}{2}(a^\mu b^\nu-a^\nu b^\mu)$. 
As no local gauge-invariant antisymmetric rank-two tensor
exists for gluons, we have $S_g(t)=0$. EMT FFs of hadrons cannot be measured directly in practice, owing to the smallness of the gravitational interaction. They can, however, be obtained from the second Mellin moments of GPDs as~\cite{Ji:1996ek}
\begin{equation}
    \label{Eq:GPD-2nd-Mellin-moments}
    \int d x\,x\,H^q(x,\xi,t)=A_q(t)+\xi^2D_q(t) \,,\qquad \int d x\,x\,E^q(x,\xi,t)=B_q(t)-\xi^2D_q(t) \,.
\end{equation}

\begin{figure}[t]
\begin{center}
\includegraphics[width=0.4\textwidth]{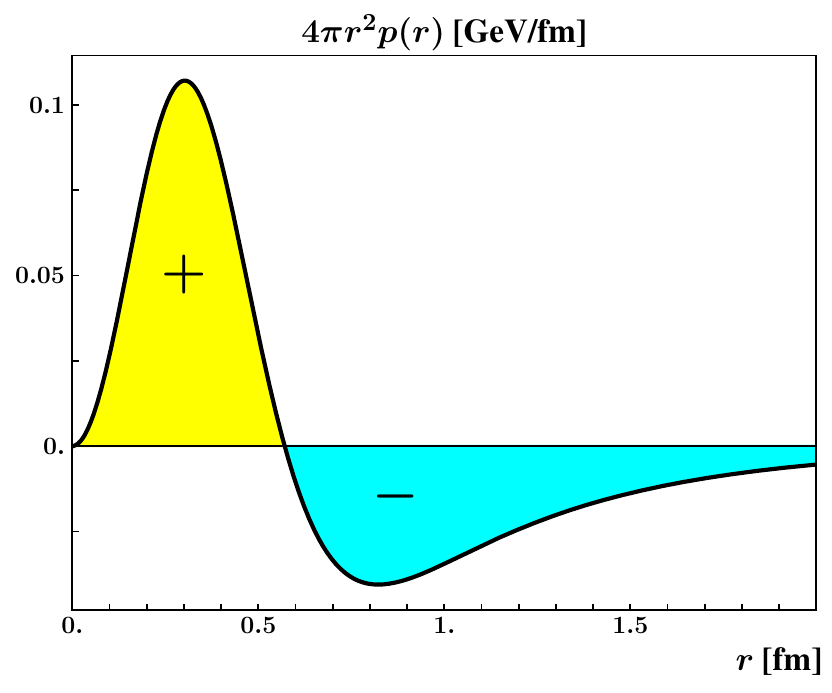} 
\caption{\footnotesize \label{Fig17} The radial distribution $4\pi r^2 p(r)$ of the nucleon from the chiral quark soliton model~\cite{Goeke:2007fp}.  The shaded regions, representing repulsive (positive) and attractive (negative) pressure contributions, have equal areas but opposite signs, and cancel each other exactly according to Eq.~(\ref{Eq:von-Laue}).}
\end{center}
\end{figure}

After 3D Fourier transform in the Breit frame, the EMT matrix elements~\eqref{EMTparameterization} can be interpreted (with similar caveats as in the electromagnetic case) in terms of 3D spatial distributions of energy, linear and angular momentum, and stress inside the hadron~\cite{Polyakov:2002yz,Polyakov:2018zvc}. 2D densities on the light front have also been introduced in~\cite{Lorce:2017wkb,Lorce:2018egm,Freese:2021czn}. Mechanical properties of the nucleon are obtained from the stress tensor, i.e.,~the spatial part of the EMT. The structure is particularly simple in the Breit frame and reads~\cite{Polyakov:2002yz}
\begin{equation}
    \int\frac{d^3\Delta}{(2\pi)^3}\,e^{-i\uvec\Delta\cdot\uvec r}\,\frac{\langle P+\frac{\Delta}{2},S'|T^{ij}(0)|P-\frac{\Delta}{2},S\rangle}{2P^0}=\delta^{ij}p(r)+\left(\frac{r^ir^j}{r^2}-\frac{1}{3}\,\delta^{ij}\right)s(r) \,,
\end{equation}
where $i,j$ are here 3D spatial indices and $r=|\uvec r\,|$ is the radial distance. By analogy with continuum mechanics, $p(r)$ is interpreted as the isotropic pressure and $s(r)$ as the shear stress or the pressure anisotropy. 
On general grounds \cite{Landau:1986aog}, they can also be understood, respectively, as the monopole and quadrupole 
contributions to the momentum flux~\cite{Ji:2021mfb}.
EMT conservation, $\partial_\mu T^{\mu\nu}=0$, implies the von Laue condition 
for mechanical equilibrium~\cite{Laue:1911lrk},
\begin{equation}\label{Eq:von-Laue}
    \int d^3r\,p(r)=0 \,,
\end{equation}
which can also be regarded as a consequence of the virial 
theorem~\cite{Lorce:2021xku}. Of particular interest is the $D$-term,
\begin{equation}
    D=M\int d^3r\,r^2\,p(r)=-\frac{4}{15}\,M\int d^3r\,r^2\,s(r),
\end{equation}
which is found
negative in systems governed by short-range interactions; 
see~\cite{Lorce:2025oot} for a recent discussion. 
An illustration how the pressure distribution inside the
nucleon may look like is shown in Fig.~\ref{Fig17} based on
the chiral quark soliton model~\cite{Goeke:2007fp}. 
In the center of the nucleon, the pressure is $p(0)=0.23$ GeV/fm$^3$ 
corresponding to an astonishing value of
$\sim 4\times 10^{29}$ atmospheric pressures. This high positive pressure in the core indicates a strong compressive stress associated with the dominance of repulsive forces between the partons. As the radial distance increases beyond $r\sim 0.6$ fm, the pressure distribution becomes negative, indicating the onset of tensile stress associated with the dominance of attractive forces that act to bind
the internal constituents of the nucleon. The presence of both compressive and tensile regions is essential to maintain the mechanical stability of the system. The mass radius (associated with the energy density in the nucleon), the scalar radius (defined through the EMT trace), and the mechanical radius (related to the normal force $p(r)+\frac23s(r)$ which is positive~definite) can also be defined from the spatial distributions of the EMT FFs~\cite{Polyakov:2018zvc}, providing new and complementary ways of characterizing the spatial extension of the nucleon beside the conventional charge radius.  Unlike the charge radius, the EMT radii also provide meaningful size estimates for the neutron. For an overview of the different aspects of the EMT FFs, we refer to~\cite{Burkert:2023wzr}.

\subsection{Observables for nucleon  GPDs}
\label{Sec-4.4-GPD-observables}

\begin{figure}[t]
\begin{center}
\includegraphics[width=0.22\textwidth]{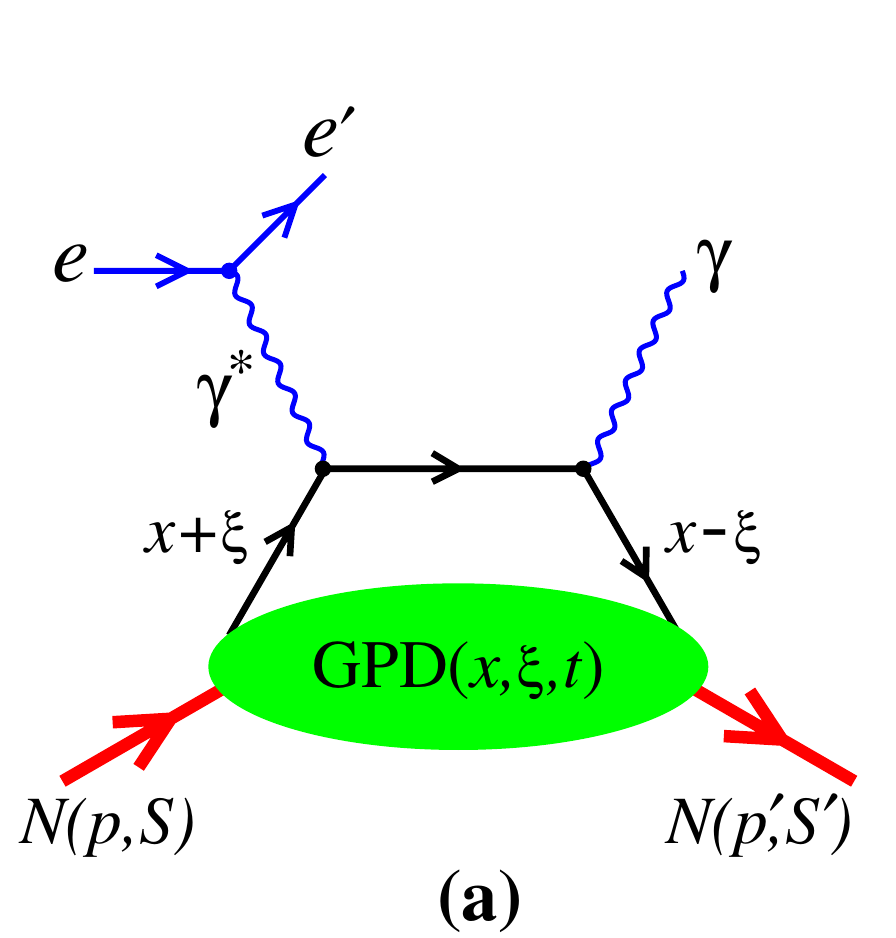} \hspace{5mm}
\includegraphics[width=0.22\textwidth]{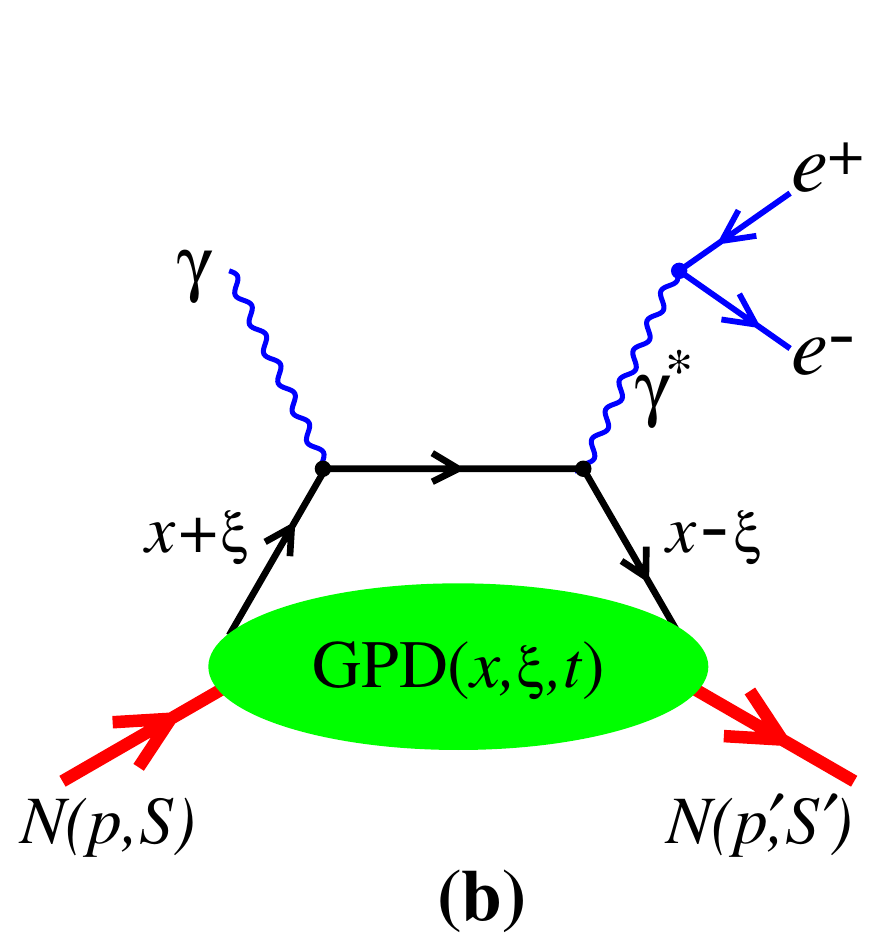}  \hspace{5mm}
\includegraphics[width=0.22\textwidth]{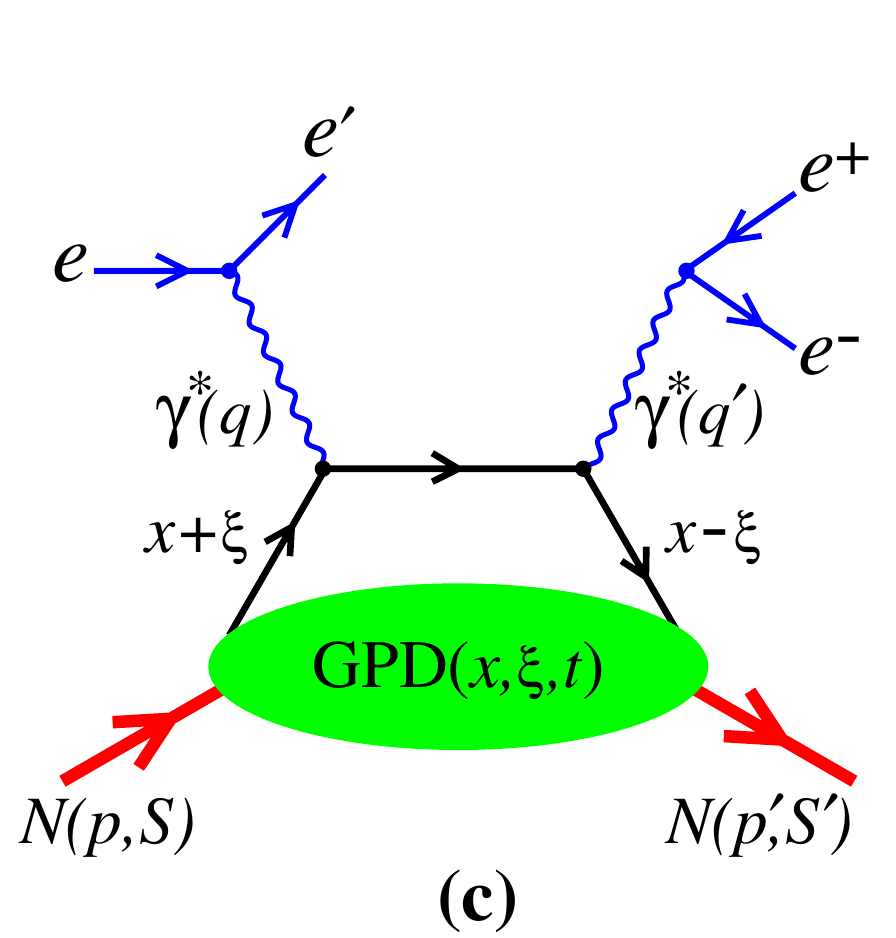}\hspace{5mm}
\includegraphics[width=0.22\textwidth]{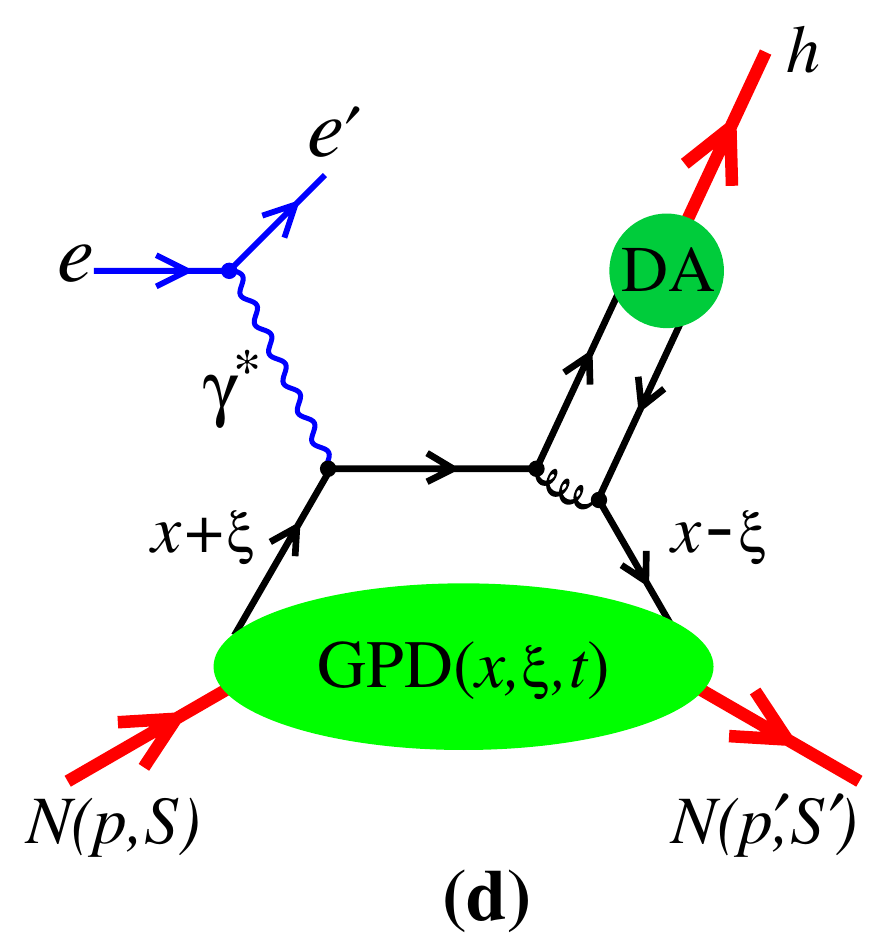}  
\caption{\footnotesize \label{Fig7} 
Amplitudes of processes described in terms of GPDs:
(a) Deeply Virtual Compton Scattering (DVCS),
(b) timelike DVCS,
(c) double DVCS,
(d) Hard Exclusive Meson Production (HEMP), 
where the darker green blob represents the distribution 
amplitude (DA) describing the formation of the meson in 
the final state. 
}
\end{center}
\end{figure}

GPDs can be experimentally accessed through various high-energy exclusive 
reactions. The possibility to probe GPDs in these reactions relies on 
factorization theorems~\cite{Collins:1998be,Collins:1996fb,Ji:1998xh}, 
as we have seen for  PDFs and TMDs. Unlike PDFs and TMDs, factorization for GPDs occurs at the level of scattering amplitudes.
A notable example is the Compton scattering amplitude with at least one virtual photon, $\gamma^{(*)}(q) +N(p)\rightarrow \gamma^{(*)}(q') + N(p')$, where factorization requires at least one of the photons to be far off-shell.
This condition is met in DVCS (see Fig.~\ref{Fig7}(a)), where $q^2$ is large and spacelike and the final photon is real, appearing as a subprocess in photon electroproduction ($eN\rightarrow e'N'\gamma$).
 Another case is timelike Compton scattering 
 (see Fig.~\ref{Fig7}(b)), where the initial photon is real and $q'^2$ is large and timelike, occurring in the photoproduction of a lepton pair  ($\gamma N \rightarrow l\bar{l}'N'$).
Finally, when both photons are off-shell, the process is known as double DVCS
(see Fig.~\ref{Fig7}(c)), which is accessed through electroproduction of a lepton pair ($eN  \rightarrow l\bar{l}'N' $).
The factorization theorems for Compton scattering are valid for $|q^2|+|q'^2|\rightarrow \infty$ at fixed $q^2/W^2$, $q'^2/W^2$, and $t$, where $W$ is the photon-nucleon center-of-mass energy.
Another privileged channel for studying GPDs is through hard exclusive electroproduction of a meson ($eN\rightarrow e' N' h$), where the factorization enters in the scattering amplitude for hard exclusive meson production (HEMP) 
when $Q^2\rightarrow \infty$ at fixed $q^2/W^2$ and $t$ (see Fig.~\ref{Fig7}(d)).

The factorized amplitudes for Compton scattering can be expressed in terms of structure functions, known as Compton FFs (CFFs), which involve the convolution of GPDs with a perturbatively calculable hard-scattering kernel. In the case of meson production, the soft part includes an additional contribution that describes the formation of the meson in the final state. At leading order, all the processes mentioned above are represented by the handbag diagrams shown in Fig.~\ref{Fig7}, where the lower blobs correspond to the relevant GPD correlators.
Reactions involving Compton scattering as a subprocess also receive contributions from the purely electromagnetic Bethe-Heitler process, which can be precisely calculated in terms of the proton electromagnetic FFs. The Bethe-Heitler process produces the same final state as the Compton process, but with the  real photon radiated by the incoming or
scattered lepton. The interference between the Bethe-Heitler and Compton amplitudes allows direct access to the CFFs at the amplitude level.
Experimental studies of GPDs have so far primarily focused on DVCS and HEMP, which provide complementary information. 
In these processes, the helicity of the produced photon or meson modulates the angular distribution of the final state, allowing the selection of various helicity combinations and, consequently, access to different CFFs. The angular structure of the cross section is particularly rich in DVCS due to its interference with the Bethe-Heitler process, which enhances sensitivity to exclusive amplitudes and makes DVCS the most favorable channel for detailed studies of GPDs. 
However, DVCS is sensitive only to CFFs associated with the 
four chiral-even GPDs and, at leading order, only to those of quarks 
and does not enable a direct flavor separation.
HEMP provides sensitivity to different quark flavor 
combinations by varying the produced meson --- at the 
price of involving an additional non-perturbative component 
to describe the dynamics of meson formation.
Chiral-odd GPDs decouple from HEMP  in Fig.~\ref{Fig7}(d) which 
proceeds through exchange of virtual 
longitudinally polarized photons \cite{Collins:1999un,Collins:1996fb}.
They can be accessed, albeit in an effective approach not 
based on collinear factorization \cite{Collins:1996fb},
in pion production mediated by the exchange of virtual transversely 
polarized photons at order $1/Q$
\cite{Goloskokov:2009ia,Ahmad:2008hp}.
Chiral-odd GPDs may also be accessed through studies of more involved 
final states like production of two vector mesons with a large 
rapidity gap \cite{Ivanov:2002jj}.

Taking as an example DVCS, the CFFs assume the form
\begin{align}\label{cff}
{\rm CFF}(\xi,t,Q^2) =\int dx \sum_a C^a\left(x,\xi,\frac{Q}{\mu},\alpha_{{\rm s}}(\mu)\right) {\rm GPD}^a (x,\xi,t,\mu) \, ,
\end{align}
where $C^a$ 
are complex-valued hard-scattering coefficient functions, systematically calculable in pQCD, and the sum is over all active partons $a$.
At leading order one has $C^q=-e^2_q\,(\frac{1}{\xi-x+i\epsilon}-\frac{1}{\xi+x-i\epsilon})$, while $C^g=0$. As a result, 
the imaginary part of the DVCS CFFs gives the GPDs along the diagonals $x=\pm \xi$, while the real part of the CFFs probes a convoluted integral of GPDs over $x$. 
Extracting GPDs from CFFs requires deconvoluting Eq.~\eqref{cff}, which is challenging because the GPD variable $x$ is integrated out and does not appear in the variables of the CFFs~\cite{Bertone:2021yyz, Moffat:2023svr}.
To tackle this issue it is essential to have data over a wide kinematic range and from different processes such as timelike Compton scattering~\cite{Berger:2001xd} and double DVCS~\cite{Guidal:2002kt,Belitsky:2002tf}.
First timelike Compton scattering measurements  have shown that this process should have a strong impact in
constraining the real part of CFFs~\cite{CLAS:2021lky}, while double DVCS gives the unique possibility to vary both quark
momenta $x$ and $\xi$ independently. 
Despite the experimental challenges of double DVCS measurements, feasibility studies at JLab and the EIC  appear promising~\cite{Alvarado:2025huq}. Other processes directly sensitive to the $x$-dependence of GPDs, which have larger count rates than double DVCS, have also been explored in Refs.~\cite{Duplancic:2018bum,Qiu:2022bpq,Qiu:2022pla,Siddikov:2022bku}. 

\begin{figure}
\centering
\includegraphics[height=0.35\textwidth]{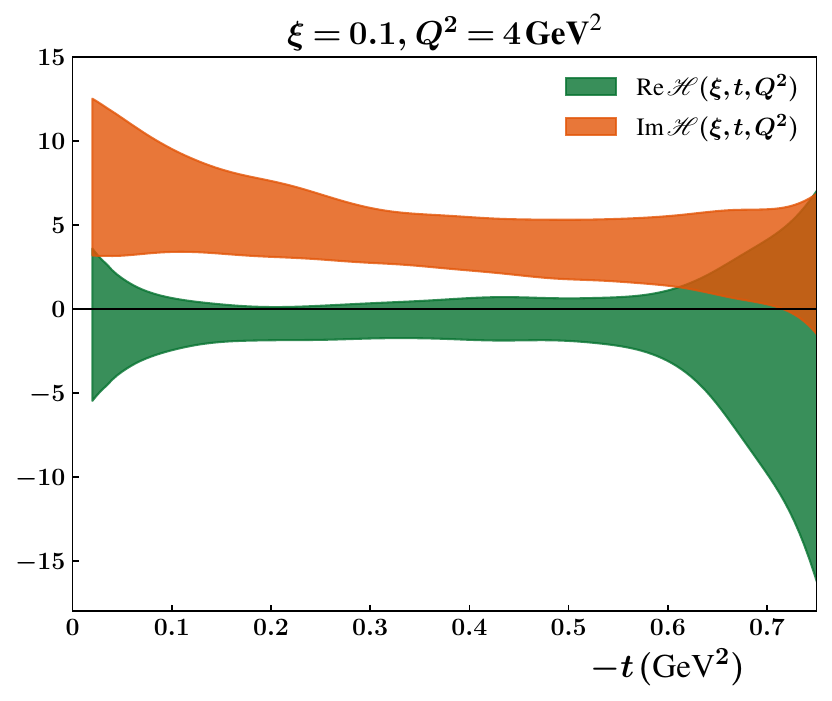}
\hspace{1 truecm}
\includegraphics[height=0.35\textwidth]{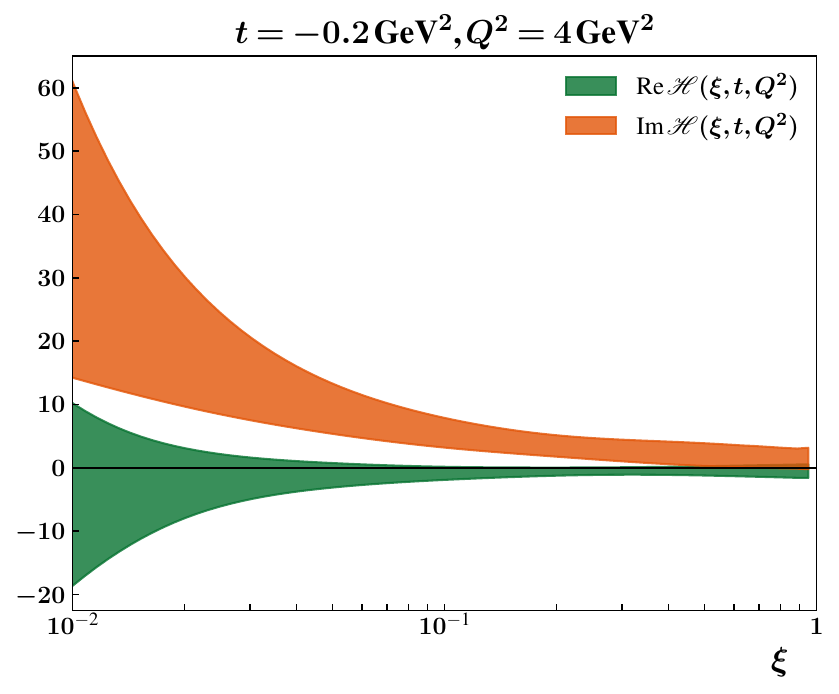}
\caption{\footnotesize Real  and imaginary parts of the CFF ${\cal H}$ at $Q^2$ = 4 GeV$^2$  as a function of $-t$ for $\xi = 0.1$ (left panel) and  as a function of $\xi$ for $t = -0.2$  GeV$^2$  (right panel), extracted from a global fit of DVCS observables using neural network techniques~\cite{Moutarde:2019tqa}. 
\label{fig:compton-ff}}
\end{figure}

The phenomenological study of GPDs is very active, leading to the first quantitative extractions of CFFs and providing valuable constraints on GPDs. Various theory-driven parameterizations  have been developed, including approaches based on the double  distribution Ansatz~\cite{Musatov:1999xp,Vanderhaeghen:1999xj,Goloskokov:2009ia}, 
models in conformal moment space~\cite{Mueller:2005ed,Kumericki:2013br,Guo:2023ahv}, 
dispersion relations~\cite{Moutarde:2019tqa,Kumericki:2009uq,Kumericki:2007sa,Diehl:2007jb}, 
and spectator models~\cite{Goldstein:2010gu,Goldstein:2013gra}. 
These frameworks provide complementary ways to describe GPDs and offer cross checks to validate the consistency of phenomenological results. Additionally, valuable open-source tools  such as PARTONS \cite{Berthou:2015oaw} and Gepard~\cite{gepard} are now available to support the modeling, analyses, and extractions of GPDs.
As an example of CFF extracted from a global analysis of DVCS, we show in Fig.~\ref{fig:compton-ff} the real and imaginary part of ${\cal H}$ obtained in the PARTONS framework using artificial neural network techniques~\cite{Moutarde:2019tqa}.
 The results are in good agreement, within uncertainties, with those from a similar neural network study in Ref.~\cite{Cuic:2020iwt}.
%, which used a smaller data set but incorporated recent DVCS data on neutrons, enabling a flavor separation of the CFF.

As discussed in Sec.~\ref{Sec-EMT-FFs}, the EMT FFs are of central 
interest for a variety of reasons. One way 
to access them involves the following steps:
(i) extracting CFFs from data on hard exclusive reactions, 
(ii) determining GPDs from CFFs through physically motivated Ans\"atze 
or GPD models, and finally
(iii) using the Mellin moment sum rules in Eq.~(\ref{Eq:GPD-2nd-Mellin-moments}). 
The steps (ii) and (iii) are at present and in the near future not yet possible 
in entirely model-independent ways. More direct approaches can be explored 
to access EMT FFs. One of them is based on a dispersion relation connecting 
the real and imaginary parts on CFFs once they are known from step (i). This
relation involves a subtraction term $\Delta(t,\mu)$ 
\cite{Teryaev:2005uj,Anikin:2007yh,Diehl:2007jb} 
which is expressed in terms of an infinite tower of FFs
related to the highest powers $\xi^{2k}$ in even polynomials
$\int dx\,x^{2k-1}H^q(x,\xi,t)$ and $\int dx\,x^{2k-1}E^q(x,\xi,t)$
\cite{Polyakov:1999gs}. The first term $k=1$ corresponds
to the EMT FF $D^q(t)$ in Eq.~(\ref{Eq:GPD-2nd-Mellin-moments}). 
Presently, to isolate this contribution, further assumptions 
are required, particularly regarding  its flavor dependence and 
the separation from higher-order terms ($k=2,\,3,\,\cdots$).
In the future, when data spanning a wide range of scales  
become available, it will be possible to distinguish the 
contributions $k=1,\,2,\,\cdots$ by exploiting their different 
evolution behavior. In particular, taking the limit $\mu\to\infty$
yields the $D$-term FF \cite{Goeke:2001tz}.
For first explorations of this approach to gain insights
on the quark EMT FF $D^q(t)$ we refer to 
\cite{Burkert:2018bqq,Kumericki:2019ddg}.
A promising approach to access information on the gluon EMT FFs
of the nucleon is threshold photoproduction of $J/\psi$ or $\Upsilon$ 
\cite{Kharzeev:2021qkd,Guo:2021ibg,Sun:2021gmi}.  (A different  phenomenological description of the latter process in terms of hadronic amplitudes  was proposed  in Ref.~\cite{Du:2020bqj}.)  For a first and, at 
the present stage, still model-dependent determination of the gluon EMT FFs based
on data we refer to \cite{Duran:2022xag}.

\subsection{GPDs for other hadrons and transition GPDs}
\label{Sec-4.5-GDS-other-hadrons-and-transitions}

In the following, we briefly discuss exclusive processes that 
allow us to extend the GPD framework to other hadrons, including 
nuclei, pions, and excited baryon states.

DVCS on nuclei can proceed via two mechanisms: coherent DVCS, where the scattered nucleus remains intact, providing access to the GPDs of the entire nucleus, 
and incoherent DVCS, where an initially bound nucleon is knocked out from the nucleus and detected in the final state, probing the in-medium nucleon GPDs. 
The combined analyses of hard exclusive reactions on free protons, nuclei, and bound protons enables the investigation of the $t$-dependence (and through the GPDs in impact-parameter space, the spatial distribution in the transverse plane) of medium-induced modifications to the parton structure of bound nucleons, which underlies also the nuclear modifications observed in DIS; see Fig.~\ref{fig4}. 
Such analyses provide stringent tests for the state-of-the-art nuclear models; we refer to~\cite{Dupre:2015jha} for a comprehensive review. 
The first measurements of nuclear GPDs were performed by the HERMES collaboration at DESY~\cite{HERMES:2009xsg}, where a direct distinction of coherent and incoherent channels was not yet possible.
More recently, the CLAS collaboration successfully achieved this separation, although with limited coverage in $x_B$ and $t$~\cite{CLAS:2017udk,CLAS:2018ddh}.
Significant experimental progress is expected 
% with data from JLab, following the 12 GeV upgrade.
with JLab12 data.
Looking further ahead, the EIC  will provide an ideal setting for studying nuclear DVCS, thanks to its collider kinematics, which facilitate the detection of recoiling nuclei and enable the polarization of incoming nuclear beams~\cite{AbdulKhalek:2021gbh}. 

Studies based on simulations for the kinematics of the future electron-ion colliders in the US and China~\cite{Chavez:2021koz,Arrington:2021biu} have also demonstrated the potential to access (off-shell) pion GPDs and, in turn, pion EMT FFs~\cite{Hatta:2025ryj}, via the Sullivan process. In this mechanism, an electron scatters off a virtual pion from the nucleon's pion cloud through the exchange of a virtual photon, while a real photon is emitted in the final state via the reaction  $\gamma^*\pi \rightarrow \gamma \pi$. Generally, EMT FFs in the timelike region $t>0$ of stable and unstable hadrons are accessible via studies of generalized distribution amplitudes in hadron-antihadron production in $e^+e^-$ annihilation processes~\cite{Diehl:1998dk}. This process was explored
to determine the EMT FFs of the $\pi^0$ \cite{Kumano:2017lhr}. 

Finally, we highlight hard exclusive processes involving quasi-elastic transitions on the hadronic side, such as DVCS or hard meson production with the transition $N\rightarrow B$, where $B$ is a different baryon or even a multi-hadron state.
Transition GPDs provide new tools for studying the resonance structure and give in addition access to the transition matrix elements of the QCD EMT, complementing the study of transition FFs.
A summary of current and planned activities on transition GPD studies can be found in Ref.~\cite{Diehl:2024bmd}, which also provides an overview of the evolving experimental program across several facilities.

\section{Generalized transverse momentum dependent parton distributions}\label{Sec-5:GTMDs}  

GTMDs were introduced as a unified framework that simultaneously extends both GPDs and TMDs~\cite{Meissner:2008ay, Meissner:2009ww}, based on a suggestion made in earlier works~\cite{Ji:2003ak,Belitsky:2003nz}. 
In the following, we briefly outline their key properties and review the main processes proposed to access them experimentally. We then discuss their connection to Wigner distributions in QCD, which represent the closest quantum analogue to classical phase-space densities. Finally, we emphasize the important role of GTMDs in providing access to the partonic orbital angular momentum.

\subsection{Definition and potential processes}
Quark GTMDs parameterize the parton correlator defined in Eq.~\eqref{GTMD-correlator}. 
For example, the leading-power GTMD correlator for unpolarized quarks is parameterized according to~\cite{Meissner:2008ay,Meissner:2009ww}
\begin{equation}
    \Phi_{\rm GTMD}^{q[\gamma^+]}(P,x,\uvec k_\perp,\xi,\uvec\Delta_\perp)=\frac{1}{2M}\,\bar u(p',S')\left[F_{1,1}^q+\frac{i\sigma^{i+}k^i_\perp}{P^+}\,F_{1,2}^q+\frac{i\sigma^{i+}\Delta^i_\perp}{P^+}\,F_{1,3}^q+\frac{i\sigma^{ij}k^i_\perp\Delta^j_\perp}{M^2}\,F_{1,4}^q\right]u(p,S) \,,
\end{equation}
where it is understood that $F_{1,i}^q=F_{1,i}^q(x,\xi,\uvec k_\perp^2,\uvec k_\perp\cdot\uvec \Delta_\perp,\uvec\Delta_\perp^2)$.
Subleading quark GTMDs have been classified as well~\cite{Meissner:2008ay,Meissner:2009ww}, alongside their gluon counterparts~\cite{Lorce:2013pza}. These distributions encode the most general information on one-body parton dynamics, representing the fully unintegrated, off-forward amplitude of a single quark (or gluon) in momentum space and at fixed light-front time.
In general, GTMDs are complex-valued functions. Their imaginary parts arise from initial- and final-state interactions, and play an essential role in generating naive \textsf{T}-odd observables, analogous to what occurs in the case of TMDs. The evolution properties of GTMDs and TMDs are closely related, because they are driven by the same QCD operator~\cite{Echevarria:2016mrc}.
Also, just as certain TMDs at small transverse separation $\uvec z_\perp$ between the parton fields can be matched onto PDFs (see Sec.~\ref{Sec-3.5-TMD-QCD}), some GTMDs under the same conditions can be matched onto GPDs~\cite{Bertone:2025vgy}.

As outlined in Sec.~\ref{sect.1.3}, GTMDs can be regarded as the ``mother'' distributions of both TMDs and GPDs.
However, not all GTMDs reduce to either TMDs or GPDs in the respective limits. 
Those GTMDs quantify correlations that are specific to off-forward kinematics such as spin-orbit couplings.
These ``GTMD-specific'' structures capture unique aspects of parton dynamics that are not contained in the simpler distributions~\cite{Meissner:2008ay,Meissner:2009ww,Kanazawa:2014nha}. The rich structure of GTMDs has motivated considerable theoretical and phenomenological interest in recent years, particularly in identifying observables and processes through which GTMDs may become experimentally accessible.
It was shown in Ref.~\cite{Hatta:2016dxp,Altinoluk:2015dpi} that gluon GTMDs at small $x$ can be probed via hard exclusive diffractive dijet production in DIS.
This study prompted  a series of related investigations~\cite{Zhou:2016rnt,Ji:2016jgn,Hagiwara:2017ofm,Boer:2018vdi,Salazar:2019ncp,Hatta:2016aoc,Mantysaari:2019csc,Boer:2023mip,Boer:2021upt,Bhattacharya:2022vvo,Bhattacharya:2024sck}, all focused on gluon GTMD observables that could potentially be accessed at the EIC.
Gluon GTMDs are expected to play a key role also in the description of exclusive $\pi^0$ production at high energies~\cite{Boussarie:2019vmk,Bhattacharya:2023yvo} and quarkonium production~\cite{Boussarie:2018zwg,Bhattacharya:2018lgm}.
In the quark sector, the first process identified to probe quark GTMDs was the exclusive pion–nucleon double Drell–Yan reaction~\cite{Bhattacharya:2017bvs,Echevarria:2022ztg}. 
In principle, this channel provides access to all leading-power quark GTMDs through suitable polarization observables. 
However, it suffers from an extremely low count rate. 
To address this challenge, exclusive $\pi^0$ production in electron–proton collisions has recently been proposed as an alternative probe~\cite{Bhattacharya:2023hbq}.

\subsection{Wigner distributions}\label{sec.-Wigner_distribution}

The phase-space formulation of quantum mechanics~\cite{Wigner:1932eb,Hillery:1983ms} allows one to describe quantum amplitudes in a way similar to statistical mechanics. If $\Psi(\uvec r)=\int \frac{d^3p}{(2\pi)^3}\,e^{i\uvec p\cdot\uvec r}\,\tilde\Psi(\uvec p)$, with $\tilde\Psi(\uvec p)=\langle p|\Psi\rangle/\sqrt{2p^0}$, is the position-space wave packet associated with the normalized physical state $|\Psi\rangle$, the expectation value of any operator $O$ can be expressed as
\begin{equation}
    \langle\Psi|O|\Psi\rangle=\int d^3R\int\frac{d^3P}{(2\pi)^3}\,\rho_\Psi(\uvec R,\uvec P)\,\langle O\rangle_{\uvec R,\uvec P} \,,
\end{equation}
where the real-valued Wigner distribution
\begin{equation}
    \rho_\Psi(\uvec R,\uvec P)=\int d^3z\,e^{-i\uvec P\cdot\uvec z}\,\Psi^*(\uvec R-\tfrac{\uvec z}{2})\Psi(\uvec R+\tfrac{\uvec z}{2})=\int\frac{d^3q}{(2\pi)^3}\,e^{-i\uvec q\cdot\uvec R}\,\tilde\Psi^*(\uvec P+\tfrac{\uvec q}{2})\tilde\Psi(\uvec P-\tfrac{\uvec q}{2})
\end{equation}
is the quantum analogue of the classical phase-space distribution. Due to Heisenberg's uncertainty relation, the Wigner distribution is in general not positive definite and admits therefore only a quasi-probabilistic interpretation. However, a probabilistic interpretation is recovered once one integrates over positions or momenta,
\begin{equation}
    \int \frac{d^3P}{(2\pi)^3}\,\rho_\Psi(\uvec R,\uvec P)=|\Psi(\uvec R)|^2\,, \qquad \int d^3R\,\rho_\Psi(\uvec R,\uvec P)=|\tilde\Psi(\uvec P)|^2 \,.
\end{equation}
The phase-space amplitude
\begin{equation}\label{Phasespaceampl}
    \langle O\rangle_{\uvec R,\uvec P}=\int\frac{d^3\Delta}{(2\pi)^3}\,e^{i\uvec\Delta\cdot\uvec R}\,\frac{\langle P+\tfrac{\Delta}{2}|O|P-\tfrac{\Delta}{2}\rangle}{\sqrt{4(P^0)^2-(\Delta^0)^2}}
\end{equation}
can be interpreted as the expectation value of the operator $O$ in a state ``localized'' in phase-space around the average position $\uvec R$ and average momentum $\uvec P$.

The phase-space formalism extends naturally to quantum field theory~\cite{Carruthers:1975sp,Bialynicki-Birula:1991jwl}, where one defines the quantum Wigner operator for fermionic fields as
\begin{equation}
    W^{q[\Gamma]}(r,k)=\int\frac{d^4z}{(2\pi)^4}\,e^{ik\cdot z}\,\,\bar\psi_q(r-\tfrac{z}{2})\;\Gamma\,{\cal W}[-\tfrac{z}{2},\tfrac{z}{2}]\;\psi_q(r+\tfrac{z}{2}) \,.
\end{equation}
Since PDF, TMD and GPD correlators are defined in terms of field operators of this form, it has been suggested in Refs.~\cite{Ji:2003ak,Belitsky:2003nz} that the corresponding distributions in position or momentum space can be regarded as particular projections of some mother distribution, namely a partonic Wigner distribution; see Fig.~\ref{Fig:functions+interpretation}. 

Similarly to the density interpretation of GPDs in impact-parameter space, the 2D Fourier transform of the GTMD correlator at $\xi=0$,
\begin{equation}
    \rho^{q[\Gamma]}(x,\uvec k_\perp,\uvec b_\perp)=\int\frac{d^2\Delta_\perp}{(2\pi)^2}\,e^{-i\uvec\Delta_\perp\cdot\uvec b_\perp}\,\Phi_{\rm GTMD}^{q[\Gamma]}(x,\uvec k_\perp,0,\uvec\Delta_\perp) =P^+\int dk^+dk^-\,\langle W^{q[\Gamma]}(r,k)\rangle_{\uvec R_\perp,P^+,\uvec 0_\perp}\,\delta(k^+-xP^+) \,,
\end{equation}
with $\uvec b_\perp=\uvec r_\perp-\uvec R_\perp$, can be interpreted as a light-front Wigner distribution~\cite{Lorce:2011kd}. The matrix element
\begin{equation}
    \langle O\rangle_{\uvec R_\perp,P^+,\uvec P_\perp}=\int\frac{d^2\Delta_\perp}{(2\pi)^2}\,e^{i\uvec\Delta_\perp\cdot\uvec R_\perp}\,\frac{\langle P^+,\uvec P_\perp+\tfrac{\uvec\Delta_\perp}{2}|O|P^+,\uvec P_\perp-\tfrac{\uvec\Delta_\perp}{2}\rangle}{2P^+}
\end{equation}
is the light-front version of Eq.~\eqref{Phasespaceampl} and indicates that the system is ``localized'' around the average transverse position $\uvec R_\perp$ (associated with the center of $P^+$) and average light-front momentum $(P^+,\uvec P_\perp)$~\cite{Lorce:2018egm}.

In total, there are 32 quark phase-space distributions at leading power for a spin-$\frac12$ hadron, half of which originate from the imaginary part of GTMDs. They encode all possible correlations
between nucleon spin ($\vec S$), quark spin ($\vec S^q$) and quark orbital angular momentum ($\vec \ell^q$)~\cite{Lorce:2015sqe}, as summarized in Table~\ref{table-wigner}. Each correlation receives both naive \textsf{T}-even and \textsf{T}-odd contributions.
As illustrated in Fig~\ref{Fig:functions+interpretation}, when integrated over $\vec b_\perp$, some of these Wigner distributions reduce to TMDs. Similarly, when integrated over $\vec k_\perp$, some of them yield GPDs in impact-parameter space, with only naive \textsf{T}-even contributions surviving in this limit. 
We emphasize that these connections hold at the bare (i.e.,~unrenormalized) level, and must be carefully revisited for renormalized quantities; see discussion in Sec.~\ref{Sec-3.5-TMD-QCD}.

\begin{table*}[h]
\centering
\caption{\footnotesize \label{tab:correlation}Correlation matrix of $\rho_{\vec S \vec S^q} $ between various nucleon polarization states (rows) and quark polarization structures (columns). \vspace{0.5cm} \label{table-wigner}}
\begin{tabular}{c|ccccc}
$\rho_{\vec S \vec S^q} $ & $U$ & $L$ & $T_x$ & $T_y$ \\
\colrule
$U$ & $\langle 1 \rangle$ & $\langle S^q_L \ell_L^q \rangle$ & $\langle S_x^q \ell_x^q \rangle$ & $\langle S_y^q \ell_y^q \rangle$ \\
$L$ & $\langle S_L \ell_L^q \rangle$ & $\langle S_L S^q_L \rangle$ & $\langle S_L \ell_L^q S_x^q \ell_x^q \rangle$ & $\langle S_L \ell_L^q S_y^q \ell_y^q \rangle$ \\
$T_x$ & $\langle S_x \ell_x^q \rangle$ & $\langle S_x \ell_x^q S^q_L \ell_L^q \rangle$ & $\langle S_x S_x^q \rangle$ & $\langle S_x \ell_x^q S_y^q \ell_y^q \rangle$ \\
$T_y$ & $\langle S_y \ell_y^q \rangle$ & $\langle S_y \ell_y^q S^q_L \ell_L^q \rangle$ & $\langle S_y \ell_y^q S_x^q \ell_x^q \rangle$ & $\langle S_y S_y^q \rangle$\\
\botrule
\end{tabular}
\end{table*}

\subsection{Orbital angular momentum}

\begin{figure}[t]
\centering
\includegraphics[width=0.35\textwidth]{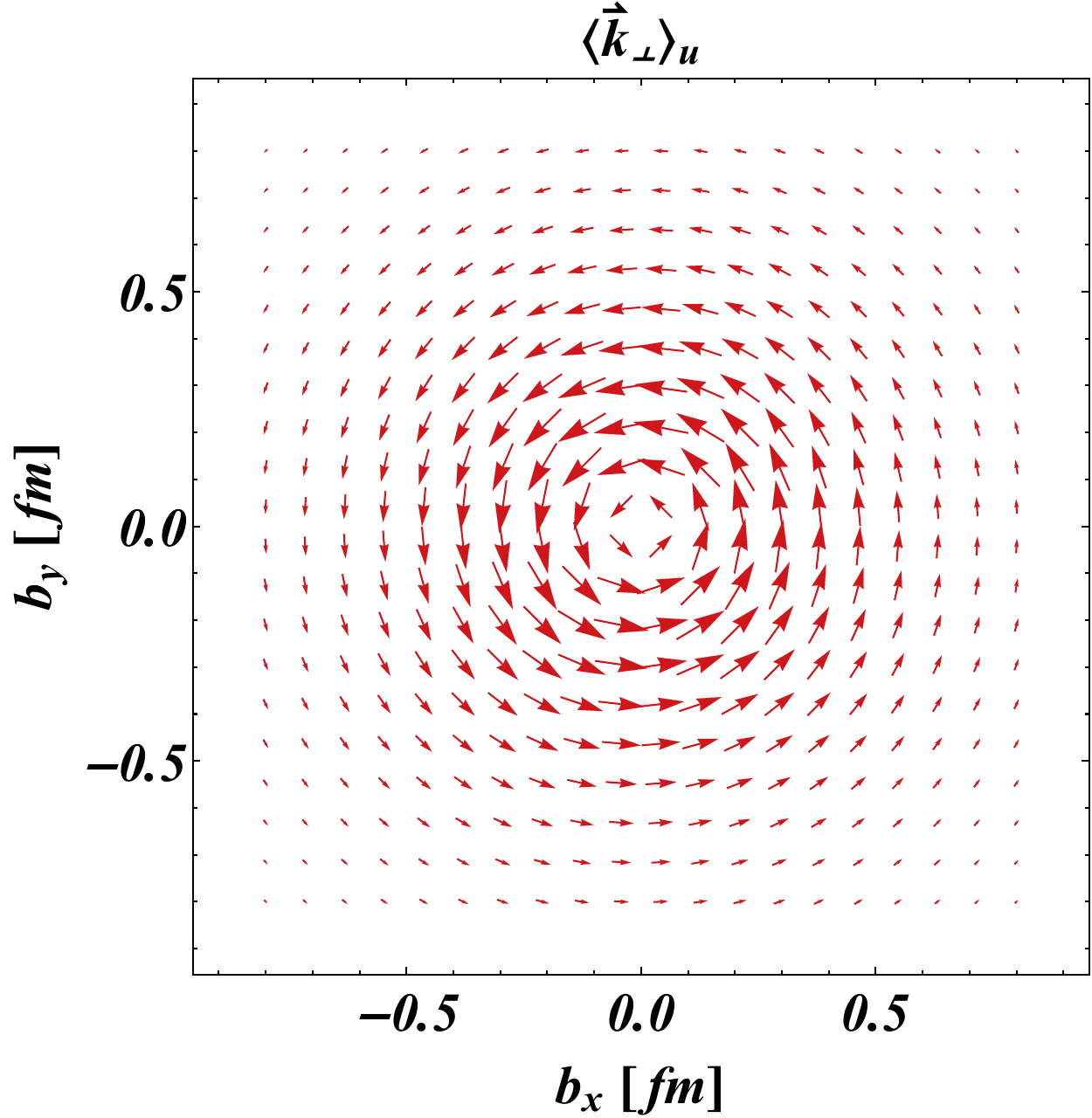}
\hspace{1cm}
\includegraphics[width=0.35\textwidth]{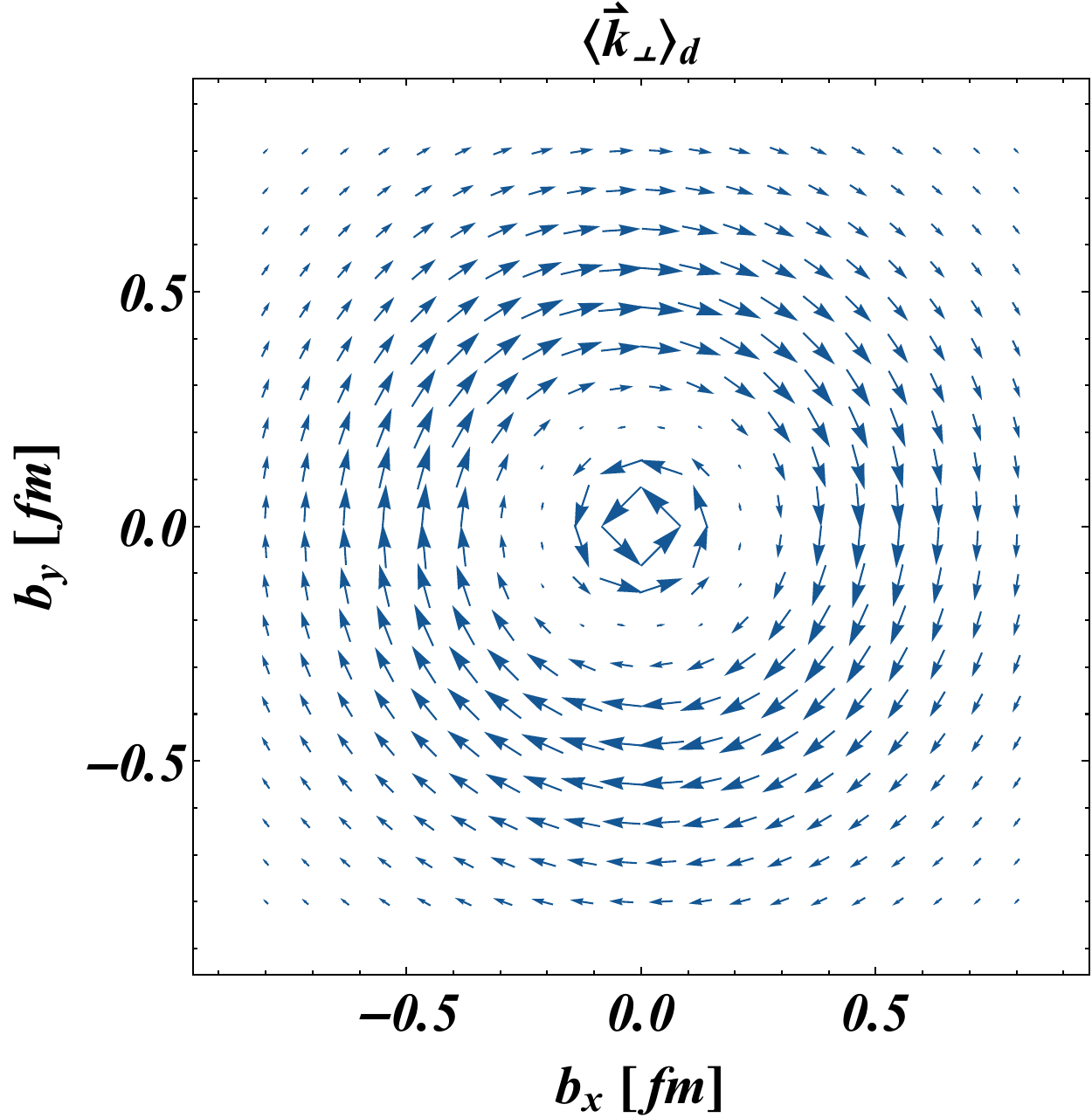}
\caption{\footnotesize Distributions in impact-parameter space of the quark mean transverse momentum, $\langle \uvec k_\perp\rangle_q(\uvec b_\perp)=\int d x\,d^2k_\perp\,\uvec k_\perp\,\rho^q_{LU}(x,\uvec k_\perp,\uvec b_\perp)$, based on a light-front constituent quark model~\cite{Lorce:2011ni}. The nucleon polarization is pointing out of the plane
and the arrows show the size and direction of the mean transverse momentum for $u$ (left panel) and $d$ (right panel) flavors. The total OAM of $u$ ($d$) quarks, $\ell^q_z=\int d^2b_\perp\,\uvec b_\perp\times\langle\uvec k_\perp\rangle_q$, is positive (negative) .\label{fig:OAM}}
\end{figure}

Orbital angular momentum (OAM) measures a correlation between position and momentum coordinates, and is therefore naturally described in terms of phase-space correlations~\cite{Shin:1992nj}. In particular, it has been observed that a specific GTMD, denoted $F^a_{1,4}$, is directly related to the quark or gluon OAM~\cite{Lorce:2011kd,Hatta:2011ku}:
\begin{equation}
    \ell^a_z=\int d x\,d^2k_\perp\int d^2b_\perp\,(\uvec b_\perp\times\uvec k_\perp)_z\,\rho^a_{LU}(x,\uvec k_\perp,\uvec b_\perp)=-\int d x\,d^2k_\perp\,\frac{\uvec k_\perp^2}{M^2}\,F^a_{1,4}(x,0,\uvec k^2_\perp,0,0) \,;\label{wigner-oam}
\end{equation}
see illustration in Fig.~\ref{fig:OAM}. Similarly, another GTMD, $G^a_{1,1}$, measures the spin-orbit correlation in an unpolarized nucleon~\cite{Lorce:2011kd}:
\begin{equation}
     C^a_z=\int d x\,d^2k_\perp\int d^2b_\perp\,(\uvec b_\perp\times\uvec k_\perp)_z\,\rho^a_{UL}(x,\uvec k_\perp,\uvec b_\perp)=\int d x\,d^2k_\perp\,\frac{\uvec k_\perp^2}{M^2}\,G^a_{1,1}(x,0,\uvec k^2_\perp,0,0) \,.
\end{equation}

The OAM depends on the path of the Wilson line in the definition of the GTMD $F_{1,4}^a$. When a straight gauge link is used in the quark sector, one obtains the kinetic OAM corresponding to the Ji decomposition (see Sec.~\ref{Sec-GPD-definition-and-all-that})~\cite{Ji:1996ek,Ji:2012sj,Ji:2012ba}. 
In contrast, a staple-shaped gauge link leads to the canonical OAM, as defined by Jaffe and Manohar~\cite{Jaffe:1989jz}.
Lattice-QCD calculations make it possible to interpolate between these two definitions by evaluating explicitly the continuous gauge-link path~\cite{Engelhardt:2017miy,Engelhardt:2020qtg}. This is achieved by varying the length of the staple, allowing for a continuous transition from the straight (Ji) to the infinitely long staple (Jaffe-Manohar) configuration. 
The lattice results for the $u-d$ flavor combination indicate that the Jaffe-Manohar OAM is larger than the Ji OAM.
A semi-classical explanation for the difference between the Jaffe-Manohar and Ji OAM was provided in Ref.~\cite{Burkardt:2012sd}. This discussion on the path dependence applies in a similar way to the spin-orbit correlation.

\section{Theoretical approaches}\label{sec6} 

In this section we give a short overview of some non-perturbative
approaches used to study the structure of hadrons, PDFs and their
generalizations with focus on the large-$N_c$ limit, models, effective theories and lattice QCD.

\subsection{\boldmath Large-$N_c$ limit in QCD}
\label{Sec-6.2-large-Nc}

In the hypothetical limit $N_c\to\infty$ in QCD,
baryons become classical solitons of meson fields \cite{Witten:1979kh}.
Also in this limit one cannot solve QCD in the non-perturbative regime.
However, the known spin-flavor symmetries of the baryon solutions can be 
explored to make predictions. Insightful examples are the anomalous 
magnetic moments of proton and neutron given by, respectively,
${\kappa_p = \frac23\,\kappa_p^u - \frac13\kappa_p^d}$ and
${\kappa_n = \frac23\,\kappa_n^u - \frac13\kappa_n^d}$
(where we neglect small strangeness and isospin symmetry breaking effects),
that yield with the values for $\kappa_p$ and $\kappa_n$ 
quoted in Sec.~\ref{Sec-1.1-pre-parton} the results
${\kappa^u \equiv \kappa_p^u = \kappa_n^d = 1.67}$ and 
${\kappa^d \equiv \kappa_p^d = \kappa_n^u = - 2.03}$.  
Large-$N_c$ QCD predicts $\kappa^u$ and $\kappa^d$ to have 
opposite signs and the same magnitude up to $1/N_c$-corrections.
This can be stated as 
$|\kappa^u+\kappa^d|/|\kappa^u-\kappa^d|={\cal O}(1/N_c)\ll 1$ and 
agrees with experiments; for more examples see \cite{Dashen:1993jt}. 
Baryon masses scale as $M\sim N_c$ while their sizes 
(e.g.~charge, mass, and mechanical radii)
scale as $N_c^0$, such that in the large-$N_c$ limit the recoil corrections discussed in Sec.~\ref{sec:IPD} become negligible. 
This implies that the 3D density interpretations of electromagnetic
(Sec.~\ref{Sec-1.1-pre-parton}) or EMT FFs
(Sec.~\ref{Sec-EMT-FFs}) become exact in the large-$N_c$ limit.
The 2D density interpretation is exact and valid for any hadron
including baryons with or without large-$N_c$ limit.

In the large-$N_c$ limit, the flavor structure of parton 
distributions of the nucleon can be predicted for $x$-values such that 
$N_cx={\cal O}(N_c^0)$~\cite{Pobylitsa:2000tt}, i.e.,\ the results are valid 
at intermediate $x$ and do not apply to the $x\to0$ or $x\to1$ limits.
Interesting results include the prediction of a large flavor asymmetry in the helicity 
sea $|(g_1^{\bar u}-g_1^{\bar d})(x)| \sim N_c^2 
 \gg |(g_1^{\bar u}+g_1^{\bar d})(x)| \sim N_c$ \cite{Diakonov:1996sr} 
or the prediction that Sivers TMDs for $u$ and $d$ quarks have the 
same magnitude and opposite signs in the large-$N_c$ limit,
${f_{1T}^{\perp u}(x,k_\perp) \approx - f_{1T}^{\perp d}(x,k_\perp)}$;
see Ref.~\cite{Pobylitsa:2000tt}.
In Ref.~\cite{Efremov:2000ar}, the large-$N_c$ prediction about 
gluon PDFs $g_1^g(x)/f_1^g(x)\sim {\cal O}(1/N_c) \ll 1$ was made.
All these predictions are supported by global fits; see 
Sec.~\ref{Sec-2.5-global-fits} and \ref{Sec-3.6-TMD-extractions}.
The large-$N_c$ approach can also relate EMT FFs
of nucleon and $\Delta$-resonance~\cite{Panteleeva:2020ejw}.

\subsection{Models}
\label{Sec6-models}

Models of hadron structure make different simplifying assumptions
about parton dynamics but typically share enough essential features
with QCD to provide meaningful insight for specific purposes. 
They can be used as exploratory tools for investigations and even 
discoveries of new phenomena, or as practical approximations to QCD 
to produce quantitative cross section estimates to help guide 
interpretations of data or prepare new experiments. 
In the following, we present a small selection of
results from the rich literature on model studies.

Soon after the first DIS measurements, the bag model was constructed
\cite{Chodos:1974je} and applied to studies of structure functions
\cite{Jaffe:1974nj}. In this model, three non-interacting quarks are
confined inside a spherical cavity by means of appropriate boundary
conditions. The model has been used through the years for numerous
exploratory studies including transversity \cite{Jaffe:1991ra}, 
 GPDs \cite{Ji:1997gm}, TMDs \cite{Avakian:2010br}, GTMDs 
 \cite{Courtoy:2016des}, EMT distributions \cite{Lorce:2022cle}, 
and a pioneering study of Wigner functions \cite{Shin:1992nj} 
that was far ahead of its time. 
One prominent result derived from the bag model is that
$h_1^q(x)$ and $g_1^q(x)$ become equal in the non-relativistic
limit at a low hadronic scale. 

The status of naive \textsf{T}-odd TMDs was unclear in QCD for some
time, when the Sivers function was proposed \cite{Sivers:1989cc} 
as a candidate to explain transverse SSAs, as \textsf{T}-odd TMDs 
were thought to be forbidden in QCD by time-reversal symmetry \cite{Collins:1992kk}. 
A spectator model study helped to clarify the issue. 
In the simplest version of the model, the on-shell spectator diquark system is 
treated as an elementary spin-0 particle and the nucleon-quark-diquark 
interaction is assumed to be point-like.  
Using this setting, Ref.~\cite{Brodsky:2002cx} calculated 
the rescattering of the quark struck by the virtual photon
with the spectator diquark by means of a one-gluon exchange; 
see Fig.~\ref{Fig:sign_change}(b) with the green ``half-blob'' 
understood as a diquark line. For a transversely polarized nucleon, 
this process generates a leading-order SSA \cite{Brodsky:2002cx}. 
Subsequently, it was recognized that this rescattering mechanism 
is entailed in the 
Wilson line in the TMD definition,
and the calculation of Ref.~\cite{Brodsky:2002cx} was de facto a model 
for the Sivers function which clarified that naive \textsf{T}-odd TMDs
are indeed allowed in QCD  
\cite{Collins:2002kn}. This model has also been used to investigate 
fundamental properties such as factorization and universality 
\cite{Collins:2007nk,Rogers:2010dm} as well as the positivity of PDFs \cite{Collins:2021vke}.
The modeling of hadron properties in the diquark picture has 
a long history \cite{Anselmino:1992vg,Barabanov:2020jvn} and has 
been explored in a variety of model versions for studies of partonic functions
\cite{Brodsky:2002cx,Jakob:1997wg,Gamberg:2003ey,Meissner:2007rx,Bacchetta:2008af}.

Light-front constituent quark models are based on a light-front
Fock expansion of the nucleon state. In its simplest version, 
the nucleon structure is modeled in terms of 3-quark or quark-diquark
light-front wave functions which encode the non-perturbative
information on the nucleon. The approach has been applied to
studies of FFs, PDFs, TMDs, GPDs, GTMDs, and Wigner functions, see e.g.~Refs.~\cite{Diehl:2000xz,Boffi:2002yy,Pasquini:2007iz,Pasquini:2007xz,Dahiya:2007is,Pasquini:2008ax,Pasquini:2010af,Lorce:2011dv,Lorce:2011ni,Lorce:2011zta,Mukherjee:2014nya,Kumar:2015fta,Choudhary:2022den,Han:2022tlh,Chakrabarti:2023djs},
providing a convenient representation of these distributions as overlap integrals of light-front wave functions corresponding to different partonic configurations. These integrals span different momentum regions, thus emphasizing the various partonic correlations encoded in each distribution.

The chiral quark soliton model is based on a low-energy chiral theory
describing the interaction of effective quark and antiquark degrees of 
freedom with Goldstone bosons of the spontaneous chiral symmetry breaking,
which is solved in the large-$N_c$ limit. The model describes consistently
quark and antiquark distributions in the nucleon with considerable success.
For instance, it naturally explains (without adjustable parameters) the
observed flavor asymmetry in the unpolarized sea $(f_1^{\bar u}-f_1^{\bar d})(x)<0$ 
\cite{Pobylitsa:1998tk}. Moreover, it predicted an even larger flavor
asymmetry in the helicity sea $(g_1^{\bar u}-g_1^{\bar d})(x)>0$
and its impact on spin asymmetries in 
$W^\pm$ 
production
\cite{Diakonov:1996sr,Dressler:1999zv}, which has been confirmed
by recent RHIC data and extractions. The model results for GPDs 
comply with polynomiality \cite{Schweitzer:2002nm}. This was 
also the first model to visualize the EMT distributions and pressure 
in the nucleon \cite{Goeke:2007fp}; see Fig.~\ref{Fig17}.
Another interesting prediction is that sea quark distributions
$f_1^{\bar q}(x,k_\perp)$ and $g_1^{\bar q}(x,k_\perp)$ fall off
with increasing $k_\perp$ significantly more slowly than the corresponding
valence quark distributions \cite{Wakamatsu:2009fn,Schweitzer:2012hh}
as a consequence of chiral symmetry breaking \cite{Schweitzer:2012hh}.
This also appears to be confirmed by  TMD extractions 
\cite{Bacchetta:2024qre}.
In addition, tensor charges \cite{Kim:1995bq,Lorce:2007fa} and
chiral-odd GPDs \cite{Kim:2024ibz} were studied in this model.
Model-independent insights into nucleon structure derived from 
chiral dynamics can be obtained for FFs and GPDs by exploring 
the impact-parameter space picture at transverse distances 
$b_\perp\sim {\cal O}(1/m_\pi)$, where $m_\pi$ denotes the pion 
mass. For studies in effective chiral theory and applications of 
this approach we refer to~\cite{Strikman:2003gz, Strikman:2009bd, 
Granados:2013moa, Granados:2015rra, Alarcon:2017lhg, Alarcon:2018zbz, Alarcon:2022adi}.

The quark target model is based on the QCD Lagrangian 
for a single flavor which is solved perturbatively.
With the spectrum of the model consisting of colored states,
it is not intended as a realistic model for hadron structure.
However, it is an interesting theoretical ground for uncovering or 
testing, e.g., relations among TMDs and GPDs \cite{Meissner:2007rx}. 
In many models, the modeling of naive \textsf{T}-odd TMDs requires 
explicit gauge field degrees of freedom, which must be introduced
through the implementation of the one-gluon exchange mechanism 
\cite{Brodsky:2002cx}. In contrast, the gauge field degrees
of freedom are present to start with in the quark target model.

Further interesting theoretical approaches to hadron structure
include dispersion relations \cite{Pasquini:2014vua,Cao:2024zlf,Cao:2025dkv,Martinez-Fernandez:2025jvk},
Nambu-Jona-Lasinio model \cite{Cloet:2007em,Freese:2019bhb}, 
Dyson-Schwinger-equation models \cite{Hecht:2000xa,Raya:2015gva},
cloudy bag model \cite{Owa:2021hnj},
holographic AdS/QCD models \cite{Brodsky:2008pf,Abidin:2009hr,Chakrabarti:2015lba,Mamo:2019mka,Mamo:2024vjh,Mamo:2024jwp},
QCD sum rule approaches \cite{Anikin:2019kwi,Azizi:2019ytx}
to name just a few; see Refs.~\cite{Burkert:2023wzr,Boussarie:2023izj,Boffi:2007yc} 
for comprehensive reviews of models and other theoretical 
approaches.

\subsection{Lattice QCD}\label{sec6:seubsec2}  

Lattice QCD, where one solves the theory in a discretized Euclidean space-time in a finite volume, offers unique opportunities to compute the non-perturbative structure of hadrons from first principles in QCD; see, for example, Ref.~\cite{Gattringer:2010zz}.
However, this machinery (formulated in Euclidean space) cannot provide direct information about light-front correlation functions (formulated in Minkowskian space) which define the different types of parton distributions.
On the other hand, one can (i) compute Mellin moments of parton distributions and (ii) approximately reconstruct their dependence on the parton momentum.
In the following, we briefly discuss both of these areas, 
while more information can be found in~\cite{Constantinou:2020hdm, FlavourLatticeAveragingGroupFLAG:2024oxs} and references therein.
The field of lattice-QCD studies of hadron structure has significantly advanced in recent years.
In fact, in some cases, lattice results have outpaced results that can currently be extracted from experiment. 

\begin{figure}[t]
\centering
\includegraphics[width=0.35\textwidth]{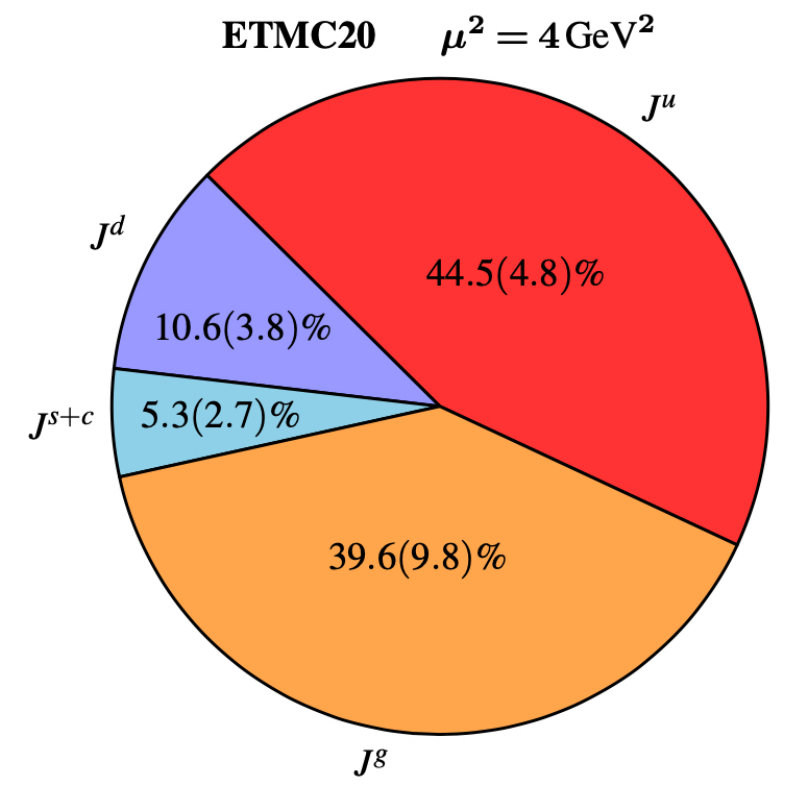} \hspace{1.0cm}
\includegraphics[width=0.43\textwidth]{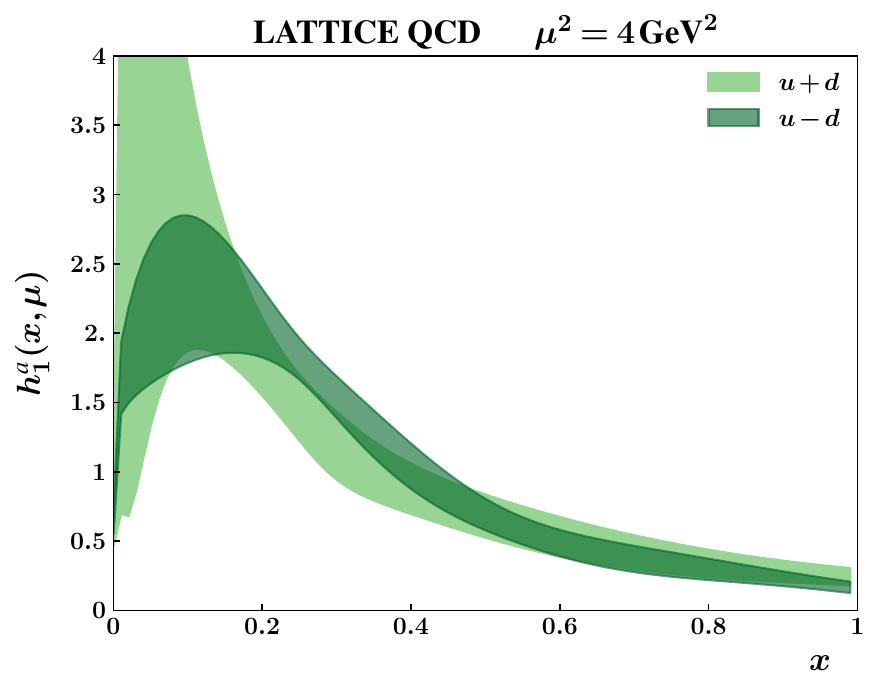}
\caption{\footnotesize Left panel: Parton angular momenta $J^a$ for the proton, along with their uncertainties, obtained from the lattice-QCD calculation discussed in Ref.~\cite{Alexandrou:2020sml}. 
The results from that paper have been rescaled as the central values there just add up to $94.6\%$.
Note that Ji's spin sum rule was not imposed as a constraint in Ref.~\cite{Alexandrou:2020sml}.
Right panel: Lattice-QCD results for the transversity distribution of the proton as a function of $x$ for the $u-d$ and $u+d$ flavor combinations from Ref.~\cite{Gao:2023ktu}, obtained in the quasi-PDF approach.
Shown are the ``NLO+LRR'' results from that paper, where ``NLO+LRR'' corresponds to a particular way of performing the matching between the quasi-PDFs and the PDFs.
The calculations have been carried out at the physical pion mass and for the nucleon momentum $|\uvec{P}\,| = 1.53 \, \textrm{GeV}$.
Generally, results from the quasi-PDF approach are most reliable for $0.2 \lesssim x \lesssim 0.8$~\cite{Gao:2023ktu}.}
\label{f:lattice}
\end{figure}

There is a long history of lattice-QCD calculations of Mellin moments of parton distributions~\cite{Hagler:2009ni}.
Those include, for instance, the lowest moment of the quark helicity PDF (axial charge) and transversity PDF (tensor charge) in Eq.~\eqref{e:moments}, and the second moment of the unpolarized PDFs in Eq.~\eqref{e:momentum_fractions}.
Such moments are defined through local operators.
Disconnected diagrams are challenging to calculate and are not always included in lattice calculations. 
They drop out from the $u-d$ flavor combination, but are needed for the complete calculation of the $u+d$ combination.
Many lattice results exist, for instance, for the tensor charges of the proton~\cite{FlavourLatticeAveragingGroupFLAG:2024oxs}, with reported uncertainties that are (much) smaller than for results extracted from experiment~\cite{Goldstein:2014aja, Radici:2018iag, Gamberg:2022kdb, Cocuzza:2023oam}.  
While combining information from experiment and lattice QCD has to be done with care, we mention that lattice results for the tensor charges have been successfully incorporated in global transversity fits~\cite{Lin:2017stx, Gamberg:2022kdb, Cocuzza:2023oam}.
Another important example of a local operator is the second moment of the GPDs $H$ and $E$ in Eq.~\eqref{ji-sumrule} that gives the angular momentum of partons $J^a$ according to Ji's spin sum rule in Eq.~\eqref{ji-sumrule}~\cite{Ji:1996ek}.
Lattice results for the $J^a$'s from the Extended Twisted Mass Collaboration (ETMC) are shown in the left panel of Fig.~\ref{f:lattice}.
The numbers indicate that at $\mu^2 = 4 \, \textrm{GeV}^2$, the spin $1/2$ of the proton can largely be described by adding $J^u$ and $J^g$.  

In principle, knowing all $x$-moments of parton distributions allows one also to reconstruct their entire $x$-dependence.
In practice, however, lattice calculations are typically limited to the lowest three moments, while higher moments suffer from complicated operator mixing and rapidly decaying signal-to-noise ratios~\cite{Hagler:2009ni}.
A breakthrough in this field occurred in 2013 in the form of the quasi-distribution approach~\cite{Ji:2013dva}.
Quasi-PDFs, for instance, are defined through non-local Euclidean correlation functions with spacelike separation of the parton fields, in contrast to the (light-cone) PDFs accessible in experiment where the parton fields are separated along the light cone; see, e.g., Eq.~\eqref{e:f1_def}.
Spacelike correlators can be computed in lattice QCD.
The key observation of Ref.~\cite{Ji:2013dva} was that the infrared behavior of quasi-PDFs and PDFs is the same, and differences occur only in the UV region where they can be computed in pQCD~\cite{Ji:2013dva}.
Schematically, the unpolarized PDF $f_1^a$ can be related to the quasi-PDF $\tilde{f}_1^a$ using a (matching) formula of the type
\begin{align}
f_1^a (x, \mu) = \int_{-\infty}^{\infty} \frac{dy}{|y|} \, C 
\left(\frac{x}{y}, \frac{\mu}{y |\uvec{P}\,|}, \alpha_{\rm s}(\mu) \right) \, 
\tilde{f}_1^a(y, |\uvec{P} \,|, \mu) + {\cal O}\left( \frac{M_{\rm had}}{|\uvec{P} \,|} \right) \,,
\label{e:matching}
\end{align}
and likewise for other PDFs, where $C$ is a perturbatively calculable matching coefficient.  
(For brevity we did not indicate mixing between parton types.)
Quasi-PDFs depend on the momentum $|\uvec{P} \,|$ of the hadron, and their support is $y \in \,(- \infty, +\infty)$.
Note the similarity of Eq.~\eqref{e:matching} and the factorization formula in Eq.~\eqref{e:QCD_factorization}.
A major goal of the lattice calculations is to reach as high momenta $|\uvec{P}\,|$ as possible in order to reduce the power corrections that are of ${\cal O}(M_{\rm had}/|\uvec{P}\,|)$.
Earlier works on how to directly address the $x$-dependence of PDFs and related quantities in lattice QCD, such as the ones in Refs.~\cite{Liu:1993cv, Detmold:2005gg, Braun:2007wv}, did not have the same impact in the community as the quasi-PDF approach.
We also note that related additional, and to some extent complementary approaches have been proposed and used for hadron structure studies~\cite{Ma:2014jla, Radyushkin:2017cyf}.
We refer to review-type papers for an overview of the multitude of theory developments and lattice simulations related to PDFs, GPDs and TMDs in this novel field of non-local Euclidean parton correlators~\cite{Cichy:2018mum, Ji:2020ect, Constantinou:2020pek, Cichy:2021lih, Lin:2025hka}.
As one example, we show in the right panel of Fig.~\ref{f:lattice} the results for the $u-d$ and $u+d$ transversity PDFs of the proton from Ref.~\cite{Gao:2023ktu}.
This calculation has been performed for $|\uvec{P} \,| = 1.53 \, \textrm{GeV}$.
Generally, it is very challenging to reach high hadron momenta in such lattice calculations which could raise concerns about large power corrections.
However, in the region $0.2 \lesssim x \lesssim 0.8$, where the present calculations are most reliable~\cite{Gamberg:2014zwa, Bhattacharya:2018zxi, Braun:2018brg}, there is reasonable agreement between the lattice results in Fig.~\ref{f:lattice} and the latest extractions of the transversity PDFs from single-hadron~\cite{Gamberg:2022kdb} and dihadron production  data~\cite{Cocuzza:2023vqs} shown in Fig.~\ref{Fig-13:TMD-extractions-f1-h1}~\cite{Gao:2023ktu}.
This and similar findings in other instances are very encouraging.
We also point out that lattice calculations of the transversity PDFs have been reported by other groups as well~\cite{Chen:2016utp, Alexandrou:2018eet, HadStruc:2021qdf}.

%%%%%%%%%%%%%%%%%%%%%%%%%%%%%%%%%%%%%%%%%
\section{Conclusions}
\label{sec:conclusions}

Our understanding of hadron structure has changed a lot over the past few decades, thanks to both new experimental tools and progress in theory. As shown by the different contributions in this section dedicated to hadron physics, what we learn about hadron structure depends on how the hadron is probed. Different features and internal dynamics become accessible depending on the resolution scale, and various theoretical approaches are needed to describe them. At high resolution, hadrons reveal their partonic content, quarks and gluons, whose interactions are governed by Quantum Chromodynamics (QCD). This picture started to take shape with Feynman's parton model and the early interpretation of Deep Inelastic Scattering (DIS) experiments, which provided the first evidence for point-like constituents inside the proton.   These developments led to a QCD-based description, where parton distribution functions (PDFs) describe hadrons in terms of their quark and gluon content.
In this chapter, we 
summarized the type of information accessible through various classes of parton distributions. 

Parton distribution functions (PDFs) describe how the longitudinal momentum of a fast-moving hadron is shared among its partons. Over the years, PDFs have been studied in great detail. A large amount of high-quality data and global fits have made this area one of the most solid and well-developed parts of hadron structure studies, playing a central role in interpreting high-energy processes at current colliders.

Beyond the one-dimensional picture provided by  PDFs, significant theoretical and experimental efforts have focused on transverse momentum dependent distributions (TMDs) and generalized parton distributions (GPDs). These functions give access to new kinds of information.
TMDs describe how partons move in the transverse plane and give rise to new effects such as transverse single-spin asymmetries, while GPDs connect momentum and spatial information and give access to  angular momentum  as well as to internal mechanical properties of hadrons, such as pressure and force distributions, through their connection to the matrix elements of the energy-momentum tensor.
Although more recent than collinear PDFs, TMDs and GPDs have become key elements in QCD phenomenology, supported by both theoretical work and targeted experimental programs.
We outlined what can be learned from these multi-dimensional distributions and how they complement the information provided by collinear PDFs, summarizing  current theoretical developments and  highlighting key experimental results.

Even more general objects are the generalized TMDs (GTMDs) and Wigner distributions, sometimes referred to as “mother distributions”, which unify and extend the information contained in TMDs and GPDs. These objects capture the full phase-space picture of partons and give the most detailed description we currently have of parton dynamics,
which includes information about orbital angular momentum and spin-orbit correlations. GTMDs are currently subject to intensive studies with a special focus on observables for these quantities.

Future experimental programs, in particular at electron-ion colliders, will provide new opportunities to test and refine these frameworks. Measurements of exclusive, semi-inclusive, and multi-dimensional observables will further constrain GPDs and TMDs, and may provide information about GTMDs. Progress in this direction will rely on the interplay between theoretical developments, phenomenological modeling, and increasingly precise data. Further key input will come from lattice QCD and models.  Overall, the field has  evolved from describing hadrons in one dimension to building a more complete, multi-dimensional picture that includes spatial, spin, and momentum correlations. These developments are gradually leading toward a more complete understanding of the partonic content of hadrons in QCD.

\begin{ack}[Acknowledgments]%
The authors thank A. Bacchetta, C.~Bissolotti, M.~Cerutti, C.~Cocuzza, 
F.~Delcarro, X. Gao, A.~Hanlon, D.~Pitonyak, L.~Polano, M.~Vanderhaeghen, and A.~Vladimirov  
for discussions and assistance in providing the phenomenological 
inputs to produce some of the figures.
The work of A.M.~was supported by the National Science Foundation under the Award No.~2110472 and Award No.~2412792.
The work of P.S.~work was supported by the National Science Foundation under 
the Award No.\ 2412625.
The authors also acknowledge partial support by the U.S. Department of Energy,
Office of Science, Office of Nuclear Physics under the umbrella of the Quark-Gluon Tomography (QGT) Topical
Collaboration with Award DE-SC0023646.
\end{ack}

%%%%%%%%%%%%%%%%%%%%%%%%%%%%%%%%%%%%%%%%%
%% Mandatory: Bibliography using bibtex 
%\bibliographystyle{Numbered-Style} %% for Numbered Reference Style

\begin{thebibliography*}{100}
\providecommand{\bibtype}[1]{}
\providecommand{\url}[1]{{\tt #1}}
\providecommand{\urlprefix}{URL }
\expandafter\ifx\csname urlstyle\endcsname\relax
  \providecommand{\doi}[1]{doi:\discretionary{}{}{}#1}\else
  \providecommand{\doi}{doi:\discretionary{}{}{}\begingroup
  \urlstyle{rm}\Url}\fi
\providecommand{\bibinfo}[2]{#2}
\providecommand{\eprint}[2][]{\url{#2}}
\makeatletter\def\@biblabel#1{\bibinfo{label}{[#1]}}\makeatother

\bibtype{Article}%
\bibitem{Frisch-Stern}
\bibinfo{author}{R. Frisch}, \bibinfo{author}{O. Stern},
  \bibinfo{title}{{\"Uber die magnetische Ablenkung von Wasserstoffmolek\"ulen
  und das magnetische Moment des Protons}}, \bibinfo{journal}{Z. Phys.}
  \bibinfo{volume}{85} (\bibinfo{year}{1933}) \bibinfo{pages}{4--16},
  \bibinfo{doi}{\doi{https://note.org/10.1007/BF01330773}}.

\bibtype{Article}%
\bibitem{Alvarez:1940zz}
\bibinfo{author}{L.~W. Alvarez}, \bibinfo{author}{F. Bloch}, \bibinfo{title}{{A
  Quantitative Determination of the Neutron Moment in Absolute Nuclear
  Magnetons}}, \bibinfo{journal}{Phys. Rev.} \bibinfo{volume}{57}
  (\bibinfo{year}{1940}) \bibinfo{pages}{111--122},
  \bibinfo{doi}{\doi{10.1103/PhysRev.57.111}}.

\bibtype{Article}%
\bibitem{ParticleDataGroup:2022pth}
\bibinfo{author}{R.~L. Workman}, et al. (\bibinfo{collaboration}{Particle Data
  Group}), \bibinfo{title}{{Review of Particle Physics}},
  \bibinfo{journal}{PTEP} \bibinfo{volume}{2022} (\bibinfo{year}{2022})
  \bibinfo{pages}{083C01}, \bibinfo{doi}{\doi{10.1093/ptep/ptac097}}.

\bibtype{Article}%
\bibitem{Mcallister:1956ng}
\bibinfo{author}{R.~W. Mcallister}, \bibinfo{author}{R. Hofstadter},
  \bibinfo{title}{{Elastic Scattering of 188-{MeV} Electrons From the Proton
  and the $\alpha$ Particle}}, \bibinfo{journal}{Phys. Rev.}
  \bibinfo{volume}{102} (\bibinfo{year}{1956}) \bibinfo{pages}{851--856},
  \bibinfo{doi}{\doi{10.1103/PhysRev.102.851}}.

\bibtype{Article}%
\bibitem{Hofstadter:1956qs}
\bibinfo{author}{R. Hofstadter}, \bibinfo{title}{{Electron scattering and
  nuclear structure}}, \bibinfo{journal}{Rev. Mod. Phys.} \bibinfo{volume}{28}
  (\bibinfo{year}{1956}) \bibinfo{pages}{214--254},
  \bibinfo{doi}{\doi{10.1103/RevModPhys.28.214}}.

\bibtype{Article}%
\bibitem{Sachs:1962zzc}
\bibinfo{author}{R.~G. Sachs}, \bibinfo{title}{{High-Energy Behavior of Nucleon
  Electromagnetic Form Factors}}, \bibinfo{journal}{Phys. Rev.}
  \bibinfo{volume}{126} (\bibinfo{year}{1962}) \bibinfo{pages}{2256--2260},
  \bibinfo{doi}{\doi{10.1103/PhysRev.126.2256}}.

\bibtype{Article}%
\bibitem{Burkardt:2000za}
\bibinfo{author}{M. Burkardt}, \bibinfo{title}{{Impact parameter dependent
  parton distributions and off forward parton distributions for $z\rightarrow
  0$}}, \bibinfo{journal}{Phys. Rev. D} \bibinfo{volume}{62}
  (\bibinfo{year}{2000}) \bibinfo{pages}{071503},
  \bibinfo{doi}{\doi{10.1103/PhysRevD.62.071503}}, \eprint{hep-ph/0005108}.

\bibtype{Article}%
\bibitem{Miller:2010nz}
\bibinfo{author}{G.~A. Miller}, \bibinfo{title}{{Transverse Charge Densities}},
  \bibinfo{journal}{Ann. Rev. Nucl. Part. Sci.} \bibinfo{volume}{60}
  (\bibinfo{year}{2010}) \bibinfo{pages}{1--25},
  \bibinfo{doi}{\doi{10.1146/annurev.nucl.012809.104508}}, \eprint{1002.0355}.

\bibtype{Article}%
\bibitem{Punjabi:2015bba}
\bibinfo{author}{V. Punjabi}, \bibinfo{author}{C.~F. Perdrisat},
  \bibinfo{author}{M.~K. Jones}, \bibinfo{author}{E.~J. Brash},
  \bibinfo{author}{C.~E. Carlson}, \bibinfo{title}{{The Structure of the
  Nucleon: Elastic Electromagnetic Form Factors}}, \bibinfo{journal}{Eur. Phys.
  J. A} \bibinfo{volume}{51} (\bibinfo{year}{2015}) \bibinfo{pages}{79},
  \bibinfo{doi}{\doi{10.1140/epja/i2015-15079-x}}, \eprint{1503.01452}.

\bibtype{Article}%
\bibitem{Gao:2021sml}
\bibinfo{author}{H. Gao}, \bibinfo{author}{M. Vanderhaeghen},
  \bibinfo{title}{{The proton charge radius}}, \bibinfo{journal}{Rev. Mod.
  Phys.} \bibinfo{volume}{94} (\bibinfo{number}{1}) (\bibinfo{year}{2022})
  \bibinfo{pages}{015002}, \bibinfo{doi}{\doi{10.1103/RevModPhys.94.015002}},
  \eprint{2105.00571}.

\bibtype{Book}%
\bibitem{Feynman:1972original}
\bibinfo{author}{R.~P. Feynman}, \bibinfo{title}{{Photon-hadron Interactions}},
  \bibinfo{publisher}{W.~A.~Benjamin} \bibinfo{year}{1972}, ISBN
  \bibinfo{isbn}{0-805-32510-7}.

\bibtype{Article}%
\bibitem{Ralston:1979ys}
\bibinfo{author}{J.~P. Ralston}, \bibinfo{author}{D.~E. Soper},
  \bibinfo{title}{{Production of Dimuons from High-Energy Polarized Proton
  Proton Collisions}}, \bibinfo{journal}{Nucl. Phys. B} \bibinfo{volume}{152}
  (\bibinfo{year}{1979}) \bibinfo{pages}{109},
  \bibinfo{doi}{\doi{10.1016/0550-3213(79)90082-8}}.

\bibtype{Article}%
\bibitem{Collins:1981uw}
\bibinfo{author}{J.~C. Collins}, \bibinfo{author}{D.~E. Soper},
  \bibinfo{title}{{Parton Distribution and Decay Functions}},
  \bibinfo{journal}{Nucl. Phys. B} \bibinfo{volume}{194} (\bibinfo{year}{1982})
  \bibinfo{pages}{445--492}, \bibinfo{doi}{\doi{10.1016/0550-3213(82)90021-9}}.

\bibtype{Article}%
\bibitem{Muller:1994ses}
\bibinfo{author}{D. M\"uller}, \bibinfo{author}{D. Robaschik},
  \bibinfo{author}{B. Geyer}, \bibinfo{author}{F.~M. Dittes},
  \bibinfo{author}{J. Ho\v{r}ej\v{s}i}, \bibinfo{title}{{Wave functions,
  evolution equations and evolution kernels from light ray operators of QCD}},
  \bibinfo{journal}{Fortsch. Phys.} \bibinfo{volume}{42} (\bibinfo{year}{1994})
  \bibinfo{pages}{101--141}, \bibinfo{doi}{\doi{10.1002/prop.2190420202}},
  \eprint{hep-ph/9812448}.

\bibtype{Article}%
\bibitem{Ji:1996ek}
\bibinfo{author}{X.-D. Ji}, \bibinfo{title}{{Gauge-Invariant Decomposition of
  Nucleon Spin}}, \bibinfo{journal}{Phys. Rev. Lett.} \bibinfo{volume}{78}
  (\bibinfo{year}{1997}) \bibinfo{pages}{610--613},
  \bibinfo{doi}{\doi{10.1103/PhysRevLett.78.610}}, \eprint{hep-ph/9603249}.

\bibtype{Article}%
\bibitem{Ji:1996nm}
\bibinfo{author}{X.-D. Ji}, \bibinfo{title}{{Deeply virtual Compton
  scattering}}, \bibinfo{journal}{Phys. Rev. D} \bibinfo{volume}{55}
  (\bibinfo{year}{1997}) \bibinfo{pages}{7114--7125},
  \bibinfo{doi}{\doi{10.1103/PhysRevD.55.7114}}, \eprint{hep-ph/9609381}.

\bibtype{Article}%
\bibitem{Radyushkin:1996nd}
\bibinfo{author}{A.~V. Radyushkin}, \bibinfo{title}{{Scaling limit of deeply
  virtual Compton scattering}}, \bibinfo{journal}{Phys. Lett. B}
  \bibinfo{volume}{380} (\bibinfo{year}{1996}) \bibinfo{pages}{417--425},
  \bibinfo{doi}{\doi{10.1016/0370-2693(96)00528-X}}, \eprint{hep-ph/9604317}.

\bibtype{Article}%
\bibitem{Radyushkin:1996ru}
\bibinfo{author}{A.~V. Radyushkin}, \bibinfo{title}{{Asymmetric gluon
  distributions and hard diffractive electroproduction}},
  \bibinfo{journal}{Phys. Lett. B} \bibinfo{volume}{385} (\bibinfo{year}{1996})
  \bibinfo{pages}{333--342}, \bibinfo{doi}{\doi{10.1016/0370-2693(96)00844-1}},
  \eprint{hep-ph/9605431}.

\bibtype{Article}%
\bibitem{Collins:1996fb}
\bibinfo{author}{J.~C. Collins}, \bibinfo{author}{L. Frankfurt},
  \bibinfo{author}{M. Strikman}, \bibinfo{title}{{Factorization for hard
  exclusive electroproduction of mesons in QCD}}, \bibinfo{journal}{Phys. Rev.
  D} \bibinfo{volume}{56} (\bibinfo{year}{1997}) \bibinfo{pages}{2982--3006},
  \bibinfo{doi}{\doi{10.1103/PhysRevD.56.2982}}, \eprint{hep-ph/9611433}.

\bibtype{Article}%
\bibitem{Ji:2003ak}
\bibinfo{author}{X.-D. Ji}, \bibinfo{title}{{Viewing the proton through 'color'
  filters}}, \bibinfo{journal}{Phys. Rev. Lett.} \bibinfo{volume}{91}
  (\bibinfo{year}{2003}) \bibinfo{pages}{062001},
  \bibinfo{doi}{\doi{10.1103/PhysRevLett.91.062001}}, \eprint{hep-ph/0304037}.

\bibtype{Article}%
\bibitem{Belitsky:2003nz}
\bibinfo{author}{A.~V. Belitsky}, \bibinfo{author}{X.-D. Ji},
  \bibinfo{author}{F. Yuan}, \bibinfo{title}{{Quark imaging in the proton via
  quantum phase space distributions}}, \bibinfo{journal}{Phys. Rev. D}
  \bibinfo{volume}{69} (\bibinfo{year}{2004}) \bibinfo{pages}{074014},
  \bibinfo{doi}{\doi{10.1103/PhysRevD.69.074014}}, \eprint{hep-ph/0307383}.

\bibtype{Article}%
\bibitem{Meissner:2008ay}
\bibinfo{author}{S. Meissner}, \bibinfo{author}{A. Metz}, \bibinfo{author}{M.
  Schlegel}, \bibinfo{author}{K. Goeke}, \bibinfo{title}{{Generalized parton
  correlation functions for a spin-0 hadron}}, \bibinfo{journal}{JHEP}
  \bibinfo{volume}{08} (\bibinfo{year}{2008}) \bibinfo{pages}{038},
  \bibinfo{doi}{\doi{10.1088/1126-6708/2008/08/038}}, \eprint{0805.3165}.

\bibtype{Article}%
\bibitem{Meissner:2009ww}
\bibinfo{author}{S. Meissner}, \bibinfo{author}{A. Metz}, \bibinfo{author}{M.
  Schlegel}, \bibinfo{title}{{Generalized parton correlation functions for a
  spin-1/2 hadron}}, \bibinfo{journal}{JHEP} \bibinfo{volume}{08}
  (\bibinfo{year}{2009}) \bibinfo{pages}{056},
  \bibinfo{doi}{\doi{10.1088/1126-6708/2009/08/056}}, \eprint{0906.5323}.

\bibtype{Article}%
\bibitem{Gross:2022hyw}
\bibinfo{author}{F. Gross}, et al., \bibinfo{title}{{50 Years of Quantum
  Chromodynamics}}, \bibinfo{journal}{Eur. Phys. J. C} \bibinfo{volume}{83}
  (\bibinfo{year}{2023}) \bibinfo{pages}{1125},
  \bibinfo{doi}{\doi{10.1140/epjc/s10052-023-11949-2}}, \eprint{2212.11107}.

\bibtype{Article}%
\bibitem{tHooft:1972tcz}
\bibinfo{author}{G. 't Hooft}, \bibinfo{author}{M.~J.~G. Veltman},
  \bibinfo{title}{{Regularization and Renormalization of Gauge Fields}},
  \bibinfo{journal}{Nucl. Phys. B} \bibinfo{volume}{44} (\bibinfo{year}{1972})
  \bibinfo{pages}{189--213}, \bibinfo{doi}{\doi{10.1016/0550-3213(72)90279-9}}.

\bibtype{Article}%
\bibitem{Deur:2025rjo}
\bibinfo{author}{A. Deur}, \bibinfo{title}{{The QCD Running Coupling}}
  (\bibinfo{year}{2025}), \eprint{2502.06535}.

\bibtype{Article}%
\bibitem{Diehl:2017wew}
\bibinfo{author}{M. Diehl}, \bibinfo{author}{J.~R. Gaunt},
  \bibinfo{title}{{Double parton scattering theory overview}},
  \bibinfo{journal}{Adv. Ser. Direct. High Energy Phys.} \bibinfo{volume}{29}
  (\bibinfo{year}{2018}) \bibinfo{pages}{7--28},
  \bibinfo{doi}{\doi{10.1142/9789813227767_0002}}, \eprint{1710.04408}.

\bibtype{Book}%
\bibitem{Bartalini:2018qje}
\bibinfo{editor}{P. Bartalini}, \bibinfo{editor}{J.~R. Gaunt} (Eds.),
  \bibinfo{title}{{Multiple Parton Interactions at the LHC}},
  \bibinfo{comment}{vol.} \bibinfo{volume}{29}, \bibinfo{publisher}{WSP}
  \bibinfo{year}{2019}, ISBN \bibinfo{isbn}{978-981-322-775-0,
  978-981-322-777-4}, \bibinfo{doi}{\doi{10.1142/10646}}.

\bibtype{Article}%
\bibitem{Bloom:1969kc}
\bibinfo{author}{E.~D. Bloom}, et al., \bibinfo{title}{{High-Energy Inelastic e
  p Scattering at 6-Degrees and 10-Degrees}}, \bibinfo{journal}{Phys. Rev.
  Lett.} \bibinfo{volume}{23} (\bibinfo{year}{1969}) \bibinfo{pages}{930--934},
  \bibinfo{doi}{\doi{10.1103/PhysRevLett.23.930}}.

\bibtype{Article}%
\bibitem{Breidenbach:1969kd}
\bibinfo{author}{M. Breidenbach}, \bibinfo{author}{J.~I. Friedman},
  \bibinfo{author}{H.~W. Kendall}, \bibinfo{author}{E.~D. Bloom},
  \bibinfo{author}{D.~H. Coward}, \bibinfo{author}{H.~C. DeStaebler},
  \bibinfo{author}{J. Drees}, \bibinfo{author}{L.~W. Mo},
  \bibinfo{author}{R.~E. Taylor}, \bibinfo{title}{{Observed behavior of highly
  inelastic electron-proton scattering}}, \bibinfo{journal}{Phys. Rev. Lett.}
  \bibinfo{volume}{23} (\bibinfo{year}{1969}) \bibinfo{pages}{935--939},
  \bibinfo{doi}{\doi{10.1103/PhysRevLett.23.935}}.

\bibtype{Article}%
\bibitem{Eichten:1973cs}
\bibinfo{author}{T. Eichten}, et al., \bibinfo{title}{{Measurement of the
  Neutrino - Nucleon Anti-neutrino - Nucleon Total Cross-sections}},
  \bibinfo{journal}{Phys. Lett. B} \bibinfo{volume}{46} (\bibinfo{year}{1973})
  \bibinfo{pages}{274--280}, \bibinfo{doi}{\doi{10.1016/0370-2693(73)90702-8}}.

\bibtype{Article}%
\bibitem{GargamelleNeutrino:1974exc}
\bibinfo{author}{H. Deden}, et al. (\bibinfo{collaboration}{Gargamelle
  Neutrino}), \bibinfo{title}{{Experimental Study of Structure Functions and
  Sum Rules in Charge Changing Interactions of Neutrinos and anti-neutrinos on
  Nucleons}}, \bibinfo{journal}{Nucl. Phys. B} \bibinfo{volume}{85}
  (\bibinfo{year}{1975}) \bibinfo{pages}{269--288},
  \bibinfo{doi}{\doi{10.1016/0550-3213(75)90008-5}}.

\bibtype{Article}%
\bibitem{Christenson:1970um}
\bibinfo{author}{J.~H. Christenson}, \bibinfo{author}{G.~S. Hicks},
  \bibinfo{author}{L.~M. Lederman}, \bibinfo{author}{P.~J. Limon},
  \bibinfo{author}{B.~G. Pope}, \bibinfo{author}{E. Zavattini},
  \bibinfo{title}{{Observation of massive muon pairs in hadron collisions}},
  \bibinfo{journal}{Phys. Rev. Lett.} \bibinfo{volume}{25}
  (\bibinfo{year}{1970}) \bibinfo{pages}{1523--1526},
  \bibinfo{doi}{\doi{10.1103/PhysRevLett.25.1523}}.

\bibtype{Article}%
\bibitem{CERN-Columbia-Oxford-Rockefeller:1979gxw}
\bibinfo{author}{A.~L.~S. Angelis}, et al.
  (\bibinfo{collaboration}{CERN-Columbia-Oxford-Rockefeller, CCOR}),
  \bibinfo{title}{{A Measurement of the Production of Massive $e^+ e^-$ Pairs
  in Proton Proton Collisions at $\sqrt{s}$=62.4 GeV}}, \bibinfo{journal}{Phys.
  Lett. B} \bibinfo{volume}{87} (\bibinfo{year}{1979})
  \bibinfo{pages}{398--402}, \bibinfo{doi}{\doi{10.1016/0370-2693(79)90563-X}}.

\bibtype{Article}%
\bibitem{Antreasyan:1980yb}
\bibinfo{author}{D. Antreasyan}, et al., \bibinfo{title}{{Measurement of Dimuon
  Production at the ISR}}, \bibinfo{journal}{Phys. Rev. Lett.}
  \bibinfo{volume}{45} (\bibinfo{year}{1980}) \bibinfo{pages}{863},
  \bibinfo{doi}{\doi{10.1103/PhysRevLett.45.863}}.

\bibtype{Article}%
\bibitem{NA3:1983ejh}
\bibinfo{author}{J. Badier}, et al. (\bibinfo{collaboration}{NA3}),
  \bibinfo{title}{{Experimental Determination of the pi Meson Structure
  Functions by the Drell-Yan Mechanism}}, \bibinfo{journal}{Z. Phys. C}
  \bibinfo{volume}{18} (\bibinfo{year}{1983}) \bibinfo{pages}{281},
  \bibinfo{doi}{\doi{10.1007/BF01573728}}.

\bibtype{Article}%
\bibitem{NA10:1986fgk}
\bibinfo{author}{S. Falciano}, et al. (\bibinfo{collaboration}{NA10}),
  \bibinfo{title}{{Angular Distributions of Muon Pairs Produced by
  194-{GeV}/$c$ Negative Pions}}, \bibinfo{journal}{Z. Phys. C}
  \bibinfo{volume}{31} (\bibinfo{year}{1986}) \bibinfo{pages}{513},
  \bibinfo{doi}{\doi{10.1007/BF01551072}}.

\bibtype{Article}%
\bibitem{NA10:1987sqk}
\bibinfo{author}{M. Guanziroli}, et al. (\bibinfo{collaboration}{NA10}),
  \bibinfo{title}{{Angular Distributions of Muon Pairs Produced by Negative
  Pions on Deuterium and Tungsten}}, \bibinfo{journal}{Z. Phys. C}
  \bibinfo{volume}{37} (\bibinfo{year}{1988}) \bibinfo{pages}{545},
  \bibinfo{doi}{\doi{10.1007/BF01549713}}.

\bibtype{Article}%
\bibitem{WA70:1987vvj}
\bibinfo{author}{M. Bonesini}, et al. (\bibinfo{collaboration}{WA70}),
  \bibinfo{title}{{Production of High Transverse Momentum Prompt Photons and
  Neutral Pions in Proton Proton Interactions at 280-GeV/c}},
  \bibinfo{journal}{Z. Phys. C} \bibinfo{volume}{38} (\bibinfo{year}{1988})
  \bibinfo{pages}{371}, \bibinfo{doi}{\doi{10.1007/BF01584385}}.

\bibtype{Article}%
\bibitem{EuropeanMuon:1983wih}
\bibinfo{author}{J.~J. Aubert}, et al. (\bibinfo{collaboration}{European
  Muon}), \bibinfo{title}{{The ratio of the nucleon structure functions $F2_n$
  for iron and deuterium}}, \bibinfo{journal}{Phys. Lett. B}
  \bibinfo{volume}{123} (\bibinfo{year}{1983}) \bibinfo{pages}{275--278},
  \bibinfo{doi}{\doi{10.1016/0370-2693(83)90437-9}}.

\bibtype{Article}%
\bibitem{EuropeanMuon:1989yki}
\bibinfo{author}{J. Ashman}, et al. (\bibinfo{collaboration}{European Muon}),
  \bibinfo{title}{{An Investigation of the Spin Structure of the Proton in Deep
  Inelastic Scattering of Polarized Muons on Polarized Protons}},
  \bibinfo{journal}{Nucl. Phys. B} \bibinfo{volume}{328} (\bibinfo{year}{1989})
  \bibinfo{pages}{1}, \bibinfo{doi}{\doi{10.1016/0550-3213(89)90089-8}}.

\bibtype{Article}%
\bibitem{SpinMuonSMC:1994met}
\bibinfo{author}{D. Adams}, et al. (\bibinfo{collaboration}{Spin Muon (SMC)}),
  \bibinfo{title}{{Measurement of the spin dependent structure function
  $g_1(x)$ of the proton}}, \bibinfo{journal}{Phys. Lett. B}
  \bibinfo{volume}{329} (\bibinfo{year}{1994}) \bibinfo{pages}{399--406},
  \bibinfo{doi}{\doi{10.1016/0370-2693(94)90793-5}}, \eprint{hep-ph/9404270}.

\bibtype{Article}%
\bibitem{D0:1999jba}
\bibinfo{author}{B. Abbott}, et al. (\bibinfo{collaboration}{D0}),
  \bibinfo{title}{{Measurement of the inclusive differential cross section for
  $Z$ bosons as a function of transverse momentum in $\bar{p}p$ collisions at
  $\sqrt{s} = 1.8$ TeV}}, \bibinfo{journal}{Phys. Rev. D} \bibinfo{volume}{61}
  (\bibinfo{year}{2000}) \bibinfo{pages}{032004},
  \bibinfo{doi}{\doi{10.1103/PhysRevD.61.032004}}, \eprint{hep-ex/9907009}.

\bibtype{Article}%
\bibitem{CDF:1999bpw}
\bibinfo{author}{T. Affolder}, et al. (\bibinfo{collaboration}{CDF}),
  \bibinfo{title}{{The transverse momentum and total cross section of $e^+e^-$
  pairs in the $Z$ boson region from $p\bar{p}$ collisions at $\sqrt{s} = 1.8$
  TeV}}, \bibinfo{journal}{Phys. Rev. Lett.} \bibinfo{volume}{84}
  (\bibinfo{year}{2000}) \bibinfo{pages}{845--850},
  \bibinfo{doi}{\doi{10.1103/PhysRevLett.84.845}}, \eprint{hep-ex/0001021}.

\bibtype{Article}%
\bibitem{D0:2010dbl}
\bibinfo{author}{V.~M. Abazov}, et al. (\bibinfo{collaboration}{D0}),
  \bibinfo{title}{{Measurement of the Normalized $Z/\gamma^* -> \mu^+\mu^-$
  Transverse Momentum Distribution in $p\bar{p}$ Collisions at $\sqrt{s}=1.96$
  TeV}}, \bibinfo{journal}{Phys. Lett. B} \bibinfo{volume}{693}
  (\bibinfo{year}{2010}) \bibinfo{pages}{522--530},
  \bibinfo{doi}{\doi{10.1016/j.physletb.2010.09.012}}, \eprint{1006.0618}.

\bibtype{Article}%
\bibitem{CDF:2012brb}
\bibinfo{author}{T. Aaltonen}, et al. (\bibinfo{collaboration}{CDF}),
  \bibinfo{title}{{Transverse momentum cross section of $e^+e^-$ pairs in the
  $Z$-boson region from $p\bar{p}$ collisions at $\sqrt{s}=1.96$ TeV}},
  \bibinfo{journal}{Phys. Rev. D} \bibinfo{volume}{86} (\bibinfo{year}{2012})
  \bibinfo{pages}{052010}, \bibinfo{doi}{\doi{10.1103/PhysRevD.86.052010}},
  \eprint{1207.7138}.

\bibtype{Article}%
\bibitem{H1:2000kis}
\bibinfo{author}{C. Adloff}, et al. (\bibinfo{collaboration}{H1}),
  \bibinfo{title}{{Elastic photoproduction of J/psi and Upsilon mesons at
  HERA}}, \bibinfo{journal}{Phys. Lett. B} \bibinfo{volume}{483}
  (\bibinfo{year}{2000}) \bibinfo{pages}{23--35},
  \bibinfo{doi}{\doi{10.1016/S0370-2693(00)00530-X}}, \eprint{hep-ex/0003020}.

\bibtype{Article}%
\bibitem{ZEUS:2001mhd}
\bibinfo{author}{S. Chekanov}, et al. (\bibinfo{collaboration}{ZEUS}),
  \bibinfo{title}{{Measurement of the neutral current cross-section and F(2)
  structure function for deep inelastic e + p scattering at HERA}},
  \bibinfo{journal}{Eur. Phys. J. C} \bibinfo{volume}{21}
  (\bibinfo{year}{2001}) \bibinfo{pages}{443--471},
  \bibinfo{doi}{\doi{10.1007/s100520100749}}, \eprint{hep-ex/0105090}.

\bibtype{Article}%
\bibitem{ZEUS:2002wfj}
\bibinfo{author}{S. Chekanov}, et al. (\bibinfo{collaboration}{ZEUS}),
  \bibinfo{title}{{Exclusive photoproduction of J / psi mesons at HERA}},
  \bibinfo{journal}{Eur. Phys. J. C} \bibinfo{volume}{24}
  (\bibinfo{year}{2002}) \bibinfo{pages}{345--360},
  \bibinfo{doi}{\doi{10.1007/s10052-002-0953-7}}, \eprint{hep-ex/0201043}.

\bibtype{Article}%
\bibitem{H1:2005dtp}
\bibinfo{author}{A. Aktas}, et al. (\bibinfo{collaboration}{H1}),
  \bibinfo{title}{{Elastic J/psi production at HERA}}, \bibinfo{journal}{Eur.
  Phys. J. C} \bibinfo{volume}{46} (\bibinfo{year}{2006})
  \bibinfo{pages}{585--603}, \bibinfo{doi}{\doi{10.1140/epjc/s2006-02519-5}},
  \eprint{hep-ex/0510016}.

\bibtype{Article}%
\bibitem{H1:2009pze}
\bibinfo{author}{F.~D. Aaron}, et al. (\bibinfo{collaboration}{H1, ZEUS}),
  \bibinfo{title}{{Combined Measurement and QCD Analysis of the Inclusive
  $e^\pm p$ Scattering Cross Sections at HERA}}, \bibinfo{journal}{JHEP}
  \bibinfo{volume}{01} (\bibinfo{year}{2010}) \bibinfo{pages}{109},
  \bibinfo{doi}{\doi{10.1007/JHEP01(2010)109}}, \eprint{0911.0884}.

\bibtype{Article}%
\bibitem{H1:2015ubc}
\bibinfo{author}{H. Abramowicz}, et al. (\bibinfo{collaboration}{H1, ZEUS}),
  \bibinfo{title}{{Combination of measurements of inclusive deep inelastic
  ${e^{\pm }p}$ scattering cross sections and QCD analysis of HERA data}},
  \bibinfo{journal}{Eur. Phys. J. C} \bibinfo{volume}{75}
  (\bibinfo{number}{12}) (\bibinfo{year}{2015}) \bibinfo{pages}{580},
  \bibinfo{doi}{\doi{10.1140/epjc/s10052-015-3710-4}}, \eprint{1506.06042}.

\bibtype{Article}%
\bibitem{HERMES:1997hjr}
\bibinfo{author}{K. Ackerstaff}, et al. (\bibinfo{collaboration}{HERMES}),
  \bibinfo{title}{{Measurement of the neutron spin structure function $g_1^n$
  with a polarized He-3 internal target}}, \bibinfo{journal}{Phys. Lett. B}
  \bibinfo{volume}{404} (\bibinfo{year}{1997}) \bibinfo{pages}{383--389},
  \bibinfo{doi}{\doi{10.1016/S0370-2693(97)00611-4}}, \eprint{hep-ex/9703005}.

\bibtype{Article}%
\bibitem{HERMES:1999ryv}
\bibinfo{author}{A. Airapetian}, et al. (\bibinfo{collaboration}{HERMES}),
  \bibinfo{title}{{Observation of a single spin azimuthal asymmetry in
  semiinclusive pion electro production}}, \bibinfo{journal}{Phys. Rev. Lett.}
  \bibinfo{volume}{84} (\bibinfo{year}{2000}) \bibinfo{pages}{4047--4051},
  \bibinfo{doi}{\doi{10.1103/PhysRevLett.84.4047}}, \eprint{hep-ex/9910062}.

\bibtype{Article}%
\bibitem{HERMES:2004mhh}
\bibinfo{author}{A. Airapetian}, et al. (\bibinfo{collaboration}{HERMES}),
  \bibinfo{title}{{Single-spin asymmetries in semi-inclusive deep-inelastic
  scattering on a transversely polarized hydrogen target}},
  \bibinfo{journal}{Phys. Rev. Lett.} \bibinfo{volume}{94}
  (\bibinfo{year}{2005}) \bibinfo{pages}{012002},
  \bibinfo{doi}{\doi{10.1103/PhysRevLett.94.012002}}, \eprint{hep-ex/0408013}.

\bibtype{Article}%
\bibitem{HERMES:2006jyl}
\bibinfo{author}{A. Airapetian}, et al. (\bibinfo{collaboration}{HERMES}),
  \bibinfo{title}{{Precise determination of the spin structure function $g_1$
  of the proton, deuteron and neutron}}, \bibinfo{journal}{Phys. Rev. D}
  \bibinfo{volume}{75} (\bibinfo{year}{2007}) \bibinfo{pages}{012007},
  \bibinfo{doi}{\doi{10.1103/PhysRevD.75.012007}}, \eprint{hep-ex/0609039}.

\bibtype{Article}%
\bibitem{HERMES:2008abz}
\bibinfo{author}{A. Airapetian}, et al. (\bibinfo{collaboration}{HERMES}),
  \bibinfo{title}{{Measurement of Azimuthal Asymmetries With Respect To Both
  Beam Charge and Transverse Target Polarization in Exclusive Electroproduction
  of Real Photons}}, \bibinfo{journal}{JHEP} \bibinfo{volume}{06}
  (\bibinfo{year}{2008}) \bibinfo{pages}{066},
  \bibinfo{doi}{\doi{10.1088/1126-6708/2008/06/066}}, \eprint{0802.2499}.

\bibtype{Article}%
\bibitem{HERMES:2009lmz}
\bibinfo{author}{A. Airapetian}, et al. (\bibinfo{collaboration}{HERMES}),
  \bibinfo{title}{{Observation of the Naive-T-odd Sivers Effect in
  Deep-Inelastic Scattering}}, \bibinfo{journal}{Phys. Rev. Lett.}
  \bibinfo{volume}{103} (\bibinfo{year}{2009}) \bibinfo{pages}{152002},
  \bibinfo{doi}{\doi{10.1103/PhysRevLett.103.152002}}, \eprint{0906.3918}.

\bibtype{Article}%
\bibitem{HERMES:2012uyd}
\bibinfo{author}{A. Airapetian}, et al. (\bibinfo{collaboration}{HERMES}),
  \bibinfo{title}{{Multiplicities of charged pions and kaons from
  semi-inclusive deep-inelastic scattering by the proton and the deuteron}},
  \bibinfo{journal}{Phys. Rev. D} \bibinfo{volume}{87} (\bibinfo{year}{2013})
  \bibinfo{pages}{074029}, \bibinfo{doi}{\doi{10.1103/PhysRevD.87.074029}},
  \eprint{1212.5407}.

\bibtype{Article}%
\bibitem{HERMES:2020ifk}
\bibinfo{author}{A. Airapetian}, et al. (\bibinfo{collaboration}{HERMES}),
  \bibinfo{title}{{Azimuthal single- and double-spin asymmetries in
  semi-inclusive deep-inelastic lepton scattering by transversely polarized
  protons}}, \bibinfo{journal}{JHEP} \bibinfo{volume}{12}
  (\bibinfo{year}{2020}) \bibinfo{pages}{010},
  \bibinfo{doi}{\doi{10.1007/JHEP12(2020)010}}, \eprint{2007.07755}.

\bibtype{Article}%
\bibitem{CLAS:2001wjj}
\bibinfo{author}{S. Stepanyan}, et al. (\bibinfo{collaboration}{CLAS}),
  \bibinfo{title}{{Observation of exclusive deeply virtual Compton scattering
  in polarized electron beam asymmetry measurements}}, \bibinfo{journal}{Phys.
  Rev. Lett.} \bibinfo{volume}{87} (\bibinfo{year}{2001})
  \bibinfo{pages}{182002}, \bibinfo{doi}{\doi{10.1103/PhysRevLett.87.182002}},
  \eprint{hep-ex/0107043}.

\bibtype{Article}%
\bibitem{CLAS:2003qum}
\bibinfo{author}{H. Avakian}, et al. (\bibinfo{collaboration}{CLAS}),
  \bibinfo{title}{{Measurement of beam-spin asymmetries for pi +
  electroproduction above the baryon resonance region}},
  \bibinfo{journal}{Phys. Rev. D} \bibinfo{volume}{69} (\bibinfo{year}{2004})
  \bibinfo{pages}{112004}, \bibinfo{doi}{\doi{10.1103/PhysRevD.69.112004}},
  \eprint{hep-ex/0301005}.

\bibtype{Article}%
\bibitem{JeffersonLabHallA:2003joy}
\bibinfo{author}{X. Zheng}, et al. (\bibinfo{collaboration}{Jefferson Lab Hall
  A}), \bibinfo{title}{{Precision measurement of the neutron spin asymmetry
  $A_1^n$ and spin flavor decomposition in the valence quark region}},
  \bibinfo{journal}{Phys. Rev. Lett.} \bibinfo{volume}{92}
  (\bibinfo{year}{2004}) \bibinfo{pages}{012004},
  \bibinfo{doi}{\doi{10.1103/PhysRevLett.92.012004}}, \eprint{nucl-ex/0308011}.

\bibtype{Article}%
\bibitem{CLAS:2007clm}
\bibinfo{author}{F.~X. Girod}, et al. (\bibinfo{collaboration}{CLAS}),
  \bibinfo{title}{{Measurement of Deeply virtual Compton scattering beam-spin
  asymmetries}}, \bibinfo{journal}{Phys. Rev. Lett.} \bibinfo{volume}{100}
  (\bibinfo{year}{2008}) \bibinfo{pages}{162002},
  \bibinfo{doi}{\doi{10.1103/PhysRevLett.100.162002}}, \eprint{0711.4805}.

\bibtype{Article}%
\bibitem{JeffersonLabHallA:2011ayy}
\bibinfo{author}{X. Qian}, et al. (\bibinfo{collaboration}{Jefferson Lab Hall
  A}), \bibinfo{title}{{Single Spin Asymmetries in Charged Pion Production from
  Semi-Inclusive Deep Inelastic Scattering on a Transversely Polarized $^3$He
  Target}}, \bibinfo{journal}{Phys. Rev. Lett.} \bibinfo{volume}{107}
  (\bibinfo{year}{2011}) \bibinfo{pages}{072003},
  \bibinfo{doi}{\doi{10.1103/PhysRevLett.107.072003}}, \eprint{1106.0363}.

\bibtype{Article}%
\bibitem{GlueX:2023pev}
\bibinfo{author}{S. Adhikari}, et al. (\bibinfo{collaboration}{GlueX}),
  \bibinfo{title}{{Measurement of the J/$\psi $ photoproduction cross section
  over the full near-threshold kinematic region}}, \bibinfo{journal}{Phys. Rev.
  C} \bibinfo{volume}{108} (\bibinfo{number}{2}) (\bibinfo{year}{2023})
  \bibinfo{pages}{025201}, \bibinfo{doi}{\doi{10.1103/PhysRevC.108.025201}},
  \eprint{2304.03845}.

\bibtype{Article}%
\bibitem{COMPASS:2005qpp}
\bibinfo{author}{E.~S. Ageev}, et al. (\bibinfo{collaboration}{COMPASS}),
  \bibinfo{title}{{Gluon polarization in the nucleon from quasi-real
  photoproduction of high-p(T) hadron pairs}}, \bibinfo{journal}{Phys. Lett. B}
  \bibinfo{volume}{633} (\bibinfo{year}{2006}) \bibinfo{pages}{25--32},
  \bibinfo{doi}{\doi{10.1016/j.physletb.2005.11.049}}, \eprint{hep-ex/0511028}.

\bibtype{Article}%
\bibitem{COMPASS:2008isr}
\bibinfo{author}{M. Alekseev}, et al. (\bibinfo{collaboration}{COMPASS}),
  \bibinfo{title}{{Collins and Sivers asymmetries for pions and kaons in
  muon-deuteron DIS}}, \bibinfo{journal}{Phys. Lett. B} \bibinfo{volume}{673}
  (\bibinfo{year}{2009}) \bibinfo{pages}{127--135},
  \bibinfo{doi}{\doi{10.1016/j.physletb.2009.01.060}}, \eprint{0802.2160}.

\bibtype{Article}%
\bibitem{COMPASS:2010wkz}
\bibinfo{author}{M.~G. Alekseev}, et al. (\bibinfo{collaboration}{COMPASS}),
  \bibinfo{title}{{The Spin-dependent Structure Function of the Proton $g_1^p$
  and a Test of the Bjorken Sum Rule}}, \bibinfo{journal}{Phys. Lett. B}
  \bibinfo{volume}{690} (\bibinfo{year}{2010}) \bibinfo{pages}{466--472},
  \bibinfo{doi}{\doi{10.1016/j.physletb.2010.05.069}}, \eprint{1001.4654}.

\bibtype{Article}%
\bibitem{COMPASS:2010hbb}
\bibinfo{author}{M.~G. Alekseev}, et al. (\bibinfo{collaboration}{COMPASS}),
  \bibinfo{title}{{Measurement of the Collins and Sivers asymmetries on
  transversely polarised protons}}, \bibinfo{journal}{Phys. Lett. B}
  \bibinfo{volume}{692} (\bibinfo{year}{2010}) \bibinfo{pages}{240--246},
  \bibinfo{doi}{\doi{10.1016/j.physletb.2010.08.001}}, \eprint{1005.5609}.

\bibtype{Article}%
\bibitem{COMPASS:2013bfs}
\bibinfo{author}{C. Adolph}, et al. (\bibinfo{collaboration}{COMPASS}),
  \bibinfo{title}{{Hadron Transverse Momentum Distributions in Muon Deep
  Inelastic Scattering at 160 GeV/$c$}}, \bibinfo{journal}{Eur. Phys. J. C}
  \bibinfo{volume}{73} (\bibinfo{number}{8}) (\bibinfo{year}{2013})
  \bibinfo{pages}{2531}, \bibinfo{doi}{\doi{10.1140/epjc/s10052-013-2531-6}},
  \eprint{1305.7317}.

\bibtype{Article}%
\bibitem{COMPASS:2014bze}
\bibinfo{author}{C. Adolph}, et al. (\bibinfo{collaboration}{COMPASS}),
  \bibinfo{title}{{Collins and Sivers asymmetries in muonproduction of pions
  and kaons off transversely polarised protons}}, \bibinfo{journal}{Phys. Lett.
  B} \bibinfo{volume}{744} (\bibinfo{year}{2015}) \bibinfo{pages}{250--259},
  \bibinfo{doi}{\doi{10.1016/j.physletb.2015.03.056}}, \eprint{1408.4405}.

\bibtype{Article}%
\bibitem{COMPASS:2017jbv}
\bibinfo{author}{M. Aghasyan}, et al. (\bibinfo{collaboration}{COMPASS}),
  \bibinfo{title}{{First measurement of transverse-spin-dependent azimuthal
  asymmetries in the Drell-Yan process}}, \bibinfo{journal}{Phys. Rev. Lett.}
  \bibinfo{volume}{119} (\bibinfo{number}{11}) (\bibinfo{year}{2017})
  \bibinfo{pages}{112002}, \bibinfo{doi}{\doi{10.1103/PhysRevLett.119.112002}},
  \eprint{1704.00488}.

\bibtype{Article}%
\bibitem{COMPASS:2017mvk}
\bibinfo{author}{M. Aghasyan}, et al. (\bibinfo{collaboration}{COMPASS}),
  \bibinfo{title}{{Transverse-momentum-dependent Multiplicities of Charged
  Hadrons in Muon-Deuteron Deep Inelastic Scattering}}, \bibinfo{journal}{Phys.
  Rev. D} \bibinfo{volume}{97} (\bibinfo{number}{3}) (\bibinfo{year}{2018})
  \bibinfo{pages}{032006}, \bibinfo{doi}{\doi{10.1103/PhysRevD.97.032006}},
  \eprint{1709.07374}.

\bibtype{Article}%
\bibitem{COMPASS:2023cgk}
\bibinfo{author}{G.~D. Alexeev}, et al. (\bibinfo{collaboration}{COMPASS}),
  \bibinfo{title}{{Transverse-spin-dependent azimuthal asymmetries of pion and
  kaon pairs produced in muon-proton and muon-deuteron semi-inclusive deep
  inelastic scattering}}, \bibinfo{journal}{Phys. Lett. B}
  \bibinfo{volume}{845} (\bibinfo{year}{2023}) \bibinfo{pages}{138155},
  \bibinfo{doi}{\doi{10.1016/j.physletb.2023.138155}}, \eprint{2301.02013}.

\bibtype{Article}%
\bibitem{PHENIX:2007kqm}
\bibinfo{author}{A. Adare}, et al. (\bibinfo{collaboration}{PHENIX}),
  \bibinfo{title}{{Inclusive cross-section and double helicity asymmetry for
  $\pi^0$ production in $p + p$ collisions at $\sqrt{s} =$ 200 GeV:
  Implications for the polarized gluon distribution in the proton}},
  \bibinfo{journal}{Phys. Rev. D} \bibinfo{volume}{76} (\bibinfo{year}{2007})
  \bibinfo{pages}{051106}, \bibinfo{doi}{\doi{10.1103/PhysRevD.76.051106}},
  \eprint{0704.3599}.

\bibtype{Article}%
\bibitem{BRAHMS:2008doi}
\bibinfo{author}{I. Arsene}, et al. (\bibinfo{collaboration}{BRAHMS}),
  \bibinfo{title}{{Single Transverse Spin Asymmetries of Identified Charged
  Hadrons in Polarized $pp$ Collisions at $\sqrt{s}$ = 62.4 GeV}},
  \bibinfo{journal}{Phys. Rev. Lett.} \bibinfo{volume}{101}
  (\bibinfo{year}{2008}) \bibinfo{pages}{042001},
  \bibinfo{doi}{\doi{10.1103/PhysRevLett.101.042001}}, \eprint{0801.1078}.

\bibtype{Article}%
\bibitem{STAR:2012ljf}
\bibinfo{author}{L. Adamczyk}, et al. (\bibinfo{collaboration}{STAR}),
  \bibinfo{title}{{Transverse Single-Spin Asymmetry and Cross-Section for
  $\pi^0$ and $\eta$ Mesons at Large Feynman-$x$ in Polarized $p+p$ Collisions
  at $\sqrt{s}=200$ GeV}}, \bibinfo{journal}{Phys. Rev. D} \bibinfo{volume}{86}
  (\bibinfo{year}{2012}) \bibinfo{pages}{051101},
  \bibinfo{doi}{\doi{10.1103/PhysRevD.86.051101}}, \eprint{1205.6826}.

\bibtype{Article}%
\bibitem{STAR:2014wox}
\bibinfo{author}{L. Adamczyk}, et al. (\bibinfo{collaboration}{STAR}),
  \bibinfo{title}{{Precision Measurement of the Longitudinal Double-spin
  Asymmetry for Inclusive Jet Production in Polarized Proton Collisions at
  $\sqrt{s}=200$ GeV}}, \bibinfo{journal}{Phys. Rev. Lett.}
  \bibinfo{volume}{115} (\bibinfo{number}{9}) (\bibinfo{year}{2015})
  \bibinfo{pages}{092002}, \bibinfo{doi}{\doi{10.1103/PhysRevLett.115.092002}},
  \eprint{1405.5134}.

\bibtype{Article}%
\bibitem{STAR:2014afm}
\bibinfo{author}{L. Adamczyk}, et al. (\bibinfo{collaboration}{STAR}),
  \bibinfo{title}{{Measurement of longitudinal spin asymmetries for weak boson
  production in polarized proton-proton collisions at RHIC}},
  \bibinfo{journal}{Phys. Rev. Lett.} \bibinfo{volume}{113}
  (\bibinfo{year}{2014}) \bibinfo{pages}{072301},
  \bibinfo{doi}{\doi{10.1103/PhysRevLett.113.072301}}, \eprint{1404.6880}.

\bibtype{Article}%
\bibitem{STAR:2015vmv}
\bibinfo{author}{L. Adamczyk}, et al. (\bibinfo{collaboration}{STAR}),
  \bibinfo{title}{{Measurement of the transverse single-spin asymmetry in
  $p^\uparrow+p \to W^{\pm}/Z^0$ at RHIC}}, \bibinfo{journal}{Phys. Rev. Lett.}
  \bibinfo{volume}{116} (\bibinfo{number}{13}) (\bibinfo{year}{2016})
  \bibinfo{pages}{132301}, \bibinfo{doi}{\doi{10.1103/PhysRevLett.116.132301}},
  \eprint{1511.06003}.

\bibtype{Article}%
\bibitem{STAR:2017wsi}
\bibinfo{author}{L. Adamczyk}, et al. (\bibinfo{collaboration}{STAR}),
  \bibinfo{title}{{Transverse spin-dependent azimuthal correlations of charged
  pion pairs measured in p$^\uparrow$+p collisions at $\sqrt{s}$ = 500 GeV}},
  \bibinfo{journal}{Phys. Lett. B} \bibinfo{volume}{780} (\bibinfo{year}{2018})
  \bibinfo{pages}{332--339},
  \bibinfo{doi}{\doi{10.1016/j.physletb.2018.02.069}}, \eprint{1710.10215}.

\bibtype{Article}%
\bibitem{PHENIX:2018dwt}
\bibinfo{author}{C. Aidala}, et al. (\bibinfo{collaboration}{PHENIX}),
  \bibinfo{title}{{Measurements of $\mu\mu$ pairs from open heavy flavor and
  Drell-Yan in $p+p$ collisions at $\sqrt{s}=200$ GeV}},
  \bibinfo{journal}{Phys. Rev. D} \bibinfo{volume}{99} (\bibinfo{number}{7})
  (\bibinfo{year}{2019}) \bibinfo{pages}{072003},
  \bibinfo{doi}{\doi{10.1103/PhysRevD.99.072003}}, \eprint{1805.02448}.

\bibtype{Article}%
\bibitem{STAR:2023jwh}
\bibinfo{author}{STAR Coll.}, \bibinfo{title}{{Measurements of
  the~$Z^{0}/\gamma^{*}$ cross section and transverse single spin asymmetry in
  510 GeV $p+p$ collisions}}, \bibinfo{journal}{Phys. Lett. B}
  \bibinfo{volume}{854} (\bibinfo{year}{2024}) \bibinfo{pages}{138715},
  \bibinfo{doi}{\doi{10.1016/j.physletb.2024.138715}}, \eprint{2308.15496}.

\bibtype{Article}%
\bibitem{LHCb:2015mad}
\bibinfo{author}{R. Aaij}, et al. (\bibinfo{collaboration}{LHCb}),
  \bibinfo{title}{{Measurement of forward W and Z boson production in $pp$
  collisions at $ \sqrt{s}=8 $ TeV}}, \bibinfo{journal}{JHEP}
  \bibinfo{volume}{01} (\bibinfo{year}{2016}) \bibinfo{pages}{155},
  \bibinfo{doi}{\doi{10.1007/JHEP01(2016)155}}, \eprint{1511.08039}.

\bibtype{Article}%
\bibitem{ATLAS:2019zci}
\bibinfo{author}{G. Aad}, et al. (\bibinfo{collaboration}{ATLAS}),
  \bibinfo{title}{{Measurement of the transverse momentum distribution of
  Drell{\textendash}Yan lepton pairs in proton{\textendash}proton collisions at
  $\sqrt{s}=13$ TeV with the ATLAS detector}}, \bibinfo{journal}{Eur. Phys. J.
  C} \bibinfo{volume}{80} (\bibinfo{number}{7}) (\bibinfo{year}{2020})
  \bibinfo{pages}{616}, \bibinfo{doi}{\doi{10.1140/epjc/s10052-020-8001-z}},
  \eprint{1912.02844}.

\bibtype{Article}%
\bibitem{CMS:2022ubq}
\bibinfo{author}{A. Tumasyan}, et al. (\bibinfo{collaboration}{CMS}),
  \bibinfo{title}{{Measurement of the mass dependence of the transverse
  momentum of lepton pairs in Drell-Yan production in proton-proton collisions
  at $\sqrt{s}$ = 13 TeV}}, \bibinfo{journal}{Eur. Phys. J. C}
  \bibinfo{volume}{83} (\bibinfo{number}{7}) (\bibinfo{year}{2023})
  \bibinfo{pages}{628}, \bibinfo{doi}{\doi{10.1140/epjc/s10052-023-11631-7}},
  \eprint{2205.04897}.

\bibtype{Article}%
\bibitem{BaBar:2014omp}
\bibinfo{author}{A.~J. Bevan}, et al. (\bibinfo{collaboration}{BaBar, Belle}),
  \bibinfo{title}{{The Physics of the B Factories}}, \bibinfo{journal}{Eur.
  Phys. J. C} \bibinfo{volume}{74} (\bibinfo{year}{2014})
  \bibinfo{pages}{3026}, \bibinfo{doi}{\doi{10.1140/epjc/s10052-014-3026-9}},
  \eprint{1406.6311}.

\bibtype{Article}%
\bibitem{Belle:2005dmx}
\bibinfo{author}{K. Abe}, et al. (\bibinfo{collaboration}{Belle}),
  \bibinfo{title}{{Measurement of azimuthal asymmetries in inclusive production
  of hadron pairs in $e^+e^-$ annihilation at Belle}}, \bibinfo{journal}{Phys.
  Rev. Lett.} \bibinfo{volume}{96} (\bibinfo{year}{2006})
  \bibinfo{pages}{232002}, \bibinfo{doi}{\doi{10.1103/PhysRevLett.96.232002}},
  \eprint{hep-ex/0507063}.

\bibtype{Article}%
\bibitem{Belle:2008fdv}
\bibinfo{author}{R. Seidl}, et al. (\bibinfo{collaboration}{Belle}),
  \bibinfo{title}{{Measurement of Azimuthal Asymmetries in Inclusive Production
  of Hadron Pairs in $e^+e^-$ Annihilation at $\sqrt{s}$ = 10.58 GeV}},
  \bibinfo{journal}{Phys. Rev. D} \bibinfo{volume}{78} (\bibinfo{year}{2008})
  \bibinfo{pages}{032011}, \bibinfo{doi}{\doi{10.1103/PhysRevD.78.032011}},
  \bibinfo{note}{[Erratum: Phys.Rev.D 86, 039905 (2012)]}, \eprint{0805.2975}.

\bibtype{Article}%
\bibitem{Belle:2011cur}
\bibinfo{author}{A. Vossen}, et al. (\bibinfo{collaboration}{Belle}),
  \bibinfo{title}{{Observation of transverse polarization asymmetries of
  charged pion pairs in $e^+e^-$ annihilation near $\sqrt{s}=10.58$ GeV}},
  \bibinfo{journal}{Phys. Rev. Lett.} \bibinfo{volume}{107}
  (\bibinfo{year}{2011}) \bibinfo{pages}{072004},
  \bibinfo{doi}{\doi{10.1103/PhysRevLett.107.072004}}, \eprint{1104.2425}.

\bibtype{Article}%
\bibitem{BaBar:2013jdt}
\bibinfo{author}{J.~P. Lees}, et al. (\bibinfo{collaboration}{BaBar}),
  \bibinfo{title}{{Measurement of Collins asymmetries in inclusive production
  of charged pion pairs in $e^+e^-$ annihilation at BABAR}},
  \bibinfo{journal}{Phys. Rev. D} \bibinfo{volume}{90} (\bibinfo{number}{5})
  (\bibinfo{year}{2014}) \bibinfo{pages}{052003},
  \bibinfo{doi}{\doi{10.1103/PhysRevD.90.052003}}, \eprint{1309.5278}.

\bibtype{Article}%
\bibitem{Belle:2017rwm}
\bibinfo{author}{R. Seidl}, et al. (\bibinfo{collaboration}{Belle}),
  \bibinfo{title}{{Invariant-mass and fractional-energy dependence of inclusive
  production of di-hadrons in $e^+e^-$ annihilation at $\sqrt{s}=$ 10.58 GeV}},
  \bibinfo{journal}{Phys. Rev. D} \bibinfo{volume}{96} (\bibinfo{number}{3})
  (\bibinfo{year}{2017}) \bibinfo{pages}{032005},
  \bibinfo{doi}{\doi{10.1103/PhysRevD.96.032005}}, \eprint{1706.08348}.

\bibtype{Article}%
\bibitem{Parisi:2025nob}
\bibinfo{author}{Giorgio Parisi}, \bibinfo{title}{{From Bjorken Scaling to
  Scaling Violations}}, \bibinfo{journal}{HiHEP} \bibinfo{volume}{1}
  (\bibinfo{number}{1}) (\bibinfo{year}{2025}) \bibinfo{pages}{17},
  \bibinfo{doi}{\doi{10.53941/hihep.2025.100017}}, \eprint{2506.03383}.

\bibtype{Article}%
\bibitem{ParticleDataGroup:2024cfk}
\bibinfo{author}{S. Navas}, et al. (\bibinfo{collaboration}{Particle Data
  Group}), \bibinfo{title}{{Review of particle physics}},
  \bibinfo{journal}{Phys. Rev. D} \bibinfo{volume}{110} (\bibinfo{number}{3})
  (\bibinfo{year}{2024}) \bibinfo{pages}{030001},
  \bibinfo{doi}{\doi{10.1103/PhysRevD.110.030001}}.

\bibtype{Article}%
\bibitem{Bjorken:1967fb}
\bibinfo{author}{J.~D. Bjorken}, \bibinfo{title}{{Current algebra at small
  distances}}, \bibinfo{journal}{Conf. Proc. C} \bibinfo{volume}{670717}
  (\bibinfo{year}{1967}) \bibinfo{pages}{55--81}.

\bibtype{Article}%
\bibitem{Bjorken:1968dy}
\bibinfo{author}{J.~D. Bjorken}, \bibinfo{title}{{Asymptotic Sum Rules at
  Infinite Momentum}}, \bibinfo{journal}{Phys. Rev.} \bibinfo{volume}{179}
  (\bibinfo{year}{1969}) \bibinfo{pages}{1547--1553},
  \bibinfo{doi}{\doi{10.1103/PhysRev.179.1547}}.

\bibtype{Article}%
\bibitem{Gross:2005nobel}
\bibinfo{author}{D.~J. Gross}, \bibinfo{title}{{Nobel Lecture: The discovery of
  asymptotic freedom and the emergence of QCD}}, \bibinfo{journal}{Rev. Mod.
  Phys.} \bibinfo{volume}{77} (\bibinfo{year}{2005}) \bibinfo{pages}{837--849},
  \bibinfo{doi}{\doi{10.1103/RevModPhys.77.837}}.

\bibtype{Article}%
\bibitem{Feynman:1969ej}
\bibinfo{author}{R.~P. Feynman}, \bibinfo{title}{{Very high-energy collisions
  of hadrons}}, \bibinfo{journal}{Phys. Rev. Lett.} \bibinfo{volume}{23}
  (\bibinfo{year}{1969}) \bibinfo{pages}{1415--1417},
  \bibinfo{doi}{\doi{[,494(1969)]}}.

\bibtype{Article}%
\bibitem{Bjorken:1969ja}
\bibinfo{author}{J.~D. Bjorken}, \bibinfo{author}{E.~A. Paschos},
  \bibinfo{title}{{Inelastic Electron-Proton and $\gamma$-Proton Scattering,
  and the Structure of the Nucleon}}, \bibinfo{journal}{Phys. Rev.}
  \bibinfo{volume}{185} (\bibinfo{year}{1969}) \bibinfo{pages}{1975--1982},
  \bibinfo{doi}{\doi{10.1103/PhysRev.185.1975}}.

\bibtype{Article}%
\bibitem{Callan:1969uq}
\bibinfo{author}{C.~G. Callan, Jr.}, \bibinfo{author}{D.~J. Gross},
  \bibinfo{title}{{High-energy electroproduction and the constitution of the
  electric current}}, \bibinfo{journal}{Phys. Rev. Lett.} \bibinfo{volume}{22}
  (\bibinfo{year}{1969}) \bibinfo{pages}{156--159},
  \bibinfo{doi}{\doi{10.1103/PhysRevLett.22.156}}.

\bibtype{Article}%
\bibitem{Drell:1970wh}
\bibinfo{author}{S.~D. Drell}, \bibinfo{author}{Tung-Mow Yan},
  \bibinfo{title}{{Massive Lepton Pair Production in Hadron-Hadron Collisions
  at High-Energies}}, \bibinfo{journal}{Phys. Rev. Lett.} \bibinfo{volume}{25}
  (\bibinfo{year}{1970}) \bibinfo{pages}{316--320},
  \bibinfo{doi}{\doi{10.1103/PhysRevLett.25.316}}.

\bibtype{Article}%
\bibitem{Entem:2025bqt}
\bibinfo{author}{D.~R. Entem}, \bibinfo{author}{F. Fern\'andez},
  \bibinfo{author}{P.~G. Ortega}, \bibinfo{author}{J. Segovia},
  \bibinfo{title}{{The Constituent Quark Model}}  (\bibinfo{year}{2025}),
  \eprint{2504.07897}.

\bibtype{Article}%
\bibitem{Collins:1989gx}
\bibinfo{author}{J.~C. Collins}, \bibinfo{author}{Davison~E. Soper},
  \bibinfo{author}{G.~F. Sterman}, \bibinfo{title}{{Factorization of Hard
  Processes in QCD}}, \bibinfo{journal}{Adv. Ser. Direct. High Energy Phys.}
  \bibinfo{volume}{5} (\bibinfo{year}{1989}) \bibinfo{pages}{1--91},
  \bibinfo{doi}{\doi{10.1142/9789814503266_0001}}, \eprint{hep-ph/0409313}.

\bibtype{Article}%
\bibitem{CTEQ:1993hwr}
\bibinfo{author}{R. Brock}, et al. (\bibinfo{collaboration}{CTEQ}),
  \bibinfo{title}{{Handbook of perturbative QCD: Version 1.0}},
  \bibinfo{journal}{Rev. Mod. Phys.} \bibinfo{volume}{67}
  (\bibinfo{year}{1995}) \bibinfo{pages}{157--248},
  \bibinfo{doi}{\doi{10.1103/RevModPhys.67.157}}.

\bibtype{Book}%
\bibitem{Collins:2011zzd}
\bibinfo{author}{J. Collins}, \bibinfo{title}{{Foundations of Perturbative
  QCD}}, \bibinfo{comment}{vol.} \bibinfo{volume}{32},
  \bibinfo{publisher}{Cambridge University Press} \bibinfo{year}{2011}, ISBN
  \bibinfo{isbn}{978-1-009-40184-5, 978-1-009-40183-8, 978-1-009-40182-1},
  \bibinfo{doi}{\doi{10.1017/9781009401845}}.

\bibtype{Article}%
\bibitem{Moch:2004pa}
\bibinfo{author}{S. Moch}, \bibinfo{author}{J.~A.~M. Vermaseren},
  \bibinfo{author}{A. Vogt}, \bibinfo{title}{{The Three loop splitting
  functions in QCD: The Nonsinglet case}}, \bibinfo{journal}{Nucl. Phys. B}
  \bibinfo{volume}{688} (\bibinfo{year}{2004}) \bibinfo{pages}{101--134},
  \bibinfo{doi}{\doi{10.1016/j.nuclphysb.2004.03.030}},
  \eprint{hep-ph/0403192}.

\bibtype{Article}%
\bibitem{Vogt:2004mw}
\bibinfo{author}{A. Vogt}, \bibinfo{author}{S. Moch}, \bibinfo{author}{J.~A.~M.
  Vermaseren}, \bibinfo{title}{{The Three-loop splitting functions in QCD: The
  Singlet case}}, \bibinfo{journal}{Nucl. Phys. B} \bibinfo{volume}{691}
  (\bibinfo{year}{2004}) \bibinfo{pages}{129--181},
  \bibinfo{doi}{\doi{10.1016/j.nuclphysb.2004.04.024}},
  \eprint{hep-ph/0404111}.

\bibtype{Article}%
\bibitem{Moch:2017uml}
\bibinfo{author}{S. Moch}, \bibinfo{author}{B. Ruijl}, \bibinfo{author}{T.
  Ueda}, \bibinfo{author}{J.~A.~M. Vermaseren}, \bibinfo{author}{A. Vogt},
  \bibinfo{title}{{Four-Loop Non-Singlet Splitting Functions in the Planar
  Limit and Beyond}}, \bibinfo{journal}{JHEP} \bibinfo{volume}{10}
  (\bibinfo{year}{2017}) \bibinfo{pages}{041},
  \bibinfo{doi}{\doi{10.1007/JHEP10(2017)041}}, \eprint{1707.08315}.

\bibtype{Article}%
\bibitem{Bollini:1972ui}
\bibinfo{author}{C.~G. Bollini}, \bibinfo{author}{J.~J. Giambiagi},
  \bibinfo{title}{{Dimensional Renormalization: The Number of Dimensions as a
  Regularizing Parameter}}, \bibinfo{journal}{Nuovo Cim. B}
  \bibinfo{volume}{12} (\bibinfo{year}{1972}) \bibinfo{pages}{20--26},
  \bibinfo{doi}{\doi{10.1007/BF02895558}}.

\bibtype{Article}%
\bibitem{Cicuta:1972jf}
\bibinfo{author}{G.~M. Cicuta}, \bibinfo{author}{E. Montaldi},
  \bibinfo{title}{{Analytic renormalization via continuous space dimension}},
  \bibinfo{journal}{Lett. Nuovo Cim.} \bibinfo{volume}{4}
  (\bibinfo{year}{1972}) \bibinfo{pages}{329--332},
  \bibinfo{doi}{\doi{10.1007/BF02756527}}.

\bibtype{Article}%
\bibitem{Bardeen:1978yd}
\bibinfo{author}{W.~A. Bardeen}, \bibinfo{author}{A.~J. Buras},
  \bibinfo{author}{D.~W. Duke}, \bibinfo{author}{T. Muta},
  \bibinfo{title}{{Deep Inelastic Scattering Beyond the Leading Order in
  Asymptotically Free Gauge Theories}}, \bibinfo{journal}{Phys. Rev. D}
  \bibinfo{volume}{18} (\bibinfo{year}{1978}) \bibinfo{pages}{3998},
  \bibinfo{doi}{\doi{10.1103/PhysRevD.18.3998}}.

\bibtype{Article}%
\bibitem{Dokshitzer:1977sg}
\bibinfo{author}{Y.~L. Dokshitzer}, \bibinfo{title}{{Calculation of the
  structure functions for deep inelastic scattering and $e^=e^-$ annihilation
  by perturbation theory in Quantum Chromodynamics.}}, \bibinfo{journal}{Sov.
  Phys. JETP} \bibinfo{volume}{46} (\bibinfo{year}{1977})
  \bibinfo{pages}{641--653}.

\bibtype{Article}%
\bibitem{Gribov:1972ri}
\bibinfo{author}{V.~N. Gribov}, \bibinfo{author}{L.~N. Lipatov},
  \bibinfo{title}{{Deep inelastic e p scattering in perturbation theory}},
  \bibinfo{journal}{Sov. J. Nucl. Phys.} \bibinfo{volume}{15}
  (\bibinfo{year}{1972}) \bibinfo{pages}{438--450}.

\bibtype{Article}%
\bibitem{Altarelli:1977zs}
\bibinfo{author}{G. Altarelli}, \bibinfo{author}{G. Parisi},
  \bibinfo{title}{{Asymptotic Freedom in Parton Language}},
  \bibinfo{journal}{Nucl. Phys. B} \bibinfo{volume}{126} (\bibinfo{year}{1977})
  \bibinfo{pages}{298--318}, \bibinfo{doi}{\doi{10.1016/0550-3213(77)90384-4}}.

\bibtype{Article}%
\bibitem{Altarelli:1979ub}
\bibinfo{author}{G. Altarelli}, \bibinfo{author}{R.~K. Ellis},
  \bibinfo{author}{G. Martinelli}, \bibinfo{title}{{Large Perturbative
  Corrections to the Drell-Yan Process in QCD}}, \bibinfo{journal}{Nucl. Phys.
  B} \bibinfo{volume}{157} (\bibinfo{year}{1979}) \bibinfo{pages}{461--497},
  \bibinfo{doi}{\doi{10.1016/0550-3213(79)90116-0}}.

\bibtype{Article}%
\bibitem{Bodwin:1984hc}
\bibinfo{author}{G.~T. Bodwin}, \bibinfo{title}{{Factorization of the Drell-Yan
  Cross-Section in Perturbation Theory}}, \bibinfo{journal}{Phys. Rev. D}
  \bibinfo{volume}{31} (\bibinfo{year}{1985}) \bibinfo{pages}{2616},
  \bibinfo{doi}{\doi{10.1103/PhysRevD.34.3932}}.

\bibtype{Article}%
\bibitem{Collins:1985ue}
\bibinfo{author}{J.~C. Collins}, \bibinfo{author}{D.~E. Soper},
  \bibinfo{author}{G.~F. Sterman}, \bibinfo{title}{{Factorization for Short
  Distance Hadron - Hadron Scattering}}, \bibinfo{journal}{Nucl. Phys. B}
  \bibinfo{volume}{261} (\bibinfo{year}{1985}) \bibinfo{pages}{104--142},
  \bibinfo{doi}{\doi{10.1016/0550-3213(85)90565-6}}.

\bibtype{Article}%
\bibitem{Hollands:2023txn}
\bibinfo{author}{S. Hollands}, \bibinfo{author}{R.~M. Wald},
  \bibinfo{title}{{The Operator Product Expansion in Quantum Field Theory}}
  (\bibinfo{year}{2023}), \eprint{2312.01096}.

\bibtype{Book}%
\bibitem{Collins:1984xc}
\bibinfo{author}{J.~C. Collins}, \bibinfo{title}{{Renormalization : An
  Introduction to Renormalization, the Renormalization Group and the
  Operator-Product Expansion}}, \bibinfo{series}{Cambridge Monographs on
  Mathematical Physics}, \bibinfo{comment}{vol.} \bibinfo{volume}{26},
  \bibinfo{publisher}{Cambridge University Press}, \bibinfo{address}{Cambridge}
  \bibinfo{year}{1984}, ISBN \bibinfo{isbn}{978-0-521-31177-9,
  978-0-511-86739-2, 978-1-009-40180-7, 978-1-009-40176-0, 978-1-009-40179-1},
  \bibinfo{doi}{\doi{10.1017/9781009401807}}.

\bibtype{Inproceedings}%
\bibitem{Jaffe:1996zw}
\bibinfo{author}{R.~L. Jaffe}, \bibinfo{title}{{Spin, twist and hadron
  structure in deep inelastic processes}}, in: \bibinfo{booktitle}{{Ettore
  Majorana International School of Nucleon Structure: 1st Course: The Spin
  Structure of the Nucleon}} \bibinfo{year}{1996}, pp.
  \bibinfo{pages}{42--129}, \eprint{hep-ph/9602236}.

\bibtype{Book}%
\bibitem{Itzykson:1980rh}
\bibinfo{author}{C. Itzykson}, \bibinfo{author}{J.~B. Zuber},
  \bibinfo{title}{{Quantum Field Theory}}, International Series In Pure and
  Applied Physics, \bibinfo{publisher}{McGraw-Hill}, \bibinfo{address}{New
  York} \bibinfo{year}{1980}, ISBN \bibinfo{isbn}{978-0-486-44568-7}.

\bibtype{Article}%
\bibitem{Jaffe:1991ra}
\bibinfo{author}{R.~L. Jaffe}, \bibinfo{author}{X.-D. Ji},
  \bibinfo{title}{{Chiral odd parton distributions and Drell-Yan processes}},
  \bibinfo{journal}{Nucl. Phys. B} \bibinfo{volume}{375} (\bibinfo{year}{1992})
  \bibinfo{pages}{527--560}, \bibinfo{doi}{\doi{10.1016/0550-3213(92)90110-W}}.

\bibtype{Article}%
\bibitem{Soper:1972xc}
\bibinfo{author}{D.~E. Soper}, \bibinfo{title}{{Infinite-momentum helicity
  states}}, \bibinfo{journal}{Phys. Rev. D} \bibinfo{volume}{5}
  (\bibinfo{year}{1972}) \bibinfo{pages}{1956--1962},
  \bibinfo{doi}{\doi{10.1103/PhysRevD.5.1956}}.

\bibtype{Article}%
\bibitem{Soffer:1994ww}
\bibinfo{author}{J. Soffer}, \bibinfo{title}{{Positivity constraints for spin
  dependent parton distributions}}, \bibinfo{journal}{Phys. Rev. Lett.}
  \bibinfo{volume}{74} (\bibinfo{year}{1995}) \bibinfo{pages}{1292--1294},
  \bibinfo{doi}{\doi{10.1103/PhysRevLett.74.1292}}, \eprint{hep-ph/9409254}.

\bibtype{Inproceedings}%
\bibitem{Ralston:2008sm}
\bibinfo{author}{J.~P. Ralston}, \bibinfo{title}{{Exploring Confinement with
  Spin}}, in: \bibinfo{booktitle}{{2nd International Workshop on Transverse
  Polarization Phenomena in Hard Processes}} \bibinfo{year}{2008}, pp.
  \bibinfo{pages}{229--236}, \bibinfo{doi}{\doi{10.1142/9789814277785_0028}},
  \eprint{0810.0871}.

\bibtype{Article}%
\bibitem{Collins:2021vke}
\bibinfo{author}{J. Collins}, \bibinfo{author}{T.~C. Rogers},
  \bibinfo{author}{N. Sato}, \bibinfo{title}{{Positivity and renormalization of
  parton densities}}, \bibinfo{journal}{Phys. Rev. D} \bibinfo{volume}{105}
  (\bibinfo{number}{7}) (\bibinfo{year}{2022}) \bibinfo{pages}{076010},
  \bibinfo{doi}{\doi{10.1103/PhysRevD.105.076010}}, \eprint{2111.01170}.

\bibtype{Article}%
\bibitem{Vogelsang:1997ak}
\bibinfo{author}{W. Vogelsang}, \bibinfo{title}{{Next-to-leading order
  evolution of transversity distributions and Soffer's inequality}},
  \bibinfo{journal}{Phys. Rev. D} \bibinfo{volume}{57} (\bibinfo{year}{1998})
  \bibinfo{pages}{1886--1894}, \bibinfo{doi}{\doi{10.1103/PhysRevD.57.1886}},
  \eprint{hep-ph/9706511}.

\bibtype{Article}%
\bibitem{Candido:2023ujx}
\bibinfo{author}{A. Candido}, \bibinfo{author}{S. Forte}, \bibinfo{author}{T.
  Giani}, \bibinfo{author}{F. Hekhorn}, \bibinfo{title}{{On the positivity of
  $\overline{\textrm{MS}}$ parton distributions}}, \bibinfo{journal}{Eur. Phys.
  J. C} \bibinfo{volume}{84} (\bibinfo{number}{3}) (\bibinfo{year}{2024})
  \bibinfo{pages}{335}, \bibinfo{doi}{\doi{10.1140/epjc/s10052-024-12681-1}},
  \eprint{2308.00025}.

\bibtype{Article}%
\bibitem{Goldberger:1958vp}
\bibinfo{author}{M.~L. Goldberger}, \bibinfo{author}{S.~B. Treiman},
  \bibinfo{title}{{Form-factors in Beta decay and muon capture}},
  \bibinfo{journal}{Phys. Rev.} \bibinfo{volume}{111} (\bibinfo{year}{1958})
  \bibinfo{pages}{354--361}, \bibinfo{doi}{\doi{10.1103/PhysRev.111.354}}.

\bibtype{Article}%
\bibitem{Adler:1965ka}
\bibinfo{author}{S.~L. Adler}, \bibinfo{title}{{Calculation of the axial vector
  coupling constant renormalization in beta decay}}, \bibinfo{journal}{Phys.
  Rev. Lett.} \bibinfo{volume}{14} (\bibinfo{year}{1965})
  \bibinfo{pages}{1051--1055},
  \bibinfo{doi}{\doi{10.1103/PhysRevLett.14.1051}}.

\bibtype{Article}%
\bibitem{Adler:1969gk}
\bibinfo{author}{S.~L. Adler}, \bibinfo{title}{{Axial vector vertex in spinor
  electrodynamics}}, \bibinfo{journal}{Phys. Rev.} \bibinfo{volume}{177}
  (\bibinfo{year}{1969}) \bibinfo{pages}{2426--2438},
  \bibinfo{doi}{\doi{10.1103/PhysRev.177.2426}}.

\bibtype{Article}%
\bibitem{Bjorken:1966jh}
\bibinfo{author}{J.~D. Bjorken}, \bibinfo{title}{{Applications of the Chiral
  U(6) x U(6) Algebra of Current Densities}}, \bibinfo{journal}{Phys. Rev.}
  \bibinfo{volume}{148} (\bibinfo{year}{1966}) \bibinfo{pages}{1467--1478},
  \bibinfo{doi}{\doi{10.1103/PhysRev.148.1467}}.

\bibtype{Article}%
\bibitem{Bjorken:1969mm}
\bibinfo{author}{J.~D. Bjorken}, \bibinfo{title}{{Inelastic Scattering of
  Polarized Leptons from Polarized Nucleons}}, \bibinfo{journal}{Phys. Rev. D}
  \bibinfo{volume}{1} (\bibinfo{year}{1970}) \bibinfo{pages}{1376--1379},
  \bibinfo{doi}{\doi{10.1103/PhysRevD.1.1376}}.

\bibtype{Article}%
\bibitem{Baikov:2010je}
\bibinfo{author}{P.~A. Baikov}, \bibinfo{author}{K.~G. Chetyrkin},
  \bibinfo{author}{J.~H. Kuhn}, \bibinfo{title}{{Adler Function, Bjorken Sum
  Rule, and the Crewther Relation to Order $\alpha^4_s$ in a General Gauge
  Theory}}, \bibinfo{journal}{Phys. Rev. Lett.} \bibinfo{volume}{104}
  (\bibinfo{year}{2010}) \bibinfo{pages}{132004},
  \bibinfo{doi}{\doi{10.1103/PhysRevLett.104.132004}}, \eprint{1001.3606}.

\bibtype{Article}%
\bibitem{Boer:2011fh}
\bibinfo{author}{D. Boer}, et al., \bibinfo{title}{{Gluons and the quark sea at
  high energies: Distributions, polarization, tomography}}
  (\bibinfo{year}{2011}), \eprint{1108.1713}.

\bibtype{Article}%
\bibitem{Altarelli:1996nm}
\bibinfo{author}{G. Altarelli}, \bibinfo{author}{R.~D. Ball},
  \bibinfo{author}{S. Forte}, \bibinfo{author}{G. Ridolfi},
  \bibinfo{title}{{Determination of the Bjorken sum and strong coupling from
  polarized structure functions}}, \bibinfo{journal}{Nucl. Phys. B}
  \bibinfo{volume}{496} (\bibinfo{year}{1997}) \bibinfo{pages}{337--357},
  \bibinfo{doi}{\doi{10.1016/S0550-3213(97)00201-0}}, \eprint{hep-ph/9701289}.

\bibtype{Article}%
\bibitem{Accardi:2012qut}
\bibinfo{author}{A. Accardi}, et al., \bibinfo{title}{{Electron Ion Collider:
  The Next QCD Frontier}: {Understanding the glue that binds us all}},
  \bibinfo{journal}{Eur. Phys. J. A} \bibinfo{volume}{52} (\bibinfo{number}{9})
  (\bibinfo{year}{2016}) \bibinfo{pages}{268},
  \bibinfo{doi}{\doi{10.1140/epja/i2016-16268-9}}, \eprint{1212.1701}.

\bibtype{Article}%
\bibitem{Gross:1973zrg}
\bibinfo{author}{D.~J. Gross}, \bibinfo{author}{Frank Wilczek},
  \bibinfo{title}{{Asymptotically free gauge theories. 2.}},
  \bibinfo{journal}{Phys. Rev. D} \bibinfo{volume}{9} (\bibinfo{year}{1974})
  \bibinfo{pages}{980--993}, \bibinfo{doi}{\doi{10.1103/PhysRevD.9.980}}.

\bibtype{Article}%
\bibitem{Politzer:1974fr}
\bibinfo{author}{H.~D. Politzer}, \bibinfo{title}{{Asymptotic Freedom: An
  Approach to Strong Interactions}}, \bibinfo{journal}{Phys. Rept.}
  \bibinfo{volume}{14} (\bibinfo{year}{1974}) \bibinfo{pages}{129--180},
  \bibinfo{doi}{\doi{10.1016/0370-1573(74)90014-3}}.

\bibtype{Article}%
\bibitem{Georgi:1974wnj}
\bibinfo{author}{H. Georgi}, \bibinfo{author}{H.~D. Politzer},
  \bibinfo{title}{{Electroproduction scaling in an asymptotically free theory
  of strong interactions}}, \bibinfo{journal}{Phys. Rev. D} \bibinfo{volume}{9}
  (\bibinfo{year}{1974}) \bibinfo{pages}{416--420},
  \bibinfo{doi}{\doi{10.1103/PhysRevD.9.416}}.

\bibtype{Article}%
\bibitem{Ellis:1982cd}
\bibinfo{author}{R.~K. Ellis}, \bibinfo{author}{W. Furmanski},
  \bibinfo{author}{R. Petronzio}, \bibinfo{title}{{Unraveling Higher Twists}},
  \bibinfo{journal}{Nucl. Phys. B} \bibinfo{volume}{212} (\bibinfo{year}{1983})
  \bibinfo{pages}{29}, \bibinfo{doi}{\doi{10.1016/0550-3213(83)90597-7}}.

\bibtype{Article}%
\bibitem{Jaffe:1983hp}
\bibinfo{author}{R.~L. Jaffe}, \bibinfo{title}{{Parton Distribution Functions
  for Twist Four}}, \bibinfo{journal}{Nucl. Phys. B} \bibinfo{volume}{229}
  (\bibinfo{year}{1983}) \bibinfo{pages}{205--230},
  \bibinfo{doi}{\doi{10.1016/0550-3213(83)90361-9}}.

\bibtype{Article}%
\bibitem{Balitsky:1987bk}
\bibinfo{author}{I.~I. Balitsky}, \bibinfo{author}{V.~M. Braun},
  \bibinfo{title}{{Evolution Equations for QCD String Operators}},
  \bibinfo{journal}{Nucl. Phys. B} \bibinfo{volume}{311} (\bibinfo{year}{1989})
  \bibinfo{pages}{541--584}, \bibinfo{doi}{\doi{10.1016/0550-3213(89)90168-5}}.

\bibtype{Article}%
\bibitem{Kanazawa:2015ajw}
\bibinfo{author}{K. Kanazawa}, \bibinfo{author}{Y. Koike}, \bibinfo{author}{A.
  Metz}, \bibinfo{author}{D. Pitonyak}, \bibinfo{author}{M. Schlegel},
  \bibinfo{title}{{Operator Constraints for Twist-3 Functions and Lorentz
  Invariance Properties of Twist-3 Observables}}, \bibinfo{journal}{Phys. Rev.
  D} \bibinfo{volume}{93} (\bibinfo{number}{5}) (\bibinfo{year}{2016})
  \bibinfo{pages}{054024}, \bibinfo{doi}{\doi{10.1103/PhysRevD.93.054024}},
  \eprint{1512.07233}.

\bibtype{Inproceedings}%
\bibitem{Caldwell:2025fjg}
\bibinfo{author}{A. Caldwell}, \bibinfo{author}{S. Dalla~Torre},
  \bibinfo{author}{R. Ent}, \bibinfo{author}{A. Levy}, \bibinfo{author}{P.
  Newman}, \bibinfo{author}{F. Olness}, \bibinfo{author}{J. Rojo},
  \bibinfo{title}{{Future Opportunities with Lepton-Hadron Collisions}}
  \bibinfo{year}{2025}, \eprint{2503.18208}.

\bibtype{Article}%
\bibitem{Accardi:2023chb}
\bibinfo{author}{A. Accardi}, et al., \bibinfo{title}{{Strong interaction
  physics at the luminosity frontier with 22 GeV electrons at Jefferson Lab}},
  \bibinfo{journal}{Eur. Phys. J. A} \bibinfo{volume}{60} (\bibinfo{number}{9})
  (\bibinfo{year}{2024}) \bibinfo{pages}{173},
  \bibinfo{doi}{\doi{10.1140/epja/s10050-024-01282-x}}, \eprint{2306.09360}.

\bibtype{Article}%
\bibitem{Ahmadova:2025vzd}
\bibinfo{author}{F. Ahmadova}, et al., \bibinfo{title}{{The Large Hadron
  electron Collider as a bridge project for CERN}}  (\bibinfo{year}{2025}),
  \eprint{2503.17727}.

\bibtype{Article}%
\bibitem{AbdulKhalek:2021gbh}
\bibinfo{author}{R. Abdul~Khalek}, et al., \bibinfo{title}{{Science
  Requirements and Detector Concepts for the Electron-Ion Collider}: {EIC
  Yellow Report}}, \bibinfo{journal}{Nucl. Phys. A} \bibinfo{volume}{1026}
  (\bibinfo{year}{2022}) \bibinfo{pages}{122447},
  \bibinfo{doi}{\doi{10.1016/j.nuclphysa.2022.122447}}, \eprint{2103.05419}.

\bibtype{Article}%
\bibitem{Anderle:2021wcy}
\bibinfo{author}{D.~P. Anderle}, et al., \bibinfo{title}{{Electron-ion collider
  in China}}, \bibinfo{journal}{Front. Phys. (Beijing)} \bibinfo{volume}{16}
  (\bibinfo{number}{6}) (\bibinfo{year}{2021}) \bibinfo{pages}{64701},
  \bibinfo{doi}{\doi{10.1007/s11467-021-1062-0}}, \eprint{2102.09222}.

\bibtype{Article}%
\bibitem{Buckley:2014ana}
\bibinfo{author}{A. Buckley}, \bibinfo{author}{J. Ferrando},
  \bibinfo{author}{Stephen Lloyd}, \bibinfo{author}{K. Nordstr\"om},
  \bibinfo{author}{B. Page}, \bibinfo{author}{M. R\"ufenacht},
  \bibinfo{author}{M. Sch\"onherr}, \bibinfo{author}{G. Watt},
  \bibinfo{title}{{LHAPDF6: parton density access in the LHC precision era}},
  \bibinfo{journal}{Eur. Phys. J. C} \bibinfo{volume}{75}
  (\bibinfo{year}{2015}) \bibinfo{pages}{132},
  \bibinfo{doi}{\doi{10.1140/epjc/s10052-015-3318-8}}, \eprint{1412.7420}.

\bibtype{Article}%
\bibitem{NNPDF:2021njg}
\bibinfo{author}{R.~D. Ball}, et al. (\bibinfo{collaboration}{NNPDF}),
  \bibinfo{title}{{The path to proton structure at 1\% accuracy}},
  \bibinfo{journal}{Eur. Phys. J. C} \bibinfo{volume}{82} (\bibinfo{number}{5})
  (\bibinfo{year}{2022}) \bibinfo{pages}{428},
  \bibinfo{doi}{\doi{10.1140/epjc/s10052-022-10328-7}}, \eprint{2109.02653}.

\bibtype{Article}%
\bibitem{Borsa:2024mss}
\bibinfo{author}{I. Borsa}, \bibinfo{author}{M. Stratmann}, \bibinfo{author}{W.
  Vogelsang}, \bibinfo{author}{D. de Florian}, \bibinfo{author}{R. Sassot},
  \bibinfo{title}{{Next-to-Next-to-Leading Order Global Analysis of Polarized
  Parton Distribution Functions}}, \bibinfo{journal}{Phys. Rev. Lett.}
  \bibinfo{volume}{133} (\bibinfo{number}{15}) (\bibinfo{year}{2024})
  \bibinfo{pages}{151901}, \bibinfo{doi}{\doi{10.1103/PhysRevLett.133.151901}},
  \eprint{2407.11635}.

\bibtype{Article}%
\bibitem{Bailey:2020ooq}
\bibinfo{author}{S. Bailey}, \bibinfo{author}{T. Cridge},
  \bibinfo{author}{L.~A. Harland-Lang}, \bibinfo{author}{A.~D. Martin},
  \bibinfo{author}{R.~S. Thorne}, \bibinfo{title}{{Parton distributions from
  LHC, HERA, Tevatron and fixed target data: MSHT20 PDFs}},
  \bibinfo{journal}{Eur. Phys. J. C} \bibinfo{volume}{81} (\bibinfo{number}{4})
  (\bibinfo{year}{2021}) \bibinfo{pages}{341},
  \bibinfo{doi}{\doi{10.1140/epjc/s10052-021-09057-0}}, \eprint{2012.04684}.

\bibtype{Article}%
\bibitem{Hou:2019qau}
\bibinfo{author}{T.-J. Hou}, et al., \bibinfo{title}{{Progress in the CTEQ-TEA
  NNLO global QCD analysis}}  (\bibinfo{year}{2019}), \eprint{1908.11394}.

\bibtype{Article}%
\bibitem{Ethier:2020way}
\bibinfo{author}{J.~J. Ethier}, \bibinfo{author}{E.~R. Nocera},
  \bibinfo{title}{{Parton Distributions in Nucleons and Nuclei}},
  \bibinfo{journal}{Ann. Rev. Nucl. Part. Sci.} \bibinfo{volume}{70}
  (\bibinfo{year}{2020}) \bibinfo{pages}{43--76},
  \bibinfo{doi}{\doi{10.1146/annurev-nucl-011720-042725}}, \eprint{2001.07722}.

\bibtype{Article}%
\bibitem{Ball:2017otu}
\bibinfo{author}{R.~D. Ball}, \bibinfo{author}{V. Bertone}, \bibinfo{author}{M.
  Bonvini}, \bibinfo{author}{S. Marzani}, \bibinfo{author}{J. Rojo},
  \bibinfo{author}{L. Rottoli}, \bibinfo{title}{{Parton distributions with
  small-x resummation: evidence for BFKL dynamics in HERA data}},
  \bibinfo{journal}{Eur. Phys. J. C} \bibinfo{volume}{78} (\bibinfo{number}{4})
  (\bibinfo{year}{2018}) \bibinfo{pages}{321},
  \bibinfo{doi}{\doi{10.1140/epjc/s10052-018-5774-4}}, \eprint{1710.05935}.

\bibtype{Article}%
\bibitem{Kumano:1997cy}
\bibinfo{author}{S. Kumano}, \bibinfo{title}{{Flavor asymmetry of anti-quark
  distributions in the nucleon}}, \bibinfo{journal}{Phys. Rept.}
  \bibinfo{volume}{303} (\bibinfo{year}{1998}) \bibinfo{pages}{183--257},
  \bibinfo{doi}{\doi{10.1016/S0370-1573(98)00016-7}}, \eprint{hep-ph/9702367}.

\bibtype{Article}%
\bibitem{Cocuzza:2021cbi}
\bibinfo{author}{C. Cocuzza}, \bibinfo{author}{W. Melnitchouk},
  \bibinfo{author}{A. Metz}, \bibinfo{author}{N. Sato}
  (\bibinfo{collaboration}{Jefferson Lab Angular Momentum (JAM)}),
  \bibinfo{title}{{Bayesian Monte~Carlo extraction of the sea asymmetry with
  SeaQuest and STAR data}}, \bibinfo{journal}{Phys. Rev. D}
  \bibinfo{volume}{104} (\bibinfo{number}{7}) (\bibinfo{year}{2021})
  \bibinfo{pages}{074031}, \bibinfo{doi}{\doi{10.1103/PhysRevD.104.074031}},
  \eprint{2109.00677}.

\bibtype{Article}%
\bibitem{Brodsky:1980pb}
\bibinfo{author}{S.~J. Brodsky}, \bibinfo{author}{P. Hoyer},
  \bibinfo{author}{C. Peterson}, \bibinfo{author}{N. Sakai},
  \bibinfo{title}{{The Intrinsic Charm of the Proton}}, \bibinfo{journal}{Phys.
  Lett. B} \bibinfo{volume}{93} (\bibinfo{year}{1980})
  \bibinfo{pages}{451--455}, \bibinfo{doi}{\doi{10.1016/0370-2693(80)90364-0}}.

\bibtype{Article}%
\bibitem{Ball:2022qks}
\bibinfo{author}{R.~D. Ball}, \bibinfo{author}{A. Candido}, \bibinfo{author}{J.
  Cruz-Martinez}, \bibinfo{author}{S. Forte}, \bibinfo{author}{T. Giani},
  \bibinfo{author}{F. Hekhorn}, \bibinfo{author}{K. Kudashkin},
  \bibinfo{author}{G. Magni}, \bibinfo{author}{J. Rojo}
  (\bibinfo{collaboration}{NNPDF}), \bibinfo{title}{{Evidence for intrinsic
  charm quarks in the proton}}, \bibinfo{journal}{Nature} \bibinfo{volume}{608}
  (\bibinfo{number}{7923}) (\bibinfo{year}{2022}) \bibinfo{pages}{483--487},
  \bibinfo{doi}{\doi{10.1038/s41586-022-04998-2}}, \eprint{2208.08372}.

\bibtype{Article}%
\bibitem{Guzzi:2022rca}
\bibinfo{author}{M. Guzzi}, \bibinfo{author}{T.~J. Hobbs}, \bibinfo{author}{K.
  Xie}, \bibinfo{author}{J. Huston}, \bibinfo{author}{P. Nadolsky},
  \bibinfo{author}{C.~P. Yuan}, \bibinfo{title}{{The persistent nonperturbative
  charm enigma}}, \bibinfo{journal}{Phys. Lett. B} \bibinfo{volume}{843}
  (\bibinfo{year}{2023}) \bibinfo{pages}{137975},
  \bibinfo{doi}{\doi{10.1016/j.physletb.2023.137975}}, \eprint{2211.01387}.

\bibtype{Article}%
\bibitem{Bertone:2024taw}
\bibinfo{author}{V. Bertone}, \bibinfo{author}{A. Chiefa},
  \bibinfo{author}{E.~R. Nocera} (\bibinfo{collaboration}{MAP
  (Multi-dimensional Analyses of Partonic distributions)}),
  \bibinfo{title}{{Helicity-dependent parton distribution functions at
  next-to-next-to-leading order accuracy from inclusive and semi-inclusive
  deep-inelastic scattering data}}, \bibinfo{journal}{Phys. Lett. B}
  \bibinfo{volume}{865} (\bibinfo{year}{2025}) \bibinfo{pages}{139497},
  \bibinfo{doi}{\doi{10.1016/j.physletb.2025.139497}}, \eprint{2404.04712}.

\bibtype{Article}%
\bibitem{Cruz-Martinez:2025ahf}
\bibinfo{author}{Juan Cruz-Martinez}, \bibinfo{author}{Toon Hasenack},
  \bibinfo{author}{Felix Hekhorn}, \bibinfo{author}{Giacomo Magni},
  \bibinfo{author}{Emanuele~R. Nocera}, \bibinfo{author}{Tanjona~R.
  Rabemananjara}, \bibinfo{author}{Juan Rojo}, \bibinfo{author}{Tanishq
  Sharma}, \bibinfo{author}{Gijs van Seeventer}, \bibinfo{title}{{NNPDFpol2.0:
  a global determination of polarised PDFs and their uncertainties at
  next-to-next-to-leading order}}, \bibinfo{journal}{JHEP} \bibinfo{volume}{07}
  (\bibinfo{year}{2025}) \bibinfo{pages}{168},
  \bibinfo{doi}{\doi{10.1007/JHEP07(2025)168}}, \eprint{2503.11814}.

\bibtype{Article}%
\bibitem{Cocuzza:2025qvf}
\bibinfo{author}{C. Cocuzza}, \bibinfo{author}{N.~T. Hunt-Smith},
  \bibinfo{author}{W. Melnitchouk}, \bibinfo{author}{N. Sato},
  \bibinfo{author}{A.~W. Thomas} (\bibinfo{collaboration}{JAM Collaboration
  (Spin PDF Analysis Group)}), \bibinfo{title}{{Global QCD analysis of spin
  PDFs in the proton with high-x and lattice constraints}},
  \bibinfo{journal}{Phys. Rev. D} \bibinfo{volume}{112} (\bibinfo{number}{11})
  (\bibinfo{year}{2025}) \bibinfo{pages}{114017},
  \bibinfo{doi}{\doi{10.1103/6fn9-1wqb}}, \eprint{2506.13616}.

\bibtype{Article}%
\bibitem{Cocuzza:2022jye}
\bibinfo{author}{C. Cocuzza}, \bibinfo{author}{W. Melnitchouk},
  \bibinfo{author}{A. Metz}, \bibinfo{author}{N. Sato}
  (\bibinfo{collaboration}{Jefferson Lab Angular Momentum (JAM)}),
  \bibinfo{title}{{Polarized antimatter in the proton from a global QCD
  analysis}}, \bibinfo{journal}{Phys. Rev. D} \bibinfo{volume}{106}
  (\bibinfo{number}{3}) (\bibinfo{year}{2022}) \bibinfo{pages}{L031502},
  \bibinfo{doi}{\doi{10.1103/PhysRevD.106.L031502}}, \eprint{2202.03372}.

\bibtype{Article}%
\bibitem{Aschenauer:2015ata}
\bibinfo{author}{E.~C. Aschenauer}, \bibinfo{author}{R. Sassot},
  \bibinfo{author}{M. Stratmann}, \bibinfo{title}{{Unveiling the Proton Spin
  Decomposition at a Future Electron-Ion Collider}}, \bibinfo{journal}{Phys.
  Rev. D} \bibinfo{volume}{92} (\bibinfo{number}{9}) (\bibinfo{year}{2015})
  \bibinfo{pages}{094030}, \bibinfo{doi}{\doi{10.1103/PhysRevD.92.094030}},
  \eprint{1509.06489}.

\bibtype{Article}%
\bibitem{Ball:2013tyh}
\bibinfo{author}{R.~D. Ball}, \bibinfo{author}{S. Forte}, \bibinfo{author}{A.
  Guffanti}, \bibinfo{author}{E.~R. Nocera}, \bibinfo{author}{G. Ridolfi},
  \bibinfo{author}{J. Rojo} (\bibinfo{collaboration}{NNPDF}),
  \bibinfo{title}{{Polarized Parton Distributions at an Electron-Ion
  Collider}}, \bibinfo{journal}{Phys. Lett. B} \bibinfo{volume}{728}
  (\bibinfo{year}{2014}) \bibinfo{pages}{524--531},
  \bibinfo{doi}{\doi{10.1016/j.physletb.2013.12.023}}, \eprint{1310.0461}.

\bibtype{Article}%
\bibitem{Barshay:1975zz}
\bibinfo{author}{S. Barshay}, \bibinfo{author}{C.~B. Dover},
  \bibinfo{author}{J.~P. Vary}, \bibinfo{title}{{Nucleus-nucleus cross sections
  and the validity of the factorization hypothesis at intermediate and high
  energies}}, \bibinfo{journal}{Phys. Rev. C} \bibinfo{volume}{11}
  (\bibinfo{year}{1975}) \bibinfo{pages}{360--369},
  \bibinfo{doi}{\doi{10.1103/PhysRevC.11.360}}.

\bibtype{Article}%
\bibitem{Klasen:2023uqj}
\bibinfo{author}{M. Klasen}, \bibinfo{author}{H. Paukkunen},
  \bibinfo{title}{{Nuclear PDFs After the First Decade of LHC Data}},
  \bibinfo{journal}{Ann. Rev. Nucl. Part. Sci.} \bibinfo{volume}{74}
  (\bibinfo{year}{2024}) \bibinfo{pages}{49--87},
  \bibinfo{doi}{\doi{10.1146/annurev-nucl-102122-022747}}, \eprint{2311.00450}.

\bibtype{Article}%
\bibitem{Ruiz:2023ozv}
\bibinfo{author}{R. Ruiz}, et al., \bibinfo{title}{{Target mass corrections in
  lepton\textendash{}nucleus DIS: Theory and applications to nuclear PDFs}},
  \bibinfo{journal}{Prog. Part. Nucl. Phys.} \bibinfo{volume}{136}
  (\bibinfo{year}{2024}) \bibinfo{pages}{104096},
  \bibinfo{doi}{\doi{10.1016/j.ppnp.2023.104096}}, \eprint{2301.07715}.

\bibtype{Article}%
\bibitem{Schienbein:2007fs}
\bibinfo{author}{I. Schienbein}, \bibinfo{author}{J.~Y. Yu},
  \bibinfo{author}{C. Keppel}, \bibinfo{author}{J.~G. Morfin},
  \bibinfo{author}{F. Olness}, \bibinfo{author}{J.~F. Owens},
  \bibinfo{title}{{Nuclear parton distribution functions from neutrino deep
  inelastic scattering}}, \bibinfo{journal}{Phys. Rev. D} \bibinfo{volume}{77}
  (\bibinfo{year}{2008}) \bibinfo{pages}{054013},
  \bibinfo{doi}{\doi{10.1103/PhysRevD.77.054013}}, \eprint{0710.4897}.

\bibtype{Article}%
\bibitem{BCDMS:1985dor}
\bibinfo{author}{G. Bari}, et al. (\bibinfo{collaboration}{BCDMS}),
  \bibinfo{title}{{A Measurement of Nuclear Effects in Deep Inelastic Muon
  Scattering on Deuterium, Nitrogen and Iron Targets}}, \bibinfo{journal}{Phys.
  Lett. B} \bibinfo{volume}{163} (\bibinfo{year}{1985}) \bibinfo{pages}{282},
  \bibinfo{doi}{\doi{10.1016/0370-2693(85)90238-2}}.

\bibtype{Article}%
\bibitem{BCDMS:1987upi}
\bibinfo{author}{A.~C. Benvenuti}, et al. (\bibinfo{collaboration}{BCDMS}),
  \bibinfo{title}{{Nuclear Effects in Deep Inelastic Muon Scattering on
  Deuterium and Iron Targets}}, \bibinfo{journal}{Phys. Lett. B}
  \bibinfo{volume}{189} (\bibinfo{year}{1987}) \bibinfo{pages}{483--487},
  \bibinfo{doi}{\doi{10.1016/0370-2693(87)90664-2}}.

\bibtype{Article}%
\bibitem{Bodek:1983qn}
\bibinfo{author}{A. Bodek}, et al., \bibinfo{title}{{Electron Scattering from
  Nuclear Targets and Quark Distributions in Nuclei}}, \bibinfo{journal}{Phys.
  Rev. Lett.} \bibinfo{volume}{50} (\bibinfo{year}{1983})
  \bibinfo{pages}{1431}, \bibinfo{doi}{\doi{10.1103/PhysRevLett.50.1431}}.

\bibtype{Article}%
\bibitem{Gomez:1993ri}
\bibinfo{author}{J. Gomez}, et al., \bibinfo{title}{{Measurement of the
  A-dependence of deep inelastic electron scattering}}, \bibinfo{journal}{Phys.
  Rev. D} \bibinfo{volume}{49} (\bibinfo{year}{1994})
  \bibinfo{pages}{4348--4372}, \bibinfo{doi}{\doi{10.1103/PhysRevD.49.4348}}.

\bibtype{Article}%
\bibitem{Dasu:1993vk}
\bibinfo{author}{S. Dasu}, et al., \bibinfo{title}{{Measurement of kinematic
  and nuclear dependence of R = sigma-L / sigma-t in deep inelastic electron
  scattering}}, \bibinfo{journal}{Phys. Rev. D} \bibinfo{volume}{49}
  (\bibinfo{year}{1994}) \bibinfo{pages}{5641--5670},
  \bibinfo{doi}{\doi{10.1103/PhysRevD.49.5641}}.

\bibtype{Article}%
\bibitem{NewMuon:1995cua}
\bibinfo{author}{P. Amaudruz}, et al. (\bibinfo{collaboration}{New Muon}),
  \bibinfo{title}{{A Reevaluation of the nuclear structure function ratios for
  D, He, Li-6, C and Ca}}, \bibinfo{journal}{Nucl. Phys. B}
  \bibinfo{volume}{441} (\bibinfo{year}{1995}) \bibinfo{pages}{3--11},
  \bibinfo{doi}{\doi{10.1016/0550-3213(94)00023-9}}, \eprint{hep-ph/9503291}.

\bibtype{Article}%
\bibitem{NewMuon:1996yuf}
\bibinfo{author}{M. Arneodo}, et al. (\bibinfo{collaboration}{New Muon}),
  \bibinfo{title}{{The A dependence of the nuclear structure function ratios}},
  \bibinfo{journal}{Nucl. Phys. B} \bibinfo{volume}{481} (\bibinfo{year}{1996})
  \bibinfo{pages}{3--22}, \bibinfo{doi}{\doi{10.1016/S0550-3213(96)90117-0}}.

\bibtype{Article}%
\bibitem{Aicher:2010cb}
\bibinfo{author}{M. Aicher}, \bibinfo{author}{A. Sch{\"a}fer},
  \bibinfo{author}{W. Vogelsang}, \bibinfo{title}{{Soft-gluon resummation and
  the valence parton distribution function of the pion}},
  \bibinfo{journal}{Phys. Rev. Lett.} \bibinfo{volume}{105}
  (\bibinfo{year}{2010}) \bibinfo{pages}{252003},
  \bibinfo{doi}{\doi{10.1103/PhysRevLett.105.252003}}, \eprint{1009.2481}.

\bibtype{Article}%
\bibitem{Barry:2021osv}
\bibinfo{author}{P.~C. Barry}, \bibinfo{author}{C.-R. Ji}, \bibinfo{author}{N.
  Sato}, \bibinfo{author}{W. Melnitchouk} (\bibinfo{collaboration}{Jefferson
  Lab Angular Momentum (JAM)}), \bibinfo{title}{{Global QCD Analysis of Pion
  Parton Distributions with Threshold Resummation}}, \bibinfo{journal}{Phys.
  Rev. Lett.} \bibinfo{volume}{127} (\bibinfo{number}{23})
  (\bibinfo{year}{2021}) \bibinfo{pages}{232001},
  \bibinfo{doi}{\doi{10.1103/PhysRevLett.127.232001}}, \eprint{2108.05822}.

\bibtype{Article}%
\bibitem{Roberts:2021nhw}
\bibinfo{author}{C.~D. Roberts}, \bibinfo{author}{D.~G. Richards},
  \bibinfo{author}{T. Horn}, \bibinfo{author}{L. Chang},
  \bibinfo{title}{{Insights into the emergence of mass from studies of pion and
  kaon structure}}, \bibinfo{journal}{Prog. Part. Nucl. Phys.}
  \bibinfo{volume}{120} (\bibinfo{year}{2021}) \bibinfo{pages}{103883},
  \bibinfo{doi}{\doi{10.1016/j.ppnp.2021.103883}}, \eprint{2102.01765}.

\bibtype{Article}%
\bibitem{Kotz:2023pbu}
\bibinfo{author}{L. Kotz}, \bibinfo{author}{A. Courtoy}, \bibinfo{author}{P.
  Nadolsky}, \bibinfo{author}{F. Olness}, \bibinfo{author}{M. Ponce-Chavez},
  \bibinfo{title}{{Analysis of parton distributions in a pion with B{\'e}zier
  parametrizations}}, \bibinfo{journal}{Phys. Rev. D} \bibinfo{volume}{109}
  (\bibinfo{number}{7}) (\bibinfo{year}{2024}) \bibinfo{pages}{074027},
  \bibinfo{doi}{\doi{10.1103/PhysRevD.109.074027}}, \eprint{2311.08447}.

\bibtype{Article}%
\bibitem{Pasquini:2023aaf}
\bibinfo{author}{B. Pasquini}, \bibinfo{author}{S. Rodini}, \bibinfo{author}{S.
  Venturini} (\bibinfo{collaboration}{MAP (Multi-dimensional Analyses of
  Partonic distributions)}), \bibinfo{title}{{Valence quark, sea, and gluon
  content of the pion from the parton distribution functions and the
  electromagnetic form factor}}, \bibinfo{journal}{Phys. Rev. D}
  \bibinfo{volume}{107} (\bibinfo{number}{11}) (\bibinfo{year}{2023})
  \bibinfo{pages}{114023}, \bibinfo{doi}{\doi{10.1103/PhysRevD.107.114023}},
  \eprint{2303.01789}.

\bibtype{Article}%
\bibitem{Saclay-CERN-CollegedeFrance-EcolePoly-Orsay:1980fhh}
\bibinfo{author}{J. Badier}, et al.
  (\bibinfo{collaboration}{Saclay-CERN-College de France-Ecole Poly-Orsay}),
  \bibinfo{title}{{Measurement of the $K^- / \pi^-$ Structure Function Ratio
  Using the {Drell-Yan} Process}}, \bibinfo{journal}{Phys. Lett. B}
  \bibinfo{volume}{93} (\bibinfo{year}{1980}) \bibinfo{pages}{354--356},
  \bibinfo{doi}{\doi{10.1016/0370-2693(80)90530-4}}.

\bibtype{Article}%
\bibitem{Bourrely:2023yzi}
\bibinfo{author}{C. Bourrely}, \bibinfo{author}{F. Buccella},
  \bibinfo{author}{W.-C. Chang}, \bibinfo{author}{J.-C. Peng},
  \bibinfo{title}{{Extraction of kaon partonic distribution functions from
  Drell-Yan and J/\ensuremath{\psi} production data}}, \bibinfo{journal}{Phys.
  Lett. B} \bibinfo{volume}{848} (\bibinfo{year}{2024})
  \bibinfo{pages}{138395}, \bibinfo{doi}{\doi{10.1016/j.physletb.2023.138395}},
  \eprint{2305.18117}.

\bibtype{Article}%
\bibitem{Chang:2024rbs}
\bibinfo{author}{W.-C. Chang}, \bibinfo{author}{J.-C. Peng},
  \bibinfo{author}{S. Platchkov}, \bibinfo{author}{T. Sawada},
  \bibinfo{title}{{Constraining kaon PDFs from Drell-Yan and
  J/\ensuremath{\psi} production}}, \bibinfo{journal}{Phys. Lett. B}
  \bibinfo{volume}{855} (\bibinfo{year}{2024}) \bibinfo{pages}{138820},
  \bibinfo{doi}{\doi{10.1016/j.physletb.2024.138820}}, \eprint{2402.02860}.

\bibtype{Article}%
\bibitem{Adams:2018pwt}
\bibinfo{author}{B. Adams}, et al., \bibinfo{title}{{Letter of Intent: A New
  QCD facility at the M2 beam line of the CERN SPS (COMPASS++/AMBER)}}
  (\bibinfo{year}{2018}), \eprint{1808.00848}.

\bibtype{Article}%
\bibitem{Arrington:2021alx}
\bibinfo{author}{J. Arrington}, et al., \bibinfo{title}{{Physics with CEBAF at
  12 GeV and future opportunities}}, \bibinfo{journal}{Prog. Part. Nucl. Phys.}
  \bibinfo{volume}{127} (\bibinfo{year}{2022}) \bibinfo{pages}{103985},
  \bibinfo{doi}{\doi{10.1016/j.ppnp.2022.103985}}, \eprint{2112.00060}.

\bibtype{Article}%
\bibitem{Arrington:2021biu}
\bibinfo{author}{J. Arrington}, et al., \bibinfo{title}{{Revealing the
  structure of light pseudoscalar mesons at the electron\textendash{}ion
  collider}}, \bibinfo{journal}{J. Phys. G} \bibinfo{volume}{48}
  (\bibinfo{number}{7}) (\bibinfo{year}{2021}) \bibinfo{pages}{075106},
  \bibinfo{doi}{\doi{10.1088/1361-6471/abf5c3}}, \eprint{2102.11788}.

\bibtype{Article}%
\bibitem{Kuraev:1977fs}
\bibinfo{author}{E.~A. Kuraev}, \bibinfo{author}{L.~N. Lipatov},
  \bibinfo{author}{V.~S. Fadin}, \bibinfo{title}{{The Pomeranchuk Singularity
  in Nonabelian Gauge Theories}}, \bibinfo{journal}{Sov. Phys. JETP}
  \bibinfo{volume}{45} (\bibinfo{year}{1977}) \bibinfo{pages}{199--204}.

\bibtype{Article}%
\bibitem{Balitsky:1978ic}
\bibinfo{author}{I.~I. Balitsky}, \bibinfo{author}{L.~N. Lipatov},
  \bibinfo{title}{{The Pomeranchuk Singularity in Quantum Chromodynamics}},
  \bibinfo{journal}{Sov. J. Nucl. Phys.} \bibinfo{volume}{28}
  (\bibinfo{year}{1978}) \bibinfo{pages}{822--829}.

\bibtype{Article}%
\bibitem{Fadin:1998py}
\bibinfo{author}{V.~S. Fadin}, \bibinfo{author}{L.~N. Lipatov},
  \bibinfo{title}{{BFKL pomeron in the next-to-leading approximation}},
  \bibinfo{journal}{Phys. Lett. B} \bibinfo{volume}{429} (\bibinfo{year}{1998})
  \bibinfo{pages}{127--134},
  \bibinfo{doi}{\doi{10.1016/S0370-2693(98)00473-0}}, \eprint{hep-ph/9802290}.

\bibtype{Article}%
\bibitem{Ciafaloni:1998gs}
\bibinfo{author}{M. Ciafaloni}, \bibinfo{author}{G. Camici},
  \bibinfo{title}{{Energy scale(s) and next-to-leading BFKL equation}},
  \bibinfo{journal}{Phys. Lett. B} \bibinfo{volume}{430} (\bibinfo{year}{1998})
  \bibinfo{pages}{349--354},
  \bibinfo{doi}{\doi{10.1016/S0370-2693(98)00551-6}}, \eprint{hep-ph/9803389}.

\bibtype{Article}%
\bibitem{Kovchegov:2015pbl}
\bibinfo{author}{Y.V. Kovchegov}, \bibinfo{author}{D. Pitonyak},
  \bibinfo{author}{M.D. Sievert}, \bibinfo{title}{{Helicity Evolution at
  Small-x}}, \bibinfo{journal}{JHEP} \bibinfo{volume}{01}
  (\bibinfo{year}{2016}) \bibinfo{pages}{072},
  \bibinfo{doi}{\doi{10.1007/JHEP01(2016)072}}, \eprint{1511.06737}.

\bibtype{Article}%
\bibitem{Kovchegov:2017lsr}
\bibinfo{author}{Y.V. Kovchegov}, \bibinfo{author}{D. Pitonyak},
  \bibinfo{author}{M.D. Sievert}, \bibinfo{title}{{Small-$x$ Asymptotics of the
  Gluon Helicity Distribution}}, \bibinfo{journal}{JHEP} \bibinfo{volume}{10}
  (\bibinfo{year}{2017}) \bibinfo{pages}{198},
  \bibinfo{doi}{\doi{10.1007/JHEP10(2017)198}}, \eprint{1706.04236}.

\bibtype{Article}%
\bibitem{Kovchegov:2018zeq}
\bibinfo{author}{Y.~V. Kovchegov}, \bibinfo{author}{M.~D. Sievert},
  \bibinfo{title}{{Valence Quark Transversity at Small $x$}},
  \bibinfo{journal}{Phys. Rev. D} \bibinfo{volume}{99} (\bibinfo{number}{5})
  (\bibinfo{year}{2019}) \bibinfo{pages}{054033},
  \bibinfo{doi}{\doi{10.1103/PhysRevD.99.054033}}, \eprint{1808.10354}.

\bibtype{Article}%
\bibitem{Cougoulic:2022gbk}
\bibinfo{author}{F. Cougoulic}, \bibinfo{author}{Y.~V. Kovchegov},
  \bibinfo{author}{A. Tarasov}, \bibinfo{author}{Y. Tawabutr},
  \bibinfo{title}{{Quark and gluon helicity evolution at small x: revised and
  updated}}, \bibinfo{journal}{JHEP} \bibinfo{volume}{07}
  (\bibinfo{year}{2022}) \bibinfo{pages}{095},
  \bibinfo{doi}{\doi{10.1007/JHEP07(2022)095}}, \eprint{2204.11898}.

\bibtype{Article}%
\bibitem{Adamiak:2023yhz}
\bibinfo{author}{D. Adamiak}, \bibinfo{author}{N. Baldonado},
  \bibinfo{author}{Y.~V. Kovchegov}, \bibinfo{author}{W. Melnitchouk},
  \bibinfo{author}{D. Pitonyak}, \bibinfo{author}{N. Sato},
  \bibinfo{author}{M.~D. Sievert}, \bibinfo{author}{A. Tarasov},
  \bibinfo{author}{Y. Tawabutr} (\bibinfo{collaboration}{Jefferson Lab Angular
  Momentum (JAM)}), \bibinfo{title}{{Global analysis of polarized DIS and SIDIS
  data with improved small-x helicity evolution}}, \bibinfo{journal}{Phys. Rev.
  D} \bibinfo{volume}{108} (\bibinfo{number}{11}) (\bibinfo{year}{2023})
  \bibinfo{pages}{114007}, \bibinfo{doi}{\doi{10.1103/PhysRevD.108.114007}},
  \eprint{2308.07461}.

\bibtype{Article}%
\bibitem{Adamiak:2025dpw}
\bibinfo{author}{Daniel Adamiak}, \bibinfo{author}{Nicholas Baldonado},
  \bibinfo{author}{Yuri~V. Kovchegov}, \bibinfo{author}{Ming Li},
  \bibinfo{author}{W. Melnitchouk}, \bibinfo{author}{Daniel Pitonyak},
  \bibinfo{author}{Nobuo Sato}, \bibinfo{author}{Matthew~D. Sievert},
  \bibinfo{author}{Andrey Tarasov}, \bibinfo{author}{Yossathorn Tawabutr}
  (\bibinfo{collaboration}{JAM Collaboration (Small-x Analysis Group)}),
  \bibinfo{title}{{First study of polarized proton-proton scattering with
  small-x helicity evolution}}, \bibinfo{journal}{Phys. Rev. D}
  \bibinfo{volume}{112} (\bibinfo{number}{9}) (\bibinfo{year}{2025})
  \bibinfo{pages}{094032}, \bibinfo{doi}{\doi{10.1103/9gnx-ycs4}},
  \eprint{2503.21006}.

\bibtype{Article}%
\bibitem{Cocuzza:2023oam}
\bibinfo{author}{C. Cocuzza}, \bibinfo{author}{A. Metz}, \bibinfo{author}{D.
  Pitonyak}, \bibinfo{author}{A. Prokudin}, \bibinfo{author}{N. Sato},
  \bibinfo{author}{R. Seidl} (\bibinfo{collaboration}{JAM}),
  \bibinfo{title}{{Transversity Distributions and Tensor Charges of the
  Nucleon: Extraction from Dihadron Production and Their Universal Nature}},
  \bibinfo{journal}{Phys. Rev. Lett.} \bibinfo{volume}{132}
  (\bibinfo{number}{9}) (\bibinfo{year}{2024}) \bibinfo{pages}{091901},
  \bibinfo{doi}{\doi{10.1103/PhysRevLett.132.091901}}, \eprint{2306.12998}.

\bibtype{Article}%
\bibitem{Cocuzza:2023vqs}
\bibinfo{author}{C. Cocuzza}, \bibinfo{author}{A. Metz}, \bibinfo{author}{D.
  Pitonyak}, \bibinfo{author}{A. Prokudin}, \bibinfo{author}{N. Sato},
  \bibinfo{author}{R. Seidl} (\bibinfo{collaboration}{Jefferson Lab Angular
  Momentum (JAM)}), \bibinfo{title}{{First simultaneous global QCD analysis of
  dihadron fragmentation functions and transversity parton distribution
  functions}}, \bibinfo{journal}{Phys. Rev. D} \bibinfo{volume}{109}
  (\bibinfo{number}{3}) (\bibinfo{year}{2024}) \bibinfo{pages}{034024},
  \bibinfo{doi}{\doi{10.1103/PhysRevD.109.034024}}, \eprint{2308.14857}.

\bibtype{Article}%
\bibitem{Froissart:1961ux}
\bibinfo{author}{M. Froissart}, \bibinfo{title}{{Asymptotic behavior and
  subtractions in the Mandelstam representation}}, \bibinfo{journal}{Phys.
  Rev.} \bibinfo{volume}{123} (\bibinfo{year}{1961})
  \bibinfo{pages}{1053--1057}, \bibinfo{doi}{\doi{10.1103/PhysRev.123.1053}}.

\bibtype{Article}%
\bibitem{Gribov:1983ivg}
\bibinfo{author}{L.~V. Gribov}, \bibinfo{author}{E.~M. Levin},
  \bibinfo{author}{M.~G. Ryskin}, \bibinfo{title}{{Semihard Processes in QCD}},
  \bibinfo{journal}{Phys. Rept.} \bibinfo{volume}{100} (\bibinfo{year}{1983})
  \bibinfo{pages}{1--150}, \bibinfo{doi}{\doi{10.1016/0370-1573(83)90022-4}}.

\bibtype{Article}%
\bibitem{McLerran:1993ni}
\bibinfo{author}{L.~D. McLerran}, \bibinfo{author}{R. Venugopalan},
  \bibinfo{title}{{Computing quark and gluon distribution functions for very
  large nuclei}}, \bibinfo{journal}{Phys. Rev. D} \bibinfo{volume}{49}
  (\bibinfo{year}{1994}) \bibinfo{pages}{2233--2241},
  \bibinfo{doi}{\doi{10.1103/PhysRevD.49.2233}}, \eprint{hep-ph/9309289}.

\bibtype{Article}%
\bibitem{McLerran:1993ka}
\bibinfo{author}{L.~D. McLerran}, \bibinfo{author}{R. Venugopalan},
  \bibinfo{title}{{Gluon distribution functions for very large nuclei at small
  transverse momentum}}, \bibinfo{journal}{Phys. Rev. D} \bibinfo{volume}{49}
  (\bibinfo{year}{1994}) \bibinfo{pages}{3352--3355},
  \bibinfo{doi}{\doi{10.1103/PhysRevD.49.3352}}, \eprint{hep-ph/9311205}.

\bibtype{Article}%
\bibitem{Gelis:2010nm}
\bibinfo{author}{F. Gelis}, \bibinfo{author}{E. Iancu}, \bibinfo{author}{J.
  Jalilian-Marian}, \bibinfo{author}{R. Venugopalan}, \bibinfo{title}{{The
  Color Glass Condensate}}, \bibinfo{journal}{Ann. Rev. Nucl. Part. Sci.}
  \bibinfo{volume}{60} (\bibinfo{year}{2010}) \bibinfo{pages}{463--489},
  \bibinfo{doi}{\doi{10.1146/annurev.nucl.010909.083629}}, \eprint{1002.0333}.

\bibtype{Article}%
\bibitem{Morreale:2021pnn}
\bibinfo{author}{A. Morreale}, \bibinfo{author}{F. Salazar},
  \bibinfo{title}{{Mining for Gluon Saturation at Colliders}},
  \bibinfo{journal}{Universe} \bibinfo{volume}{7} (\bibinfo{number}{8})
  (\bibinfo{year}{2021}) \bibinfo{pages}{312},
  \bibinfo{doi}{\doi{10.3390/universe7080312}}, \eprint{2108.08254}.

\bibtype{Article}%
\bibitem{Jalilian-Marian:1997qno}
\bibinfo{author}{J. Jalilian-Marian}, \bibinfo{author}{A. Kovner},
  \bibinfo{author}{A. Leonidov}, \bibinfo{author}{H. Weigert},
  \bibinfo{title}{{The BFKL equation from the Wilson renormalization group}},
  \bibinfo{journal}{Nucl. Phys. B} \bibinfo{volume}{504} (\bibinfo{year}{1997})
  \bibinfo{pages}{415--431},
  \bibinfo{doi}{\doi{10.1016/S0550-3213(97)00440-9}}, \eprint{hep-ph/9701284}.

\bibtype{Article}%
\bibitem{Jalilian-Marian:1997jhx}
\bibinfo{author}{J. Jalilian-Marian}, \bibinfo{author}{A. Kovner},
  \bibinfo{author}{A. Leonidov}, \bibinfo{author}{H. Weigert},
  \bibinfo{title}{{The Wilson renormalization group for low x physics: Towards
  the high density regime}}, \bibinfo{journal}{Phys. Rev. D}
  \bibinfo{volume}{59} (\bibinfo{year}{1998}) \bibinfo{pages}{014014},
  \bibinfo{doi}{\doi{10.1103/PhysRevD.59.014014}}, \eprint{hep-ph/9706377}.

\bibtype{Article}%
\bibitem{Weigert:2000gi}
\bibinfo{author}{Heribert Weigert}, \bibinfo{title}{{Unitarity at small Bjorken
  x}}, \bibinfo{journal}{Nucl. Phys. A} \bibinfo{volume}{703}
  (\bibinfo{year}{2002}) \bibinfo{pages}{823--860},
  \bibinfo{doi}{\doi{10.1016/S0375-9474(01)01668-2}}, \eprint{hep-ph/0004044}.

\bibtype{Article}%
\bibitem{Iancu:2000hn}
\bibinfo{author}{E. Iancu}, \bibinfo{author}{A. Leonidov},
  \bibinfo{author}{L.~D. McLerran}, \bibinfo{title}{{Nonlinear gluon evolution
  in the color glass condensate. 1.}}, \bibinfo{journal}{Nucl. Phys. A}
  \bibinfo{volume}{692} (\bibinfo{year}{2001}) \bibinfo{pages}{583--645},
  \bibinfo{doi}{\doi{10.1016/S0375-9474(01)00642-X}}, \eprint{hep-ph/0011241}.

\bibtype{Article}%
\bibitem{Iancu:2001ad}
\bibinfo{author}{E. Iancu}, \bibinfo{author}{A. Leonidov},
  \bibinfo{author}{L.~D. McLerran}, \bibinfo{title}{{The Renormalization group
  equation for the color glass condensate}}, \bibinfo{journal}{Phys. Lett. B}
  \bibinfo{volume}{510} (\bibinfo{year}{2001}) \bibinfo{pages}{133--144},
  \bibinfo{doi}{\doi{10.1016/S0370-2693(01)00524-X}}, \eprint{hep-ph/0102009}.

\bibtype{Article}%
\bibitem{Kovchegov:1999yj}
\bibinfo{author}{Y.~V. Kovchegov}, \bibinfo{title}{{Small x F(2) structure
  function of a nucleus including multiple pomeron exchanges}},
  \bibinfo{journal}{Phys. Rev. D} \bibinfo{volume}{60} (\bibinfo{year}{1999})
  \bibinfo{pages}{034008}, \bibinfo{doi}{\doi{10.1103/PhysRevD.60.034008}},
  \eprint{hep-ph/9901281}.

\bibtype{Article}%
\bibitem{Kowalski:2007rw}
\bibinfo{author}{H. Kowalski}, \bibinfo{author}{T. Lappi}, \bibinfo{author}{R.
  Venugopalan}, \bibinfo{title}{{Nuclear enhancement of universal dynamics of
  high parton densities}}, \bibinfo{journal}{Phys. Rev. Lett.}
  \bibinfo{volume}{100} (\bibinfo{year}{2008}) \bibinfo{pages}{022303},
  \bibinfo{doi}{\doi{10.1103/PhysRevLett.100.022303}}, \eprint{0705.3047}.

\bibtype{Article}%
\bibitem{Kowalski:2006hc}
\bibinfo{author}{H. Kowalski}, \bibinfo{author}{L. Motyka}, \bibinfo{author}{G.
  Watt}, \bibinfo{title}{{Exclusive diffractive processes at HERA within the
  dipole picture}}, \bibinfo{journal}{Phys. Rev. D} \bibinfo{volume}{74}
  (\bibinfo{year}{2006}) \bibinfo{pages}{074016},
  \bibinfo{doi}{\doi{10.1103/PhysRevD.74.074016}}, \eprint{hep-ph/0606272}.

\bibtype{Article}%
\bibitem{Kowalski:2003hm}
\bibinfo{author}{H. Kowalski}, \bibinfo{author}{D. Teaney}, \bibinfo{title}{{An
  Impact parameter dipole saturation model}}, \bibinfo{journal}{Phys. Rev. D}
  \bibinfo{volume}{68} (\bibinfo{year}{2003}) \bibinfo{pages}{114005},
  \bibinfo{doi}{\doi{10.1103/PhysRevD.68.114005}}, \eprint{hep-ph/0304189}.

\bibtype{Article}%
\bibitem{Matveev:1973ra}
\bibinfo{author}{V.~A. Matveev}, \bibinfo{author}{R.~M. Muradian},
  \bibinfo{author}{A.~N. Tavkhelidze}, \bibinfo{title}{{Automodellism in the
  large-angle elastic scattering and structure of hadrons}},
  \bibinfo{journal}{Lett. Nuovo Cim.} \bibinfo{volume}{7}
  (\bibinfo{year}{1973}) \bibinfo{pages}{719--723},
  \bibinfo{doi}{\doi{10.1007/BF02728133}}.

\bibtype{Article}%
\bibitem{Brodsky:1973kr}
\bibinfo{author}{S.~J. Brodsky}, \bibinfo{author}{G.~R. Farrar},
  \bibinfo{title}{{Scaling Laws at Large Transverse Momentum}},
  \bibinfo{journal}{Phys. Rev. Lett.} \bibinfo{volume}{31}
  (\bibinfo{year}{1973}) \bibinfo{pages}{1153--1156},
  \bibinfo{doi}{\doi{10.1103/PhysRevLett.31.1153}}.

\bibtype{Article}%
\bibitem{Brodsky:1994kg}
\bibinfo{author}{S.~J. Brodsky}, \bibinfo{author}{M. Burkardt},
  \bibinfo{author}{I. Schmidt}, \bibinfo{title}{{Perturbative QCD constraints
  on the shape of polarized quark and gluon distributions}},
  \bibinfo{journal}{Nucl. Phys. B} \bibinfo{volume}{441} (\bibinfo{year}{1995})
  \bibinfo{pages}{197--214}, \bibinfo{doi}{\doi{10.1016/0550-3213(95)00009-H}},
  \eprint{hep-ph/9401328}.

\bibtype{Article}%
\bibitem{Ball:2016spl}
\bibinfo{author}{R.~D. Ball}, \bibinfo{author}{E.~R. Nocera},
  \bibinfo{author}{J. Rojo}, \bibinfo{title}{{The asymptotic behaviour of
  parton distributions at small and large $x$}}, \bibinfo{journal}{Eur. Phys.
  J. C} \bibinfo{volume}{76} (\bibinfo{number}{7}) (\bibinfo{year}{2016})
  \bibinfo{pages}{383}, \bibinfo{doi}{\doi{10.1140/epjc/s10052-016-4240-4}},
  \eprint{1604.00024}.

\bibtype{Article}%
\bibitem{Bloom:1970xb}
\bibinfo{author}{E.~D. Bloom}, \bibinfo{author}{F.~J. Gilman},
  \bibinfo{title}{{Scaling, Duality, and the Behavior of Resonances in
  Inelastic electron-Proton Scattering}}, \bibinfo{journal}{Phys. Rev. Lett.}
  \bibinfo{volume}{25} (\bibinfo{year}{1970}) \bibinfo{pages}{1140},
  \bibinfo{doi}{\doi{10.1103/PhysRevLett.25.1140}}.

\bibtype{Article}%
\bibitem{Accardi:2009br}
\bibinfo{author}{A. Accardi}, \bibinfo{author}{M.~E. Christy},
  \bibinfo{author}{C.~E. Keppel}, \bibinfo{author}{W. Melnitchouk},
  \bibinfo{author}{P. Monaghan}, \bibinfo{author}{J.~G. Morf\'\i{}n},
  \bibinfo{author}{J.~F. Owens}, \bibinfo{title}{{New parton distributions from
  large-$x$ and low-$Q^2$ data}}, \bibinfo{journal}{Phys. Rev. D}
  \bibinfo{volume}{81} (\bibinfo{year}{2010}) \bibinfo{pages}{034016},
  \bibinfo{doi}{\doi{10.1103/PhysRevD.81.034016}}, \eprint{0911.2254}.

\bibtype{Article}%
\bibitem{Accardi:2016qay}
\bibinfo{author}{A. Accardi}, \bibinfo{author}{L.~T. Brady},
  \bibinfo{author}{W. Melnitchouk}, \bibinfo{author}{J.~F. Owens},
  \bibinfo{author}{N. Sato}, \bibinfo{title}{{Constraints on large-$x$ parton
  distributions from new weak boson production and deep-inelastic scattering
  data}}, \bibinfo{journal}{Phys. Rev. D} \bibinfo{volume}{93}
  (\bibinfo{number}{11}) (\bibinfo{year}{2016}) \bibinfo{pages}{114017},
  \bibinfo{doi}{\doi{10.1103/PhysRevD.93.114017}}, \eprint{1602.03154}.

\bibtype{Article}%
\bibitem{Sterman:1986aj}
\bibinfo{author}{G.~F. Sterman}, \bibinfo{title}{{Summation of Large
  Corrections to Short Distance Hadronic Cross-Sections}},
  \bibinfo{journal}{Nucl. Phys. B} \bibinfo{volume}{281} (\bibinfo{year}{1987})
  \bibinfo{pages}{310--364}, \bibinfo{doi}{\doi{10.1016/0550-3213(87)90258-6}}.

\bibtype{Article}%
\bibitem{Catani:1989ne}
\bibinfo{author}{S. Catani}, \bibinfo{author}{L. Trentadue},
  \bibinfo{title}{{Resummation of the QCD Perturbative Series for Hard
  Processes}}, \bibinfo{journal}{Nucl. Phys. B} \bibinfo{volume}{327}
  (\bibinfo{year}{1989}) \bibinfo{pages}{323--352},
  \bibinfo{doi}{\doi{10.1016/0550-3213(89)90273-3}}.

\bibtype{Article}%
\bibitem{Aicher:2011ai}
\bibinfo{author}{M. Aicher}, \bibinfo{author}{A. Sch{\"a}fer},
  \bibinfo{author}{W. Vogelsang}, \bibinfo{title}{{Threshold-Resummed Cross
  Section for the Drell-Yan Process in Pion-Nucleon Collisions at COMPASS}},
  \bibinfo{journal}{Phys. Rev. D} \bibinfo{volume}{83} (\bibinfo{year}{2011})
  \bibinfo{pages}{114023}, \bibinfo{doi}{\doi{10.1103/PhysRevD.83.114023}},
  \eprint{1104.3512}.

\bibtype{Article}%
\bibitem{Corcella:2005us}
\bibinfo{author}{G. Corcella}, \bibinfo{author}{L. Magnea},
  \bibinfo{title}{{Soft-gluon resummation effects on parton distributions}},
  \bibinfo{journal}{Phys. Rev. D} \bibinfo{volume}{72} (\bibinfo{year}{2005})
  \bibinfo{pages}{074017}, \bibinfo{doi}{\doi{10.1103/PhysRevD.72.074017}},
  \eprint{hep-ph/0506278}.

\bibtype{Article}%
\bibitem{Ravndal:1973kt}
\bibinfo{author}{F. Ravndal}, \bibinfo{title}{{On the azimuthal dependence of
  semiinclusive, deep inelastic electroproduction cross-sections}},
  \bibinfo{journal}{Phys. Lett. B} \bibinfo{volume}{43} (\bibinfo{year}{1973})
  \bibinfo{pages}{301--303}, \bibinfo{doi}{\doi{10.1016/0370-2693(73)90445-0}}.

\bibtype{Article}%
\bibitem{Cahn:1978se}
\bibinfo{author}{R.~N. Cahn}, \bibinfo{title}{{Azimuthal Dependence in
  Leptoproduction: A Simple Parton Model Calculation}}, \bibinfo{journal}{Phys.
  Lett. B} \bibinfo{volume}{78} (\bibinfo{year}{1978})
  \bibinfo{pages}{269--273}, \bibinfo{doi}{\doi{10.1016/0370-2693(78)90020-5}}.

\bibtype{Article}%
\bibitem{Collins:1984kg}
\bibinfo{author}{J.~C. Collins}, \bibinfo{author}{D.~E. Soper},
  \bibinfo{author}{G.~F. Sterman}, \bibinfo{title}{{Transverse Momentum
  Distribution in Drell-Yan Pair and W and Z Boson Production}},
  \bibinfo{journal}{Nucl. Phys. B} \bibinfo{volume}{250} (\bibinfo{year}{1985})
  \bibinfo{pages}{199--224}, \bibinfo{doi}{\doi{10.1016/0550-3213(85)90479-1}}.

\bibtype{Article}%
\bibitem{Sivers:1989cc}
\bibinfo{author}{D.~W. Sivers}, \bibinfo{title}{{Single Spin Production
  Asymmetries from the Hard Scattering of Point-Like Constituents}},
  \bibinfo{journal}{Phys. Rev. D} \bibinfo{volume}{41} (\bibinfo{year}{1990})
  \bibinfo{pages}{83}, \bibinfo{doi}{\doi{10.1103/PhysRevD.41.83}}.

\bibtype{Article}%
\bibitem{Collins:1992kk}
\bibinfo{author}{J.~C. Collins}, \bibinfo{title}{{Fragmentation of transversely
  polarized quarks probed in transverse momentum distributions}},
  \bibinfo{journal}{Nucl. Phys. B} \bibinfo{volume}{396} (\bibinfo{year}{1993})
  \bibinfo{pages}{161--182}, \bibinfo{doi}{\doi{10.1016/0550-3213(93)90262-N}},
  \eprint{hep-ph/9208213}.

\bibtype{Article}%
\bibitem{Anselmino:1994tv}
\bibinfo{author}{M. Anselmino}, \bibinfo{author}{M. Boglione},
  \bibinfo{author}{F. Murgia}, \bibinfo{title}{{Single spin asymmetry for
  $p\uparrow p\rightarrow \pi X$ in perturbative QCD}}, \bibinfo{journal}{Phys.
  Lett. B} \bibinfo{volume}{362} (\bibinfo{year}{1995})
  \bibinfo{pages}{164--172}, \bibinfo{doi}{\doi{10.1016/0370-2693(95)01168-P}},
  \eprint{hep-ph/9503290}.

\bibtype{Article}%
\bibitem{Tangerman:1994eh}
\bibinfo{author}{R.~D. Tangerman}, \bibinfo{author}{P.~J. Mulders},
  \bibinfo{title}{{Intrinsic transverse momentum and the polarized Drell-Yan
  process}}, \bibinfo{journal}{Phys. Rev. D} \bibinfo{volume}{51}
  (\bibinfo{year}{1995}) \bibinfo{pages}{3357--3372},
  \bibinfo{doi}{\doi{10.1103/PhysRevD.51.3357}}, \eprint{hep-ph/9403227}.

\bibtype{Article}%
\bibitem{Kotzinian:1994dv}
\bibinfo{author}{A. Kotzinian}, \bibinfo{title}{{New quark distributions and
  semiinclusive electroproduction on the polarized nucleons}},
  \bibinfo{journal}{Nucl. Phys. B} \bibinfo{volume}{441} (\bibinfo{year}{1995})
  \bibinfo{pages}{234--248}, \bibinfo{doi}{\doi{10.1016/0550-3213(95)00098-D}},
  \eprint{hep-ph/9412283}.

\bibtype{Article}%
\bibitem{Mulders:1995dh}
\bibinfo{author}{P.~J. Mulders}, \bibinfo{author}{R.~D. Tangerman},
  \bibinfo{title}{{The Complete tree level result up to order 1/Q for polarized
  deep inelastic leptoproduction}}, \bibinfo{journal}{Nucl. Phys. B}
  \bibinfo{volume}{461} (\bibinfo{year}{1996}) \bibinfo{pages}{197--237},
  \bibinfo{doi}{\doi{10.1016/0550-3213(95)00632-X}}, \eprint{hep-ph/9510301}.

\bibtype{Article}%
\bibitem{Boer:1997nt}
\bibinfo{author}{D. Boer}, \bibinfo{author}{P.~J. Mulders},
  \bibinfo{title}{{Time reversal odd distribution functions in
  leptoproduction}}, \bibinfo{journal}{Phys. Rev. D} \bibinfo{volume}{57}
  (\bibinfo{year}{1998}) \bibinfo{pages}{5780--5786},
  \bibinfo{doi}{\doi{10.1103/PhysRevD.57.5780}}, \eprint{hep-ph/9711485}.

\bibtype{Article}%
\bibitem{Boussarie:2023izj}
\bibinfo{author}{R. Boussarie}, et al., \bibinfo{title}{{TMD Handbook}}
  (\bibinfo{year}{2023}), \eprint{2304.03302}.

\bibtype{Article}%
\bibitem{DAlesio:2007bjf}
\bibinfo{author}{U. D'Alesio}, \bibinfo{author}{F. Murgia},
  \bibinfo{title}{{Azimuthal and Single Spin Asymmetries in Hard Scattering
  Processes}}, \bibinfo{journal}{Prog. Part. Nucl. Phys.} \bibinfo{volume}{61}
  (\bibinfo{year}{2008}) \bibinfo{pages}{394--454},
  \bibinfo{doi}{\doi{10.1016/j.ppnp.2008.01.001}}, \eprint{0712.4328}.

\bibtype{Article}%
\bibitem{Anselmino:2005nn}
\bibinfo{author}{M. Anselmino}, \bibinfo{author}{M. Boglione},
  \bibinfo{author}{U. D'Alesio}, \bibinfo{author}{A. Kotzinian},
  \bibinfo{author}{F. Murgia}, \bibinfo{author}{A. Prokudin},
  \bibinfo{title}{{The Role of Cahn and sivers effects in deep inelastic
  scattering}}, \bibinfo{journal}{Phys. Rev. D} \bibinfo{volume}{71}
  (\bibinfo{year}{2005}) \bibinfo{pages}{074006},
  \bibinfo{doi}{\doi{10.1103/PhysRevD.71.074006}}, \eprint{hep-ph/0501196}.

\bibtype{Article}%
\bibitem{DAlesio:2004eso}
\bibinfo{author}{U. D'Alesio}, \bibinfo{author}{F. Murgia},
  \bibinfo{title}{{Parton intrinsic motion in inclusive particle production:
  Unpolarized cross sections, single spin asymmetries and the Sivers effect}},
  \bibinfo{journal}{Phys. Rev. D} \bibinfo{volume}{70} (\bibinfo{year}{2004})
  \bibinfo{pages}{074009}, \bibinfo{doi}{\doi{10.1103/PhysRevD.70.074009}},
  \eprint{hep-ph/0408092}.

\bibtype{Article}%
\bibitem{Schweitzer:2010tt}
\bibinfo{author}{P. Schweitzer}, \bibinfo{author}{T. Teckentrup},
  \bibinfo{author}{A. Metz}, \bibinfo{title}{{Intrinsic transverse parton
  momenta in deeply inelastic reactions}}, \bibinfo{journal}{Phys. Rev. D}
  \bibinfo{volume}{81} (\bibinfo{year}{2010}) \bibinfo{pages}{094019},
  \bibinfo{doi}{\doi{10.1103/PhysRevD.81.094019}}, \eprint{1003.2190}.

\bibtype{Article}%
\bibitem{Kane:1978nd}
\bibinfo{author}{G.~L. Kane}, \bibinfo{author}{J. Pumplin}, \bibinfo{author}{W.
  Repko}, \bibinfo{title}{{Transverse Quark Polarization in Large $p_T$
  Reactions, $e^+ e^-$ Jets, and Leptoproduction: A Test of QCD}},
  \bibinfo{journal}{Phys. Rev. Lett.} \bibinfo{volume}{41}
  (\bibinfo{year}{1978}) \bibinfo{pages}{1689},
  \bibinfo{doi}{\doi{10.1103/PhysRevLett.41.1689}}.

\bibtype{Article}%
\bibitem{Dick:1975ty}
\bibinfo{author}{L. Dick}, et al., \bibinfo{title}{{Spin effects in the
  inclusive reactions $\pi^\pm +p(\uparrow)\rightarrow \pi^\pm +$ anything at 8
  GeV/c}}, \bibinfo{journal}{Phys. Lett. B} \bibinfo{volume}{57}
  (\bibinfo{year}{1975}) \bibinfo{pages}{93--96},
  \bibinfo{doi}{\doi{10.1016/0370-2693(75)90252-X}}.

\bibtype{Article}%
\bibitem{Lam:1980uc}
\bibinfo{author}{C.~S. Lam}, \bibinfo{author}{Wu-Ki Tung}, \bibinfo{title}{{A
  Parton Model Relation Sans {QCD} Modifications in Lepton Pair Productions}},
  \bibinfo{journal}{Phys. Rev. D} \bibinfo{volume}{21} (\bibinfo{year}{1980})
  \bibinfo{pages}{2712}, \bibinfo{doi}{\doi{10.1103/PhysRevD.21.2712}}.

\bibtype{Article}%
\bibitem{Collins:1977iv}
\bibinfo{author}{J.~C. Collins}, \bibinfo{author}{D.~E. Soper},
  \bibinfo{title}{{Angular Distribution of Dileptons in High-Energy Hadron
  Collisions}}, \bibinfo{journal}{Phys. Rev. D} \bibinfo{volume}{16}
  (\bibinfo{year}{1977}) \bibinfo{pages}{2219},
  \bibinfo{doi}{\doi{10.1103/PhysRevD.16.2219}}.

\bibtype{Article}%
\bibitem{Boer:1999mm}
\bibinfo{author}{D. Boer}, \bibinfo{title}{{Investigating the origins of
  transverse spin asymmetries at RHIC}}, \bibinfo{journal}{Phys. Rev. D}
  \bibinfo{volume}{60} (\bibinfo{year}{1999}) \bibinfo{pages}{014012},
  \bibinfo{doi}{\doi{10.1103/PhysRevD.60.014012}}, \eprint{hep-ph/9902255}.

\bibtype{Article}%
\bibitem{Peng:2015spa}
\bibinfo{author}{J.-C. Peng}, \bibinfo{author}{W.-C. Chang},
  \bibinfo{author}{R.~E. McClellan}, \bibinfo{author}{O. Teryaev},
  \bibinfo{title}{{Interpretation of Angular Distributions of $Z$-boson
  Production at Colliders}}, \bibinfo{journal}{Phys. Lett. B}
  \bibinfo{volume}{758} (\bibinfo{year}{2016}) \bibinfo{pages}{384--388},
  \bibinfo{doi}{\doi{10.1016/j.physletb.2016.05.035}}, \eprint{1511.08932}.

\bibtype{Article}%
\bibitem{Lambertsen:2016wgj}
\bibinfo{author}{M. Lambertsen}, \bibinfo{author}{W. Vogelsang},
  \bibinfo{title}{{Drell-Yan lepton angular distributions in perturbative
  QCD}}, \bibinfo{journal}{Phys. Rev. D} \bibinfo{volume}{93}
  (\bibinfo{number}{11}) (\bibinfo{year}{2016}) \bibinfo{pages}{114013},
  \bibinfo{doi}{\doi{10.1103/PhysRevD.93.114013}}, \eprint{1605.02625}.

\bibtype{Article}%
\bibitem{Bacchetta:2006tn}
\bibinfo{author}{A. Bacchetta}, \bibinfo{author}{M. Diehl}, \bibinfo{author}{K.
  Goeke}, \bibinfo{author}{A. Metz}, \bibinfo{author}{P.J. Mulders},
  \bibinfo{author}{M. Schlegel}, \bibinfo{title}{{Semi-inclusive deep inelastic
  scattering at small transverse momentum}}, \bibinfo{journal}{JHEP}
  \bibinfo{volume}{02} (\bibinfo{year}{2007}) \bibinfo{pages}{093},
  \bibinfo{doi}{\doi{10.1088/1126-6708/2007/02/093}}, \eprint{hep-ph/0611265}.

\bibtype{Article}%
\bibitem{Mulders:2000sh}
\bibinfo{author}{P.~J. Mulders}, \bibinfo{author}{J. Rodrigues},
  \bibinfo{title}{{Transverse momentum dependence in gluon distribution and
  fragmentation functions}}, \bibinfo{journal}{Phys. Rev. D}
  \bibinfo{volume}{63} (\bibinfo{year}{2001}) \bibinfo{pages}{094021},
  \bibinfo{doi}{\doi{10.1103/PhysRevD.63.094021}}, \eprint{hep-ph/0009343}.

\bibtype{Article}%
\bibitem{Meissner:2007rx}
\bibinfo{author}{S. Meissner}, \bibinfo{author}{A. Metz}, \bibinfo{author}{K.
  Goeke}, \bibinfo{title}{{Relations between generalized and transverse
  momentum dependent parton distributions}}, \bibinfo{journal}{Phys. Rev. D}
  \bibinfo{volume}{76} (\bibinfo{year}{2007}) \bibinfo{pages}{034002},
  \bibinfo{doi}{\doi{10.1103/PhysRevD.76.034002}}, \eprint{hep-ph/0703176}.

\bibtype{Article}%
\bibitem{Efremov:1984ip}
\bibinfo{author}{A.~V. Efremov}, \bibinfo{author}{O.~V. Teryaev},
  \bibinfo{title}{{QCD Asymmetry and Polarized Hadron Structure Functions}},
  \bibinfo{journal}{Phys. Lett. B} \bibinfo{volume}{150} (\bibinfo{year}{1985})
  \bibinfo{pages}{383}, \bibinfo{doi}{\doi{10.1016/0370-2693(85)90999-2}}.

\bibtype{Article}%
\bibitem{Qiu:1991pp}
\bibinfo{author}{J.-W. Qiu}, \bibinfo{author}{G.~F. Sterman},
  \bibinfo{title}{{Single transverse spin asymmetries}},
  \bibinfo{journal}{Phys. Rev. Lett.} \bibinfo{volume}{67}
  (\bibinfo{year}{1991}) \bibinfo{pages}{2264--2267},
  \bibinfo{doi}{\doi{10.1103/PhysRevLett.67.2264}}.

\bibtype{Article}%
\bibitem{Kouvaris:2006zy}
\bibinfo{author}{C. Kouvaris}, \bibinfo{author}{J.-W. Qiu}, \bibinfo{author}{W.
  Vogelsang}, \bibinfo{author}{F. Yuan}, \bibinfo{title}{{Single
  transverse-spin asymmetry in high transverse momentum pion production in pp
  collisions}}, \bibinfo{journal}{Phys. Rev. D} \bibinfo{volume}{74}
  (\bibinfo{year}{2006}) \bibinfo{pages}{114013},
  \bibinfo{doi}{\doi{10.1103/PhysRevD.74.114013}}, \eprint{hep-ph/0609238}.

\bibtype{Article}%
\bibitem{Koike:2009ge}
\bibinfo{author}{Y. Koike}, \bibinfo{author}{T. Tomita},
  \bibinfo{title}{{Soft-fermion-pole contribution to single-spin asymmetry for
  pion production in pp collisions}}, \bibinfo{journal}{Phys. Lett. B}
  \bibinfo{volume}{675} (\bibinfo{year}{2009}) \bibinfo{pages}{181--189},
  \bibinfo{doi}{\doi{10.1016/j.physletb.2009.04.017}}, \eprint{0903.1923}.

\bibtype{Article}%
\bibitem{Kang:2010zzb}
\bibinfo{author}{Z.-B. Kang}, \bibinfo{author}{F. Yuan}, \bibinfo{author}{J.
  Zhou}, \bibinfo{title}{{Twist-three fragmentation function contribution to
  the single spin asymmetry in $p p$ collisions}}, \bibinfo{journal}{Phys.
  Lett. B} \bibinfo{volume}{691} (\bibinfo{year}{2010})
  \bibinfo{pages}{243--248},
  \bibinfo{doi}{\doi{10.1016/j.physletb.2010.07.003}}, \eprint{1002.0399}.

\bibtype{Article}%
\bibitem{Metz:2012ct}
\bibinfo{author}{A. Metz}, \bibinfo{author}{D. Pitonyak},
  \bibinfo{title}{{Fragmentation contribution to the transverse single-spin
  asymmetry in proton-proton collisions}}, \bibinfo{journal}{Phys. Lett. B}
  \bibinfo{volume}{723} (\bibinfo{year}{2013}) \bibinfo{pages}{365--370},
  \bibinfo{doi}{\doi{10.1016/j.physletb.2013.05.043}}, \eprint{1212.5037}.

\bibtype{Article}%
\bibitem{Anselmino:2005sh}
\bibinfo{author}{M. Anselmino}, \bibinfo{author}{M. Boglione},
  \bibinfo{author}{U. D'Alesio}, \bibinfo{author}{E. Leader},
  \bibinfo{author}{S. Melis}, \bibinfo{author}{F. Murgia}, \bibinfo{title}{{The
  general partonic structure for hadronic spin asymmetries}},
  \bibinfo{journal}{Phys. Rev. D} \bibinfo{volume}{73} (\bibinfo{year}{2006})
  \bibinfo{pages}{014020}, \bibinfo{doi}{\doi{10.1103/PhysRevD.73.014020}},
  \eprint{hep-ph/0509035}.

\bibtype{Article}%
\bibitem{Bacchetta:1999kz}
\bibinfo{author}{A. Bacchetta}, \bibinfo{author}{M. Boglione},
  \bibinfo{author}{A. Henneman}, \bibinfo{author}{P.~J. Mulders},
  \bibinfo{title}{{Bounds on transverse momentum dependent distribution and
  fragmentation functions}}, \bibinfo{journal}{Phys. Rev. Lett.}
  \bibinfo{volume}{85} (\bibinfo{year}{2000}) \bibinfo{pages}{712--715},
  \bibinfo{doi}{\doi{10.1103/PhysRevLett.85.712}}, \eprint{hep-ph/9912490}.

\bibtype{Article}%
\bibitem{Metz:2016swz}
\bibinfo{author}{A. Metz}, \bibinfo{author}{A. Vossen}, \bibinfo{title}{{Parton
  Fragmentation Functions}}, \bibinfo{journal}{Prog. Part. Nucl. Phys.}
  \bibinfo{volume}{91} (\bibinfo{year}{2016}) \bibinfo{pages}{136--202},
  \bibinfo{doi}{\doi{10.1016/j.ppnp.2016.08.003}}, \eprint{1607.02521}.

\bibtype{Article}%
\bibitem{Arnold:2008kf}
\bibinfo{author}{S. Arnold}, \bibinfo{author}{A. Metz}, \bibinfo{author}{M.
  Schlegel}, \bibinfo{title}{{Dilepton production from polarized hadron hadron
  collisions}}, \bibinfo{journal}{Phys. Rev. D} \bibinfo{volume}{79}
  (\bibinfo{year}{2009}) \bibinfo{pages}{034005},
  \bibinfo{doi}{\doi{10.1103/PhysRevD.79.034005}}, \eprint{0809.2262}.

\bibtype{Article}%
\bibitem{Liu:2018trl}
\bibinfo{author}{X. Liu}, \bibinfo{author}{F. Ringer}, \bibinfo{author}{W.
  Vogelsang}, \bibinfo{author}{F. Yuan}, \bibinfo{title}{{Lepton-jet
  Correlations in Deep Inelastic Scattering at the Electron-Ion Collider}},
  \bibinfo{journal}{Phys. Rev. Lett.} \bibinfo{volume}{122}
  (\bibinfo{number}{19}) (\bibinfo{year}{2019}) \bibinfo{pages}{192003},
  \bibinfo{doi}{\doi{10.1103/PhysRevLett.122.192003}}, \eprint{1812.08077}.

\bibtype{Article}%
\bibitem{Boer:2011kf}
\bibinfo{author}{D. Boer}, \bibinfo{author}{W.~J. den Dunnen},
  \bibinfo{author}{C. Pisano}, \bibinfo{author}{M. Schlegel},
  \bibinfo{author}{W. Vogelsang}, \bibinfo{title}{{Linearly Polarized Gluons
  and the Higgs Transverse Momentum Distribution}}, \bibinfo{journal}{Phys.
  Rev. Lett.} \bibinfo{volume}{108} (\bibinfo{year}{2012})
  \bibinfo{pages}{032002}, \bibinfo{doi}{\doi{10.1103/PhysRevLett.108.032002}},
  \eprint{1109.1444}.

\bibtype{Article}%
\bibitem{Echevarria:2015uaa}
\bibinfo{author}{M.~G. Echevarria}, \bibinfo{author}{T. Kasemets},
  \bibinfo{author}{P.~J. Mulders}, \bibinfo{author}{C. Pisano},
  \bibinfo{title}{{QCD evolution of (un)polarized gluon TMDPDFs and the Higgs
  $q_T$-distribution}}, \bibinfo{journal}{JHEP} \bibinfo{volume}{07}
  (\bibinfo{year}{2015}) \bibinfo{pages}{158},
  \bibinfo{doi}{\doi{10.1007/JHEP07(2015)158}}, \eprint{1502.05354}.

\bibtype{Article}%
\bibitem{Echevarria:2019ynx}
\bibinfo{author}{M.~G. Echevarria}, \bibinfo{title}{{Proper TMD factorization
  for quarkonia production: $pp\to\eta_{c,b}$ as a study case}},
  \bibinfo{journal}{JHEP} \bibinfo{volume}{10} (\bibinfo{year}{2019})
  \bibinfo{pages}{144}, \bibinfo{doi}{\doi{10.1007/JHEP10(2019)144}},
  \eprint{1907.06494}.

\bibtype{Article}%
\bibitem{Metz:2011wb}
\bibinfo{author}{A. Metz}, \bibinfo{author}{J. Zhou},
  \bibinfo{title}{{Distribution of linearly polarized gluons inside a large
  nucleus}}, \bibinfo{journal}{Phys. Rev. D} \bibinfo{volume}{84}
  (\bibinfo{year}{2011}) \bibinfo{pages}{051503},
  \bibinfo{doi}{\doi{10.1103/PhysRevD.84.051503}}, \eprint{1105.1991}.

\bibtype{Article}%
\bibitem{Dominguez:2011br}
\bibinfo{author}{F. Dominguez}, \bibinfo{author}{J.-W. Qiu},
  \bibinfo{author}{B.-W. Xiao}, \bibinfo{author}{F. Yuan}, \bibinfo{title}{{On
  the linearly polarized gluon distributions in the color dipole model}},
  \bibinfo{journal}{Phys. Rev. D} \bibinfo{volume}{85} (\bibinfo{year}{2012})
  \bibinfo{pages}{045003}, \bibinfo{doi}{\doi{10.1103/PhysRevD.85.045003}},
  \eprint{1109.6293}.

\bibtype{Article}%
\bibitem{Collins:2002kn}
\bibinfo{author}{J.~C. Collins}, \bibinfo{title}{{Leading twist single
  transverse-spin asymmetries: Drell-Yan and deep inelastic scattering}},
  \bibinfo{journal}{Phys. Lett. B} \bibinfo{volume}{536} (\bibinfo{year}{2002})
  \bibinfo{pages}{43--48}, \bibinfo{doi}{\doi{10.1016/S0370-2693(02)01819-1}},
  \eprint{hep-ph/0204004}.

\bibtype{Article}%
\bibitem{Bomhof:2006dp}
\bibinfo{author}{C.~J. Bomhof}, \bibinfo{author}{P.~J. Mulders},
  \bibinfo{author}{F. Pijlman}, \bibinfo{title}{{The Construction of
  gauge-links in arbitrary hard processes}}, \bibinfo{journal}{Eur. Phys. J. C}
  \bibinfo{volume}{47} (\bibinfo{year}{2006}) \bibinfo{pages}{147--162},
  \bibinfo{doi}{\doi{10.1140/epjc/s2006-02554-2}}, \eprint{hep-ph/0601171}.

\bibtype{Article}%
\bibitem{Collins:2007nk}
\bibinfo{author}{J. Collins}, \bibinfo{author}{J.-W. Qiu},
  \bibinfo{title}{{$k_{T}$ factorization is violated in production of
  high-transverse-momentum particles in hadron-hadron collisions}},
  \bibinfo{journal}{Phys. Rev. D} \bibinfo{volume}{75} (\bibinfo{year}{2007})
  \bibinfo{pages}{114014}, \bibinfo{doi}{\doi{10.1103/PhysRevD.75.114014}},
  \eprint{0705.2141}.

\bibtype{Article}%
\bibitem{Rogers:2010dm}
\bibinfo{author}{T.~C. Rogers}, \bibinfo{author}{P.~J. Mulders},
  \bibinfo{title}{{No Generalized TMD-Factorization in Hadro-Production of High
  Transverse Momentum Hadrons}}, \bibinfo{journal}{Phys. Rev. D}
  \bibinfo{volume}{81} (\bibinfo{year}{2010}) \bibinfo{pages}{094006},
  \bibinfo{doi}{\doi{10.1103/PhysRevD.81.094006}}, \eprint{1001.2977}.

\bibtype{Article}%
\bibitem{Metz:2002iz}
\bibinfo{author}{A. Metz}, \bibinfo{title}{{Gluon-exchange in spin-dependent
  fragmentation}}, \bibinfo{journal}{Phys. Lett. B} \bibinfo{volume}{549}
  (\bibinfo{year}{2002}) \bibinfo{pages}{139--145},
  \bibinfo{doi}{\doi{10.1016/S0370-2693(02)02899-X}}, \eprint{hep-ph/0209054}.

\bibtype{Article}%
\bibitem{Collins:2004nx}
\bibinfo{author}{J.~C. Collins}, \bibinfo{author}{A. Metz},
  \bibinfo{title}{{Universality of soft and collinear factors in
  hard-scattering factorization}}, \bibinfo{journal}{Phys. Rev. Lett.}
  \bibinfo{volume}{93} (\bibinfo{year}{2004}) \bibinfo{pages}{252001},
  \bibinfo{doi}{\doi{10.1103/PhysRevLett.93.252001}}, \eprint{hep-ph/0408249}.

\bibtype{Article}%
\bibitem{Yuan:2008yv}
\bibinfo{author}{F. Yuan}, \bibinfo{title}{{Collins Asymmetry at Hadron
  Colliders}}, \bibinfo{journal}{Phys. Rev. D} \bibinfo{volume}{77}
  (\bibinfo{year}{2008}) \bibinfo{pages}{074019},
  \bibinfo{doi}{\doi{10.1103/PhysRevD.77.074019}}, \eprint{0801.3441}.

\bibtype{Article}%
\bibitem{Gamberg:2008yt}
\bibinfo{author}{L.~P. Gamberg}, \bibinfo{author}{A. Mukherjee},
  \bibinfo{author}{P.~J. Mulders}, \bibinfo{title}{{Spectral analysis of
  gluonic pole matrix elements for fragmentation}}, \bibinfo{journal}{Phys.
  Rev. D} \bibinfo{volume}{77} (\bibinfo{year}{2008}) \bibinfo{pages}{114026},
  \bibinfo{doi}{\doi{10.1103/PhysRevD.77.114026}}, \eprint{0803.2632}.

\bibtype{Article}%
\bibitem{Meissner:2008yf}
\bibinfo{author}{S. Meissner}, \bibinfo{author}{A. Metz},
  \bibinfo{title}{{Partonic pole matrix elements for fragmentation}},
  \bibinfo{journal}{Phys. Rev. Lett.} \bibinfo{volume}{102}
  (\bibinfo{year}{2009}) \bibinfo{pages}{172003},
  \bibinfo{doi}{\doi{10.1103/PhysRevLett.102.172003}}, \eprint{0812.3783}.

\bibtype{Article}%
\bibitem{Collins:1981uk}
\bibinfo{author}{J.~C. Collins}, \bibinfo{author}{D.~E. Soper},
  \bibinfo{title}{{Back-To-Back Jets in QCD}}, \bibinfo{journal}{Nucl. Phys. B}
  \bibinfo{volume}{193} (\bibinfo{year}{1981}) \bibinfo{pages}{381},
  \bibinfo{doi}{\doi{10.1016/0550-3213(81)90339-4}}.

\bibtype{Article}%
\bibitem{Ji:2004wu}
\bibinfo{author}{X.-D. Ji}, \bibinfo{author}{J.-P. Ma}, \bibinfo{author}{F.
  Yuan}, \bibinfo{title}{{QCD factorization for semi-inclusive deep-inelastic
  scattering at low transverse momentum}}, \bibinfo{journal}{Phys. Rev. D}
  \bibinfo{volume}{71} (\bibinfo{year}{2005}) \bibinfo{pages}{034005},
  \bibinfo{doi}{\doi{10.1103/PhysRevD.71.034005}}, \eprint{hep-ph/0404183}.

\bibtype{Article}%
\bibitem{Ebert:2021jhy}
\bibinfo{author}{M.~A. Ebert}, \bibinfo{author}{A. Gao}, \bibinfo{author}{I.~W.
  Stewart}, \bibinfo{title}{{Factorization for azimuthal asymmetries in SIDIS
  at next-to-leading power}}, \bibinfo{journal}{JHEP} \bibinfo{volume}{06}
  (\bibinfo{year}{2022}) \bibinfo{pages}{007},
  \bibinfo{doi}{\doi{10.1007/JHEP06(2022)007}}, \eprint{2112.07680}.

\bibtype{Article}%
\bibitem{Gamberg:2022lju}
\bibinfo{author}{L. Gamberg}, \bibinfo{author}{Z.-B. Kang},
  \bibinfo{author}{D.~Y. Shao}, \bibinfo{author}{J. Terry}, \bibinfo{author}{F.
  Zhao}, \bibinfo{title}{{Transverse-momentum-dependent factorization at
  next-to-leading power}}  (\bibinfo{year}{2022}), \eprint{2211.13209}.

\bibtype{Article}%
\bibitem{Rodini:2023plb}
\bibinfo{author}{S. Rodini}, \bibinfo{author}{A. Vladimirov},
  \bibinfo{title}{{Transverse momentum dependent factorization for SIDIS at
  next-to-leading power}}, \bibinfo{journal}{Phys. Rev. D}
  \bibinfo{volume}{110} (\bibinfo{number}{3}) (\bibinfo{year}{2024})
  \bibinfo{pages}{034009}, \bibinfo{doi}{\doi{10.1103/PhysRevD.110.034009}},
  \eprint{2306.09495}.

\bibtype{Article}%
\bibitem{Dokshitzer:1978yd}
\bibinfo{author}{Y.~L. Dokshitzer}, \bibinfo{author}{D. Diakonov},
  \bibinfo{author}{S.~I. Troian}, \bibinfo{title}{{On the Transverse Momentum
  Distribution of Massive Lepton Pairs}}, \bibinfo{journal}{Phys. Lett. B}
  \bibinfo{volume}{79} (\bibinfo{year}{1978}) \bibinfo{pages}{269--272},
  \bibinfo{doi}{\doi{10.1016/0370-2693(78)90240-X}}.

\bibtype{Article}%
\bibitem{Parisi:1979se}
\bibinfo{author}{G. Parisi}, \bibinfo{author}{R. Petronzio},
  \bibinfo{title}{{Small Transverse Momentum Distributions in Hard Processes}},
  \bibinfo{journal}{Nucl. Phys. B} \bibinfo{volume}{154} (\bibinfo{year}{1979})
  \bibinfo{pages}{427--440}, \bibinfo{doi}{\doi{10.1016/0550-3213(79)90040-3}}.

\bibtype{Article}%
\bibitem{Ji:2006br}
\bibinfo{author}{X.-D. Ji}, \bibinfo{author}{J.-W. Qiu}, \bibinfo{author}{W.
  Vogelsang}, \bibinfo{author}{F. Yuan}, \bibinfo{title}{{Single-transverse
  spin asymmetry in semi-inclusive deep inelastic scattering}},
  \bibinfo{journal}{Phys. Lett. B} \bibinfo{volume}{638} (\bibinfo{year}{2006})
  \bibinfo{pages}{178--186},
  \bibinfo{doi}{\doi{10.1016/j.physletb.2006.05.044}}, \eprint{hep-ph/0604128}.

\bibtype{Article}%
\bibitem{Bacchetta:2008xw}
\bibinfo{author}{A. Bacchetta}, \bibinfo{author}{D. Boer}, \bibinfo{author}{M.
  Diehl}, \bibinfo{author}{P.J. Mulders}, \bibinfo{title}{{Matches and
  mismatches in the descriptions of semi-inclusive processes at low and high
  transverse momentum}}, \bibinfo{journal}{JHEP} \bibinfo{volume}{08}
  (\bibinfo{year}{2008}) \bibinfo{pages}{023},
  \bibinfo{doi}{\doi{10.1088/1126-6708/2008/08/023}}, \eprint{0803.0227}.

\bibtype{Article}%
\bibitem{Boglione:2014oea}
\bibinfo{author}{M. Boglione}, \bibinfo{author}{J.~O. Gonzalez~Hernandez},
  \bibinfo{author}{S. Melis}, \bibinfo{author}{A. Prokudin}, \bibinfo{title}{{A
  study on the interplay between perturbative QCD and CSS/TMD formalism in
  SIDIS processes}}, \bibinfo{journal}{JHEP} \bibinfo{volume}{02}
  (\bibinfo{year}{2015}) \bibinfo{pages}{095},
  \bibinfo{doi}{\doi{10.1007/JHEP02(2015)095}}, \eprint{1412.1383}.

\bibtype{Article}%
\bibitem{Collins:2016hqq}
\bibinfo{author}{J. Collins}, \bibinfo{author}{L. Gamberg}, \bibinfo{author}{A.
  Prokudin}, \bibinfo{author}{T.~C. Rogers}, \bibinfo{author}{N. Sato},
  \bibinfo{author}{B. Wang}, \bibinfo{title}{{Relating Transverse Momentum
  Dependent and Collinear Factorization Theorems in a Generalized Formalism}},
  \bibinfo{journal}{Phys. Rev. D} \bibinfo{volume}{94} (\bibinfo{number}{3})
  (\bibinfo{year}{2016}) \bibinfo{pages}{034014},
  \bibinfo{doi}{\doi{10.1103/PhysRevD.94.034014}}, \eprint{1605.00671}.

\bibtype{Article}%
\bibitem{Gonzalez-Hernandez:2023iso}
\bibinfo{author}{J.~O. Gonzalez-Hernandez}, \bibinfo{author}{T. Rainaldi},
  \bibinfo{author}{T.~C. Rogers}, \bibinfo{title}{{Resolution to the problem of
  consistent large transverse momentum in TMDs}}, \bibinfo{journal}{Phys. Rev.
  D} \bibinfo{volume}{107} (\bibinfo{number}{9}) (\bibinfo{year}{2023})
  \bibinfo{pages}{094029}, \bibinfo{doi}{\doi{10.1103/PhysRevD.107.094029}},
  \eprint{2303.04921}.

\bibtype{Article}%
\bibitem{Collins:2003fm}
\bibinfo{author}{J.~C. Collins}, \bibinfo{title}{{What exactly is a parton
  density?}}, \bibinfo{journal}{Acta Phys. Polon. B} \bibinfo{volume}{34}
  (\bibinfo{year}{2003}) \bibinfo{pages}{3103}, \eprint{hep-ph/0304122}.

\bibtype{Article}%
\bibitem{Aybat:2011zv}
\bibinfo{author}{S.~M. Aybat}, \bibinfo{author}{T.~C. Rogers},
  \bibinfo{title}{{TMD Parton Distribution and Fragmentation Functions with QCD
  Evolution}}, \bibinfo{journal}{Phys. Rev. D} \bibinfo{volume}{83}
  (\bibinfo{year}{2011}) \bibinfo{pages}{114042},
  \bibinfo{doi}{\doi{10.1103/PhysRevD.83.114042}}, \eprint{1101.5057}.

\bibtype{Article}%
\bibitem{Gehrmann:2014yya}
\bibinfo{author}{T. Gehrmann}, \bibinfo{author}{T. Luebbert},
  \bibinfo{author}{L.~L. Yang}, \bibinfo{title}{{Calculation of the transverse
  parton distribution functions at next-to-next-to-leading order}},
  \bibinfo{journal}{JHEP} \bibinfo{volume}{06} (\bibinfo{year}{2014})
  \bibinfo{pages}{155}, \bibinfo{doi}{\doi{10.1007/JHEP06(2014)155}},
  \eprint{1403.6451}.

\bibtype{Article}%
\bibitem{Kang:2011mr}
\bibinfo{author}{Z.-B. Kang}, \bibinfo{author}{B.-W. Xiao}, \bibinfo{author}{F.
  Yuan}, \bibinfo{title}{{QCD Resummation for Single Spin Asymmetries}},
  \bibinfo{journal}{Phys. Rev. Lett.} \bibinfo{volume}{107}
  (\bibinfo{year}{2011}) \bibinfo{pages}{152002},
  \bibinfo{doi}{\doi{10.1103/PhysRevLett.107.152002}}, \eprint{1106.0266}.

\bibtype{Article}%
\bibitem{Aybat:2011ge}
\bibinfo{author}{S.~M. Aybat}, \bibinfo{author}{J.~C. Collins},
  \bibinfo{author}{J.-W. Qiu}, \bibinfo{author}{T.~C. Rogers},
  \bibinfo{title}{{The QCD Evolution of the Sivers Function}},
  \bibinfo{journal}{Phys. Rev. D} \bibinfo{volume}{85} (\bibinfo{year}{2012})
  \bibinfo{pages}{034043}, \bibinfo{doi}{\doi{10.1103/PhysRevD.85.034043}},
  \eprint{1110.6428}.

\bibtype{Article}%
\bibitem{Ebert:2022cku}
\bibinfo{author}{M.~A. Ebert}, \bibinfo{author}{J.~K.~L. Michel},
  \bibinfo{author}{I.~W. Stewart}, \bibinfo{author}{Z. Sun},
  \bibinfo{title}{{Disentangling long and short distances in momentum-space
  TMDs}}, \bibinfo{journal}{JHEP} \bibinfo{volume}{07} (\bibinfo{year}{2022})
  \bibinfo{pages}{129}, \bibinfo{doi}{\doi{10.1007/JHEP07(2022)129}},
  \eprint{2201.07237}.

\bibtype{Article}%
\bibitem{delRio:2024vvq}
\bibinfo{author}{O. del Rio}, \bibinfo{author}{A. Prokudin},
  \bibinfo{author}{I. Scimemi}, \bibinfo{author}{A. Vladimirov},
  \bibinfo{title}{{Transverse momentum moments}}, \bibinfo{journal}{Phys. Rev.
  D} \bibinfo{volume}{110} (\bibinfo{number}{1}) (\bibinfo{year}{2024})
  \bibinfo{pages}{016003}, \bibinfo{doi}{\doi{10.1103/PhysRevD.110.016003}},
  \eprint{2402.01836}.

\bibtype{Article}%
\bibitem{Bacchetta:2025ara}
\bibinfo{author}{A. Bacchetta}, \bibinfo{author}{V. Bertone},
  \bibinfo{author}{C. Bissolotti}, \bibinfo{author}{M. Cerutti},
  \bibinfo{author}{M. Radici}, \bibinfo{author}{S. Rodini}, \bibinfo{author}{L.
  Rossi} (\bibinfo{collaboration}{MAP (Multi-dimensional Analyses of Partonic
  distributions)}), \bibinfo{title}{{Neural-Network Extraction of Unpolarized
  Transverse-Momentum-Dependent Distributions}}, \bibinfo{journal}{Phys. Rev.
  Lett.} \bibinfo{volume}{135} (\bibinfo{number}{2}) (\bibinfo{year}{2025})
  \bibinfo{pages}{021904}, \bibinfo{doi}{\doi{10.1103/csc2-bj91}},
  \eprint{2502.04166}.

\bibtype{Article}%
\bibitem{Moos:2025sal}
\bibinfo{author}{Valentin Moos}, \bibinfo{author}{Ignazio Scimemi},
  \bibinfo{author}{Alexey Vladimirov}, \bibinfo{author}{Pia Zurita},
  \bibinfo{title}{{Determination of unpolarized TMD distributions from the fit
  of Drell-Yan and SIDIS data at N$^{4}$LL}}, \bibinfo{journal}{JHEP}
  \bibinfo{volume}{11} (\bibinfo{year}{2025}) \bibinfo{pages}{134},
  \bibinfo{doi}{\doi{10.1007/JHEP11(2025)134}}, \eprint{2503.11201}.

\bibtype{Article}%
\bibitem{Cammarota:2020qcw}
\bibinfo{author}{J. Cammarota}, \bibinfo{author}{L. Gamberg},
  \bibinfo{author}{Z.-B. Kang}, \bibinfo{author}{J.~A. Miller},
  \bibinfo{author}{D. Pitonyak}, \bibinfo{author}{A. Prokudin},
  \bibinfo{author}{T.~C. Rogers}, \bibinfo{author}{N. Sato}
  (\bibinfo{collaboration}{Jefferson Lab Angular Momentum}),
  \bibinfo{title}{{Origin of single transverse-spin asymmetries in high-energy
  collisions}}, \bibinfo{journal}{Phys. Rev. D} \bibinfo{volume}{102}
  (\bibinfo{number}{5}) (\bibinfo{year}{2020}) \bibinfo{pages}{054002},
  \bibinfo{doi}{\doi{10.1103/PhysRevD.102.054002}}, \eprint{2002.08384}.

\bibtype{Article}%
\bibitem{Gamberg:2022kdb}
\bibinfo{author}{L. Gamberg}, \bibinfo{author}{M. Malda},
  \bibinfo{author}{J.~A. Miller}, \bibinfo{author}{D. Pitonyak},
  \bibinfo{author}{A. Prokudin}, \bibinfo{author}{N. Sato}
  (\bibinfo{collaboration}{Jefferson Lab Angular Momentum (JAM), Jefferson Lab
  Angular Momentum}), \bibinfo{title}{{Updated QCD global analysis of single
  transverse-spin asymmetries: Extracting $\tilde{H}$, and the role of the
  Soffer bound and lattice QCD}}, \bibinfo{journal}{Phys. Rev. D}
  \bibinfo{volume}{106} (\bibinfo{number}{3}) (\bibinfo{year}{2022})
  \bibinfo{pages}{034014}, \bibinfo{doi}{\doi{10.1103/PhysRevD.106.034014}},
  \eprint{2205.00999}.

\bibtype{Article}%
\bibitem{Boer:1997mf}
\bibinfo{author}{D. Boer}, \bibinfo{author}{R. Jakob}, \bibinfo{author}{P.~J.
  Mulders}, \bibinfo{title}{{Asymmetries in polarized hadron production in
  $e^+e^-$ annihilation up to order $1/Q$}}, \bibinfo{journal}{Nucl. Phys. B}
  \bibinfo{volume}{504} (\bibinfo{year}{1997}) \bibinfo{pages}{345--380},
  \bibinfo{doi}{\doi{10.1016/S0550-3213(97)00456-2}}, \eprint{hep-ph/9702281}.

\bibtype{Article}%
\bibitem{Collins:1993kq}
\bibinfo{author}{J.~C. Collins}, \bibinfo{author}{S.~F. Heppelmann},
  \bibinfo{author}{G.~A. Ladinsky}, \bibinfo{title}{{Measuring transversity
  densities in singly polarized hadron hadron and lepton - hadron collisions}},
  \bibinfo{journal}{Nucl. Phys. B} \bibinfo{volume}{420} (\bibinfo{year}{1994})
  \bibinfo{pages}{565--582}, \bibinfo{doi}{\doi{10.1016/0550-3213(94)90078-7}},
  \eprint{hep-ph/9305309}.

\bibtype{Article}%
\bibitem{Bianconi:1999cd}
\bibinfo{author}{A. Bianconi}, \bibinfo{author}{S. Boffi}, \bibinfo{author}{R.
  Jakob}, \bibinfo{author}{M. Radici}, \bibinfo{title}{{Two hadron interference
  fragmentation functions. Part 1. General framework}}, \bibinfo{journal}{Phys.
  Rev. D} \bibinfo{volume}{62} (\bibinfo{year}{2000}) \bibinfo{pages}{034008},
  \bibinfo{doi}{\doi{10.1103/PhysRevD.62.034008}}, \eprint{hep-ph/9907475}.

\bibtype{Article}%
\bibitem{Bacchetta:2004it}
\bibinfo{author}{A. Bacchetta}, \bibinfo{author}{M. Radici},
  \bibinfo{title}{{Dihadron interference fragmentation functions in
  proton-proton collisions}}, \bibinfo{journal}{Phys. Rev. D}
  \bibinfo{volume}{70} (\bibinfo{year}{2004}) \bibinfo{pages}{094032},
  \bibinfo{doi}{\doi{10.1103/PhysRevD.70.094032}}, \eprint{hep-ph/0409174}.

\bibtype{Article}%
\bibitem{Artru:1995zu}
\bibinfo{author}{X. Artru}, \bibinfo{author}{J.~C. Collins},
  \bibinfo{title}{{Measuring transverse spin correlations by 4 particle
  correlations in $e^+e^- \rightarrow $ 2 jets}}, \bibinfo{journal}{Z. Phys. C}
  \bibinfo{volume}{69} (\bibinfo{year}{1996}) \bibinfo{pages}{277--286},
  \bibinfo{doi}{\doi{10.1007/s002880050028}}, \eprint{hep-ph/9504220}.

\bibtype{Article}%
\bibitem{Boer:2003ya}
\bibinfo{author}{D. Boer}, \bibinfo{author}{R. Jakob}, \bibinfo{author}{M.
  Radici}, \bibinfo{title}{{Interference fragmentation functions in electron
  positron annihilation}}, \bibinfo{journal}{Phys. Rev. D} \bibinfo{volume}{67}
  (\bibinfo{year}{2003}) \bibinfo{pages}{094003},
  \bibinfo{doi}{\doi{10.1103/PhysRevD.67.094003}}, \eprint{hep-ph/0302232}.

\bibtype{Article}%
\bibitem{Matevosyan:2018icf}
\bibinfo{author}{H.~H. Matevosyan}, \bibinfo{author}{A. Bacchetta},
  \bibinfo{author}{D. Boer}, \bibinfo{author}{A. Courtoy}, \bibinfo{author}{A.
  Kotzinian}, \bibinfo{author}{M. Radici}, \bibinfo{author}{A.~W. Thomas},
  \bibinfo{title}{{Semi-inclusive production of two back-to-back hadron pairs
  in $e^+e^-$ annihilation revisited}}, \bibinfo{journal}{Phys. Rev. D}
  \bibinfo{volume}{97} (\bibinfo{number}{7}) (\bibinfo{year}{2018})
  \bibinfo{pages}{074019}, \bibinfo{doi}{\doi{10.1103/PhysRevD.97.074019}},
  \eprint{1802.01578}.

\bibtype{Article}%
\bibitem{Pitonyak:2023gjx}
\bibinfo{author}{D. Pitonyak}, \bibinfo{author}{C. Cocuzza},
  \bibinfo{author}{A. Metz}, \bibinfo{author}{A. Prokudin}, \bibinfo{author}{N.
  Sato}, \bibinfo{title}{{Number Density Interpretation of Dihadron
  Fragmentation Functions}}, \bibinfo{journal}{Phys. Rev. Lett.}
  \bibinfo{volume}{132} (\bibinfo{number}{1}) (\bibinfo{year}{2024})
  \bibinfo{pages}{011902}, \bibinfo{doi}{\doi{10.1103/PhysRevLett.132.011902}},
  \eprint{2305.11995}.

\bibtype{Article}%
\bibitem{HERMES:2008mcr}
\bibinfo{author}{A. Airapetian}, et al. (\bibinfo{collaboration}{HERMES}),
  \bibinfo{title}{{Evidence for a Transverse Single-Spin Asymmetry in
  Leptoproduction of $\pi^+\pi^-$ Pairs}}, \bibinfo{journal}{JHEP}
  \bibinfo{volume}{06} (\bibinfo{year}{2008}) \bibinfo{pages}{017},
  \bibinfo{doi}{\doi{10.1088/1126-6708/2008/06/017}}, \eprint{0803.2367}.

\bibtype{Article}%
\bibitem{Metz:2012ui}
\bibinfo{author}{A. Metz}, \bibinfo{author}{D. Pitonyak}, \bibinfo{author}{A.
  Sch{\"a}fer}, \bibinfo{author}{M. Schlegel}, \bibinfo{author}{W. Vogelsang},
  \bibinfo{author}{J. Zhou}, \bibinfo{title}{{Single-spin asymmetries in
  inclusive deep inelastic scattering and multiparton correlations in the
  nucleon}}, \bibinfo{journal}{Phys. Rev. D} \bibinfo{volume}{86}
  (\bibinfo{year}{2012}) \bibinfo{pages}{094039},
  \bibinfo{doi}{\doi{10.1103/PhysRevD.86.094039}}, \eprint{1209.3138}.

\bibtype{Article}%
\bibitem{Gamberg:2013kla}
\bibinfo{author}{L. Gamberg}, \bibinfo{author}{Z.-B. Kang}, \bibinfo{author}{A.
  Prokudin}, \bibinfo{title}{{Indication on the process-dependence of the
  Sivers effect}}, \bibinfo{journal}{Phys. Rev. Lett.} \bibinfo{volume}{110}
  (\bibinfo{number}{23}) (\bibinfo{year}{2013}) \bibinfo{pages}{232301},
  \bibinfo{doi}{\doi{10.1103/PhysRevLett.110.232301}}, \eprint{1302.3218}.

\bibtype{Article}%
\bibitem{Bacchetta:2020gko}
\bibinfo{author}{A. Bacchetta}, \bibinfo{author}{F. Delcarro},
  \bibinfo{author}{C. Pisano}, \bibinfo{author}{M. Radici},
  \bibinfo{title}{{The 3-dimensional distribution of quarks in momentum
  space}}, \bibinfo{journal}{Phys. Lett. B} \bibinfo{volume}{827}
  (\bibinfo{year}{2022}) \bibinfo{pages}{136961},
  \bibinfo{doi}{\doi{10.1016/j.physletb.2022.136961}}, \eprint{2004.14278}.

\bibtype{Article}%
\bibitem{Barone:2009hw}
\bibinfo{author}{V. Barone}, \bibinfo{author}{S. Melis}, \bibinfo{author}{A.
  Prokudin}, \bibinfo{title}{{The Boer-Mulders effect in unpolarized SIDIS: An
  Analysis of the COMPASS and HERMES data on the cos 2 phi asymmetry}},
  \bibinfo{journal}{Phys. Rev. D} \bibinfo{volume}{81} (\bibinfo{year}{2010})
  \bibinfo{pages}{114026}, \bibinfo{doi}{\doi{10.1103/PhysRevD.81.114026}},
  \eprint{0912.5194}.

\bibtype{Article}%
\bibitem{Lefky:2014eia}
\bibinfo{author}{C. Lefky}, \bibinfo{author}{A. Prokudin},
  \bibinfo{title}{{Extraction of the distribution function $h^{\perp}_{1T}$
  from experimental data}}, \bibinfo{journal}{Phys. Rev. D}
  \bibinfo{volume}{91} (\bibinfo{number}{3}) (\bibinfo{year}{2015})
  \bibinfo{pages}{034010}, \bibinfo{doi}{\doi{10.1103/PhysRevD.91.034010}},
  \eprint{1411.0580}.

\bibtype{Article}%
\bibitem{Bhattacharya:2021twu}
\bibinfo{author}{S. Bhattacharya}, \bibinfo{author}{Z.-B. Kang},
  \bibinfo{author}{A. Metz}, \bibinfo{author}{G. Penn}, \bibinfo{author}{D.
  Pitonyak}, \bibinfo{title}{{First global QCD analysis of the TMD $g_{1T}$
  from semi-inclusive DIS data}}, \bibinfo{journal}{Phys. Rev. D}
  \bibinfo{volume}{105} (\bibinfo{number}{3}) (\bibinfo{year}{2022})
  \bibinfo{pages}{034007}, \bibinfo{doi}{\doi{10.1103/PhysRevD.105.034007}},
  \eprint{2110.10253}.

\bibtype{Article}%
\bibitem{Horstmann:2022xkk}
\bibinfo{author}{M. Horstmann}, \bibinfo{author}{A. Sch{\"a}fer},
  \bibinfo{author}{A. Vladimirov}, \bibinfo{title}{{Study of the worm-gear-T
  function $g_{1T}$ with semi-inclusive DIS data}}, \bibinfo{journal}{Phys.
  Rev. D} \bibinfo{volume}{107} (\bibinfo{number}{3}) (\bibinfo{year}{2023})
  \bibinfo{pages}{034016}, \bibinfo{doi}{\doi{10.1103/PhysRevD.107.034016}},
  \eprint{2210.07268}.

\bibtype{Article}%
\bibitem{Yang:2024bfz}
\bibinfo{author}{K. Yang}, \bibinfo{author}{T. Liu}, \bibinfo{author}{P. Sun},
  \bibinfo{author}{Y. Zhao}, \bibinfo{author}{B.-Q. Ma},
  \bibinfo{title}{{Extraction of transhelicity worm-gear distributions and
  opportunities at the Electron-Ion Collider in China}},
  \bibinfo{journal}{Phys. Rev. D} \bibinfo{volume}{110} (\bibinfo{number}{3})
  (\bibinfo{year}{2024}) \bibinfo{pages}{034036},
  \bibinfo{doi}{\doi{10.1103/PhysRevD.110.034036}}, \eprint{2403.12795}.

\bibtype{Article}%
\bibitem{Bacchetta:2024yzl}
\bibinfo{author}{A. Bacchetta}, \bibinfo{author}{A. Bongallino},
  \bibinfo{author}{M. Cerutti}, \bibinfo{author}{M. Radici},
  \bibinfo{author}{L. Rossi} (\bibinfo{collaboration}{MAP (Multi-dimensional
  Analyses of Partonic distributions)}), \bibinfo{title}{{Exploring the
  Three-Dimensional Momentum Distribution of Longitudinally Polarized Quarks in
  the Proton}}, \bibinfo{journal}{Phys. Rev. Lett.} \bibinfo{volume}{134}
  (\bibinfo{number}{12}) (\bibinfo{year}{2025}) \bibinfo{pages}{121901},
  \bibinfo{doi}{\doi{10.1103/PhysRevLett.134.121901}}, \eprint{2409.18078}.

\bibtype{Article}%
\bibitem{Soper:1976jc}
\bibinfo{author}{D.~E. Soper}, \bibinfo{title}{{The Parton Model and the
  Bethe-Salpeter Wave Function}}, \bibinfo{journal}{Phys. Rev. D}
  \bibinfo{volume}{15} (\bibinfo{year}{1977}) \bibinfo{pages}{1141},
  \bibinfo{doi}{\doi{10.1103/PhysRevD.15.1141}}.

\bibtype{Article}%
\bibitem{Burkardt:2002hr}
\bibinfo{author}{M. Burkardt}, \bibinfo{title}{{Impact parameter space
  interpretation for generalized parton distributions}}, \bibinfo{journal}{Int.
  J. Mod. Phys. A} \bibinfo{volume}{18} (\bibinfo{year}{2003})
  \bibinfo{pages}{173--208}, \bibinfo{doi}{\doi{10.1142/S0217751X03012370}},
  \eprint{hep-ph/0207047}.

\bibtype{Article}%
\bibitem{Ralston:2001xs}
\bibinfo{author}{J.~P. Ralston}, \bibinfo{author}{B. Pire},
  \bibinfo{title}{{Femtophotography of protons to nuclei with deeply virtual
  Compton scattering}}, \bibinfo{journal}{Phys. Rev. D} \bibinfo{volume}{66}
  (\bibinfo{year}{2002}) \bibinfo{pages}{111501},
  \bibinfo{doi}{\doi{10.1103/PhysRevD.66.111501}}, \eprint{hep-ph/0110075}.

\bibtype{Article}%
\bibitem{Diehl:2002he}
\bibinfo{author}{M. Diehl}, \bibinfo{title}{{Generalized parton distributions
  in impact parameter space}}, \bibinfo{journal}{Eur. Phys. J. C}
  \bibinfo{volume}{25} (\bibinfo{year}{2002}) \bibinfo{pages}{223--232},
  \bibinfo{doi}{\doi{10.1007/s10052-002-1016-9}}, \eprint{hep-ph/0205208}.

\bibtype{Article}%
\bibitem{Polyakov:2002yz}
\bibinfo{author}{M.~V. Polyakov}, \bibinfo{title}{{Generalized parton
  distributions and strong forces inside nucleons and nuclei}},
  \bibinfo{journal}{Phys. Lett. B} \bibinfo{volume}{555} (\bibinfo{year}{2003})
  \bibinfo{pages}{57--62}, \bibinfo{doi}{\doi{10.1016/S0370-2693(03)00036-4}},
  \eprint{hep-ph/0210165}.

\bibtype{Article}%
\bibitem{Diehl:2003ny}
\bibinfo{author}{M. Diehl}, \bibinfo{title}{{Generalized parton
  distributions}}, \bibinfo{journal}{Phys. Rept.} \bibinfo{volume}{388}
  (\bibinfo{year}{2003}) \bibinfo{pages}{41--277},
  \bibinfo{doi}{\doi{10.1016/j.physrep.2003.08.002}}, \eprint{hep-ph/0307382}.

\bibtype{Article}%
\bibitem{Ji:1998pc}
\bibinfo{author}{X.-D. Ji}, \bibinfo{title}{{Off forward parton
  distributions}}, \bibinfo{journal}{J. Phys. G} \bibinfo{volume}{24}
  (\bibinfo{year}{1998}) \bibinfo{pages}{1181--1205},
  \bibinfo{doi}{\doi{10.1088/0954-3899/24/7/002}}, \eprint{hep-ph/9807358}.

\bibtype{Article}%
\bibitem{Goeke:2001tz}
\bibinfo{author}{K. Goeke}, \bibinfo{author}{M.~V. Polyakov},
  \bibinfo{author}{M. Vanderhaeghen}, \bibinfo{title}{{Hard exclusive reactions
  and the structure of hadrons}}, \bibinfo{journal}{Prog. Part. Nucl. Phys.}
  \bibinfo{volume}{47} (\bibinfo{year}{2001}) \bibinfo{pages}{401--515},
  \bibinfo{doi}{\doi{10.1016/S0146-6410(01)00158-2}}, \eprint{hep-ph/0106012}.

\bibtype{Article}%
\bibitem{Belitsky:2005qn}
\bibinfo{author}{A.~V. Belitsky}, \bibinfo{author}{A.~V. Radyushkin},
  \bibinfo{title}{{Unraveling hadron structure with generalized parton
  distributions}}, \bibinfo{journal}{Phys. Rept.} \bibinfo{volume}{418}
  (\bibinfo{year}{2005}) \bibinfo{pages}{1--387},
  \bibinfo{doi}{\doi{10.1016/j.physrep.2005.06.002}}, \eprint{hep-ph/0504030}.

\bibtype{Article}%
\bibitem{Boffi:2007yc}
\bibinfo{author}{S. Boffi}, \bibinfo{author}{B. Pasquini},
  \bibinfo{title}{{Generalized parton distributions and the structure of the
  nucleon}}, \bibinfo{journal}{Riv. Nuovo Cim.} \bibinfo{volume}{30}
  (\bibinfo{number}{9}) (\bibinfo{year}{2007}) \bibinfo{pages}{387--448},
  \bibinfo{doi}{\doi{10.1393/ncr/i2007-10025-7}}, \eprint{0711.2625}.

\bibtype{Article}%
\bibitem{Mezrag:2022pqk}
\bibinfo{author}{C\'edric Mezrag}, \bibinfo{title}{{An Introductory Lecture on
  Generalised Parton Distributions}}, \bibinfo{journal}{Few Body Syst.}
  \bibinfo{volume}{63} (\bibinfo{number}{3}) (\bibinfo{year}{2022})
  \bibinfo{pages}{62}, \bibinfo{doi}{\doi{10.1007/s00601-022-01765-x}},
  \eprint{2207.13584}.

\bibtype{Article}%
\bibitem{Guidal:2013rya}
\bibinfo{author}{M. Guidal}, \bibinfo{author}{H. Moutarde}, \bibinfo{author}{M.
  Vanderhaeghen}, \bibinfo{title}{{Generalized Parton Distributions in the
  valence region from Deeply Virtual Compton Scattering}},
  \bibinfo{journal}{Rept. Prog. Phys.} \bibinfo{volume}{76}
  (\bibinfo{year}{2013}) \bibinfo{pages}{066202},
  \bibinfo{doi}{\doi{10.1088/0034-4885/76/6/066202}}, \eprint{1303.6600}.

\bibtype{Article}%
\bibitem{Favart:2015umi}
\bibinfo{author}{L. Favart}, \bibinfo{author}{M. Guidal}, \bibinfo{author}{T.
  Horn}, \bibinfo{author}{P. Kroll}, \bibinfo{title}{{Deeply Virtual Meson
  Production on the nucleon}}, \bibinfo{journal}{Eur. Phys. J. A}
  \bibinfo{volume}{52} (\bibinfo{number}{6}) (\bibinfo{year}{2016})
  \bibinfo{pages}{158}, \bibinfo{doi}{\doi{10.1140/epja/i2016-16158-2}},
  \eprint{1511.04535}.

\bibtype{Article}%
\bibitem{Kumericki:2016ehc}
\bibinfo{author}{K. Kumeri\v{c}ki}, \bibinfo{author}{S. Liuti},
  \bibinfo{author}{H. Moutarde}, \bibinfo{title}{{GPD phenomenology and DVCS
  fitting}: {Entering the high-precision era}}, \bibinfo{journal}{Eur. Phys. J.
  A} \bibinfo{volume}{52} (\bibinfo{number}{6}) (\bibinfo{year}{2016})
  \bibinfo{pages}{157}, \bibinfo{doi}{\doi{10.1140/epja/i2016-16157-3}},
  \eprint{1602.02763}.

\bibtype{Article}%
\bibitem{Diehl:2023nmm}
\bibinfo{author}{S. Diehl}, \bibinfo{title}{{Experimental exploration of the 3D
  nucleon structure}}, \bibinfo{journal}{Prog. Part. Nucl. Phys.}
  \bibinfo{volume}{133} (\bibinfo{year}{2023}) \bibinfo{pages}{104069},
  \bibinfo{doi}{\doi{10.1016/j.ppnp.2023.104069}}.

\bibtype{Article}%
\bibitem{Penttinen:2000dg}
\bibinfo{author}{M. Penttinen}, \bibinfo{author}{M.~V. Polyakov},
  \bibinfo{author}{A.~G. Shuvaev}, \bibinfo{author}{M. Strikman},
  \bibinfo{title}{{DVCS amplitude in the parton model}},
  \bibinfo{journal}{Phys. Lett. B} \bibinfo{volume}{491} (\bibinfo{year}{2000})
  \bibinfo{pages}{96--100}, \bibinfo{doi}{\doi{10.1016/S0370-2693(00)01035-2}},
  \eprint{hep-ph/0006321}.

\bibtype{Article}%
\bibitem{Kiptily:2002nx}
\bibinfo{author}{D.~V. Kiptily}, \bibinfo{author}{M.~V. Polyakov},
  \bibinfo{title}{{Genuine twist three contributions to the generalized parton
  distributions from instantons}}, \bibinfo{journal}{Eur. Phys. J. C}
  \bibinfo{volume}{37} (\bibinfo{year}{2004}) \bibinfo{pages}{105--114},
  \bibinfo{doi}{\doi{10.1140/epjc/s2004-01957-3}}, \eprint{hep-ph/0212372}.

\bibtype{Article}%
\bibitem{Hatta:2012cs}
\bibinfo{author}{Y. Hatta}, \bibinfo{author}{S. Yoshida},
  \bibinfo{title}{{Twist analysis of the nucleon spin in QCD}},
  \bibinfo{journal}{JHEP} \bibinfo{volume}{10} (\bibinfo{year}{2012})
  \bibinfo{pages}{080}, \bibinfo{doi}{\doi{10.1007/JHEP10(2012)080}},
  \eprint{1207.5332}.

\bibtype{Article}%
\bibitem{Leader:2013jra}
\bibinfo{author}{E. Leader}, \bibinfo{author}{C. Lorcé}, \bibinfo{title}{{The
  angular momentum controversy: What\textquoteright{}s it all about and does it
  matter?}}, \bibinfo{journal}{Phys. Rept.} \bibinfo{volume}{541}
  (\bibinfo{number}{3}) (\bibinfo{year}{2014}) \bibinfo{pages}{163--248},
  \bibinfo{doi}{\doi{10.1016/j.physrep.2014.02.010}}, \eprint{1309.4235}.

\bibtype{Article}%
\bibitem{Ji-encyclopedia}
\bibinfo{author}{X. Ji}, \bibinfo{author}{Y. Guo}, \bibinfo{title}{{Nucleon
  spin decomposition}}, \bibinfo{journal}{Encyclopedia of Particle Physics}
  (\bibinfo{year}{2025}).

\bibtype{Article}%
\bibitem{Hagler:2004yt}
\bibinfo{author}{Ph. Hägler}, \bibinfo{title}{{Form-factor decomposition of
  generalized parton distributions at leading twist}}, \bibinfo{journal}{Phys.
  Lett. B} \bibinfo{volume}{594} (\bibinfo{year}{2004})
  \bibinfo{pages}{164--170},
  \bibinfo{doi}{\doi{10.1016/j.physletb.2004.05.014}}, \eprint{hep-ph/0404138}.

\bibtype{Article}%
\bibitem{Efremov:1979qk}
\bibinfo{author}{A.~V. Efremov}, \bibinfo{author}{A.~V. Radyushkin},
  \bibinfo{title}{{Factorization and Asymptotical Behavior of Pion Form-Factor
  in QCD}}, \bibinfo{journal}{Phys. Lett. B} \bibinfo{volume}{94}
  (\bibinfo{year}{1980}) \bibinfo{pages}{245--250},
  \bibinfo{doi}{\doi{10.1016/0370-2693(80)90869-2}}.

\bibtype{Article}%
\bibitem{Lepage:1980fj}
\bibinfo{author}{G.~P. Lepage}, \bibinfo{author}{S.~J. Brodsky},
  \bibinfo{title}{{Exclusive Processes in Perturbative Quantum
  Chromodynamics}}, \bibinfo{journal}{Phys. Rev. D} \bibinfo{volume}{22}
  (\bibinfo{year}{1980}) \bibinfo{pages}{2157},
  \bibinfo{doi}{\doi{10.1103/PhysRevD.22.2157}}.

\bibtype{Article}%
\bibitem{Lorce:2020onh}
\bibinfo{author}{C. Lorcé}, \bibinfo{title}{{Charge Distributions of Moving
  Nucleons}}, \bibinfo{journal}{Phys. Rev. Lett.} \bibinfo{volume}{125}
  (\bibinfo{number}{23}) (\bibinfo{year}{2020}) \bibinfo{pages}{232002},
  \bibinfo{doi}{\doi{10.1103/PhysRevLett.125.232002}}, \eprint{2007.05318}.

\bibtype{Inbook}%
\bibitem{Fleming:1974af}
\bibinfo{author}{G.~N. Fleming}, \bibinfo{title}{{Charge Distributions from
  Relativistic Form-Factors}} \bibinfo{year}{1974} pp.
  \bibinfo{pages}{357--374}, \bibinfo{doi}{\doi{10.1007/978-94-010-2274-3_22}}.

\bibtype{Article}%
\bibitem{Jaffe:2020ebz}
\bibinfo{author}{R.~L. Jaffe}, \bibinfo{title}{{Ambiguities in the definition
  of local spatial densities in light hadrons}}, \bibinfo{journal}{Phys. Rev.
  D} \bibinfo{volume}{103} (\bibinfo{number}{1}) (\bibinfo{year}{2021})
  \bibinfo{pages}{016017}, \bibinfo{doi}{\doi{10.1103/PhysRevD.103.016017}},
  \eprint{2010.15887}.

\bibtype{Article}%
\bibitem{Freese:2021mzg}
\bibinfo{author}{A. Freese}, \bibinfo{author}{G.~A. Miller},
  \bibinfo{title}{{Unified formalism for electromagnetic and gravitational
  probes: Densities}}, \bibinfo{journal}{Phys. Rev. D} \bibinfo{volume}{105}
  (\bibinfo{number}{1}) (\bibinfo{year}{2022}) \bibinfo{pages}{014003},
  \bibinfo{doi}{\doi{10.1103/PhysRevD.105.014003}}, \eprint{2108.03301}.

\bibtype{Article}%
\bibitem{Epelbaum:2022fjc}
\bibinfo{author}{E. Epelbaum}, \bibinfo{author}{J. Gegelia},
  \bibinfo{author}{N. Lange}, \bibinfo{author}{U.~G. Mei{\ss}ner},
  \bibinfo{author}{M.~V. Polyakov}, \bibinfo{title}{{Definition of Local
  Spatial Densities in Hadrons}}, \bibinfo{journal}{Phys. Rev. Lett.}
  \bibinfo{volume}{129} (\bibinfo{number}{1}) (\bibinfo{year}{2022})
  \bibinfo{pages}{012001}, \bibinfo{doi}{\doi{10.1103/PhysRevLett.129.012001}},
  \eprint{2201.02565}.

\bibtype{Article}%
\bibitem{Miller:2019ysh}
\bibinfo{author}{G.~A. Miller}, \bibinfo{author}{S.~J. Brodsky},
  \bibinfo{title}{{Frame-independent spatial coordinate $\tilde{z}$:
  Implications for light-front wave functions, deep inelastic scattering,
  light-front holography, and lattice QCD calculations}},
  \bibinfo{journal}{Phys. Rev. C} \bibinfo{volume}{102} (\bibinfo{number}{2})
  (\bibinfo{year}{2020}) \bibinfo{pages}{022201},
  \bibinfo{doi}{\doi{10.1103/PhysRevC.102.022201}}, \eprint{1912.08911}.

\bibtype{Article}%
\bibitem{Pire:1998nw}
\bibinfo{author}{B. Pire}, \bibinfo{author}{J. Soffer}, \bibinfo{author}{O.
  Teryaev}, \bibinfo{title}{{Positivity constraints for off - forward parton
  distributions}}, \bibinfo{journal}{Eur. Phys. J. C} \bibinfo{volume}{8}
  (\bibinfo{year}{1999}) \bibinfo{pages}{103--106},
  \bibinfo{doi}{\doi{10.1007/s100529901063}}, \eprint{hep-ph/9804284}.

\bibtype{Article}%
\bibitem{Kirch:2005in}
\bibinfo{author}{M. Kirch}, \bibinfo{author}{P.~V. Pobylitsa},
  \bibinfo{author}{K. Goeke}, \bibinfo{title}{{Inequalities for nucleon
  generalized parton distributions with helicity flip}},
  \bibinfo{journal}{Phys. Rev. D} \bibinfo{volume}{72} (\bibinfo{year}{2005})
  \bibinfo{pages}{054019}, \bibinfo{doi}{\doi{10.1103/PhysRevD.72.054019}},
  \eprint{hep-ph/0507048}.

\bibtype{Article}%
\bibitem{Pobylitsa:2002gw}
\bibinfo{author}{P.~V. Pobylitsa}, \bibinfo{title}{{Disentangling positivity
  constraints for generalized parton distributions}}, \bibinfo{journal}{Phys.
  Rev. D} \bibinfo{volume}{65} (\bibinfo{year}{2002}) \bibinfo{pages}{114015},
  \bibinfo{doi}{\doi{10.1103/PhysRevD.65.114015}}, \eprint{hep-ph/0201030}.

\bibtype{Article}%
\bibitem{Diehl:2005jf}
\bibinfo{author}{M. Diehl}, \bibinfo{author}{Ph. Hägler},
  \bibinfo{title}{{Spin densities in the transverse plane and generalized
  transversity distributions}}, \bibinfo{journal}{Eur. Phys. J. C}
  \bibinfo{volume}{44} (\bibinfo{year}{2005}) \bibinfo{pages}{87--101},
  \bibinfo{doi}{\doi{10.1140/epjc/s2005-02342-6}}, \eprint{hep-ph/0504175}.

\bibtype{Article}%
\bibitem{Lorce:2011dv}
\bibinfo{author}{C. Lorcé}, \bibinfo{author}{B. Pasquini}, \bibinfo{author}{M.
  Vanderhaeghen}, \bibinfo{title}{{Unified framework for generalized and
  transverse-momentum dependent parton distributions within a 3Q light-cone
  picture of the nucleon}}, \bibinfo{journal}{JHEP} \bibinfo{volume}{05}
  (\bibinfo{year}{2011}) \bibinfo{pages}{041},
  \bibinfo{doi}{\doi{10.1007/JHEP05(2011)041}}, \eprint{1102.4704}.

\bibtype{Article}%
\bibitem{Burkardt:2003je}
\bibinfo{author}{M. Burkardt}, \bibinfo{author}{D.~S. Hwang},
  \bibinfo{title}{{Sivers asymmetry and generalized parton distributions in
  impact parameter space}}, \bibinfo{journal}{Phys. Rev. D}
  \bibinfo{volume}{69} (\bibinfo{year}{2004}) \bibinfo{pages}{074032},
  \bibinfo{doi}{\doi{10.1103/PhysRevD.69.074032}}, \eprint{hep-ph/0309072}.

\bibtype{Article}%
\bibitem{Burkardt:2003uw}
\bibinfo{author}{M. Burkardt}, \bibinfo{title}{{Chromodynamic lensing and
  transverse single spin asymmetries}}, \bibinfo{journal}{Nucl. Phys. A}
  \bibinfo{volume}{735} (\bibinfo{year}{2004}) \bibinfo{pages}{185--199},
  \bibinfo{doi}{\doi{10.1016/j.nuclphysa.2004.02.008}},
  \eprint{hep-ph/0302144}.

\bibtype{Article}%
\bibitem{Pasquini:2019evu}
\bibinfo{author}{B. Pasquini}, \bibinfo{author}{S. Rodini}, \bibinfo{author}{A.
  Bacchetta}, \bibinfo{title}{{Revisiting model relations between T-odd
  transverse-momentum dependent parton distributions and generalized parton
  distributions}}, \bibinfo{journal}{Phys. Rev. D} \bibinfo{volume}{100}
  (\bibinfo{number}{5}) (\bibinfo{year}{2019}) \bibinfo{pages}{054039},
  \bibinfo{doi}{\doi{10.1103/PhysRevD.100.054039}}, \eprint{1907.06960}.

\bibtype{Article}%
\bibitem{Dupre:2017hfs}
\bibinfo{author}{R. Dupr\'e}, \bibinfo{author}{M. Guidal}, \bibinfo{author}{S.
  Niccolai}, \bibinfo{author}{M. Vanderhaeghen}, \bibinfo{title}{{Analysis of
  Deeply Virtual Compton Scattering Data at Jefferson Lab and Proton
  Tomography}}, \bibinfo{journal}{Eur. Phys. J. A} \bibinfo{volume}{53}
  (\bibinfo{number}{8}) (\bibinfo{year}{2017}) \bibinfo{pages}{171},
  \bibinfo{doi}{\doi{10.1140/epja/i2017-12356-8}}, \eprint{1704.07330}.

\bibtype{Article}%
\bibitem{Kobzarev:1962wt}
\bibinfo{author}{I.~Yu. Kobzarev}, \bibinfo{author}{L.~B. Okun},
  \bibinfo{title}{{Gravitational interaction of fermions}},
  \bibinfo{journal}{Zh. Eksp. Teor. Fiz.} \bibinfo{volume}{43}
  (\bibinfo{year}{1962}) \bibinfo{pages}{1904--1909}.

\bibtype{Article}%
\bibitem{Pagels:1966zza}
\bibinfo{author}{H. Pagels}, \bibinfo{title}{{Energy-Momentum Structure Form
  Factors of Particles}}, \bibinfo{journal}{Phys. Rev.} \bibinfo{volume}{144}
  (\bibinfo{year}{1966}) \bibinfo{pages}{1250--1260},
  \bibinfo{doi}{\doi{10.1103/PhysRev.144.1250}}.

\bibtype{Article}%
\bibitem{Bakker:2004ib}
\bibinfo{author}{B.~L.~G. Bakker}, \bibinfo{author}{E. Leader},
  \bibinfo{author}{T.~L. Trueman}, \bibinfo{title}{{A Critique of the angular
  momentum sum rules and a new angular momentum sum rule}},
  \bibinfo{journal}{Phys. Rev. D} \bibinfo{volume}{70} (\bibinfo{year}{2004})
  \bibinfo{pages}{114001}, \bibinfo{doi}{\doi{10.1103/PhysRevD.70.114001}},
  \eprint{hep-ph/0406139}.

\bibtype{Article}%
\bibitem{Polyakov:2018zvc}
\bibinfo{author}{M.~V. Polyakov}, \bibinfo{author}{P. Schweitzer},
  \bibinfo{title}{{Forces inside hadrons: pressure, surface tension, mechanical
  radius, and all that}}, \bibinfo{journal}{Int. J. Mod. Phys. A}
  \bibinfo{volume}{33} (\bibinfo{number}{26}) (\bibinfo{year}{2018})
  \bibinfo{pages}{1830025}, \bibinfo{doi}{\doi{10.1142/S0217751X18300259}},
  \eprint{1805.06596}.

\bibtype{Article}%
\bibitem{Lorce:2017wkb}
\bibinfo{author}{C. Lorcé}, \bibinfo{author}{L. Mantovani},
  \bibinfo{author}{B. Pasquini}, \bibinfo{title}{{Spatial distribution of
  angular momentum inside the nucleon}}, \bibinfo{journal}{Phys. Lett. B}
  \bibinfo{volume}{776} (\bibinfo{year}{2018}) \bibinfo{pages}{38--47},
  \bibinfo{doi}{\doi{10.1016/j.physletb.2017.11.018}}, \eprint{1704.08557}.

\bibtype{Article}%
\bibitem{Lorce:2018egm}
\bibinfo{author}{C. Lorcé}, \bibinfo{author}{H. Moutarde},
  \bibinfo{author}{A.~P. Trawi\'nski}, \bibinfo{title}{{Revisiting the
  mechanical properties of the nucleon}}, \bibinfo{journal}{Eur. Phys. J. C}
  \bibinfo{volume}{79} (\bibinfo{number}{1}) (\bibinfo{year}{2019})
  \bibinfo{pages}{89}, \bibinfo{doi}{\doi{10.1140/epjc/s10052-019-6572-3}},
  \eprint{1810.09837}.

\bibtype{Article}%
\bibitem{Freese:2021czn}
\bibinfo{author}{A. Freese}, \bibinfo{author}{G.~A. Miller},
  \bibinfo{title}{{Forces within hadrons on the light front}},
  \bibinfo{journal}{Phys. Rev. D} \bibinfo{volume}{103} (\bibinfo{year}{2021})
  \bibinfo{pages}{094023}, \bibinfo{doi}{\doi{10.1103/PhysRevD.103.094023}},
  \eprint{2102.01683}.

\bibtype{Book}%
\bibitem{Landau:1986aog}
\bibinfo{author}{L.~D. Landau}, \bibinfo{author}{E.~M. Lifshitz},
  \bibinfo{title}{{Theory of Elasticity}}, \bibinfo{series}{Course of
  Theoretical Physics}, \bibinfo{comment}{vol.} \bibinfo{volume}{7},
  \bibinfo{publisher}{Elsevier Butterworth-Heinemann}, \bibinfo{address}{New
  York} \bibinfo{year}{1986}, ISBN \bibinfo{isbn}{978-0-7506-2633-0},
  \bibinfo{doi}{\doi{10.1016/C2009-0-25521-8}}.

\bibtype{Article}%
\bibitem{Ji:2021mfb}
\bibinfo{author}{X.-D. Ji}, \bibinfo{author}{Y. Liu},
  \bibinfo{title}{{Momentum-Current Gravitational Multipoles of Hadrons}},
  \bibinfo{journal}{Phys. Rev. D} \bibinfo{volume}{106} (\bibinfo{number}{3})
  (\bibinfo{year}{2022}) \bibinfo{pages}{034028},
  \bibinfo{doi}{\doi{10.1103/PhysRevD.106.034028}}, \eprint{2110.14781}.

\bibtype{Article}%
\bibitem{Laue:1911lrk}
\bibinfo{author}{M. Laue}, \bibinfo{title}{{Zur Dynamik der
  Relativit{\"a}tstheorie}}, \bibinfo{journal}{Annalen Phys.}
  \bibinfo{volume}{340} (\bibinfo{number}{8}) (\bibinfo{year}{1911})
  \bibinfo{pages}{524--542}, \bibinfo{doi}{\doi{10.1002/andp.19113400808}}.

\bibtype{Article}%
\bibitem{Lorce:2021xku}
\bibinfo{author}{C. Lorcé}, \bibinfo{author}{A. Metz}, \bibinfo{author}{B.
  Pasquini}, \bibinfo{author}{S. Rodini}, \bibinfo{title}{{Energy-momentum
  tensor in QCD: nucleon mass decomposition and mechanical equilibrium}},
  \bibinfo{journal}{JHEP} \bibinfo{volume}{11} (\bibinfo{year}{2021})
  \bibinfo{pages}{121}, \bibinfo{doi}{\doi{10.1007/JHEP11(2021)121}},
  \eprint{2109.11785}.

\bibtype{Article}%
\bibitem{Lorce:2025oot}
\bibinfo{author}{C. Lorc{\'e}}, \bibinfo{author}{Peter Schweitzer},
  \bibinfo{title}{{Pressure inside hadrons: criticism, conjectures, and all
  that}}, \bibinfo{journal}{Acta Phys. Polon. B} \bibinfo{volume}{56}
  (\bibinfo{year}{2025}) \bibinfo{pages}{3--A17},
  \bibinfo{doi}{\doi{10.5506/APhysPolB.56.3-A17}}, \eprint{2501.04622}.

\bibtype{Article}%
\bibitem{Goeke:2007fp}
\bibinfo{author}{K. Goeke}, \bibinfo{author}{J. Grabis}, \bibinfo{author}{J.
  Ossmann}, \bibinfo{author}{M.~V. Polyakov}, \bibinfo{author}{P. Schweitzer},
  \bibinfo{author}{A. Silva}, \bibinfo{author}{D. Urbano},
  \bibinfo{title}{{Nucleon form-factors of the energy momentum tensor in the
  chiral quark-soliton model}}, \bibinfo{journal}{Phys. Rev. D}
  \bibinfo{volume}{75} (\bibinfo{year}{2007}) \bibinfo{pages}{094021},
  \bibinfo{doi}{\doi{10.1103/PhysRevD.75.094021}}, \eprint{hep-ph/0702030}.

\bibtype{Article}%
\bibitem{Burkert:2023wzr}
\bibinfo{author}{V.~D. Burkert}, \bibinfo{author}{L. Elouadrhiri},
  \bibinfo{author}{F.~X. Girod}, \bibinfo{author}{C. Lorcé},
  \bibinfo{author}{P. Schweitzer}, \bibinfo{author}{P.~E. Shanahan},
  \bibinfo{title}{{Colloquium: Gravitational form factors of the proton}},
  \bibinfo{journal}{Rev. Mod. Phys.} \bibinfo{volume}{95} (\bibinfo{number}{4})
  (\bibinfo{year}{2023}) \bibinfo{pages}{041002},
  \bibinfo{doi}{\doi{10.1103/RevModPhys.95.041002}}, \eprint{2303.08347}.

\bibtype{Article}%
\bibitem{Collins:1998be}
\bibinfo{author}{J.~C. Collins}, \bibinfo{author}{A. Freund},
  \bibinfo{title}{{Proof of factorization for deeply virtual Compton scattering
  in QCD}}, \bibinfo{journal}{Phys. Rev. D} \bibinfo{volume}{59}
  (\bibinfo{year}{1999}) \bibinfo{pages}{074009},
  \bibinfo{doi}{\doi{10.1103/PhysRevD.59.074009}}, \eprint{hep-ph/9801262}.

\bibtype{Article}%
\bibitem{Ji:1998xh}
\bibinfo{author}{X.-D. Ji}, \bibinfo{author}{J. Osborne}, \bibinfo{title}{{One
  loop corrections and all order factorization in deeply virtual Compton
  scattering}}, \bibinfo{journal}{Phys. Rev. D} \bibinfo{volume}{58}
  (\bibinfo{year}{1998}) \bibinfo{pages}{094018},
  \bibinfo{doi}{\doi{10.1103/PhysRevD.58.094018}}, \eprint{hep-ph/9801260}.

\bibtype{Article}%
\bibitem{Collins:1999un}
\bibinfo{author}{J.~C. Collins}, \bibinfo{author}{M. Diehl},
  \bibinfo{title}{{Transversity distribution does not contribute to hard
  exclusive electroproduction of mesons}}, \bibinfo{journal}{Phys. Rev. D}
  \bibinfo{volume}{61} (\bibinfo{year}{2000}) \bibinfo{pages}{114015},
  \bibinfo{doi}{\doi{10.1103/PhysRevD.61.114015}}, \eprint{hep-ph/9907498}.

\bibtype{Article}%
\bibitem{Goloskokov:2009ia}
\bibinfo{author}{S.~V. Goloskokov}, \bibinfo{author}{P. Kroll},
  \bibinfo{title}{{An Attempt to understand exclusive $\pi^+$
  electroproduction}}, \bibinfo{journal}{Eur. Phys. J. C} \bibinfo{volume}{65}
  (\bibinfo{year}{2010}) \bibinfo{pages}{137--151},
  \bibinfo{doi}{\doi{10.1140/epjc/s10052-009-1178-9}}, \eprint{0906.0460}.

\bibtype{Article}%
\bibitem{Ahmad:2008hp}
\bibinfo{author}{S. Ahmad}, \bibinfo{author}{G.~R. Goldstein},
  \bibinfo{author}{S. Liuti}, \bibinfo{title}{{Nucleon Tensor Charge from
  Exclusive $\pi^0$ Electroproduction}}, \bibinfo{journal}{Phys. Rev. D}
  \bibinfo{volume}{79} (\bibinfo{year}{2009}) \bibinfo{pages}{054014},
  \bibinfo{doi}{\doi{10.1103/PhysRevD.79.054014}}, \eprint{0805.3568}.

\bibtype{Article}%
\bibitem{Ivanov:2002jj}
\bibinfo{author}{D.~Yu. Ivanov}, \bibinfo{author}{B. Pire}, \bibinfo{author}{L.
  Szymanowski}, \bibinfo{author}{O.~V. Teryaev}, \bibinfo{title}{{Probing
  chiral odd GPD's in diffractive electroproduction of two vector mesons}},
  \bibinfo{journal}{Phys. Lett. B} \bibinfo{volume}{550} (\bibinfo{year}{2002})
  \bibinfo{pages}{65--76}, \bibinfo{doi}{\doi{10.1016/S0370-2693(02)02856-3}},
  \eprint{hep-ph/0209300}.

\bibtype{Article}%
\bibitem{Bertone:2021yyz}
\bibinfo{author}{V. Bertone}, \bibinfo{author}{H. Dutrieux},
  \bibinfo{author}{C. Mezrag}, \bibinfo{author}{H. Moutarde},
  \bibinfo{author}{P. Sznajder}, \bibinfo{title}{{Deconvolution problem of
  deeply virtual Compton scattering}}, \bibinfo{journal}{Phys. Rev. D}
  \bibinfo{volume}{103} (\bibinfo{number}{11}) (\bibinfo{year}{2021})
  \bibinfo{pages}{114019}, \bibinfo{doi}{\doi{10.1103/PhysRevD.103.114019}},
  \eprint{2104.03836}.

\bibtype{Article}%
\bibitem{Moffat:2023svr}
\bibinfo{author}{E. Moffat}, \bibinfo{author}{A. Freese}, \bibinfo{author}{I.
  Clo\"et}, \bibinfo{author}{T. Donohoe}, \bibinfo{author}{L. Gamberg},
  \bibinfo{author}{W. Melnitchouk}, \bibinfo{author}{A. Metz},
  \bibinfo{author}{A. Prokudin}, \bibinfo{author}{N. Sato},
  \bibinfo{title}{{Shedding light on shadow generalized parton distributions}},
  \bibinfo{journal}{Phys. Rev. D} \bibinfo{volume}{108} (\bibinfo{number}{3})
  (\bibinfo{year}{2023}) \bibinfo{pages}{036027},
  \bibinfo{doi}{\doi{10.1103/PhysRevD.108.036027}}, \eprint{2303.12006}.

\bibtype{Article}%
\bibitem{Berger:2001xd}
\bibinfo{author}{E.~R. Berger}, \bibinfo{author}{M. Diehl}, \bibinfo{author}{B.
  Pire}, \bibinfo{title}{{Time - like Compton scattering: Exclusive
  photoproduction of lepton pairs}}, \bibinfo{journal}{Eur. Phys. J. C}
  \bibinfo{volume}{23} (\bibinfo{year}{2002}) \bibinfo{pages}{675--689},
  \bibinfo{doi}{\doi{10.1007/s100520200917}}, \eprint{hep-ph/0110062}.

\bibtype{Article}%
\bibitem{Guidal:2002kt}
\bibinfo{author}{M. Guidal}, \bibinfo{author}{M. Vanderhaeghen},
  \bibinfo{title}{{Double deeply virtual Compton scattering off the nucleon}},
  \bibinfo{journal}{Phys. Rev. Lett.} \bibinfo{volume}{90}
  (\bibinfo{year}{2003}) \bibinfo{pages}{012001},
  \bibinfo{doi}{\doi{10.1103/PhysRevLett.90.012001}}, \eprint{hep-ph/0208275}.

\bibtype{Article}%
\bibitem{Belitsky:2002tf}
\bibinfo{author}{{Belitsky, A. V. and M\"uller, D.}},
  \bibinfo{title}{{Exclusive electroproduction of lepton pairs as a probe of
  nucleon structure}}, \bibinfo{journal}{Phys. Rev. Lett.} \bibinfo{volume}{90}
  (\bibinfo{year}{2003}) \bibinfo{pages}{022001},
  \bibinfo{doi}{\doi{10.1103/PhysRevLett.90.022001}}, \eprint{hep-ph/0210313}.

\bibtype{Article}%
\bibitem{CLAS:2021lky}
\bibinfo{author}{P. Chatagnon}, et al. (\bibinfo{collaboration}{CLAS}),
  \bibinfo{title}{{First Measurement of Timelike Compton Scattering}},
  \bibinfo{journal}{Phys. Rev. Lett.} \bibinfo{volume}{127}
  (\bibinfo{number}{26}) (\bibinfo{year}{2021}) \bibinfo{pages}{262501},
  \bibinfo{doi}{\doi{10.1103/PhysRevLett.127.262501}}, \eprint{2108.11746}.

\bibtype{Article}%
\bibitem{Alvarado:2025huq}
\bibinfo{author}{J.~S. Alvarado}, \bibinfo{author}{M. Hoballah},
  \bibinfo{author}{E. Voutier}, \bibinfo{title}{{Sensitivity of double deeply
  virtual Compton-scattering observables to generalized parton distributions}},
  \bibinfo{journal}{Phys. Rev. C} \bibinfo{volume}{111} (\bibinfo{number}{6})
  (\bibinfo{year}{2025}) \bibinfo{pages}{065205},
  \bibinfo{doi}{\doi{10.1103/PhysRevC.111.065205}}, \eprint{2502.02346}.

\bibtype{Article}%
\bibitem{Duplancic:2018bum}
\bibinfo{author}{G. Duplan\v{c}i\'c}, \bibinfo{author}{K.
  Passek-Kumeri\v{c}ki}, \bibinfo{author}{B. Pire}, \bibinfo{author}{L.
  Szymanowski}, \bibinfo{author}{S. Wallon}, \bibinfo{title}{{Probing axial
  quark generalized parton distributions through exclusive photoproduction of a
  $\gamma\,\pi^\pm$ pair with a large invariant mass}}, \bibinfo{journal}{JHEP}
  \bibinfo{volume}{11} (\bibinfo{year}{2018}) \bibinfo{pages}{179},
  \bibinfo{doi}{\doi{10.1007/JHEP11(2018)179}}, \eprint{1809.08104}.

\bibtype{Article}%
\bibitem{Qiu:2022bpq}
\bibinfo{author}{J.-W. Qiu}, \bibinfo{author}{Z. Yu},
  \bibinfo{title}{{Exclusive production of a pair of high transverse momentum
  photons in pion-nucleon collisions for extracting generalized parton
  distributions}}, \bibinfo{journal}{JHEP} \bibinfo{volume}{08}
  (\bibinfo{year}{2022}) \bibinfo{pages}{103},
  \bibinfo{doi}{\doi{10.1007/JHEP08(2022)103}}, \eprint{2205.07846}.

\bibtype{Article}%
\bibitem{Qiu:2022pla}
\bibinfo{author}{J.-W. Qiu}, \bibinfo{author}{Z. Yu}, \bibinfo{title}{{Single
  diffractive hard exclusive processes for the study of generalized parton
  distributions}}, \bibinfo{journal}{Phys. Rev. D} \bibinfo{volume}{107}
  (\bibinfo{number}{1}) (\bibinfo{year}{2023}) \bibinfo{pages}{014007},
  \bibinfo{doi}{\doi{10.1103/PhysRevD.107.014007}}, \eprint{2210.07995}.

\bibtype{Article}%
\bibitem{Siddikov:2022bku}
\bibinfo{author}{M. Siddikov}, \bibinfo{author}{I. Schmidt},
  \bibinfo{title}{{Exclusive production of quarkonia pairs in collinear
  factorization framework}}, \bibinfo{journal}{Phys. Rev. D}
  \bibinfo{volume}{107} (\bibinfo{number}{3}) (\bibinfo{year}{2023})
  \bibinfo{pages}{034037}, \bibinfo{doi}{\doi{10.1103/PhysRevD.107.034037}},
  \eprint{2212.14019}.

\bibtype{Article}%
\bibitem{Musatov:1999xp}
\bibinfo{author}{I.~V. Musatov}, \bibinfo{author}{A.~V. Radyushkin},
  \bibinfo{title}{{Evolution and models for skewed parton distributions}},
  \bibinfo{journal}{Phys. Rev. D} \bibinfo{volume}{61} (\bibinfo{year}{2000})
  \bibinfo{pages}{074027}, \bibinfo{doi}{\doi{10.1103/PhysRevD.61.074027}},
  \eprint{hep-ph/9905376}.

\bibtype{Article}%
\bibitem{Vanderhaeghen:1999xj}
\bibinfo{author}{M. Vanderhaeghen}, \bibinfo{author}{P.~A.~M. Guichon},
  \bibinfo{author}{M. Guidal}, \bibinfo{title}{{Deeply virtual
  electroproduction of photons and mesons on the nucleon: Leading order
  amplitudes and power corrections}}, \bibinfo{journal}{Phys. Rev. D}
  \bibinfo{volume}{60} (\bibinfo{year}{1999}) \bibinfo{pages}{094017},
  \bibinfo{doi}{\doi{10.1103/PhysRevD.60.094017}}, \eprint{hep-ph/9905372}.

\bibtype{Article}%
\bibitem{Mueller:2005ed}
\bibinfo{author}{{M\"uller, D. and Sch{\"a}fer, A.}}, \bibinfo{title}{{Complex
  conformal spin partial wave expansion of generalized parton distributions and
  distribution amplitudes}}, \bibinfo{journal}{Nucl. Phys. B}
  \bibinfo{volume}{739} (\bibinfo{year}{2006}) \bibinfo{pages}{1--59},
  \bibinfo{doi}{\doi{10.1016/j.nuclphysb.2006.01.019}},
  \eprint{hep-ph/0509204}.

\bibtype{Article}%
\bibitem{Kumericki:2013br}
\bibinfo{author}{K. Kumeri{\v{c}}ki}, \bibinfo{author}{D. M\"uller},
  \bibinfo{author}{M. Murray}, \bibinfo{title}{{HERMES impact for the access of
  Compton form factors}}, \bibinfo{journal}{Phys. Part. Nucl.}
  \bibinfo{volume}{45} (\bibinfo{number}{4}) (\bibinfo{year}{2014})
  \bibinfo{pages}{723--755}, \bibinfo{doi}{\doi{10.1134/S1063779614040108}},
  \eprint{1301.1230}.

\bibtype{Article}%
\bibitem{Guo:2023ahv}
\bibinfo{author}{Y. Guo}, \bibinfo{author}{X.-D. Ji}, \bibinfo{author}{M.~G.
  Santiago}, \bibinfo{author}{K. Shiells}, \bibinfo{author}{J. Yang},
  \bibinfo{title}{{Generalized parton distributions through universal moment
  parameterization: non-zero skewness case}}, \bibinfo{journal}{JHEP}
  \bibinfo{volume}{05} (\bibinfo{year}{2023}) \bibinfo{pages}{150},
  \bibinfo{doi}{\doi{10.1007/JHEP05(2023)150}}, \eprint{2302.07279}.

\bibtype{Article}%
\bibitem{Moutarde:2019tqa}
\bibinfo{author}{H. Moutarde}, \bibinfo{author}{P. Sznajder},
  \bibinfo{author}{J. Wagner}, \bibinfo{title}{{Unbiased determination of DVCS
  Compton Form Factors}}, \bibinfo{journal}{Eur. Phys. J. C}
  \bibinfo{volume}{79} (\bibinfo{number}{7}) (\bibinfo{year}{2019})
  \bibinfo{pages}{614}, \bibinfo{doi}{\doi{10.1140/epjc/s10052-019-7117-5}},
  \eprint{1905.02089}.

\bibtype{Article}%
\bibitem{Kumericki:2009uq}
\bibinfo{author}{{Kumeri{\v{c}}ki, K. and M\"uller, D.}},
  \bibinfo{title}{{Deeply virtual Compton scattering at small $x_B$ and the
  access to the GPD H}}, \bibinfo{journal}{Nucl. Phys. B} \bibinfo{volume}{841}
  (\bibinfo{year}{2010}) \bibinfo{pages}{1--58},
  \bibinfo{doi}{\doi{10.1016/j.nuclphysb.2010.07.015}}, \eprint{0904.0458}.

\bibtype{Article}%
\bibitem{Kumericki:2007sa}
\bibinfo{author}{{Kumeri\v{c}ki, K. and M\"uller, D. and Passek-Kumeri\v{c}ki,
  K.}}, \bibinfo{title}{{Towards a fitting procedure for deeply virtual Compton
  scattering at next-to-leading order and beyond}}, \bibinfo{journal}{Nucl.
  Phys. B} \bibinfo{volume}{794} (\bibinfo{year}{2008})
  \bibinfo{pages}{244--323},
  \bibinfo{doi}{\doi{10.1016/j.nuclphysb.2007.10.029}},
  \eprint{hep-ph/0703179}.

\bibtype{Article}%
\bibitem{Diehl:2007jb}
\bibinfo{author}{M. Diehl}, \bibinfo{author}{D.~Yu. Ivanov},
  \bibinfo{title}{{Dispersion representations for hard exclusive processes:
  beyond the Born approximation}}, \bibinfo{journal}{Eur. Phys. J. C}
  \bibinfo{volume}{52} (\bibinfo{year}{2007}) \bibinfo{pages}{919--932},
  \bibinfo{doi}{\doi{10.1140/epjc/s10052-007-0401-9}}, \eprint{0707.0351}.

\bibtype{Article}%
\bibitem{Goldstein:2010gu}
\bibinfo{author}{G.~R. Goldstein}, \bibinfo{author}{J.~O. Gonzalez~Hernandez},
  \bibinfo{author}{S. Liuti}, \bibinfo{title}{{Flexible Parametrization of
  Generalized Parton Distributions from Deeply Virtual Compton Scattering
  Observables}}, \bibinfo{journal}{Phys. Rev. D} \bibinfo{volume}{84}
  (\bibinfo{year}{2011}) \bibinfo{pages}{034007},
  \bibinfo{doi}{\doi{10.1103/PhysRevD.84.034007}}, \eprint{1012.3776}.

\bibtype{Article}%
\bibitem{Goldstein:2013gra}
\bibinfo{author}{G.~R. Goldstein}, \bibinfo{author}{J.~O. Gonzalez~Hernandez},
  \bibinfo{author}{S. Liuti}, \bibinfo{title}{{Flexible Parametrization of
  Generalized Parton Distributions: The Chiral-Odd Sector}},
  \bibinfo{journal}{Phys. Rev. D} \bibinfo{volume}{91} (\bibinfo{number}{11})
  (\bibinfo{year}{2015}) \bibinfo{pages}{114013},
  \bibinfo{doi}{\doi{10.1103/PhysRevD.91.114013}}, \eprint{1311.0483}.

\bibtype{Article}%
\bibitem{Berthou:2015oaw}
\bibinfo{author}{B. Berthou}, et al., \bibinfo{title}{{PARTONS: PARtonic
  Tomography Of Nucleon Software}: {A computing framework for the phenomenology
  of Generalized Parton Distributions}}, \bibinfo{journal}{Eur. Phys. J. C}
  \bibinfo{volume}{78} (\bibinfo{number}{6}) (\bibinfo{year}{2018})
  \bibinfo{pages}{478}, \bibinfo{doi}{\doi{10.1140/epjc/s10052-018-5948-0}},
  \eprint{1512.06174}.

\bibtype{Misc}%
\bibitem{gepard}
\bibinfo{author}{{K. Kumerički}}, \bibinfo{title}{Gepard: Tool for studying
  the 3d quark and gluon distributions in the nucleon},
  \bibinfo{howpublished}{\url{https://gepard.phy.hr/credits.html}}
  \bibinfo{year}{2006}.

\bibtype{Article}%
\bibitem{Cuic:2020iwt}
\bibinfo{author}{M. \v{C}ui\'c}, \bibinfo{author}{K. Kumeri\v{c}ki},
  \bibinfo{author}{A. Schäfer}, \bibinfo{title}{{Separation of Quark Flavors
  Using Deeply Virtual Compton Scattering Data}}, \bibinfo{journal}{Phys. Rev.
  Lett.} \bibinfo{volume}{125} (\bibinfo{number}{23}) (\bibinfo{year}{2020})
  \bibinfo{pages}{232005}, \bibinfo{doi}{\doi{10.1103/PhysRevLett.125.232005}},
  \eprint{2007.00029}.

\bibtype{Inproceedings}%
\bibitem{Teryaev:2005uj}
\bibinfo{author}{O.~V. Teryaev}, \bibinfo{title}{{Analytic properties of hard
  exclusive amplitudes}}, in: \bibinfo{booktitle}{{11th International
  Conference on Elastic and Diffractive Scattering: Towards High Energy
  Frontiers: The 20th Anniversary of the Blois Workshops, 17th Rencontre de
  Blois}} \bibinfo{year}{2005}, \eprint{hep-ph/0510031}.

\bibtype{Article}%
\bibitem{Anikin:2007yh}
\bibinfo{author}{I.~V. Anikin}, \bibinfo{author}{O.~V. Teryaev},
  \bibinfo{title}{{Dispersion relations and subtractions in hard exclusive
  processes}}, \bibinfo{journal}{Phys. Rev. D} \bibinfo{volume}{76}
  (\bibinfo{year}{2007}) \bibinfo{pages}{056007},
  \bibinfo{doi}{\doi{10.1103/PhysRevD.76.056007}}, \eprint{0704.2185}.

\bibtype{Article}%
\bibitem{Polyakov:1999gs}
\bibinfo{author}{M.~V. Polyakov}, \bibinfo{author}{C. Weiss},
  \bibinfo{title}{{Skewed and double distributions in pion and nucleon}},
  \bibinfo{journal}{Phys. Rev. D} \bibinfo{volume}{60} (\bibinfo{year}{1999})
  \bibinfo{pages}{114017}, \bibinfo{doi}{\doi{10.1103/PhysRevD.60.114017}},
  \eprint{hep-ph/9902451}.

\bibtype{Article}%
\bibitem{Burkert:2018bqq}
\bibinfo{author}{V.~D. Burkert}, \bibinfo{author}{L. Elouadrhiri},
  \bibinfo{author}{F.~X. Girod}, \bibinfo{title}{{The pressure distribution
  inside the proton}}, \bibinfo{journal}{Nature} \bibinfo{volume}{557}
  (\bibinfo{number}{7705}) (\bibinfo{year}{2018}) \bibinfo{pages}{396--399},
  \bibinfo{doi}{\doi{10.1038/s41586-018-0060-z}}.

\bibtype{Article}%
\bibitem{Kumericki:2019ddg}
\bibinfo{author}{K. Kumeri{\v{c}}ki}, \bibinfo{title}{{Measurability of
  pressure inside the proton}}, \bibinfo{journal}{Nature} \bibinfo{volume}{570}
  (\bibinfo{number}{7759}) (\bibinfo{year}{2019}) \bibinfo{pages}{E1--E2},
  \bibinfo{doi}{\doi{10.1038/s41586-019-1211-6}}.

\bibtype{Article}%
\bibitem{Kharzeev:2021qkd}
\bibinfo{author}{D.~E. Kharzeev}, \bibinfo{title}{{Mass radius of the proton}},
  \bibinfo{journal}{Phys. Rev. D} \bibinfo{volume}{104} (\bibinfo{number}{5})
  (\bibinfo{year}{2021}) \bibinfo{pages}{054015},
  \bibinfo{doi}{\doi{10.1103/PhysRevD.104.054015}}, \eprint{2102.00110}.

\bibtype{Article}%
\bibitem{Guo:2021ibg}
\bibinfo{author}{Y. Guo}, \bibinfo{author}{X.-D. Ji}, \bibinfo{author}{Y. Liu},
  \bibinfo{title}{{QCD Analysis of Near-Threshold Photon-Proton Production of
  Heavy Quarkonium}}, \bibinfo{journal}{Phys. Rev. D} \bibinfo{volume}{103}
  (\bibinfo{number}{9}) (\bibinfo{year}{2021}) \bibinfo{pages}{096010},
  \bibinfo{doi}{\doi{10.1103/PhysRevD.103.096010}}, \eprint{2103.11506}.

\bibtype{Article}%
\bibitem{Sun:2021gmi}
\bibinfo{author}{P. Sun}, \bibinfo{author}{X.-B. Tong}, \bibinfo{author}{F.
  Yuan}, \bibinfo{title}{{Perturbative QCD analysis of near threshold heavy
  quarkonium photoproduction at large momentum transfer}},
  \bibinfo{journal}{Phys. Lett. B} \bibinfo{volume}{822} (\bibinfo{year}{2021})
  \bibinfo{pages}{136655}, \bibinfo{doi}{\doi{10.1016/j.physletb.2021.136655}},
  \eprint{2103.12047}.

\bibtype{Article}%
\bibitem{Du:2020bqj}
\bibinfo{author}{M.-L. Du}, \bibinfo{author}{V. Baru}, \bibinfo{author}{F.-K.
  Guo}, \bibinfo{author}{C. Hanhart}, \bibinfo{author}{U.-G. Mei{\ss}ner},
  \bibinfo{author}{A. Nefediev}, \bibinfo{author}{I. Strakovsky},
  \bibinfo{title}{{Deciphering the mechanism of near-threshold $J/\psi$
  photoproduction}}, \bibinfo{journal}{Eur. Phys. J. C} \bibinfo{volume}{80}
  (\bibinfo{number}{11}) (\bibinfo{year}{2020}) \bibinfo{pages}{1053},
  \bibinfo{doi}{\doi{10.1140/epjc/s10052-020-08620-5}}, \eprint{2009.08345}.

\bibtype{Article}%
\bibitem{Duran:2022xag}
\bibinfo{author}{B. Duran}, et al., \bibinfo{title}{{Determining the gluonic
  gravitational form factors of the proton}}, \bibinfo{journal}{Nature}
  \bibinfo{volume}{615} (\bibinfo{number}{7954}) (\bibinfo{year}{2023})
  \bibinfo{pages}{813--816}, \bibinfo{doi}{\doi{10.1038/s41586-023-05730-4}},
  \eprint{2207.05212}.

\bibtype{Article}%
\bibitem{Dupre:2015jha}
\bibinfo{author}{R. Dupr\'e}, \bibinfo{author}{S. Scopetta},
  \bibinfo{title}{{3D Structure and Nuclear Targets}}, \bibinfo{journal}{Eur.
  Phys. J. A} \bibinfo{volume}{52} (\bibinfo{number}{6}) (\bibinfo{year}{2016})
  \bibinfo{pages}{159}, \bibinfo{doi}{\doi{10.1140/epja/i2016-16159-1}},
  \eprint{1510.00794}.

\bibtype{Article}%
\bibitem{HERMES:2009xsg}
\bibinfo{author}{A. Airapetian}, et al. (\bibinfo{collaboration}{HERMES}),
  \bibinfo{title}{{Nuclear-mass dependence of azimuthal beam-helicity and
  beam-charge asymmetries in deeply virtual Compton scattering}},
  \bibinfo{journal}{Phys. Rev. C} \bibinfo{volume}{81} (\bibinfo{year}{2010})
  \bibinfo{pages}{035202}, \bibinfo{doi}{\doi{10.1103/PhysRevC.81.035202}},
  \eprint{0911.0091}.

\bibtype{Article}%
\bibitem{CLAS:2017udk}
\bibinfo{author}{M. Hattawy}, et al. (\bibinfo{collaboration}{CLAS}),
  \bibinfo{title}{{First Exclusive Measurement of Deeply Virtual Compton
  Scattering off $^4$He: Toward the 3D Tomography of Nuclei}},
  \bibinfo{journal}{Phys. Rev. Lett.} \bibinfo{volume}{119}
  (\bibinfo{number}{20}) (\bibinfo{year}{2017}) \bibinfo{pages}{202004},
  \bibinfo{doi}{\doi{10.1103/PhysRevLett.119.202004}}, \eprint{1707.03361}.

\bibtype{Article}%
\bibitem{CLAS:2018ddh}
\bibinfo{author}{M. Hattawy}, et al. (\bibinfo{collaboration}{CLAS}),
  \bibinfo{title}{{Exploring the Structure of the Bound Proton with Deeply
  Virtual Compton Scattering}}, \bibinfo{journal}{Phys. Rev. Lett.}
  \bibinfo{volume}{123} (\bibinfo{number}{3}) (\bibinfo{year}{2019})
  \bibinfo{pages}{032502}, \bibinfo{doi}{\doi{10.1103/PhysRevLett.123.032502}},
  \eprint{1812.07628}.

\bibtype{Article}%
\bibitem{Chavez:2021koz}
\bibinfo{author}{J.~M.~M. Ch\'avez}, \bibinfo{author}{V. Bertone},
  \bibinfo{author}{F. De~Soto~Borrero}, \bibinfo{author}{M. Defurne},
  \bibinfo{author}{C. Mezrag}, \bibinfo{author}{H. Moutarde},
  \bibinfo{author}{J. Rodr\'\i{}guez-Quintero}, \bibinfo{author}{J. Segovia},
  \bibinfo{title}{{Accessing the Pion 3D Structure at US and China Electron-Ion
  Colliders}}, \bibinfo{journal}{Phys. Rev. Lett.} \bibinfo{volume}{128}
  (\bibinfo{number}{20}) (\bibinfo{year}{2022}) \bibinfo{pages}{202501},
  \bibinfo{doi}{\doi{10.1103/PhysRevLett.128.202501}}, \eprint{2110.09462}.

\bibtype{Article}%
\bibitem{Hatta:2025ryj}
\bibinfo{author}{Y. Hatta}, \bibinfo{author}{J. Schoenleber},
  \bibinfo{title}{{Sullivan Process near Threshold and the Pion Gravitational
  Form Factors}}, \bibinfo{journal}{Phys. Rev. Lett.} \bibinfo{volume}{134}
  (\bibinfo{number}{25}) (\bibinfo{year}{2025}) \bibinfo{pages}{251901},
  \bibinfo{doi}{\doi{10.1103/y9fq-y84c}}, \eprint{2502.12061}.

\bibtype{Article}%
\bibitem{Diehl:1998dk}
\bibinfo{author}{M. Diehl}, \bibinfo{author}{T. Gousset}, \bibinfo{author}{B.
  Pire}, \bibinfo{author}{O. Teryaev}, \bibinfo{title}{{Probing partonic
  structure in $\gamma^*\gamma \rightarrow \pi \pi $ near threshold}},
  \bibinfo{journal}{Phys. Rev. Lett.} \bibinfo{volume}{81}
  (\bibinfo{year}{1998}) \bibinfo{pages}{1782--1785},
  \bibinfo{doi}{\doi{10.1103/PhysRevLett.81.1782}}, \eprint{hep-ph/9805380}.

\bibtype{Article}%
\bibitem{Kumano:2017lhr}
\bibinfo{author}{S. Kumano}, \bibinfo{author}{Q.-T. Song},
  \bibinfo{author}{O.~V. Teryaev}, \bibinfo{title}{{Hadron tomography by
  generalized distribution amplitudes in pion-pair production process $\gamma^*
  \gamma \rightarrow \pi^0 \pi^0 $ and gravitational form factors for pion}},
  \bibinfo{journal}{Phys. Rev. D} \bibinfo{volume}{97} (\bibinfo{number}{1})
  (\bibinfo{year}{2018}) \bibinfo{pages}{014020},
  \bibinfo{doi}{\doi{10.1103/PhysRevD.97.014020}}, \eprint{1711.08088}.

\bibtype{Article}%
\bibitem{Diehl:2024bmd}
\bibinfo{author}{S. Diehl}, et al., \bibinfo{title}{{Exploring baryon
  resonances with transition generalized parton distributions: status and
  perspectives}}, \bibinfo{journal}{Eur. Phys. J. A} \bibinfo{volume}{61}
  (\bibinfo{number}{6}) (\bibinfo{year}{2025}) \bibinfo{pages}{131},
  \bibinfo{doi}{\doi{10.1140/epja/s10050-025-01552-2}}, \eprint{2405.15386}.

\bibtype{Article}%
\bibitem{Lorce:2013pza}
\bibinfo{author}{C. Lorcé}, \bibinfo{author}{B. Pasquini},
  \bibinfo{title}{{Structure analysis of the generalized correlator of quark
  and gluon for a spin-1/2 target}}, \bibinfo{journal}{JHEP}
  \bibinfo{volume}{09} (\bibinfo{year}{2013}) \bibinfo{pages}{138},
  \bibinfo{doi}{\doi{10.1007/JHEP09(2013)138}}, \eprint{1307.4497}.

\bibtype{Article}%
\bibitem{Echevarria:2016mrc}
\bibinfo{author}{M.~G. Echevarria}, \bibinfo{author}{A. Idilbi},
  \bibinfo{author}{K. Kanazawa}, \bibinfo{author}{C. Lorcé},
  \bibinfo{author}{A. Metz}, \bibinfo{author}{B. Pasquini}, \bibinfo{author}{M.
  Schlegel}, \bibinfo{title}{{Proper definition and evolution of generalized
  transverse momentum dependent distributions}}, \bibinfo{journal}{Phys. Lett.
  B} \bibinfo{volume}{759} (\bibinfo{year}{2016}) \bibinfo{pages}{336--341},
  \bibinfo{doi}{\doi{10.1016/j.physletb.2016.05.086}}, \eprint{1602.06953}.

\bibtype{Article}%
\bibitem{Bertone:2025vgy}
\bibinfo{author}{V. Bertone}, \bibinfo{author}{M.~G. Echevarria},
  \bibinfo{author}{O. del Rio}, \bibinfo{author}{S. Rodini},
  \bibinfo{title}{{One-loop matching for leading-twist generalised
  transverse-momentum-dependent distributions}}, \bibinfo{journal}{JHEP}
  \bibinfo{volume}{05} (\bibinfo{year}{2025}) \bibinfo{pages}{183},
  \bibinfo{doi}{\doi{10.1007/JHEP05(2025)183}}, \eprint{2502.07576}.

\bibtype{Article}%
\bibitem{Kanazawa:2014nha}
\bibinfo{author}{K. Kanazawa}, \bibinfo{author}{C. Lorcé}, \bibinfo{author}{A.
  Metz}, \bibinfo{author}{B. Pasquini}, \bibinfo{author}{M. Schlegel},
  \bibinfo{title}{{Twist-2 generalized transverse-momentum dependent parton
  distributions and the spin/orbital structure of the nucleon}},
  \bibinfo{journal}{Phys. Rev. D} \bibinfo{volume}{90} (\bibinfo{number}{1})
  (\bibinfo{year}{2014}) \bibinfo{pages}{014028},
  \bibinfo{doi}{\doi{10.1103/PhysRevD.90.014028}}, \eprint{1403.5226}.

\bibtype{Article}%
\bibitem{Hatta:2016dxp}
\bibinfo{author}{Y. Hatta}, \bibinfo{author}{B.-W. Xiao}, \bibinfo{author}{F.
  Yuan}, \bibinfo{title}{{Probing the Small- x Gluon Tomography in Correlated
  Hard Diffractive Dijet Production in Deep Inelastic Scattering}},
  \bibinfo{journal}{Phys. Rev. Lett.} \bibinfo{volume}{116}
  (\bibinfo{number}{20}) (\bibinfo{year}{2016}) \bibinfo{pages}{202301},
  \bibinfo{doi}{\doi{10.1103/PhysRevLett.116.202301}}, \eprint{1601.01585}.

\bibtype{Article}%
\bibitem{Altinoluk:2015dpi}
\bibinfo{author}{T. Altinoluk}, \bibinfo{author}{N. Armesto},
  \bibinfo{author}{G. Beuf}, \bibinfo{author}{A.~H. Rezaeian},
  \bibinfo{title}{{Diffractive Dijet Production in Deep Inelastic Scattering
  and Photon-Hadron Collisions in the Color Glass Condensate}},
  \bibinfo{journal}{Phys. Lett. B} \bibinfo{volume}{758} (\bibinfo{year}{2016})
  \bibinfo{pages}{373--383},
  \bibinfo{doi}{\doi{10.1016/j.physletb.2016.05.032}}, \eprint{1511.07452}.

\bibtype{Article}%
\bibitem{Zhou:2016rnt}
\bibinfo{author}{J. Zhou}, \bibinfo{title}{{Elliptic gluon generalized
  transverse-momentum-dependent distribution inside a large nucleus}},
  \bibinfo{journal}{Phys. Rev. D} \bibinfo{volume}{94} (\bibinfo{number}{11})
  (\bibinfo{year}{2016}) \bibinfo{pages}{114017},
  \bibinfo{doi}{\doi{10.1103/PhysRevD.94.114017}}, \eprint{1611.02397}.

\bibtype{Article}%
\bibitem{Ji:2016jgn}
\bibinfo{author}{X.-D. Ji}, \bibinfo{author}{F. Yuan}, \bibinfo{author}{Y.
  Zhao}, \bibinfo{title}{{Hunting the Gluon Orbital Angular Momentum at the
  Electron-Ion Collider}}, \bibinfo{journal}{Phys. Rev. Lett.}
  \bibinfo{volume}{118} (\bibinfo{number}{19}) (\bibinfo{year}{2017})
  \bibinfo{pages}{192004}, \bibinfo{doi}{\doi{10.1103/PhysRevLett.118.192004}},
  \eprint{1612.02438}.

\bibtype{Article}%
\bibitem{Hagiwara:2017ofm}
\bibinfo{author}{Y. Hagiwara}, \bibinfo{author}{Y. Hatta},
  \bibinfo{author}{B.-W. Xiao}, \bibinfo{author}{F. Yuan},
  \bibinfo{title}{{Elliptic Flow in Small Systems due to Elliptic Gluon
  Distributions?}}, \bibinfo{journal}{Phys. Lett. B} \bibinfo{volume}{771}
  (\bibinfo{year}{2017}) \bibinfo{pages}{374--378},
  \bibinfo{doi}{\doi{10.1016/j.physletb.2017.05.083}}, \eprint{1701.04254}.

\bibtype{Article}%
\bibitem{Boer:2018vdi}
\bibinfo{author}{D. Boer}, \bibinfo{author}{T. Van~Daal},
  \bibinfo{author}{P.~J. Mulders}, \bibinfo{author}{E. Petreska},
  \bibinfo{title}{{Directed flow from C-odd gluon correlations at small $x$}},
  \bibinfo{journal}{JHEP} \bibinfo{volume}{07} (\bibinfo{year}{2018})
  \bibinfo{pages}{140}, \bibinfo{doi}{\doi{10.1007/JHEP07(2018)140}},
  \eprint{1805.05219}.

\bibtype{Article}%
\bibitem{Salazar:2019ncp}
\bibinfo{author}{F. Salazar}, \bibinfo{author}{B. Schenke},
  \bibinfo{title}{{Diffractive dijet production in impact parameter dependent
  saturation models}}, \bibinfo{journal}{Phys. Rev. D} \bibinfo{volume}{100}
  (\bibinfo{number}{3}) (\bibinfo{year}{2019}) \bibinfo{pages}{034007},
  \bibinfo{doi}{\doi{10.1103/PhysRevD.100.034007}}, \eprint{1905.03763}.

\bibtype{Article}%
\bibitem{Hatta:2016aoc}
\bibinfo{author}{Y. Hatta}, \bibinfo{author}{Y. Nakagawa}, \bibinfo{author}{F.
  Yuan}, \bibinfo{author}{Y. Zhao}, \bibinfo{author}{B. Xiao},
  \bibinfo{title}{{Gluon orbital angular momentum at small-$x$}},
  \bibinfo{journal}{Phys. Rev. D} \bibinfo{volume}{95} (\bibinfo{number}{11})
  (\bibinfo{year}{2017}) \bibinfo{pages}{114032},
  \bibinfo{doi}{\doi{10.1103/PhysRevD.95.114032}}, \eprint{1612.02445}.

\bibtype{Article}%
\bibitem{Mantysaari:2019csc}
\bibinfo{author}{H. M\"antysaari}, \bibinfo{author}{N. Mueller},
  \bibinfo{author}{B. Schenke}, \bibinfo{title}{{Diffractive Dijet Production
  and Wigner Distributions from the Color Glass Condensate}},
  \bibinfo{journal}{Phys. Rev. D} \bibinfo{volume}{99} (\bibinfo{number}{7})
  (\bibinfo{year}{2019}) \bibinfo{pages}{074004},
  \bibinfo{doi}{\doi{10.1103/PhysRevD.99.074004}}, \eprint{1902.05087}.

\bibtype{Article}%
\bibitem{Boer:2023mip}
\bibinfo{author}{D. Boer}, \bibinfo{author}{C. Setyadi},
  \bibinfo{title}{{Probing gluon GTMDs through exclusive coherent diffractive
  processes}}, \bibinfo{journal}{Eur. Phys. J. C} \bibinfo{volume}{83}
  (\bibinfo{number}{10}) (\bibinfo{year}{2023}) \bibinfo{pages}{890},
  \bibinfo{doi}{\doi{10.1140/epjc/s10052-023-12040-6}}, \eprint{2301.07980}.

\bibtype{Article}%
\bibitem{Boer:2021upt}
\bibinfo{author}{D. Boer}, \bibinfo{author}{C. Setyadi}, \bibinfo{title}{{GTMD
  model predictions for diffractive dijet production at EIC}},
  \bibinfo{journal}{Phys. Rev. D} \bibinfo{volume}{104} (\bibinfo{number}{7})
  (\bibinfo{year}{2021}) \bibinfo{pages}{074006},
  \bibinfo{doi}{\doi{10.1103/PhysRevD.104.074006}}, \eprint{2106.15148}.

\bibtype{Article}%
\bibitem{Bhattacharya:2022vvo}
\bibinfo{author}{S. Bhattacharya}, \bibinfo{author}{R. Boussarie},
  \bibinfo{author}{Y. Hatta}, \bibinfo{title}{{Signature of the Gluon Orbital
  Angular Momentum}}, \bibinfo{journal}{Phys. Rev. Lett.} \bibinfo{volume}{128}
  (\bibinfo{number}{18}) (\bibinfo{year}{2022}) \bibinfo{pages}{182002},
  \bibinfo{doi}{\doi{10.1103/PhysRevLett.128.182002}}, \eprint{2201.08709}.

\bibtype{Article}%
\bibitem{Bhattacharya:2024sck}
\bibinfo{author}{S. Bhattacharya}, \bibinfo{author}{R. Boussarie},
  \bibinfo{author}{Y. Hatta}, \bibinfo{title}{{Exploring orbital angular
  momentum and spin-orbit correlations for gluons at the Electron-Ion
  Collider}}, \bibinfo{journal}{Phys. Rev. D} \bibinfo{volume}{111}
  (\bibinfo{number}{3}) (\bibinfo{year}{2025}) \bibinfo{pages}{034019},
  \bibinfo{doi}{\doi{10.1103/PhysRevD.111.034019}}, \eprint{2404.04209}.

\bibtype{Article}%
\bibitem{Boussarie:2019vmk}
\bibinfo{author}{R. Boussarie}, \bibinfo{author}{Y. Hatta}, \bibinfo{author}{L.
  Szymanowski}, \bibinfo{author}{S. Wallon}, \bibinfo{title}{{Probing the Gluon
  Sivers Function with an Unpolarized Target: GTMD Distributions and the
  Odderons}}, \bibinfo{journal}{Phys. Rev. Lett.} \bibinfo{volume}{124}
  (\bibinfo{number}{17}) (\bibinfo{year}{2020}) \bibinfo{pages}{172501},
  \bibinfo{doi}{\doi{10.1103/PhysRevLett.124.172501}}, \eprint{1912.08182}.

\bibtype{Article}%
\bibitem{Bhattacharya:2023yvo}
\bibinfo{author}{S. Bhattacharya}, \bibinfo{author}{D. Zheng},
  \bibinfo{author}{J. Zhou}, \bibinfo{title}{{Accessing the gluon GTMD F1,4 in
  exclusive \ensuremath{\pi}0 production in ep collisions}},
  \bibinfo{journal}{Phys. Rev. D} \bibinfo{volume}{109} (\bibinfo{number}{9})
  (\bibinfo{year}{2024}) \bibinfo{pages}{096029},
  \bibinfo{doi}{\doi{10.1103/PhysRevD.109.096029}}, \eprint{2304.05784}.

\bibtype{Article}%
\bibitem{Boussarie:2018zwg}
\bibinfo{author}{R. Boussarie}, \bibinfo{author}{Y. Hatta},
  \bibinfo{author}{B.-W. Xiao}, \bibinfo{author}{F. Yuan},
  \bibinfo{title}{{Probing the Weizsäcker-Williams gluon Wigner distribution
  in $pp$ collisions}}, \bibinfo{journal}{Phys. Rev. D} \bibinfo{volume}{98}
  (\bibinfo{number}{7}) (\bibinfo{year}{2018}) \bibinfo{pages}{074015},
  \bibinfo{doi}{\doi{10.1103/PhysRevD.98.074015}}, \eprint{1807.08697}.

\bibtype{Article}%
\bibitem{Bhattacharya:2018lgm}
\bibinfo{author}{S. Bhattacharya}, \bibinfo{author}{A. Metz},
  \bibinfo{author}{V.~K. Ojha}, \bibinfo{author}{J.-Y. Tsai},
  \bibinfo{author}{J. Zhou}, \bibinfo{title}{{Exclusive double quarkonium
  production and generalized TMDs of gluons}}, \bibinfo{journal}{Phys. Lett. B}
  \bibinfo{volume}{833} (\bibinfo{year}{2022}) \bibinfo{pages}{137383},
  \bibinfo{doi}{\doi{10.1016/j.physletb.2022.137383}}, \eprint{1802.10550}.

\bibtype{Article}%
\bibitem{Bhattacharya:2017bvs}
\bibinfo{author}{S. Bhattacharya}, \bibinfo{author}{A. Metz},
  \bibinfo{author}{J. Zhou}, \bibinfo{title}{{Generalized TMDs and the
  exclusive double Drell\textendash{}Yan process}}, \bibinfo{journal}{Phys.
  Lett. B} \bibinfo{volume}{771} (\bibinfo{year}{2017})
  \bibinfo{pages}{396--400},
  \bibinfo{doi}{\doi{10.1016/j.physletb.2017.05.081}}, \eprint{1702.04387}.

\bibtype{Article}%
\bibitem{Echevarria:2022ztg}
\bibinfo{author}{M.~G. Echevarria}, \bibinfo{author}{P.~A. Gutierrez~Garcia},
  \bibinfo{author}{I. Scimemi}, \bibinfo{title}{{GTMDs and the factorization of
  exclusive double Drell-Yan}}, \bibinfo{journal}{Phys. Lett. B}
  \bibinfo{volume}{840} (\bibinfo{year}{2023}) \bibinfo{pages}{137881},
  \bibinfo{doi}{\doi{10.1016/j.physletb.2023.137881}}, \eprint{2208.00021}.

\bibtype{Article}%
\bibitem{Bhattacharya:2023hbq}
\bibinfo{author}{S. Bhattacharya}, \bibinfo{author}{D. Zheng},
  \bibinfo{author}{J. Zhou}, \bibinfo{title}{{Probing the Quark Orbital Angular
  Momentum at Electron-Ion Colliders Using Exclusive \ensuremath{\pi}0
  Production}}, \bibinfo{journal}{Phys. Rev. Lett.} \bibinfo{volume}{133}
  (\bibinfo{number}{5}) (\bibinfo{year}{2024}) \bibinfo{pages}{051901},
  \bibinfo{doi}{\doi{10.1103/PhysRevLett.133.051901}}, \eprint{2312.01309}.

\bibtype{Article}%
\bibitem{Wigner:1932eb}
\bibinfo{author}{E.~P. Wigner}, \bibinfo{title}{{On the quantum correction for
  thermodynamic equilibrium}}, \bibinfo{journal}{Phys. Rev.}
  \bibinfo{volume}{40} (\bibinfo{year}{1932}) \bibinfo{pages}{749--760},
  \bibinfo{doi}{\doi{10.1103/PhysRev.40.749}}.

\bibtype{Article}%
\bibitem{Hillery:1983ms}
\bibinfo{author}{M. Hillery}, \bibinfo{author}{R.~F. O'Connell},
  \bibinfo{author}{M.~O. Scully}, \bibinfo{author}{E.~P. Wigner},
  \bibinfo{title}{{Distribution functions in physics: Fundamentals}},
  \bibinfo{journal}{Phys. Rept.} \bibinfo{volume}{106} (\bibinfo{year}{1984})
  \bibinfo{pages}{121--167}, \bibinfo{doi}{\doi{10.1016/0370-1573(84)90160-1}}.

\bibtype{Article}%
\bibitem{Carruthers:1975sp}
\bibinfo{author}{P. Carruthers}, \bibinfo{author}{F. Zachariasen},
  \bibinfo{title}{{Relativistic Quantum Transport Theory Approach to
  Multiparticle Production}}, \bibinfo{journal}{Phys. Rev. D}
  \bibinfo{volume}{13} (\bibinfo{year}{1976}) \bibinfo{pages}{950},
  \bibinfo{doi}{\doi{10.1103/PhysRevD.13.950}}.

\bibtype{Article}%
\bibitem{Bialynicki-Birula:1991jwl}
\bibinfo{author}{I. Bialynicki-Birula}, \bibinfo{author}{P. Gornicki},
  \bibinfo{author}{J. Rafelski}, \bibinfo{title}{{Phase space structure of the
  Dirac vacuum}}, \bibinfo{journal}{Phys. Rev. D} \bibinfo{volume}{44}
  (\bibinfo{year}{1991}) \bibinfo{pages}{1825--1835},
  \bibinfo{doi}{\doi{10.1103/PhysRevD.44.1825}}.

\bibtype{Article}%
\bibitem{Lorce:2011kd}
\bibinfo{author}{C. Lorcé}, \bibinfo{author}{B. Pasquini},
  \bibinfo{title}{{Quark Wigner Distributions and Orbital Angular Momentum}},
  \bibinfo{journal}{Phys. Rev. D} \bibinfo{volume}{84} (\bibinfo{year}{2011})
  \bibinfo{pages}{014015}, \bibinfo{doi}{\doi{10.1103/PhysRevD.84.014015}},
  \eprint{1106.0139}.

\bibtype{Article}%
\bibitem{Lorce:2015sqe}
\bibinfo{author}{C. Lorcé}, \bibinfo{author}{B. Pasquini},
  \bibinfo{title}{{Multipole decomposition of the nucleon transverse phase
  space}}, \bibinfo{journal}{Phys. Rev. D} \bibinfo{volume}{93}
  (\bibinfo{number}{3}) (\bibinfo{year}{2016}) \bibinfo{pages}{034040},
  \bibinfo{doi}{\doi{10.1103/PhysRevD.93.034040}}, \eprint{1512.06744}.

\bibtype{Article}%
\bibitem{Shin:1992nj}
\bibinfo{author}{G.~R. Shin}, \bibinfo{author}{I. Bialynicki-Birula},
  \bibinfo{author}{J. Rafelski}, \bibinfo{title}{{Wigner function of polarized,
  localized relativistic spin 1/2 particles}}, \bibinfo{journal}{Phys. Rev. A}
  \bibinfo{volume}{46} (\bibinfo{year}{1992}) \bibinfo{pages}{645},
  \bibinfo{doi}{\doi{10.1103/PhysRevA.46.645}}.

\bibtype{Article}%
\bibitem{Hatta:2011ku}
\bibinfo{author}{Y. Hatta}, \bibinfo{title}{{Notes on the orbital angular
  momentum of quarks in the nucleon}}, \bibinfo{journal}{Phys. Lett. B}
  \bibinfo{volume}{708} (\bibinfo{year}{2012}) \bibinfo{pages}{186--190},
  \bibinfo{doi}{\doi{10.1016/j.physletb.2012.01.024}}, \eprint{1111.3547}.

\bibtype{Article}%
\bibitem{Lorce:2011ni}
\bibinfo{author}{C. Lorcé}, \bibinfo{author}{B. Pasquini}, \bibinfo{author}{X.
  Xiong}, \bibinfo{author}{F. Yuan}, \bibinfo{title}{{The quark orbital angular
  momentum from Wigner distributions and light-cone wave functions}},
  \bibinfo{journal}{Phys. Rev. D} \bibinfo{volume}{85} (\bibinfo{year}{2012})
  \bibinfo{pages}{114006}, \bibinfo{doi}{\doi{10.1103/PhysRevD.85.114006}},
  \eprint{1111.4827}.

\bibtype{Article}%
\bibitem{Ji:2012sj}
\bibinfo{author}{X.-D. Ji}, \bibinfo{author}{X. Xiong}, \bibinfo{author}{F.
  Yuan}, \bibinfo{title}{{Proton Spin Structure from Measurable Parton
  Distributions}}, \bibinfo{journal}{Phys. Rev. Lett.} \bibinfo{volume}{109}
  (\bibinfo{year}{2012}) \bibinfo{pages}{152005},
  \bibinfo{doi}{\doi{10.1103/PhysRevLett.109.152005}}, \eprint{1202.2843}.

\bibtype{Article}%
\bibitem{Ji:2012ba}
\bibinfo{author}{X.-D. Ji}, \bibinfo{author}{X. Xiong}, \bibinfo{author}{F.
  Yuan}, \bibinfo{title}{{Probing Parton Orbital Angular Momentum in
  Longitudinally Polarized Nucleon}}, \bibinfo{journal}{Phys. Rev. D}
  \bibinfo{volume}{88} (\bibinfo{number}{1}) (\bibinfo{year}{2013})
  \bibinfo{pages}{014041}, \bibinfo{doi}{\doi{10.1103/PhysRevD.88.014041}},
  \eprint{1207.5221}.

\bibtype{Article}%
\bibitem{Jaffe:1989jz}
\bibinfo{author}{R.~L. Jaffe}, \bibinfo{author}{A. Manohar},
  \bibinfo{title}{{The $g_1$ Problem: Fact and Fantasy on the Spin of the
  Proton}}, \bibinfo{journal}{Nucl. Phys. B} \bibinfo{volume}{337}
  (\bibinfo{year}{1990}) \bibinfo{pages}{509--546},
  \bibinfo{doi}{\doi{10.1016/0550-3213(90)90506-9}}.

\bibtype{Article}%
\bibitem{Engelhardt:2017miy}
\bibinfo{author}{M. Engelhardt}, \bibinfo{title}{{Quark orbital dynamics in the
  proton from Lattice QCD -- from Ji to Jaffe-Manohar orbital angular
  momentum}}, \bibinfo{journal}{Phys. Rev. D} \bibinfo{volume}{95}
  (\bibinfo{number}{9}) (\bibinfo{year}{2017}) \bibinfo{pages}{094505},
  \bibinfo{doi}{\doi{10.1103/PhysRevD.95.094505}}, \eprint{1701.01536}.

\bibtype{Article}%
\bibitem{Engelhardt:2020qtg}
\bibinfo{author}{M. Engelhardt}, \bibinfo{author}{J.~R. Green},
  \bibinfo{author}{N. Hasan}, \bibinfo{author}{S. Krieg}, \bibinfo{author}{S.
  Meinel}, \bibinfo{author}{J. Negele}, \bibinfo{author}{A. Pochinsky},
  \bibinfo{author}{S. Syritsyn}, \bibinfo{title}{{From Ji to Jaffe-Manohar
  orbital angular momentum in lattice QCD using a direct derivative method}},
  \bibinfo{journal}{Phys. Rev. D} \bibinfo{volume}{102} (\bibinfo{number}{7})
  (\bibinfo{year}{2020}) \bibinfo{pages}{074505},
  \bibinfo{doi}{\doi{10.1103/PhysRevD.102.074505}}, \eprint{2008.03660}.

\bibtype{Article}%
\bibitem{Burkardt:2012sd}
\bibinfo{author}{M. Burkardt}, \bibinfo{title}{{Parton Orbital Angular Momentum
  and Final State Interactions}}, \bibinfo{journal}{Phys. Rev. D}
  \bibinfo{volume}{88} (\bibinfo{number}{1}) (\bibinfo{year}{2013})
  \bibinfo{pages}{014014}, \bibinfo{doi}{\doi{10.1103/PhysRevD.88.014014}},
  \eprint{1205.2916}.

\bibtype{Article}%
\bibitem{Witten:1979kh}
\bibinfo{author}{E. Witten}, \bibinfo{title}{{Baryons in the 1/n Expansion}},
  \bibinfo{journal}{Nucl. Phys. B} \bibinfo{volume}{160} (\bibinfo{year}{1979})
  \bibinfo{pages}{57--115}, \bibinfo{doi}{\doi{10.1016/0550-3213(79)90232-3}}.

\bibtype{Article}%
\bibitem{Dashen:1993jt}
\bibinfo{author}{R.~F. Dashen}, \bibinfo{author}{E.~E. Jenkins},
  \bibinfo{author}{A.~V. Manohar}, \bibinfo{title}{{The $1/N_c$ expansion for
  baryons}}, \bibinfo{journal}{Phys. Rev. D} \bibinfo{volume}{49}
  (\bibinfo{year}{1994}) \bibinfo{pages}{4713},
  \bibinfo{doi}{\doi{10.1103/PhysRevD.51.2489}}, \eprint{hep-ph/9310379}.

\bibtype{Article}%
\bibitem{Pobylitsa:2000tt}
\bibinfo{author}{P.~V. Pobylitsa}, \bibinfo{author}{M.~V. Polyakov},
  \bibinfo{title}{{New positivity bounds on parton distributions in
  multicolored QCD}}, \bibinfo{journal}{Phys. Rev. D} \bibinfo{volume}{62}
  (\bibinfo{year}{2000}) \bibinfo{pages}{097502},
  \bibinfo{doi}{\doi{10.1103/PhysRevD.62.097502}}, \eprint{hep-ph/0004094}.

\bibtype{Article}%
\bibitem{Diakonov:1996sr}
\bibinfo{author}{D. Diakonov}, \bibinfo{author}{V. Petrov}, \bibinfo{author}{P.
  Pobylitsa}, \bibinfo{author}{M.~V. Polyakov}, \bibinfo{author}{C. Weiss},
  \bibinfo{title}{{Nucleon parton distributions at low normalization point in
  the large-$N_c$ limit}}, \bibinfo{journal}{Nucl. Phys. B}
  \bibinfo{volume}{480} (\bibinfo{year}{1996}) \bibinfo{pages}{341--380},
  \bibinfo{doi}{\doi{10.1016/S0550-3213(96)00486-5}}, \eprint{hep-ph/9606314}.

\bibtype{Article}%
\bibitem{Efremov:2000ar}
\bibinfo{author}{A.~V. Efremov}, \bibinfo{author}{K. Goeke},
  \bibinfo{author}{P.~V. Pobylitsa}, \bibinfo{title}{{Gluon and quark
  distributions in large-$N_c$ QCD: Theory versus phenomenology}},
  \bibinfo{journal}{Phys. Lett. B} \bibinfo{volume}{488} (\bibinfo{year}{2000})
  \bibinfo{pages}{182--186},
  \bibinfo{doi}{\doi{10.1016/S0370-2693(00)00858-3}}, \eprint{hep-ph/0004196}.

\bibtype{Article}%
\bibitem{Panteleeva:2020ejw}
\bibinfo{author}{J.~Y. Panteleeva}, \bibinfo{author}{M.~V. Polyakov},
  \bibinfo{title}{{Quadrupole pressure and shear forces inside baryons in the
  large-$N_c$ limit}}, \bibinfo{journal}{Phys. Lett. B} \bibinfo{volume}{809}
  (\bibinfo{year}{2020}) \bibinfo{pages}{135707},
  \bibinfo{doi}{\doi{10.1016/j.physletb.2020.135707}}, \eprint{2004.02912}.

\bibtype{Article}%
\bibitem{Chodos:1974je}
\bibinfo{author}{A. Chodos}, \bibinfo{author}{R.~L. Jaffe}, \bibinfo{author}{K.
  Johnson}, \bibinfo{author}{Charles~B. Thorn}, \bibinfo{author}{V.~F.
  Weisskopf}, \bibinfo{title}{{A New Extended Model of Hadrons}},
  \bibinfo{journal}{Phys. Rev. D} \bibinfo{volume}{9} (\bibinfo{year}{1974})
  \bibinfo{pages}{3471--3495}, \bibinfo{doi}{\doi{10.1103/PhysRevD.9.3471}}.

\bibtype{Article}%
\bibitem{Jaffe:1974nj}
\bibinfo{author}{R.~L. Jaffe}, \bibinfo{title}{{Deep Inelastic Structure
  Functions in an Approximation to the Bag Theory}}, \bibinfo{journal}{Phys.
  Rev. D} \bibinfo{volume}{11} (\bibinfo{year}{1975}) \bibinfo{pages}{1953},
  \bibinfo{doi}{\doi{10.1103/PhysRevD.11.1953}}.

\bibtype{Article}%
\bibitem{Ji:1997gm}
\bibinfo{author}{X.-D. Ji}, \bibinfo{author}{W. Melnitchouk},
  \bibinfo{author}{X. Song}, \bibinfo{title}{{A Study of off forward parton
  distributions}}, \bibinfo{journal}{Phys. Rev. D} \bibinfo{volume}{56}
  (\bibinfo{year}{1997}) \bibinfo{pages}{5511--5523},
  \bibinfo{doi}{\doi{10.1103/PhysRevD.56.5511}}, \eprint{hep-ph/9702379}.

\bibtype{Article}%
\bibitem{Avakian:2010br}
\bibinfo{author}{H. Avakian}, \bibinfo{author}{A.~V. Efremov},
  \bibinfo{author}{P. Schweitzer}, \bibinfo{author}{F. Yuan},
  \bibinfo{title}{{The transverse momentum dependent distribution functions in
  the bag model}}, \bibinfo{journal}{Phys. Rev. D} \bibinfo{volume}{81}
  (\bibinfo{year}{2010}) \bibinfo{pages}{074035},
  \bibinfo{doi}{\doi{10.1103/PhysRevD.81.074035}}, \eprint{1001.5467}.

\bibtype{Article}%
\bibitem{Courtoy:2016des}
\bibinfo{author}{A. Courtoy}, \bibinfo{author}{A.~S. Miramontes},
  \bibinfo{title}{{Quark Orbital Angular Momentum in the MIT Bag Model}},
  \bibinfo{journal}{Phys. Rev. D} \bibinfo{volume}{95} (\bibinfo{number}{1})
  (\bibinfo{year}{2017}) \bibinfo{pages}{014027},
  \bibinfo{doi}{\doi{10.1103/PhysRevD.95.014027}}, \eprint{1611.03375}.

\bibtype{Article}%
\bibitem{Lorce:2022cle}
\bibinfo{author}{C. Lorcé}, \bibinfo{author}{P. Schweitzer},
  \bibinfo{author}{K. Tezgin}, \bibinfo{title}{{2D energy-momentum tensor
  distributions of nucleon in a large-Nc quark model from ultrarelativistic to
  nonrelativistic limit}}, \bibinfo{journal}{Phys. Rev. D}
  \bibinfo{volume}{106} (\bibinfo{number}{1}) (\bibinfo{year}{2022})
  \bibinfo{pages}{014012}, \bibinfo{doi}{\doi{10.1103/PhysRevD.106.014012}},
  \eprint{2202.01192}.

\bibtype{Article}%
\bibitem{Brodsky:2002cx}
\bibinfo{author}{S.~J. Brodsky}, \bibinfo{author}{D.~S. Hwang},
  \bibinfo{author}{I. Schmidt}, \bibinfo{title}{{Final state interactions and
  single spin asymmetries in semiinclusive deep inelastic scattering}},
  \bibinfo{journal}{Phys. Lett. B} \bibinfo{volume}{530} (\bibinfo{year}{2002})
  \bibinfo{pages}{99--107}, \bibinfo{doi}{\doi{10.1016/S0370-2693(02)01320-5}},
  \eprint{hep-ph/0201296}.

\bibtype{Article}%
\bibitem{Anselmino:1992vg}
\bibinfo{author}{M. Anselmino}, \bibinfo{author}{E. Predazzi},
  \bibinfo{author}{S. Ekelin}, \bibinfo{author}{S. Fredriksson},
  \bibinfo{author}{D.~B. Lichtenberg}, \bibinfo{title}{{Diquarks}},
  \bibinfo{journal}{Rev. Mod. Phys.} \bibinfo{volume}{65}
  (\bibinfo{year}{1993}) \bibinfo{pages}{1199--1234},
  \bibinfo{doi}{\doi{10.1103/RevModPhys.65.1199}}.

\bibtype{Article}%
\bibitem{Barabanov:2020jvn}
\bibinfo{author}{M.~Yu. Barabanov}, et al., \bibinfo{title}{{Diquark
  correlations in hadron physics: Origin, impact and evidence}},
  \bibinfo{journal}{Prog. Part. Nucl. Phys.} \bibinfo{volume}{116}
  (\bibinfo{year}{2021}) \bibinfo{pages}{103835},
  \bibinfo{doi}{\doi{10.1016/j.ppnp.2020.103835}}, \eprint{2008.07630}.

\bibtype{Article}%
\bibitem{Jakob:1997wg}
\bibinfo{author}{R. Jakob}, \bibinfo{author}{P.~J. Mulders},
  \bibinfo{author}{J. Rodrigues}, \bibinfo{title}{{Modeling quark distribution
  and fragmentation functions}}, \bibinfo{journal}{Nucl. Phys. A}
  \bibinfo{volume}{626} (\bibinfo{year}{1997}) \bibinfo{pages}{937--965},
  \bibinfo{doi}{\doi{10.1016/S0375-9474(97)00588-5}}, \eprint{hep-ph/9704335}.

\bibtype{Article}%
\bibitem{Gamberg:2003ey}
\bibinfo{author}{L.~P. Gamberg}, \bibinfo{author}{G.~R. Goldstein},
  \bibinfo{author}{K.~A. Oganessyan}, \bibinfo{title}{{Novel transversity
  properties in semiinclusive deep inelastic scattering}},
  \bibinfo{journal}{Phys. Rev. D} \bibinfo{volume}{67} (\bibinfo{year}{2003})
  \bibinfo{pages}{071504}, \bibinfo{doi}{\doi{10.1103/PhysRevD.67.071504}},
  \eprint{hep-ph/0301018}.

\bibtype{Article}%
\bibitem{Bacchetta:2008af}
\bibinfo{author}{A. Bacchetta}, \bibinfo{author}{F. Conti}, \bibinfo{author}{M.
  Radici}, \bibinfo{title}{{Transverse-momentum distributions in a diquark
  spectator model}}, \bibinfo{journal}{Phys. Rev. D} \bibinfo{volume}{78}
  (\bibinfo{year}{2008}) \bibinfo{pages}{074010},
  \bibinfo{doi}{\doi{10.1103/PhysRevD.78.074010}}, \eprint{0807.0323}.

\bibtype{Article}%
\bibitem{Diehl:2000xz}
\bibinfo{author}{M. Diehl}, \bibinfo{author}{T. Feldmann}, \bibinfo{author}{R.
  Jakob}, \bibinfo{author}{P. Kroll}, \bibinfo{title}{{The overlap
  representation of skewed quark and gluon distributions}},
  \bibinfo{journal}{Nucl. Phys. B} \bibinfo{volume}{596} (\bibinfo{year}{2001})
  \bibinfo{pages}{33--65}, \bibinfo{doi}{\doi{10.1016/S0550-3213(00)00684-2}},
  \eprint{hep-ph/0009255}.

\bibtype{Article}%
\bibitem{Boffi:2002yy}
\bibinfo{author}{S. Boffi}, \bibinfo{author}{B. Pasquini}, \bibinfo{author}{M.
  Traini}, \bibinfo{title}{{Linking generalized parton distributions to
  constituent quark models}}, \bibinfo{journal}{Nucl. Phys. B}
  \bibinfo{volume}{649} (\bibinfo{year}{2003}) \bibinfo{pages}{243--262},
  \bibinfo{doi}{\doi{10.1016/S0550-3213(02)01016-7}}, \eprint{hep-ph/0207340}.

\bibtype{Article}%
\bibitem{Pasquini:2007iz}
\bibinfo{author}{B. Pasquini}, \bibinfo{author}{S. Boffi},
  \bibinfo{title}{{Electroweak structure of the nucleon, meson cloud and
  light-cone wavefunctions}}, \bibinfo{journal}{Phys. Rev. D}
  \bibinfo{volume}{76} (\bibinfo{year}{2007}) \bibinfo{pages}{074011},
  \bibinfo{doi}{\doi{10.1103/PhysRevD.76.074011}}, \eprint{0707.2897}.

\bibtype{Article}%
\bibitem{Pasquini:2007xz}
\bibinfo{author}{B. Pasquini}, \bibinfo{author}{S. Boffi},
  \bibinfo{title}{{Nucleon spin densities in a light-front constituent quark
  model}}, \bibinfo{journal}{Phys. Lett. B} \bibinfo{volume}{653}
  (\bibinfo{year}{2007}) \bibinfo{pages}{23--28},
  \bibinfo{doi}{\doi{10.1016/j.physletb.2007.07.037}}, \eprint{0705.4345}.

\bibtype{Article}%
\bibitem{Dahiya:2007is}
\bibinfo{author}{H. Dahiya}, \bibinfo{author}{A. Mukherjee},
  \bibinfo{author}{S. Ray}, \bibinfo{title}{{Parton Distributions in Impact
  Parameter Space}}, \bibinfo{journal}{Phys. Rev. D} \bibinfo{volume}{76}
  (\bibinfo{year}{2007}) \bibinfo{pages}{034010},
  \bibinfo{doi}{\doi{10.1103/PhysRevD.76.034010}}, \eprint{0705.3580}.

\bibtype{Article}%
\bibitem{Pasquini:2008ax}
\bibinfo{author}{B. Pasquini}, \bibinfo{author}{S. Cazzaniga},
  \bibinfo{author}{S. Boffi}, \bibinfo{title}{{Transverse momentum dependent
  parton distributions in a light-cone quark model}}, \bibinfo{journal}{Phys.
  Rev. D} \bibinfo{volume}{78} (\bibinfo{year}{2008}) \bibinfo{pages}{034025},
  \bibinfo{doi}{\doi{10.1103/PhysRevD.78.034025}}, \eprint{0806.2298}.

\bibtype{Article}%
\bibitem{Pasquini:2010af}
\bibinfo{author}{B. Pasquini}, \bibinfo{author}{F. Yuan},
  \bibinfo{title}{{Sivers and Boer-Mulders functions in Light-Cone Quark
  Models}}, \bibinfo{journal}{Phys. Rev. D} \bibinfo{volume}{81}
  (\bibinfo{year}{2010}) \bibinfo{pages}{114013},
  \bibinfo{doi}{\doi{10.1103/PhysRevD.81.114013}}, \eprint{1001.5398}.

\bibtype{Article}%
\bibitem{Lorce:2011zta}
\bibinfo{author}{C. Lorcé}, \bibinfo{author}{B. Pasquini}, \bibinfo{title}{{On
  the Origin of Model Relations among Transverse-Momentum Dependent Parton
  Distributions}}, \bibinfo{journal}{Phys. Rev. D} \bibinfo{volume}{84}
  (\bibinfo{year}{2011}) \bibinfo{pages}{034039},
  \bibinfo{doi}{\doi{10.1103/PhysRevD.84.034039}}, \eprint{1104.5651}.

\bibtype{Article}%
\bibitem{Mukherjee:2014nya}
\bibinfo{author}{A. Mukherjee}, \bibinfo{author}{S. Nair},
  \bibinfo{author}{V.~K. Ojha}, \bibinfo{title}{{Quark Wigner Distributions and
  Orbital Angular Momentum in Light-front Dressed Quark Model}},
  \bibinfo{journal}{Phys. Rev. D} \bibinfo{volume}{90} (\bibinfo{number}{1})
  (\bibinfo{year}{2014}) \bibinfo{pages}{014024},
  \bibinfo{doi}{\doi{10.1103/PhysRevD.90.014024}}, \eprint{1403.6233}.

\bibtype{Article}%
\bibitem{Kumar:2015fta}
\bibinfo{author}{N. Kumar}, \bibinfo{author}{H. Dahiya},
  \bibinfo{title}{{Generalized Parton Distributions of proton for nonzero
  skewness in transverse and longitudinal position spaces}},
  \bibinfo{journal}{Int. J. Mod. Phys. A} \bibinfo{volume}{30}
  (\bibinfo{number}{02}) (\bibinfo{year}{2015}) \bibinfo{pages}{1550010},
  \bibinfo{doi}{\doi{10.1142/S0217751X15500104}}, \eprint{1501.04745}.

\bibtype{Article}%
\bibitem{Choudhary:2022den}
\bibinfo{author}{P. Choudhary}, \bibinfo{author}{B. Gurjar},
  \bibinfo{author}{D. Chakrabarti}, \bibinfo{author}{A. Mukherjee},
  \bibinfo{title}{{Gravitational form factors and mechanical properties of the
  proton: Connections between distributions in 2D and 3D}},
  \bibinfo{journal}{Phys. Rev. D} \bibinfo{volume}{106} (\bibinfo{number}{7})
  (\bibinfo{year}{2022}) \bibinfo{pages}{076004},
  \bibinfo{doi}{\doi{10.1103/PhysRevD.106.076004}}, \eprint{2206.12206}.

\bibtype{Article}%
\bibitem{Han:2022tlh}
\bibinfo{author}{Y. Han}, \bibinfo{author}{T. Liu}, \bibinfo{author}{B.-Q. Ma},
  \bibinfo{title}{{Six-dimensional light-front Wigner distribution of
  hadrons}}, \bibinfo{journal}{Phys. Lett. B} \bibinfo{volume}{830}
  (\bibinfo{year}{2022}) \bibinfo{pages}{137127},
  \bibinfo{doi}{\doi{10.1016/j.physletb.2022.137127}}, \eprint{2202.10359}.

\bibtype{Article}%
\bibitem{Chakrabarti:2023djs}
\bibinfo{author}{D. Chakrabarti}, \bibinfo{author}{P. Choudhary},
  \bibinfo{author}{B. Gurjar}, \bibinfo{author}{R. Kishore},
  \bibinfo{author}{T. Maji}, \bibinfo{author}{C. Mondal}, \bibinfo{author}{A.
  Mukherjee}, \bibinfo{title}{{Gluon distributions in the proton in a
  light-front spectator model}}, \bibinfo{journal}{Phys. Rev. D}
  \bibinfo{volume}{108} (\bibinfo{number}{1}) (\bibinfo{year}{2023})
  \bibinfo{pages}{014009}, \bibinfo{doi}{\doi{10.1103/PhysRevD.108.014009}},
  \eprint{2304.09908}.

\bibtype{Article}%
\bibitem{Pobylitsa:1998tk}
\bibinfo{author}{P.~V. Pobylitsa}, \bibinfo{author}{M.~V. Polyakov},
  \bibinfo{author}{K. Goeke}, \bibinfo{author}{T. Watabe}, \bibinfo{author}{C.
  Weiss}, \bibinfo{title}{{Isovector unpolarized quark distribution in the
  nucleon in the large-$N_c$ limit}}, \bibinfo{journal}{Phys. Rev. D}
  \bibinfo{volume}{59} (\bibinfo{year}{1999}) \bibinfo{pages}{034024},
  \bibinfo{doi}{\doi{10.1103/PhysRevD.59.034024}}, \eprint{hep-ph/9804436}.

\bibtype{Article}%
\bibitem{Dressler:1999zv}
\bibinfo{author}{B. Dressler}, \bibinfo{author}{K. Goeke},
  \bibinfo{author}{M.~V. Polyakov}, \bibinfo{author}{P. Schweitzer},
  \bibinfo{author}{M. Strikman}, \bibinfo{author}{C. Weiss},
  \bibinfo{title}{{Polarized anti-quark flavor asymmetry in Drell-Yan pair
  production}}, \bibinfo{journal}{Eur. Phys. J. C} \bibinfo{volume}{18}
  (\bibinfo{year}{2001}) \bibinfo{pages}{719--722},
  \bibinfo{doi}{\doi{10.1007/s100520100567}}, \eprint{hep-ph/9910464}.

\bibtype{Article}%
\bibitem{Schweitzer:2002nm}
\bibinfo{author}{P. Schweitzer}, \bibinfo{author}{S. Boffi},
  \bibinfo{author}{M. Radici}, \bibinfo{title}{{Polynomiality of unpolarized
  off forward distribution functions and the D term in the chiral quark soliton
  model}}, \bibinfo{journal}{Phys. Rev. D} \bibinfo{volume}{66}
  (\bibinfo{year}{2002}) \bibinfo{pages}{114004},
  \bibinfo{doi}{\doi{10.1103/PhysRevD.66.114004}}, \eprint{hep-ph/0207230}.

\bibtype{Article}%
\bibitem{Wakamatsu:2009fn}
\bibinfo{author}{M. Wakamatsu}, \bibinfo{title}{{Transverse momentum
  distributions of quarks in the nucleon from the Chiral Quark Soliton Model}},
  \bibinfo{journal}{Phys. Rev. D} \bibinfo{volume}{79} (\bibinfo{year}{2009})
  \bibinfo{pages}{094028}, \bibinfo{doi}{\doi{10.1103/PhysRevD.79.094028}},
  \eprint{0903.1886}.

\bibtype{Article}%
\bibitem{Schweitzer:2012hh}
\bibinfo{author}{P. Schweitzer}, \bibinfo{author}{M. Strikman},
  \bibinfo{author}{C. Weiss}, \bibinfo{title}{{Intrinsic transverse momentum
  and parton correlations from dynamical chiral symmetry breaking}},
  \bibinfo{journal}{JHEP} \bibinfo{volume}{01} (\bibinfo{year}{2013})
  \bibinfo{pages}{163}, \bibinfo{doi}{\doi{10.1007/JHEP01(2013)163}},
  \eprint{1210.1267}.

\bibtype{Article}%
\bibitem{Bacchetta:2024qre}
\bibinfo{author}{A. Bacchetta}, \bibinfo{author}{V. Bertone},
  \bibinfo{author}{C. Bissolotti}, \bibinfo{author}{G. Bozzi},
  \bibinfo{author}{M. Cerutti}, \bibinfo{author}{F. Delcarro},
  \bibinfo{author}{M. Radici}, \bibinfo{author}{L. Rossi}, \bibinfo{author}{A.
  Signori} (\bibinfo{collaboration}{MAP (Multi-dimensional Analyses of Partonic
  distributions)}), \bibinfo{title}{{Flavor dependence of unpolarized quark
  transverse momentum distributions from a global fit}},
  \bibinfo{journal}{JHEP} \bibinfo{volume}{08} (\bibinfo{year}{2024})
  \bibinfo{pages}{232}, \bibinfo{doi}{\doi{10.1007/JHEP08(2024)232}},
  \eprint{2405.13833}.

\bibtype{Article}%
\bibitem{Kim:1995bq}
\bibinfo{author}{H.-C. Kim}, \bibinfo{author}{M.~V. Polyakov},
  \bibinfo{author}{K. Goeke}, \bibinfo{title}{{Nucleon tensor charges in the
  SU(2) chiral quark - soliton model}}, \bibinfo{journal}{Phys. Rev. D}
  \bibinfo{volume}{53} (\bibinfo{year}{1996}) \bibinfo{pages}{4715--4718},
  \bibinfo{doi}{\doi{10.1103/PhysRevD.53.R4715}}, \eprint{hep-ph/9509283}.

\bibtype{Article}%
\bibitem{Lorce:2007fa}
\bibinfo{author}{C. Lorcé}, \bibinfo{title}{{Tensor charges of light baryons
  in the Infinite Momentum Frame}}, \bibinfo{journal}{Phys. Rev. D}
  \bibinfo{volume}{79} (\bibinfo{year}{2009}) \bibinfo{pages}{074027},
  \bibinfo{doi}{\doi{10.1103/PhysRevD.79.074027}}, \eprint{0708.4168}.

\bibtype{Article}%
\bibitem{Kim:2024ibz}
\bibinfo{author}{J.-Y. Kim}, \bibinfo{author}{C. Weiss},
  \bibinfo{title}{{Chiral-odd generalized parton distributions in the large-Nc
  limit of QCD: Spin-flavor structure, polynomiality, and sum rules}},
  \bibinfo{journal}{Phys. Rev. D} \bibinfo{volume}{111} (\bibinfo{number}{7})
  (\bibinfo{year}{2025}) \bibinfo{pages}{074007},
  \bibinfo{doi}{\doi{10.1103/PhysRevD.111.074007}}, \eprint{2411.17634}.

\bibtype{Article}%
\bibitem{Strikman:2003gz}
\bibinfo{author}{M. Strikman}, \bibinfo{author}{C. Weiss},
  \bibinfo{title}{{Chiral dynamics and the growth of the nucleon's gluonic
  transverse size at small x}}, \bibinfo{journal}{Phys. Rev. D}
  \bibinfo{volume}{69} (\bibinfo{year}{2004}) \bibinfo{pages}{054012},
  \bibinfo{doi}{\doi{10.1103/PhysRevD.69.054012}}, \eprint{hep-ph/0308191}.

\bibtype{Article}%
\bibitem{Strikman:2009bd}
\bibinfo{author}{M. Strikman}, \bibinfo{author}{C. Weiss},
  \bibinfo{title}{{Chiral dynamics and partonic structure at large transverse
  distances}}, \bibinfo{journal}{Phys. Rev. D} \bibinfo{volume}{80}
  (\bibinfo{year}{2009}) \bibinfo{pages}{114029},
  \bibinfo{doi}{\doi{10.1103/PhysRevD.80.114029}}, \eprint{0906.3267}.

\bibtype{Article}%
\bibitem{Granados:2013moa}
\bibinfo{author}{C. Granados}, \bibinfo{author}{C. Weiss},
  \bibinfo{title}{{Chiral dynamics and peripheral transverse densities}},
  \bibinfo{journal}{JHEP} \bibinfo{volume}{01} (\bibinfo{year}{2014})
  \bibinfo{pages}{092}, \bibinfo{doi}{\doi{10.1007/JHEP01(2014)092}},
  \eprint{1308.1634}.

\bibtype{Article}%
\bibitem{Granados:2015rra}
\bibinfo{author}{C. Granados}, \bibinfo{author}{C. Weiss},
  \bibinfo{title}{{Light-front representation of chiral dynamics in peripheral
  transverse densities}}, \bibinfo{journal}{JHEP} \bibinfo{volume}{07}
  (\bibinfo{year}{2015}) \bibinfo{pages}{170},
  \bibinfo{doi}{\doi{10.1007/JHEP07(2015)170}}, \eprint{1503.04839}.

\bibtype{Article}%
\bibitem{Alarcon:2017lhg}
\bibinfo{author}{J.~M. Alarc{\'o}n}, \bibinfo{author}{C. Weiss},
  \bibinfo{title}{{Nucleon form factors in dispersively improved Chiral
  Effective Field Theory II: Electromagnetic form factors}},
  \bibinfo{journal}{Phys. Rev. C} \bibinfo{volume}{97} (\bibinfo{number}{5})
  (\bibinfo{year}{2018}) \bibinfo{pages}{055203},
  \bibinfo{doi}{\doi{10.1103/PhysRevC.97.055203}}, \eprint{1710.06430}.

\bibtype{Article}%
\bibitem{Alarcon:2018zbz}
\bibinfo{author}{J.~M. Alarc{\'o}n}, \bibinfo{author}{D.~W. Higinbotham},
  \bibinfo{author}{C. Weiss}, \bibinfo{author}{Z. Ye}, \bibinfo{title}{{Proton
  charge radius extraction from electron scattering data using dispersively
  improved chiral effective field theory}}, \bibinfo{journal}{Phys. Rev. C}
  \bibinfo{volume}{99} (\bibinfo{number}{4}) (\bibinfo{year}{2019})
  \bibinfo{pages}{044303}, \bibinfo{doi}{\doi{10.1103/PhysRevC.99.044303}},
  \eprint{1809.06373}.

\bibtype{Article}%
\bibitem{Alarcon:2022adi}
\bibinfo{author}{J.~M. Alarc{\'o}n}, \bibinfo{author}{C. Weiss},
  \bibinfo{title}{{Transverse charge and current densities in the nucleon from
  dispersively improved chiral effective field theory}},
  \bibinfo{journal}{Phys. Rev. D} \bibinfo{volume}{106} (\bibinfo{number}{5})
  (\bibinfo{year}{2022}) \bibinfo{pages}{054005},
  \bibinfo{doi}{\doi{10.1103/PhysRevD.106.054005}}, \eprint{2204.11863}.

\bibtype{Article}%
\bibitem{Pasquini:2014vua}
\bibinfo{author}{B. Pasquini}, \bibinfo{author}{M.~V. Polyakov},
  \bibinfo{author}{M. Vanderhaeghen}, \bibinfo{title}{{Dispersive evaluation of
  the D-term form factor in deeply virtual Compton scattering}},
  \bibinfo{journal}{Phys. Lett. B} \bibinfo{volume}{739} (\bibinfo{year}{2014})
  \bibinfo{pages}{133--138},
  \bibinfo{doi}{\doi{10.1016/j.physletb.2014.10.047}}, \eprint{1407.5960}.

\bibtype{Article}%
\bibitem{Cao:2024zlf}
\bibinfo{author}{Xiong-Hui Cao}, \bibinfo{author}{Feng-Kun Guo},
  \bibinfo{author}{Q.-Z. Li}, \bibinfo{author}{D.-L. Yao},
  \bibinfo{title}{{Dispersive determination of nucleon gravitational form
  factors}}, \bibinfo{journal}{Nature Commun.} \bibinfo{volume}{16}
  (\bibinfo{number}{1}) (\bibinfo{year}{2025}) \bibinfo{pages}{6979},
  \bibinfo{doi}{\doi{10.1038/s41467-025-62278-9}}, \eprint{2411.13398}.

\bibtype{Article}%
\bibitem{Cao:2025dkv}
\bibinfo{author}{X.-H. Cao}, \bibinfo{author}{F.-K. Guo},
  \bibinfo{author}{Q.-Z. Li}, \bibinfo{author}{B.-W. Wu},
  \bibinfo{author}{D.-L. Yao}, \bibinfo{title}{{Gravitational form factors of
  pions, kaons and nucleons from dispersion relations}}
  (\bibinfo{year}{2025}), \eprint{2507.05375}.

\bibtype{Article}%
\bibitem{Martinez-Fernandez:2025jvk}
\bibinfo{author}{V. Mart{\'\i}nez-Fern{\'a}ndez}, \bibinfo{author}{D. Binosi},
  \bibinfo{author}{C. Mezrag}, \bibinfo{author}{Z.-Q. Yao},
  \bibinfo{title}{{Constraining the Energy Momentum Tensor through DVCS
  Dispersion Relation beyond Leading Power}}  (\bibinfo{year}{2025}),
  \eprint{2509.06669}.

\bibtype{Article}%
\bibitem{Cloet:2007em}
\bibinfo{author}{I.~C. Cloet}, \bibinfo{author}{W. Bentz},
  \bibinfo{author}{A.~W. Thomas}, \bibinfo{title}{{Transversity quark
  distributions in a covariant quark-diquark model}}, \bibinfo{journal}{Phys.
  Lett. B} \bibinfo{volume}{659} (\bibinfo{year}{2008})
  \bibinfo{pages}{214--220},
  \bibinfo{doi}{\doi{10.1016/j.physletb.2007.09.071}}, \eprint{0708.3246}.

\bibtype{Article}%
\bibitem{Freese:2019bhb}
\bibinfo{author}{A. Freese}, \bibinfo{author}{I.~C. Clo{\"e}t},
  \bibinfo{title}{{Gravitational form factors of light mesons}},
  \bibinfo{journal}{Phys. Rev. C} \bibinfo{volume}{100} (\bibinfo{number}{1})
  (\bibinfo{year}{2019}) \bibinfo{pages}{015201},
  \bibinfo{doi}{\doi{10.1103/PhysRevC.100.015201}}, \eprint{1903.09222}.

\bibtype{Article}%
\bibitem{Hecht:2000xa}
\bibinfo{author}{M.~B. Hecht}, \bibinfo{author}{Craig~D. Roberts},
  \bibinfo{author}{S.~M. Schmidt}, \bibinfo{title}{{Valence quark distributions
  in the pion}}, \bibinfo{journal}{Phys. Rev. C} \bibinfo{volume}{63}
  (\bibinfo{year}{2001}) \bibinfo{pages}{025213},
  \bibinfo{doi}{\doi{10.1103/PhysRevC.63.025213}}, \eprint{nucl-th/0008049}.

\bibtype{Article}%
\bibitem{Raya:2015gva}
\bibinfo{author}{K. Raya}, \bibinfo{author}{L. Chang}, \bibinfo{author}{A.
  Bashir}, \bibinfo{author}{J.~J. Cobos-Martinez}, \bibinfo{author}{L.~X.
  Guti{\'e}rrez-Guerrero}, \bibinfo{author}{C.~D. Roberts},
  \bibinfo{author}{P.~C. Tandy}, \bibinfo{title}{{Structure of the neutral pion
  and its electromagnetic transition form factor}}, \bibinfo{journal}{Phys.
  Rev. D} \bibinfo{volume}{93} (\bibinfo{number}{7}) (\bibinfo{year}{2016})
  \bibinfo{pages}{074017}, \bibinfo{doi}{\doi{10.1103/PhysRevD.93.074017}},
  \eprint{1510.02799}.

\bibtype{Article}%
\bibitem{Owa:2021hnj}
\bibinfo{author}{S. Owa}, \bibinfo{author}{A.~W. Thomas},
  \bibinfo{author}{X.~G. Wang}, \bibinfo{title}{{Effect of the pion field on
  the distributions of pressure and shear in the proton}},
  \bibinfo{journal}{Phys. Lett. B} \bibinfo{volume}{829} (\bibinfo{year}{2022})
  \bibinfo{pages}{137136}, \bibinfo{doi}{\doi{10.1016/j.physletb.2023.138396}},
  \eprint{2106.00929}.

\bibtype{Article}%
\bibitem{Brodsky:2008pf}
\bibinfo{author}{S.~J. Brodsky}, \bibinfo{author}{G.~F. de Teramond},
  \bibinfo{title}{{Light-Front Dynamics and AdS/QCD Correspondence:
  Gravitational Form Factors of Composite Hadrons}}, \bibinfo{journal}{Phys.
  Rev. D} \bibinfo{volume}{78} (\bibinfo{year}{2008}) \bibinfo{pages}{025032},
  \bibinfo{doi}{\doi{10.1103/PhysRevD.78.025032}}, \eprint{0804.0452}.

\bibtype{Article}%
\bibitem{Abidin:2009hr}
\bibinfo{author}{Z. Abidin}, \bibinfo{author}{C.~E. Carlson},
  \bibinfo{title}{{Nucleon electromagnetic and gravitational form factors from
  holography}}, \bibinfo{journal}{Phys. Rev. D} \bibinfo{volume}{79}
  (\bibinfo{year}{2009}) \bibinfo{pages}{115003},
  \bibinfo{doi}{\doi{10.1103/PhysRevD.79.115003}}, \eprint{0903.4818}.

\bibtype{Article}%
\bibitem{Chakrabarti:2015lba}
\bibinfo{author}{D. Chakrabarti}, \bibinfo{author}{C. Mondal},
  \bibinfo{author}{A. Mukherjee}, \bibinfo{title}{{Gravitational form factors
  and transverse spin sum rule in a light front quark-diquark model in
  AdS/QCD}}, \bibinfo{journal}{Phys. Rev. D} \bibinfo{volume}{91}
  (\bibinfo{number}{11}) (\bibinfo{year}{2015}) \bibinfo{pages}{114026},
  \bibinfo{doi}{\doi{10.1103/PhysRevD.91.114026}}, \eprint{1505.02013}.

\bibtype{Article}%
\bibitem{Mamo:2019mka}
\bibinfo{author}{K.~A. Mamo}, \bibinfo{author}{I. Zahed},
  \bibinfo{title}{{Diffractive photoproduction of $J/\psi$ and $\Upsilon$ using
  holographic QCD: gravitational form factors and GPD of gluons in the
  proton}}, \bibinfo{journal}{Phys. Rev. D} \bibinfo{volume}{101}
  (\bibinfo{number}{8}) (\bibinfo{year}{2020}) \bibinfo{pages}{086003},
  \bibinfo{doi}{\doi{10.1103/PhysRevD.101.086003}}, \eprint{1910.04707}.

\bibtype{Article}%
\bibitem{Mamo:2024vjh}
\bibinfo{author}{K.~A. Mamo}, \bibinfo{author}{I. Zahed},
  \bibinfo{title}{{String-based parametrization of nucleon GPDs at any
  skewness: A comparison to lattice QCD}}, \bibinfo{journal}{Phys. Rev. D}
  \bibinfo{volume}{110} (\bibinfo{number}{11}) (\bibinfo{year}{2024})
  \bibinfo{pages}{114016}, \bibinfo{doi}{\doi{10.1103/PhysRevD.110.114016}},
  \eprint{2404.13245}.

\bibtype{Article}%
\bibitem{Mamo:2024jwp}
\bibinfo{author}{K.A. Mamo}, \bibinfo{author}{I. Zahed},
  \bibinfo{title}{{Parametrization of Generalized Parton Distributions from
  t-Channel String Exchange in AdS Spaces}}, \bibinfo{journal}{Phys. Rev.
  Lett.} \bibinfo{volume}{133} (\bibinfo{number}{24}) (\bibinfo{year}{2024})
  \bibinfo{pages}{241901}, \bibinfo{doi}{\doi{10.1103/PhysRevLett.133.241901}},
  \eprint{2411.04162}.

\bibtype{Article}%
\bibitem{Anikin:2019kwi}
\bibinfo{author}{I.~V. Anikin}, \bibinfo{title}{{Gravitational form factors
  within light-cone sum rules at leading order}}, \bibinfo{journal}{Phys. Rev.
  D} \bibinfo{volume}{99} (\bibinfo{number}{9}) (\bibinfo{year}{2019})
  \bibinfo{pages}{094026}, \bibinfo{doi}{\doi{10.1103/PhysRevD.99.094026}},
  \eprint{1902.00094}.

\bibtype{Article}%
\bibitem{Azizi:2019ytx}
\bibinfo{author}{K. Azizi}, \bibinfo{author}{U. {\"O}zdem},
  \bibinfo{title}{{Nucleon{\textquoteright}s energy{\textendash}momentum tensor
  form factors in light-cone QCD}}, \bibinfo{journal}{Eur. Phys. J. C}
  \bibinfo{volume}{80} (\bibinfo{number}{2}) (\bibinfo{year}{2020})
  \bibinfo{pages}{104}, \bibinfo{doi}{\doi{10.1140/epjc/s10052-020-7676-5}},
  \eprint{1908.06143}.

\bibtype{Book}%
\bibitem{Gattringer:2010zz}
\bibinfo{author}{C. Gattringer}, \bibinfo{author}{C.~B. Lang},
  \bibinfo{title}{{Quantum chromodynamics on the lattice}},
  \bibinfo{comment}{vol.} \bibinfo{volume}{788}, \bibinfo{publisher}{Springer},
  \bibinfo{address}{Berlin} \bibinfo{year}{2010}, ISBN
  \bibinfo{isbn}{978-3-642-01849-7, 978-3-642-01850-3},
  \bibinfo{doi}{\doi{10.1007/978-3-642-01850-3}}.

\bibtype{Article}%
\bibitem{Constantinou:2020hdm}
\bibinfo{author}{M. Constantinou}, et al., \bibinfo{title}{{Parton
  distributions and lattice-QCD calculations: Toward 3D structure}},
  \bibinfo{journal}{Prog. Part. Nucl. Phys.} \bibinfo{volume}{121}
  (\bibinfo{year}{2021}) \bibinfo{pages}{103908},
  \bibinfo{doi}{\doi{10.1016/j.ppnp.2021.103908}}, \eprint{2006.08636}.

\bibtype{Article}%
\bibitem{FlavourLatticeAveragingGroupFLAG:2024oxs}
\bibinfo{author}{Y. Aoki}, et al. (\bibinfo{collaboration}{Flavour Lattice
  Averaging Group (FLAG)}), \bibinfo{title}{{FLAG Review 2024}}
  (\bibinfo{year}{2024}), \eprint{2411.04268}.

\bibtype{Article}%
\bibitem{Hagler:2009ni}
\bibinfo{author}{Ph. H{\"a}gler}, \bibinfo{title}{{Hadron structure from
  lattice quantum chromodynamics}}, \bibinfo{journal}{Phys. Rept.}
  \bibinfo{volume}{490} (\bibinfo{year}{2010}) \bibinfo{pages}{49--175},
  \bibinfo{doi}{\doi{10.1016/j.physrep.2009.12.008}}, \eprint{0912.5483}.

\bibtype{Article}%
\bibitem{Goldstein:2014aja}
\bibinfo{author}{G.~R. Goldstein}, \bibinfo{author}{J.~O. Gonzalez~Hernandez},
  \bibinfo{author}{S. Liuti}, \bibinfo{title}{{Flavor dependence of chiral odd
  generalized parton distributions and the tensor charge from the analysis of
  combined $\pi^0$ and $\eta$ exclusive electroproduction data}}
  (\bibinfo{year}{2014}), \eprint{1401.0438}.

\bibtype{Article}%
\bibitem{Radici:2018iag}
\bibinfo{author}{M. Radici}, \bibinfo{author}{A. Bacchetta},
  \bibinfo{title}{{First Extraction of Transversity from a Global Analysis of
  Electron-Proton and Proton-Proton Data}}, \bibinfo{journal}{Phys. Rev. Lett.}
  \bibinfo{volume}{120} (\bibinfo{number}{19}) (\bibinfo{year}{2018})
  \bibinfo{pages}{192001}, \bibinfo{doi}{\doi{10.1103/PhysRevLett.120.192001}},
  \eprint{1802.05212}.

\bibtype{Article}%
\bibitem{Lin:2017stx}
\bibinfo{author}{H.-W. Lin}, \bibinfo{author}{W. Melnitchouk},
  \bibinfo{author}{A. Prokudin}, \bibinfo{author}{N. Sato}, \bibinfo{author}{H.
  Shows}, \bibinfo{title}{{First Monte Carlo Global Analysis of Nucleon
  Transversity with Lattice QCD Constraints}}, \bibinfo{journal}{Phys. Rev.
  Lett.} \bibinfo{volume}{120} (\bibinfo{number}{15}) (\bibinfo{year}{2018})
  \bibinfo{pages}{152502}, \bibinfo{doi}{\doi{10.1103/PhysRevLett.120.152502}},
  \eprint{1710.09858}.

\bibtype{Article}%
\bibitem{Ji:2013dva}
\bibinfo{author}{X.-D. Ji}, \bibinfo{title}{{Parton Physics on a Euclidean
  Lattice}}, \bibinfo{journal}{Phys. Rev. Lett.} \bibinfo{volume}{110}
  (\bibinfo{year}{2013}) \bibinfo{pages}{262002},
  \bibinfo{doi}{\doi{10.1103/PhysRevLett.110.262002}}, \eprint{1305.1539}.

\bibtype{Article}%
\bibitem{Liu:1993cv}
\bibinfo{author}{K.-F. Liu}, \bibinfo{author}{S.-J. Dong},
  \bibinfo{title}{{Origin of difference between anti-d and anti-u partons in
  the nucleon}}, \bibinfo{journal}{Phys. Rev. Lett.} \bibinfo{volume}{72}
  (\bibinfo{year}{1994}) \bibinfo{pages}{1790--1793},
  \bibinfo{doi}{\doi{10.1103/PhysRevLett.72.1790}}, \eprint{hep-ph/9306299}.

\bibtype{Article}%
\bibitem{Detmold:2005gg}
\bibinfo{author}{W. Detmold}, \bibinfo{author}{C.~J.~D. Lin},
  \bibinfo{title}{{Deep-inelastic scattering and the operator product expansion
  in lattice QCD}}, \bibinfo{journal}{Phys. Rev. D} \bibinfo{volume}{73}
  (\bibinfo{year}{2006}) \bibinfo{pages}{014501},
  \bibinfo{doi}{\doi{10.1103/PhysRevD.73.014501}}, \eprint{hep-lat/0507007}.

\bibtype{Article}%
\bibitem{Braun:2007wv}
\bibinfo{author}{V. Braun}, \bibinfo{author}{D. M\"uller},
  \bibinfo{title}{{Exclusive processes in position space and the pion
  distribution amplitude}}, \bibinfo{journal}{Eur. Phys. J. C}
  \bibinfo{volume}{55} (\bibinfo{year}{2008}) \bibinfo{pages}{349--361},
  \bibinfo{doi}{\doi{10.1140/epjc/s10052-008-0608-4}}, \eprint{0709.1348}.

\bibtype{Article}%
\bibitem{Ma:2014jla}
\bibinfo{author}{Yan-Qing Ma}, \bibinfo{author}{Jian-Wei Qiu},
  \bibinfo{title}{{Extracting Parton Distribution Functions from Lattice QCD
  Calculations}}, \bibinfo{journal}{Phys. Rev. D} \bibinfo{volume}{98}
  (\bibinfo{number}{7}) (\bibinfo{year}{2018}) \bibinfo{pages}{074021},
  \bibinfo{doi}{\doi{10.1103/PhysRevD.98.074021}}, \eprint{1404.6860}.

\bibtype{Article}%
\bibitem{Radyushkin:2017cyf}
\bibinfo{author}{A.~V. Radyushkin}, \bibinfo{title}{{Quasi-parton distribution
  functions, momentum distributions, and pseudo-parton distribution
  functions}}, \bibinfo{journal}{Phys. Rev. D} \bibinfo{volume}{96}
  (\bibinfo{number}{3}) (\bibinfo{year}{2017}) \bibinfo{pages}{034025},
  \bibinfo{doi}{\doi{10.1103/PhysRevD.96.034025}}, \eprint{1705.01488}.

\bibtype{Article}%
\bibitem{Cichy:2018mum}
\bibinfo{author}{K. Cichy}, \bibinfo{author}{M. Constantinou},
  \bibinfo{title}{{A guide to light-cone PDFs from Lattice QCD: an overview of
  approaches, techniques and results}}, \bibinfo{journal}{Adv. High Energy
  Phys.} \bibinfo{volume}{2019} (\bibinfo{year}{2019})
  \bibinfo{pages}{3036904}, \bibinfo{doi}{\doi{10.1155/2019/3036904}},
  \eprint{1811.07248}.

\bibtype{Article}%
\bibitem{Ji:2020ect}
\bibinfo{author}{X.-D. Ji}, \bibinfo{author}{Y.-S. Liu}, \bibinfo{author}{Y.
  Liu}, \bibinfo{author}{J.-H. Zhang}, \bibinfo{author}{Y. Zhao},
  \bibinfo{title}{{Large-momentum effective theory}}, \bibinfo{journal}{Rev.
  Mod. Phys.} \bibinfo{volume}{93} (\bibinfo{number}{3}) (\bibinfo{year}{2021})
  \bibinfo{pages}{035005}, \bibinfo{doi}{\doi{10.1103/RevModPhys.93.035005}},
  \eprint{2004.03543}.

\bibtype{Article}%
\bibitem{Constantinou:2020pek}
\bibinfo{author}{M. Constantinou}, \bibinfo{title}{{The x-dependence of
  hadronic parton distributions: A review on the progress of lattice QCD}},
  \bibinfo{journal}{Eur. Phys. J. A} \bibinfo{volume}{57} (\bibinfo{number}{2})
  (\bibinfo{year}{2021}) \bibinfo{pages}{77},
  \bibinfo{doi}{\doi{10.1140/epja/s10050-021-00353-7}}, \eprint{2010.02445}.

\bibtype{Article}%
\bibitem{Cichy:2021lih}
\bibinfo{author}{K. Cichy}, \bibinfo{title}{{Progress in $x$-dependent partonic
  distributions from lattice QCD}}, \bibinfo{journal}{PoS}
  \bibinfo{volume}{LATTICE2021} (\bibinfo{year}{2022}) \bibinfo{pages}{017},
  \bibinfo{doi}{\doi{10.22323/1.396.0017}}, \eprint{2110.07440}.

\bibtype{Article}%
\bibitem{Lin:2025hka}
\bibinfo{author}{H.-W. Lin}, \bibinfo{title}{{Mapping parton distributions of
  hadrons with lattice QCD}}, \bibinfo{journal}{Prog. Part. Nucl. Phys.}
  \bibinfo{volume}{144} (\bibinfo{year}{2025}) \bibinfo{pages}{104177},
  \bibinfo{doi}{\doi{10.1016/j.ppnp.2025.104177}}, \eprint{2506.05025}.

\bibtype{Article}%
\bibitem{Alexandrou:2020sml}
\bibinfo{author}{C. Alexandrou}, \bibinfo{author}{S. Bacchio},
  \bibinfo{author}{M. Constantinou}, \bibinfo{author}{J. Finkenrath},
  \bibinfo{author}{K. Hadjiyiannakou}, \bibinfo{author}{K. Jansen},
  \bibinfo{author}{G. Koutsou}, \bibinfo{author}{H. Panagopoulos},
  \bibinfo{author}{G. Spanoudes}, \bibinfo{title}{{Complete flavor
  decomposition of the spin and momentum fraction of the proton using lattice
  QCD simulations at physical pion mass}}, \bibinfo{journal}{Phys. Rev. D}
  \bibinfo{volume}{101} (\bibinfo{number}{9}) (\bibinfo{year}{2020})
  \bibinfo{pages}{094513}, \bibinfo{doi}{\doi{10.1103/PhysRevD.101.094513}},
  \eprint{2003.08486}.

\bibtype{Article}%
\bibitem{Gao:2023ktu}
\bibinfo{author}{X. Gao}, \bibinfo{author}{A.D. Hanlon}, \bibinfo{author}{S.
  Mukherjee}, \bibinfo{author}{P. Petreczky}, \bibinfo{author}{Q. Shi},
  \bibinfo{author}{S. Syritsyn}, \bibinfo{author}{Y. Zhao},
  \bibinfo{title}{{Transversity PDFs of the proton from lattice QCD with
  physical quark masses}}, \bibinfo{journal}{Phys. Rev. D}
  \bibinfo{volume}{109} (\bibinfo{number}{5}) (\bibinfo{year}{2024})
  \bibinfo{pages}{054506}, \bibinfo{doi}{\doi{10.1103/PhysRevD.109.054506}},
  \eprint{2310.19047}.

\bibtype{Article}%
\bibitem{Gamberg:2014zwa}
\bibinfo{author}{L. Gamberg}, \bibinfo{author}{Z.-B. Kang}, \bibinfo{author}{I.
  Vitev}, \bibinfo{author}{H. Xing}, \bibinfo{title}{{Quasi-parton distribution
  functions: a study in the diquark spectator model}}, \bibinfo{journal}{Phys.
  Lett. B} \bibinfo{volume}{743} (\bibinfo{year}{2015})
  \bibinfo{pages}{112--120},
  \bibinfo{doi}{\doi{10.1016/j.physletb.2015.02.021}}, \eprint{1412.3401}.

\bibtype{Article}%
\bibitem{Bhattacharya:2018zxi}
\bibinfo{author}{S. Bhattacharya}, \bibinfo{author}{C. Cocuzza},
  \bibinfo{author}{A. Metz}, \bibinfo{title}{{Generalized quasi parton
  distributions in a diquark spectator model}}, \bibinfo{journal}{Phys. Lett.
  B} \bibinfo{volume}{788} (\bibinfo{year}{2019}) \bibinfo{pages}{453--463},
  \bibinfo{doi}{\doi{10.1016/j.physletb.2018.09.061}}, \eprint{1808.01437}.

\bibtype{Article}%
\bibitem{Braun:2018brg}
\bibinfo{author}{V.~M. Braun}, \bibinfo{author}{A. Vladimirov},
  \bibinfo{author}{J.-H. Zhang}, \bibinfo{title}{{Power corrections and
  renormalons in parton quasidistributions}}, \bibinfo{journal}{Phys. Rev. D}
  \bibinfo{volume}{99} (\bibinfo{number}{1}) (\bibinfo{year}{2019})
  \bibinfo{pages}{014013}, \bibinfo{doi}{\doi{10.1103/PhysRevD.99.014013}},
  \eprint{1810.00048}.

\bibtype{Article}%
\bibitem{Chen:2016utp}
\bibinfo{author}{J.-W. Chen}, \bibinfo{author}{S.~D. Cohen},
  \bibinfo{author}{X.-D. Ji}, \bibinfo{author}{H.-W. Lin},
  \bibinfo{author}{J.-H. Zhang}, \bibinfo{title}{{Nucleon Helicity and
  Transversity Parton Distributions from Lattice QCD}}, \bibinfo{journal}{Nucl.
  Phys. B} \bibinfo{volume}{911} (\bibinfo{year}{2016})
  \bibinfo{pages}{246--273},
  \bibinfo{doi}{\doi{10.1016/j.nuclphysb.2016.07.033}}, \eprint{1603.06664}.

\bibtype{Article}%
\bibitem{Alexandrou:2018eet}
\bibinfo{author}{C. Alexandrou}, \bibinfo{author}{K. Cichy},
  \bibinfo{author}{M. Constantinou}, \bibinfo{author}{K. Jansen},
  \bibinfo{author}{A. Scapellato}, \bibinfo{author}{F. Steffens},
  \bibinfo{title}{{Transversity parton distribution functions from lattice
  QCD}}, \bibinfo{journal}{Phys. Rev. D} \bibinfo{volume}{98}
  (\bibinfo{number}{9}) (\bibinfo{year}{2018}) \bibinfo{pages}{091503},
  \bibinfo{doi}{\doi{10.1103/PhysRevD.98.091503}}, \eprint{1807.00232}.

\bibtype{Article}%
\bibitem{HadStruc:2021qdf}
\bibinfo{author}{C. Egerer}, et al. (\bibinfo{collaboration}{HadStruc}),
  \bibinfo{title}{{Transversity parton distribution function of the nucleon
  using the pseudodistribution approach}}, \bibinfo{journal}{Phys. Rev. D}
  \bibinfo{volume}{105} (\bibinfo{number}{3}) (\bibinfo{year}{2022})
  \bibinfo{pages}{034507}, \bibinfo{doi}{\doi{10.1103/PhysRevD.105.034507}},
  \eprint{2111.01808}.

\end{thebibliography*}

%\bibliography{main-reference.bib}
\end{document}